\documentclass[twocolumn,trackchanges]{aastex62}

\usepackage{amsmath}

\newcommand{\unitstyle}[1]{\ensuremath{\mathrm{#1}}}
\newcommand{\power}[2]{\ensuremath{{#1}^{#2}}}

\newcommand{\Mega}{\unitstyle{M}}
\newcommand{\Giga}{\unitstyle{G}}

\newcommand{\second}{\unitstyle{s}}

\newcommand{\Kelvin}{\unitstyle{K}}
\newcommand{\K}{\Kelvin}  

\newcommand{\Msun}{\ensuremath{\unitstyle{M}_\odot}}
\newcommand{\Lsun}{\ensuremath{\unitstyle{L}_{\odot}}}
\newcommand{\Rsun}{\ensuremath{\unitstyle{R}_{\odot}}}

\newcommand{\Myr}{\Mega\yr}
\newcommand{\Gyr}{\Giga\yr}

\newcommand{\Msunyr}{\Msun\,\power{\yr}{-1}}

\newcommand{\yr}{\unitstyle{yr}}        
\newcommand{\kms}{\ensuremath{\mathrm{km}\,\second^{-1}}}

\newcommand{\code}[1]{\texttt{#1}}
\newcommand{\mesa}{\code{MESA}}
\newcommand{\MESA}{\mesa}

\newcommand{\GYRE}{\code{GYRE}}
\newcommand{\stella}{\code{STELLA}}
\newcommand{\STELLA}{\stella}
\newcommand{\rsp}{\code{RSP}}
\newcommand{\RSP}{\rsp}

\newcommand{\mesastar}{\mesa\code{star}}
\newcommand{\mesabinary}{\mesa\code{binary}}
\newcommand{\MESAstar}{\mesastar}

\newcommand{\eos}{\code{eos}}

\newcommand{\net}{\code{net}}

\newcommand{\kB}{\ensuremath{k_\mathrm{B}}} 
\newcommand{\NA}{\ensuremath{N_\mathrm{\!A}}} 

\RequirePackage{bm}

\newcommand{\Dif}{\ensuremath{\mathrm{D}}}

\newcommand{\DDt}[1]{\frac{\Dif #1}{\Dif t}} 
\newcommand{\ddm}[1]{\frac{\partial #1}{\partial m}} 

\newcommand{\dxdy}[2]{{\frac{\partial{#1}}{\partial{#2}}}}

\newcommand{\dm}[1]{\ensuremath{dm_{#1}}} 
\newcommand{\dr}[1]{\ensuremath{dr_{#1}}} 

\newcommand{\timestep}{\ensuremath{\delta t}} 

\newcommand{\dt}{\ensuremath{\delta t}}
\newcommand{\area}{\ensuremath{\mathcal{A}}}

\newcommand{\epsnuc}{\ensuremath{\epsilon_{\mathrm{nuc}}}} 
\newcommand{\epsgrav}{\ensuremath{\epsilon_{\mathrm{grav}}}} 
\newcommand{\epsnu}{\ensuremath{\epsilon_{\mathrm{\nu}}}} 
\newcommand{\vhat}{\ensuremath{\hat{v}}}

\newcommand{\SNEC}{\texttt{SNEC}}
\newcommand{\Ni}{^{56}{\rm Ni}}
\newcommand{\Eexp}{E_{\rm exp}}
\newcommand{\MNi}{M_{\rm Ni}}

\newcommand{\grada}{\ensuremath{\nabla_{{\rm ad}}}}

\newcommand{\gradL}{\ensuremath{\nabla_{{\rm L}}}}
\newcommand{\gradr}{\ensuremath{\nabla_{{\rm rad}}}}

\newcommand{\Teff}{\ensuremath{T_{\rm eff}}}	
\newcommand{\Dconv}{\ensuremath{D_{\rm conv}}}
\newcommand{\vconv}{\ensuremath{v_{\rm conv}}}

\newcommand{\mabund}{\ensuremath{m_{\rm a}}}
\newcommand{\mcore}{\ensuremath{m_{\rm c}}}
\newcommand{\BV}{Brunt-V\"{a}is\"{a}l\"{a}}

\newcommand{\Xhyd}{\ensuremath{X}}
\newcommand{\Yhel}{\ensuremath{Y}}

\newcommand{\Xcore}{\ensuremath{\Xhyd_{\rm c}}}
\newcommand{\Ycore}{\ensuremath{\Yhel_{\rm c}}}
\newcommand{\Tc}{\ensuremath{T_{\mathrm{c}}}} 

\newcommand{\logT}{\ensuremath{\log(T/\mathrm{K})}} 

\newcommand{\zbar}{\ensuremath{\bar{Z}}} 

\newcommand{\eface}{\mathcal{E}_{\rm face}} 
\newcommand{\efacek}[1]{\mathcal{E}_{{\rm face},#1}} 
\newcommand{\Mstart}{{\rm start}} 
\newcommand{\Mmid}{{\rm mid}} 
\newcommand{\pass}{\rm pass} 
\newcommand{\dLdm}{the \code{dLdm}-form}
\newcommand{\dedt}{the \code{dedt}-form}

\newcommand{\mesaone}{Paper~I}  
\newcommand{\mesatwo}{Paper~II} 
\newcommand{\mesathree}{Paper~III} 
\newcommand{\mesafour}{Paper~IV} 

\newcommand{\revision}[1]{#1}

\newlength{\apjcolwidth}
\newlength{\figwidth}
\newlength{\doublewide}
\DeclareMathAlphabet{\mathpzc}{OT1}{pzc}{m}{it}

\begin{document}
\title{Modules for Experiments in Stellar Astrophysics (MESA): Pulsating Variable Stars, Rotation, Convective Boundaries, and Energy Conservation}

\author{Bill Paxton}
\affiliation{Kavli Institute for Theoretical Physics, University of California, Santa Barbara, CA 93106, USA}

\author[0000-0001-7217-4884]{R.~Smolec}
\affiliation{Nicolaus Copernicus Astronomical Center of the Polish Academy of Sciences, Bartycka 18, PL-00-716 Warszawa, Poland}  

\author[0000-0002-4870-8855]{Josiah Schwab}
\altaffiliation{Hubble Fellow}
\affiliation{Department of Astronomy and Astrophysics, University of California, Santa Cruz, CA 95064, USA}

\author{A.~Gautschy}
\affiliation{CBmA, 4410 Liestal, Switzerland}

\author{Lars Bildsten}
\affiliation{Kavli Institute for Theoretical Physics, University of California, Santa Barbara, CA 93106, USA}
\affiliation{Department of Physics, University of California, Santa Barbara, CA 93106, USA}

\author[0000-0002-8171-8596]{Matteo Cantiello}
\affiliation{Center for Computational Astrophysics, Flatiron Institute, 162 5th Avenue, New York, NY 10010, \
USA}
\affiliation{Department of Astrophysical Sciences, Princeton University, Princeton, NJ 08544, USA}

\author[0000-0002-4442-5700]{Aaron Dotter}
\affiliation{Harvard-Smithsonian Center for Astrophysics, Cambridge, MA 02138, USA}

\author[0000-0003-3441-7624]{R. Farmer}
\affiliation{Anton Pannenkoek Institute for Astronomy, University of Amsterdam, NL-1090 GE Amsterdam, The Netherlands}
\affiliation{GRAPPA, University of Amsterdam, Science Park 904, 1098 XH Amsterdam, The Netherlands}

\author[0000-0003-1012-3031]{Jared A. Goldberg}
\affiliation{Department of Physics, University of California, Santa Barbara, CA 93106, USA}

\author[0000-0001-5048-9973]{Adam S. Jermyn}
\affiliation{Kavli Institute for Theoretical Physics, University of California, Santa Barbara, CA 93106, USA}

\author{S.M.~Kanbur}
\affiliation{Department of Physics, SUNY Oswego, NY 13126, USA}

\author[0000-0002-0338-8181]{Pablo Marchant}
\affiliation{Department of Physics and Astronomy, Northwestern University, 2145 Sheridan Road, Evanston, IL 60208, USA}

\author[0000-0002-8107-118X]{Anne Thoul}
\affiliation{Space sciences, Technologies and Astrophysics Research (STAR) Institute, Universit\'e de Li\`ege, All\'ee du 6 Ao$\hat{u}$t 19C, Bat. B5C, 4000 Li\`ege, Belgium}

\author[0000-0002-2522-8605]{Richard H. D. Townsend}
\affiliation{Department of Astronomy, University of Wisconsin-Madison, Madison, WI 53706, USA}

\author[0000-0002-6828-0630]{William M.~Wolf}
\affiliation{Department of Physics and Astronomy, University of Wisconsin-Eau Claire, Eau Claire, WI 54701, USA}
\affiliation{School of Earth and Space Exploration, Arizona State University, Tempe, AZ 85287, USA}

\author[0000-0002-0659-1783]{Michael Zhang}
\affil{Department of Astronomy, California Institute of Technology, Pasadena, CA 91125, USA}

\author[0000-0002-0474-159X]{F.X.~Timmes}
\affiliation{School of Earth and Space Exploration, Arizona State University, Tempe, AZ 85287, USA}

\correspondingauthor{F.X.~Timmes}
\email{fxtimmes@gmail.com}

\begin{abstract}
%
%
\noindent
We update the capabilities of the open-knowledge software instrument Modules for Experiments in Stellar Astrophysics (\MESA).
\RSP\ is a new functionality in \MESAstar\ that models the non-linear radial stellar pulsations that
characterize RR~Lyrae, Cepheids, and other classes of variable stars.
We significantly enhance numerical energy conservation capabilities, including during mass changes.
For example, this enables calculations through the He flash that conserve energy to better than 0.001\%.
To improve the modeling of rotating stars in \MESA, 
we introduce a new approach to modifying the pressure and temperature equations of 
stellar structure, and a formulation of the projection effects of gravity darkening.
A new scheme for tracking convective boundaries yields reliable values of the
convective-core mass, and allows the natural emergence of adiabatic semiconvection regions
during both core hydrogen- and helium-burning phases.
We quantify the parallel performance of \MESA\ on current generation multicore architectures
and demonstrate improvements in the computational efficiency of radiative levitation. 
We report updates to the equation of state and nuclear reaction physics modules. 
We briefly discuss the current treatment of fallback in core-collapse supernova models
and the thermodynamic evolution of supernova explosions.
We close by discussing the new \MESA\ Testhub software infrastructure to enhance source-code development.

\end{abstract}




\section{Introduction}\label{s.intro}

One of the foundations upon which modern astrophysics rests is the
fundamental properties of stars throughout their evolution. The
advent of transformative capabilities in space- and ground-based
hardware instruments is providing an unprecedented volume of high-quality measurements of stars, significantly strengthening and
extending the observational data upon which all of stellar
astrophysics ultimately depends.  
For example, the {\it Parker Solar Probe} will provide new information 
on the flow of energy, structure, and dynamics of the closest star
\citep{parker_1958_aa, feng_2010_aa, cranmer_2018_aa, gombosi_2018_aa}
and the {\it Daniel K. Inouye Solar Telescope} will provide high
temporal and spatial resolution with adaptive optics to reveal 
the nature of the the outer layers of the Sun 
\citep{parker_1958_ab, snow_2018_aa, mccomas_2018_aa}.

The exceptional precision of stellar brightness measurements
achieved by the planet-hunting space telescopes {\it Kepler/K2}
\citep{borucki_2010_aa,howell_2014_aa} and 
{\it CoRoT} \citep{auvergne_2009_aa} ushered in a new era in stellar
photometric variability investigations.  
The {\it Transiting Exoplanet Survey Satellite} is building upon this
legacy by surveying most of the sky in roughly month-long sectors
covering four 24$^{\circ}$\,$\times$\,24$^{\circ}$ areas from the
ecliptic poles to near the ecliptic plane \citep{ricker_2016_aa}. The
mission will produce light curves for about 200,000 nearby late-type
stars sampled at a 2 minute cadence to open a new era of stellar
variability exploration \citep[e.g.,][]{dragomir_2019_aa,huang_2018_aa,ball_2018_aa,shen_2018_aa,wang_2019_aa}.
The {\it Characterizing Exoplanets Satellite} will complement these
surveys by providing a unique, large sample of high precision
photometric monitoring of selected bright target stars
\citep{broeg_2013_aa,gaidos_2017_aa}.

The {\it Gaia} Data Release 2, containing about 1.7 billion stars, begins
the process of converting the spectrophotometric measurements to
distances, proper motions, luminosities, effective temperatures,
surface gravities, and elemental compositions \citep{gaia-collaboration_2018_aa,bailer-jones_2018_aa,lindegren_2018_aa,luri_2018_aa}.  
This stellar census will revolutionize a range of questions related to the origin,
structure, and evolutionary history of stars in the Milky Way
\citep[e.g.,][]{gaia-collaboration_2018_ab,gaia-collaboration_2018_ac,riess_2018_aa}.
The infrared instruments aboard 
the {\it  James Webb Space Telescope} \citep{gardner_2006_aa,beichman_2012_aa,artigau_2014_aa,rieke_2015_aa} 
will search for the first and second generation stars \citep{rydberg_2013_aa,kelly_2018_aa,windhorst_2018_aa}, 
assess how galaxies evolved from their formation \citep{zackrisson_2011_aa},
observe the formation of stars from the initial stages of collapse onwards \citep{senarath_2018_aa},
and measure the physical and chemical properties of stellar-planetary systems \citep{deming_2009_aa}.
The {\it Laser Interferometer Gravitational-Wave Observatory} and 
{\it Virgo} interferometers have 
demonstrated the existence of binary stellar-mass black hole systems
\citep{abbott_2017_ab,abbott_2017_ac,abbott_2017_ad}
and neutron star mergers \citep{abbott_2017_aa,abbott_2017_ae,abbott_2017_af}, 
and will continue to monitor the sky with improved broadband detectors
for gravitational waves from compact binary inspirals and asymmetrical
exploding massive stars.

In partnership with this ongoing explosion of activity in stellar
astrophysics, community-driven software instruments are transforming 
how stellar theory, modeling, and simulations interact with observations 
\citep[e.g.,][]{turk_2011_aa, foreman-mackey_2013_aa, ness_2015_aa, choi_2016_aa, astropy-collaboration_2018_aa}.
Modules for Experiments in Stellar
Astrophysics (\MESA) was introduced in
\citealt{paxton_2011_aa} (\mesaone) and significantly expanded its
range of capabilities in 
\citealt{paxton_2013_aa} (\mesatwo),
\citealt{paxton_2015_aa} (\mesathree), and
\citealt{paxton_2018_aa} (\mesafour).
These prior papers, as well as this one, are software instrument papers 
that describe the capabilities and limitations of \MESA\ while
also comparing to other available numerical or analytic results.

This instrument paper describes the major new advances to \MESA\ for 
variable stars, numerical energy conservation, rotation, and convective boundaries.
We do not fully explore the science results and
their implications in this paper.  The scientific potential of these
new capabilities will be unlocked in future work via the efforts of
the growing, 1,000-strong \MESA\ research community.

\begin{figure}[!htb]
\centering
\includegraphics[width=1.0\columnwidth]{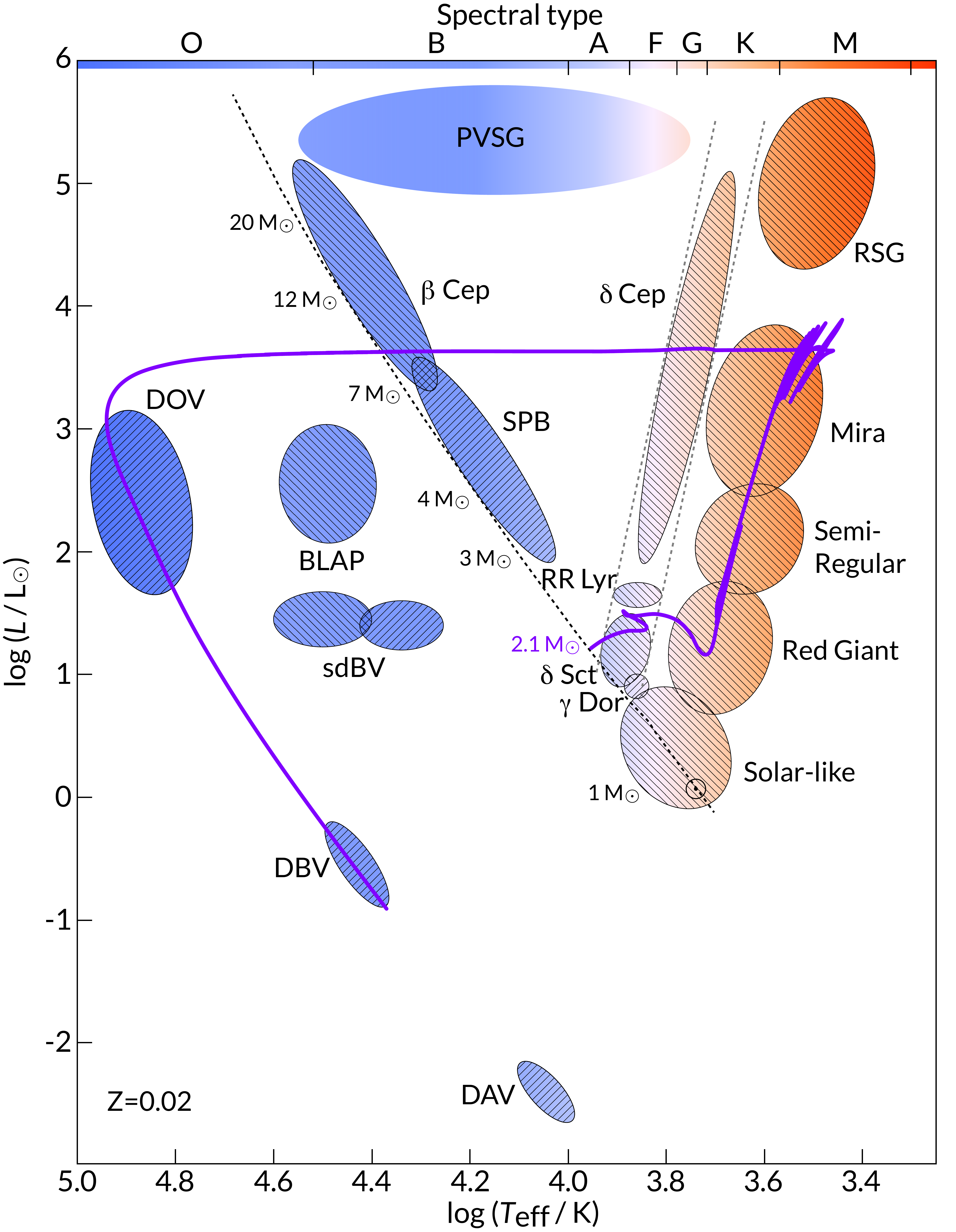}
\caption{
Classes of pulsating variable stars in the Hertzsprung-Russell (HR) diagram, including 
regions driven by the \ion{He}{2} bump ($\delta$ Ceph, $\delta$ Sct, RR~Lyrae) 
and Fe bump ($\beta$ Ceph, SPB) in the opacity.
Backslash (\textbackslash) fills represent pressure modes and slash (/) fills
represent gravity modes.  The zero age main-sequence (ZAMS, black dashed curve) 
is labeled with the locations of selected masses.  The
classical instability strip for radial
pulsations is shown by the gray dashed curves. Evolution of a
2.1\,\Msun \ \MESA \ model (at $Z = 0.02$) from ZAMS to a white dwarf (WD) is shown by the
purple curve.
\revision{Figure design from \revision{\citet{papics_2013_aa}}.}
}
\label{f.hrpulse}
\end{figure}

Millions of variable stars have been discovered in the Milky Way and Magellanic Clouds,
the Local Group
(e.g., {\it Optical Gravitational Lensing Experiment}, OGLE, \citealt{rsp_ogle4};
{\it MACHO Project}, \citealt{alcock_2003_aa}; {\it Palomar Transient Factory}, \citealt{soraisam_2018_aa})
and beyond \citep[e.g.,][]{conroy_2018_aa}.
Figure~\ref{f.hrpulse} shows the broad classifications of these pulsating stars.
Pulsating stars such as RR~Lyrae and the brighter $\delta$ Cephei
(the classical Cepheids) are common, and a strong direct relationship
between their luminosities and pulsation periods
established Cepheids
\citep{leavitt_1908_aa,freedman_2001_aa,majaess_2009_aa,riess_2016_aa,riess_2018_aa}
and RR Lyrae in infrared bands
\citep{clementini_2001_aa, benedict_2002_aa, klein_2014_aa, muraveva_2018_aa, muraveva_2018_ab}
as key distance indicators.
New classes of variable stars are still being discovered:
Blue Large-Amplitude Pulsators (BLAPs) are a new family of pulsating variable stars
\citep{pietrukowicz_2017_aa}.
BLAPs are rare;  
only 14 variable stars are attributed by OGLE to this class after examining 
$\simeq$\,10$^9$ stars. They vary in brightness by 
$\simeq$\,20\% on $\simeq$\,30~min timescales \citep{pietrukowicz_2013_aa}.
An important new addition to \MESA\ is 
the capability to model radially-pulsating variable stars.

Numerical energy conservation
is rarely discussed by stellar evolution software instrument papers, 
or shown in science papers as part of establishing robustness of 
the solutions obtained with the software instrument.
Yet stellar
evolution calculations generally use low-order, implicit time
integration with potentially poorly conditioned matrices whose matrix elements
contain limited-precision partial derivatives that can severely limit
the quality of solutions. The cumulative effect of such errors 
can be substantial \citep{reiter_1995_aa}.
We implement a set of changes in \MESA \ which, when
applicable, can significantly improve the energy conservation properties of
stellar evolution models at both global and local levels.  This can reduce
cumulative errors in energy conservation to 1\% or less for applications
such as the evolution of a 1~\Msun\ model from the pre-main sequence to the end of core He-burning 
or a core-collapse supernova from soon after explosion to shock breakout.

Rotation modifies a star's structure
\citep{von-zeipel_1924_aa,tassoul_2000_aa,maeder_2000_aa}.
We present a new approach in \MESA\ for calculating the factors 
that modify the pressure and temperature equations of stellar structure 
within the shellular approximation.  
A rotating star is also oblate, with a larger
radius at its equator than at its poles. As a result, the equator has
a lower surface gravity
and thus a lower effective temperature \Teff\ \citep{von-zeipel_1924_ab,chandrasekhar_1933_aa}.
Hence, the equator is ``gravity darkened'',
the poles ``gravity brightened'', and this effect can play an important role 
in the classification of stars. 
The new extensions to \MESA\ open a pathway 
for correcting \Teff\ and $L$ for aspect-dependent effects.

Stars transport energy by convection, whether within a
core, an envelope, or throughout the interior.  These
convection regions showcase the interplay between composition mixing,
gradients, and diffusion, and the transport of energy
through the radial exchange of matter.
It is necessary to ensure that convective boundaries are properly
positioned because their placement can strongly influence
the evolution of the stellar model
\citep[][\mesafour]{gabriel_2014_aa,salaris_2017_aa}.  
We implement an improved algorithm for
correctly locating the convective boundaries and 
naturally allowing the emergence of adiabatic semiconvection regions during core H and He burning.

The paper is organized as follows. 
Section~\ref{s.rsp} introduces a new capability to model large-amplitude radially-pulsating variable stars.
Section~\ref{s.energy} highlights energy conservation in \MESA.
Section~\ref{s.rot} describes new rotation and gravity darkening factors,
Section~\ref{s.cpm} explores a new treatment of convective boundaries, and 
Section~\ref{s.parallel} examines the parallel performance of \mesastar.
Appendix~\ref{s.updates} reports updates to the equation of state (EOS) and nuclear reaction modules.
Appendix~\ref{a.rot} details properties of the rotation factors.
Appendix~\ref{s.snec} discusses the current treatment of fallback in core-collapse supernovae (SN),  
and the thermodynamic evolution from massive star explosions.
Appendix~\ref{s.testhub} introduces the \MESA\ Testhub for source code development.

Important symbols are defined in Table 1.  Acronyms are defined
in Table~\ref{tab:acronym}. Components of \MESA, such as
modules and routines, are in typewriter font e.g., \texttt{eos}.

\startlongtable
\begin{deluxetable}{p{0.06\apjcolwidth}p{0.60\apjcolwidth}r}
  \tablecolumns{3}
  \tablewidth{1.0\apjcolwidth}
  \tablecaption{Important symbols.
    Single character symbols are listed first, 
    symbols with modifiers are listed second, and
    symbols for the \RSP\ convection model are listed third.
    Some symbols may be further subscripted, for example, 
    by $\mathrm{c}$ (indicating a central quantity),
    by a cell index $k$, or
    by  species index $i$.
\label{t.list-of-symbols}}
  \tablehead{\colhead{Name} & \colhead{Description} & \colhead{Appears}}
  \startdata
\area          & $4 \pi r^2$ \ \ Area of face           & \ref{s.rspeq}   \\
$e$                &  Specific internal energy                       & \ref{s.rspeq} \\ 
$E$                &  Energy                                         & \ref{s.energy}  \\
$F$                & Flux                                            & \ref{s.rspeq}  \\ 
$L$                & Luminosity                                      & \ref{s.intro}\\
$m$            & Mass coordinate               & \ref{s.rspeq} \\ 
$M$            & Stellar mass                           & \ref{s.intro} \\ 
$\Phi$         & Roche potential                        & \ref{s.rot}   \\ 
$p$            &  Pressure                                       & \ref{s.rspeq}       \\  
$P$            & Period                                 & \ref{s.rspeq} \\
$\rho$         & Mass density                           & \ref{s.rspeq}       \\ 
$r$            & Radial coordinate           & \ref{s.rspeq} \\ 
$s$            & Specific entropy                                & \ref{s.eos} \\
$T$            & Temperature                                     & \ref{s.rsp.action} \\
$u$            & Velocity                    &  \ref{s.rspeq} \\ 
$V$            & 1/$\rho$ \ \ Specific volume               &  \ref{s.rspeq} \\ 
$\Omega$       & Rotation angular frequency                     & \ref{s.rot} \\
$X$            & Hydrogen mass fraction                      & \ref{s.rsp.action} \\
$Y$            & Helium mass fraction                        & \ref{s.cpm} \\
$Z$            & Metal mass fraction                         & \ref{s.rsp.action} \\
  \hline
$c_p$              &  Specific heat at constant pressure             & \ref{s.intro}   \\ 
$c_V$              &  Specific heat at constant volume               & \ref{s.eos}   \\ 
\timestep          &  Numerical time step                             & \ref{s.mdot}  \\ 
\dm{}              &  Mass of cell                                   & \ref{s.energy}  \\ 
\grada             &  Adiabatic temperature gradient                 & \ref{s.mdot}  \\ 
\gradL             &  Ledoux temperature gradient                    & \ref{s.cpm}    \\ 
\gradr             &  Radiative temperature gradient                 & \ref{s.cpm}    \\ 
$i_{\rm rot}$      & Specific moment of inertia                      & \ref{s.rot} \\
$j_{\rm rot}$      & Specific angular momentum                       & \ref{s.rot} \\
\Teff              & Effective temperature                           & \ref{s.intro} \\
  \hline
$\alpha$           & Mixing length parameter                         & \ref{s.rspeq} \\
$\alpha_{\rm c}$   & Convective flux parameter                       & \ref{s.rspeq} \\
$\alpha_{\rm cut}$ & Artificial viscosity parameter                   & \ref{s.rspeq} \\
$\alpha_{\rm d}$   & Turbulent dissipation parameter                 & \ref{s.rspeq} \\
$\alpha_{\rm m}$   & Eddy-viscous dissipation parameter              & \ref{s.rspeq} \\
$\alpha_{\rm p}$   & Turbulent pressure parameter                    & \ref{s.rspeq} \\
$\alpha_{\rm s}$   & Turbulent source parameter                      & \ref{s.rspeq} \\
$\alpha_{\rm t}$   & Turbulent flux parameter                        & \ref{s.rspeq} \\
\revision{$C$}     & \revision{$S - D - D_{\rm r}$ \quad convective coupling}  & \revision{\ref{s.rspeq}} \\
$C_{\rm q}$        & Artificial viscosity parameter                  & \ref{s.rspeq} \\
\revision{$\Delta u$} & \revision{Change in velocity across a cell}   & \revision{\ref{s.intro}} \\
$D$          &  $\alpha_{\rm d} ( e_{\rm t}^{3/2} / \alpha H_{\rm p} )$  \ Turbulent dissipation    & \ref{s.rspeq} \\
$D_{\rm r}$  & ($4\sigma\gamma_{\rm r}^2 / \alpha^2$)  ($T^3V^2 / c_p\kappa H_{\rm p}^2) e_{\rm t}$  & \ref{s.rspeq} \\
             & Radiative cooling  &  \\
$\epsilon_{\rm q}$  & $(4/3)  (q/\rho) \left( \partial u/\partial r - u/r\right)^2$   & \ref{s.rspeq} \\
             & Viscous energy transfer rate  &  \\
$e_{\rm t}$    & Specific turbulent energy              & \ref{s.rspeq} \\
$F_{\rm c}$  & $\alpha\alpha_{\rm c}\rho T c_p e_{\rm t}^{1/2} Y_{\rm sag}$  \quad  Convective flux  & \ref{s.rspeq} \\
$F_{\rm r}$  & $-$$(4acT^3/3\kappa\rho) \ \partial T/\partial r$ \ Radiative flux  & \ref{s.rspeq} \\
$F_{\rm t}$  & $-$$(\alpha\alpha_{\rm t}\rho H_{\rm p} e_{\rm t}^{1/2}) \partial e_{\rm t}/\partial r$\,\,Turbulent~flux & \ref{s.rspeq} \\
$\gamma_{\rm r}$  & Radiative cooling parameter                     & \ref{s.rspeq} \\
$H_{\rm p}$   & Pressure scale height & \ref{t.list-of-symbols} \\
$\kappa$       & Opacity                                & \ref{s.intro} \\
$p_{\rm av}$  & $C_{\rm q} p \left[{\rm min}\left( \Delta u/\sqrt{pV} + \alpha_{\rm cut},\, 0\right)\right]^2$  & \ref{s.rspeq} \\
              & Artificial viscosity pressure  &  \\
$p_{\rm t}$   & $\alpha_{\rm p}\rho e_{\rm t}$  \quad Turbulent pressure   & \ref{s.rspeq} \\
$q$           & $\alpha_{\rm m}\rho\alpha H_{\rm p} e_{\rm t}^{1/2}$\,\,Kinetic\,\,turbulent\,\,viscosity   & \ref{s.rspeq} \\
$Q$           & ($\partial V / \partial T)|_{p}$\,\,Thermal\,expansion\,coefficient  & \ref{s.intro} \\
$\revision{s}$        & \revision{Specifc entropy}                                    & \revision{\ref{t.list-of-symbols}} \\
$S$           & $\alpha\alpha_{\rm s}e_{\rm t}^{1/2} (TpQ/H_{\rm p}) Y_{\rm sag}$\,\,Source~function & \ref{s.rspeq} \\
$U_{\rm q}$   & $(1/\rho r^3) \ \partial/\partial r \left[\frac{4}{3}qr^3 \left(\partial u/\partial r - u/r\right)\right]$ & \ref{s.rspeq} \\
              & Viscous momentum transfer rate  &  \\
$Y_{\rm sag}$ & \revision{$-H_{\rm p}/c_{\rm p} \ \partial s / \partial r$}\,\,\,\,superadiabatic~gradient & \revision{\ref{t.list-of-symbols}} \\
  \enddata
\end{deluxetable}

\startlongtable
\begin{deluxetable}{clr}
  \tablecolumns{3}
  \tablewidth{0.9\apjcolwidth}
  \tablecaption{Acronyms used in this paper.\label{tab:acronym}}
  \tablehead{ \colhead{Acronym} & \colhead{Description} & \colhead{Appears} }
  \startdata 
1O     & First Overtone                 & \ref{s.rsp.action.lna} \\
2O     & Second Overtone                & \ref{s.rsp.action.lna} \\
BEP    & Binary Evolution Pulsators     & \ref{sssec:rsp_BEP} \\
BLAP   & Blue Large-Amplitude Pulsators & \ref{s.intro} \\
CHeB   & Core Helium Burning            & \ref{s.cpm} \\
CPM    & Convective Premixing          & \ref{s.cpm} \\
EOS    & Equation of State              & \ref{s.intro} \\  
HADS   & High Amplitude Delta Scuti     & \ref{sssec:rsp_DSCT} \\
HR     & Hertzsprung Russell            & \ref{s.intro} \\      
LNA    & Linear Non-Adiabatic           & \ref{s.rsp.action} \\
MLT    & Mixing Length Theory           & \ref{s.cpm} \\  
MS     & Main Sequence                  & \ref{sssec:rsp_DSCT}  \\
RSP    & Radial Stellar Pulsations      & \ref{s.rspeq} \\
TAMS   & Terminal Age Main Sequence     & \ref{sec:roche} \\
WD     & White Dwarf                    & \ref{s.intro} \\
ZAMS   & Zero Age Main Sequence         & \ref{s.intro}  \\
  \enddata
\end{deluxetable}

\section{Radial Stellar Pulsations}\label{s.rsp}

Cepheids, RR Lyrae, and other classes of variable stars
are observed to brighten and dim periodically.
They can be modeled as radially symmetric, large amplitude, 
nonlinear oscillations of self-gravitating gas spheres.
Software instruments for precision asteroseismology such as \GYRE\ \citep{townsend_2013_aa,townsend_2018_aa}
model the small amplitude, linear oscillations of stars. 
Software instruments such as \RSP, described below, are necessary to
model the time evolution of large amplitude,
self-excited, nonlinear pulsations over many cycles 
to produce luminosity and radial velocity histories that can be compared to observations.

Early nonlinear radial pulsation models
considered purely radiative envelopes
\citep[e.g.,][]{rsp_christy, rsp_stel75, rsp_DYN, rsp_TGRID}.
Later, radiation hydrodynamic treatments followed 
with implicit adaptive grids \citep{rsp_dd,rsp_df91}.
While these purely radiative models qualitatively reproduced
light and radial velocity curves, it was clear 
that convection driven by partial ionization of H and He 
carries most of the flux in the envelopes of RR~Lyrae and Cepheids. 
Prescriptions for coupling convection with pulsations were developed
\citep[e.g.,][]{rsp_stel82,rsp_kuhfuss} that reside, with modifications, in 
modern software instruments \citep{rsp_IC, rsp_FB1, rsp_kbsc02, rsp_sm08a}.
Models from these software instruments can reproduce 
the overall morphology of light and radial velocity curves of classical pulsators 
\citep[e.g.,][]{rsp_fbk00,rsp_Marconi15}, features of specific 
objects \citep[e.g.,][]{rsp_kw06,rsp_Marconi13,smolec_bep2013}, and 
dynamical phenomena such as the Hertzsprung progression 
\citep[e.g.,][]{hertzsprung_1926_aa,bono_2000_aa}. 
Unsolved problems include double-mode pulsations \citep{rsp_kbsc02,rsp_sm10}
and the cyclic modulations of RR~Lyrae light curves
\citep[e.g., the Blazhko effect,][]{blazko_1907_aa,szabo_2010_aa}.
For background material we refer the reader to 
\citet{gautschy_1995_aa, gautschy_1996_aa}, \citet{rsp_buchlerAIP09}, and \citet{rsp_MarcellaRev}.

\subsection{Radial Stellar Pulsations - \RSP}\label{s.rspeq}

\RSP\ is a new functionality in \MESAstar\ 
that models large amplitude, self-excited, nonlinear pulsations that stars develop 
when they cross instability domains in the HR diagram (see Figure~\ref{f.hrpulse}).
\RSP\ is closely integrated with the \MESA\ environment.
Instead of calling the standard \mesastar\ routine to evaluate equations
and solve for a new model using Newton-Raphson iterations
(see Section~\ref{s.energy}), a separate
routine does the same for \RSP\ using a different set of equations and a
different Newton-Raphson solver.  The different equations include time-dependent
convection in a form appropriate for modelling nonlinear pulsations,
and the different solver uses a band diagonal matrix approach since
the equations as currently implemented do not fit into a
three-block stencil needed for the standard block tridiagonal solver.  Moreover, instead of calling the usual
\mesastar\ routine to get a starting model, a separate routine creates an \RSP\
model envelope that is consistent with the \RSP\ set of equations.
\RSP\ uses the same \MESA\ 
opacity and equation of state (EOS) modules, 
inlist structure, profile and history output files, 
photo files for saving and restarting runs, \code{run\_star\_extras} extensions,
and hooks for using externally supplied routines.

\RSP\ follows \cite{rsp_sm08a}, where the momentum and specific internal
energy equations are
\begin{align}
& \DDt{u}=-\area\ddm{}\left(p+p{\rm _t} + p_{\rm av}\right)+U_{\rm q} -\frac{Gm}{r^2}\,,\label{eq:mom} \\
& \DDt{e}+\left(p+p_{\rm av}\right)\DDt{V}=-\ddm{}\left[\area\left(F_{\rm r} +F_{\rm c}\right)\right]-C\,,\label{eq:ie}
\end{align}
where $\Dif/\Dif t$ is the Lagrangian time derivative.
The generation of a specific turbulent energy, $e_{\rm t}$, is described by the one-equation \cite{rsp_kuhfuss} model
\begin{equation}
 \DDt{e_{\rm t}}+p_{\rm t}\DDt{V} =- \ddm{}\left(\area F_{\rm t}\right)+\epsilon_{\rm q} + C\,.\label{eq:te}
\end{equation}
The latter two equations are added to give an equation for the 
specific internal and turbulent energies
\begin{multline}
\DDt{}\left(e+e_{\rm t}\right)+\left(p+p_{\rm t}+p_{\rm av}\right)\DDt{V}=\\-\ddm{}\left[\area\left(F_{\rm r} +F_{\rm c}+F_{\rm t}\right)\right]+\epsilon_{\rm q}\,.\label{eq:tote}
\end{multline}
Definitions for all terms entering these equations are given in
Table~\ref{t.list-of-symbols}.
\RSP\ solves Equations \eqref{eq:mom},  \eqref{eq:te}, and \eqref{eq:tote}.
The diffusion approximation is used for the radiative flux $F_{\rm r}$ and
its numerical implementation follows \cite{rsp_stel75}. 
\revision{Numerical implementation of the superadiabtic gradient follows \cite{rsp_stel82}. 
All equations are discretized on a Lagrangian mesh.}

Several quantities enter the convection model. 
For the momentum equation these are the
turbulent pressure $p_{\rm t}$ (Table~\ref{t.list-of-symbols} lists the relationship 
with the specific turbulent energy $e_{\rm t}$)
and viscous momentum transfer rate $U_{\rm q}$.
For the turbulent energy equation these are the
work done by turbulent pressure, the divergence of the turbulent flux $F_{\rm t}$, 
and the viscous energy transfer rate $\epsilon_{\rm q}$.
The convective coupling term $C = S - D - D_{\rm r}$ 
appears with opposite sign in the internal and turbulent energy equations. 
Generation of the turbulent
energy is driven by the source function $S$, while turbulent dissipation $D$ and
radiative cooling $D_{\rm r}$ contribute to its decay. 
Radiative cooling of convective eddies follows \cite{rsp_wh98}. Details
of the turbulent convection model are discussed in \cite{rsp_kuhfuss}, \cite{rsp_wh98} and \cite{rsp_sm08a}.

\begin{deluxetable}{lll}[!ht]
  \tablecolumns{3}
  \tablewidth{1.0\apjcolwidth}
  \tablecaption{Free parameters of the \RSP\ convection model, their base values, and associated \MESA\ controls that multiply the base values. \label{tab:rsp_alphas}}
  \tablehead{ \colhead{Parameter \quad $=$} & \colhead{Base Value \quad $\times$} & \colhead{Control Value} }
  \startdata 
    $\alpha$        & $1$    & \code{RSP\_alfa}  \\
    $\alpha_{\rm m}$ & $1$    & \code{RSP\_alfam} \\
    $\alpha_{\rm s}$ & $(1/2)\sqrt{2/3}$ & \code{RSP\_alfas} \\
    $\alpha_{\rm c}$ & $(1/2)\sqrt{2/3}$ & \code{RSP\_alfac} \\
    $\alpha_{\rm d}$ & $(8/3)\sqrt{2/3}$ & \code{RSP\_alfad} \\
    $\alpha_{\rm p}$ & $2/3$           & \code{RSP\_alfap} \\
    $\alpha_{\rm t}$ & $1$   & \code{RSP\_alfat} \\
    $\gamma_{\rm r}$ & $2\sqrt{3}$     & \code{RSP\_gammar} 
  \enddata
\end{deluxetable}

These terms in the convection model depend on the free parameters listed in
Table~\ref{tab:rsp_alphas}.
If radiative cooling and turbulent pressure are neglected, the
time-independent version of the \cite{rsp_kuhfuss} convection model
reduces to standard mixing length theory provided base values
are used for $\alpha_{\rm s}$, $\alpha_{\rm c}$ and $\alpha_{\rm d}$ (associated controls set to 1).
Base values for $\alpha_{\rm p}$ and $\gamma_{\rm r}$ follow \cite{rsp_FB1} and \cite{rsp_wh98}, respectively. 
Experience suggests $\alpha_{\rm t} \simeq 0.01$, $\alpha_{\rm m} \lesssim 1$, and $\alpha \lesssim 2$ 
are useful starting choices.

Periods of pulsation modes depend weakly on the values of these free
parameters.  Pulsation growth rates and light and radial velocity curves are, however,
sensitive to the free parameters. Calibration 
with multiple observational constraints is unlikely to yield a unique
set of parameters that gives satisfactory results across the HR
diagram for all pulsation modes. We stress that parameter surveys are
an essential part of any science application of \RSP.

In Equations \eqref{eq:mom} and \eqref{eq:ie}, $p_{\rm av}$ is the artificial viscosity pressure \citep{rsp_richtmyer}
for numerically handling shocks that may develop during pulsations.
We adopt the \cite{rsp_stel75} two-parameter formulation as the default. 
The \cite{rsp_TW} artificial pressure-tensor form, which was 
implemented in \mesathree, can also be used in \RSP. 

\begin{figure}[hbt!]
\centering
\includegraphics[width=1.0\columnwidth]{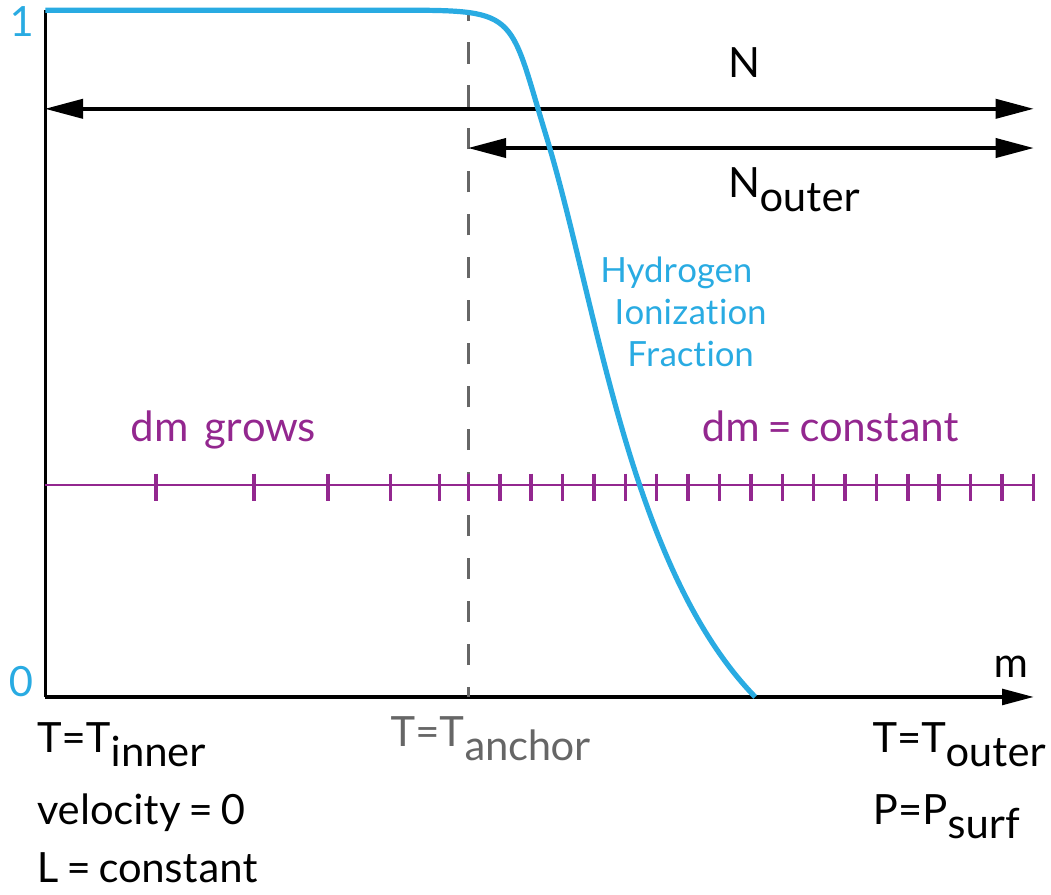}
\caption{
Grid structure in \RSP.
The inner boundary at the base of the static envelope is defined by a chosen temperature 
(\code{RSP\_T\_inner}). The model surface has 
a fixed temperature $T_{\rm outer}$, derived from $\Teff$, and pressure (\code{RSP\_Psurf}).
The anchor temperature (\code{RSP\_T\_anchor}) is usually located where H ionizes, shown by the blue curve.
The envelope is divided into $N$ Lagrangian mass cells (\code{RSP\_nz}).
Between the anchor and the surface are
$N_{\rm outer}$ cells (\code{RSP\_nz\_outer}), each with a constant mass.
Between the inner boundary and the anchor the mass of each cell increases.
}
\label{f.rsp.grid}
\end{figure}

The numerical scheme to solve discrete versions of the equations
is based on the intrinsically energy conserving method given by
\cite{rsp_fraley}.  Details of the numerical implementation, along with \RSP's lineage \citep{rsp_stel75,rsp_kb88}, 
are discussed in \cite{rsp_sm08a}.
\revision{During the nonlinear integration, 
$u=0$, $L={\rm constant}$ and $e_t=0$ at the inner boundary (See Figure~\ref{f.rsp.grid}).
The latter condition holds also for the outermost boundary, i.e., outermost cell is radiative.
External pressure is fixed and zero by default.}

\subsection{\RSP\ in Action}\label{s.rsp.action}

\RSP\ performs three operations: 
building an initial model;
conducting a linear non-adiabatic (LNA) stability analysis on that model; and 
integrating the time-dependent nonlinear equations.

\subsubsection{Building an Initial Model}\label{s.rsp.action.grid}

Since the energy density of radial pulsations drops rapidly going inward from a
star's surface, a full stellar model reaching to the center is frequently not necessary.  The use of \RSP\ is
currently restricted to cases in which pulsations are determined by the
structure of the envelope and are independent of the detailed structure of the core.
\RSP\ begins by building a chemically homogeneous envelope from given stellar parameters ($M$, $L$, \Teff, $X$, and $Z$).
These parameters can be freely chosen and need not originate from a \mesastar\ model.  
It is not yet possible to directly import an envelope
from \mesastar\ into \RSP\, primarily because of the different treatments of 
convection (a version of mixing length theory in \mesastar\ versus
detailed time evolution of turbulence in \RSP).  
Tighter integration of $\MESAstar$ and \RSP\ is a future project.


Specifications for the initial model include the number of cells and
the temperature at the base (see Figure~\ref{f.rsp.grid}).
This inner boundary temperature is defined by a chosen temperature  (\code{RSP\_T\_inner}\,$\simeq$\,2$\times$10$^6$\,\Kelvin)
that should be set hot enough so that the eigenvector amplitudes
generated in the following stability analysis go
to zero, and cool enough to exclude regions of nuclear burning and 
justify the assumption of chemical homogenity.  
The model is divided into inner and outer regions at a specified anchor temperature.  
In the outer region, cells have the same mass; in the inner region cell masses grow by a constant factor 
so that the innermost cells are significantly larger than the ones at the surface.
The anchor temperature should
be in the part of the model driving the pulsations.
For example, for pulsations in the classical instability strip a value of $T_{\rm anchor}$=11,000\,\Kelvin\ is typical.
In the case of $Z$-bump pulsations a higher temperature would be appropriate. 
Proper choice of the number of outer cells and placement of the anchor are necessary 
to ensure that the driving region is well resolved. 

The initial model builder iteratively constructs an envelope in hydrostatic equilibrium that satisfies the
\RSP\ equations.  
Starting from the outer radius determined by $L$ and \Teff\ this process involves selection
of a cell mass to be used in the outer part of the envelope and a scale factor that
is used to progressively increase cell masses in the inner region.  Those choices must match
the desired number of cells, both $N$ and $N_{\rm outer}$, and also satisfy the surface boundary conditions
and the required temperatures at the anchor location and at the inner boundary.  
The model builder is a complex multistage iterative procedure that works well for the range of cases presented in the following
but may fail when applied outside of that range.

\subsubsection{Stability Analysis}\label{s.rsp.action.lna}

The LNA analysis is performed on the initial model using a full linearization of the \RSP\ equations 
\revision{\citep[for details see][]{rsp_phd_smolec}}.
These include time-dependent convection, moving beyond the frozen-in convection approximation made in software instruments like \code{GYRE}.
This yields the eigenmodes, periods, and growth rates.
The eigenvectors are used to perturb the initial model for the time evolution.

\begin{figure}[ht!]
\begin{center}
\includegraphics[width=\columnwidth]{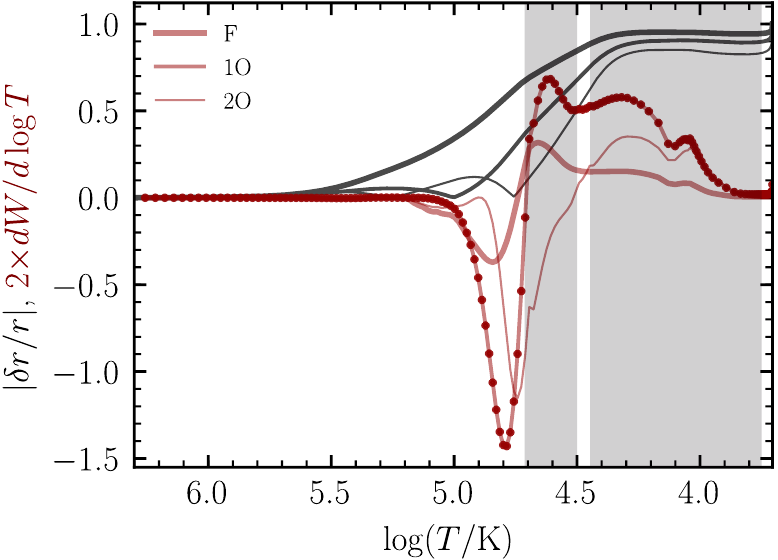}
\end{center}
\vspace{-0.15in}
\caption{\RSP\ LNA analysis of a classical Cepheid model. 
Shown are the displacement amplitudes (black) and the differential work
(red) done by the lowest three radial eigenmodes. 
The thickest curves are for the fundamental (F) mode, medium thickness for the 
first overtone (1O) mode and the thinnest curves for the second overtone (2O) mode.
The dots indicate the cell locations.
\revision{The grey areas show the extent
of the convection zones around H ionization, first He and second He ionization, respectively.}
\label{fig:rsp_LINA}}
\end{figure}

Figure~\ref{fig:rsp_LINA} shows amplitudes of radial displacements and
differential work for the first three eigenmodes of the classical
Cepheid model in the \mesa\ \code{test\_suite}.
A common resolution for exploratory model surveys, $N=150$ and $N_{\rm outer} = 40$, is adopted.
For $T \gtrsim 5 \times 10^{5}\,\Kelvin$, 
the displacements and differential work of all three radial eigenmodes are negligible,
indicating that the extent of the computational domain is sufficient.

\begin{figure}[ht!]
\begin{center}
\includegraphics[width=\columnwidth]{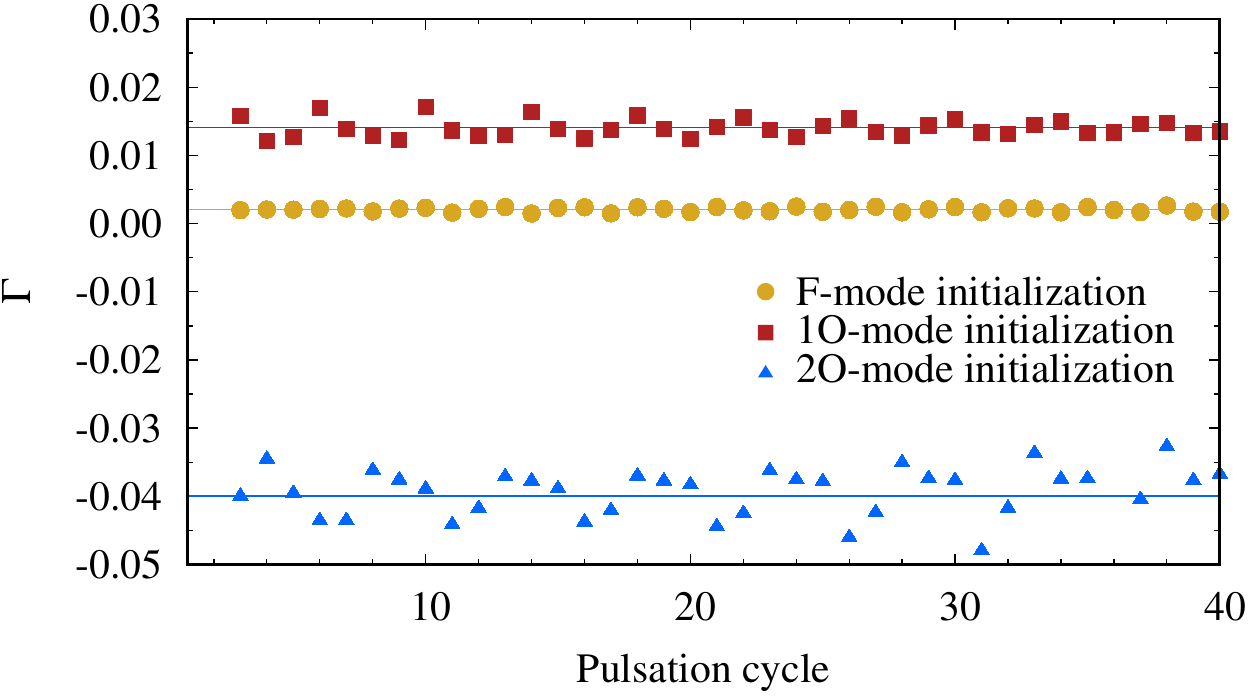}
\end{center}
\vspace{-0.15in}
\caption{Comparison of the fractional growth rate $\Gamma$ during the initial cycles of the time integration.
Horizontal lines show the LNA predictions. 
An RR~Lyrae model 
($M$\,=\,0.65\,\Msun, $L$\,=\,45\,\Lsun, \Teff\,=\,7,100\,\Kelvin, $X$\,=\,0.75, $Z$\,=\,0.0014) 
was initialized with a 0.1~\kms\ amplitude pure F-mode (circles), 1O-mode (squares), or
2O-mode (triangles) and evolved.
\label{fig:rsp_grtest}}
\end{figure}

\subsubsection{Evolution in the Linear Regime}\label{s.rsp.action.verify}

The initial static model is perturbed with a linear combination of 
the velocity eigenvectors of the three lowest order radial modes.
More specifically, the velocity eigenvectors are scaled to have a surface value of 1.  
\code{RSP\_fraction\_1st\_overtone} and \code{RSP\_fraction\_2nd\_overtone} 
multiply the 1O and 2O eigenvectors, respectively. The F-mode eigenvector is then multiplied by
(1 - \code{RSP\_fraction\_1st\_overtone} - \code{RSP\_fraction\_2nd\_overtone}).  
The linear combination of these three scaled eigenvectors is then multiplied by
the surface velocity \code{RSP\_kick\_vsurf\_km\_per\_sec}.

The time integration commences with a 
constant time step (\code{RSP\_target\_steps\_per\_cycle})
and continues for a specified number of pulsation cycles (\code{RSP\_max\_num\_periods}).
A new cycle begins when the model passes through a maximum radius. 
Controls allow filtering out secondary maxima in the radial velocity curve.

Figure~\ref{fig:rsp_grtest} shows $\Gamma$ the fractional growth of the kinetic energy per pulsation period
near the start of a time integration, where 
\begin{equation}
\Gamma=2(E_{\rm k,\,max}^{i+1}-E_{\rm k,\,max}^{i})/(E_{\rm k,\,max}^{i+1}+E_{\rm k,\,max}^{i}),
\label{eq.rsp.gamma}
\end{equation}
and $E_{\rm k,\,max}^{i}$ is the maximum kinetic energy of the envelope during pulsation cycle $i$.
Agreement between these three time integrations and the corresponding LNA analyses is satisfactory.
Similarly, the pulsation periods match the linear values
during the low-amplitude phase of development. 
Consistency between the time integrations and LNA analyses 
form the basis for interpreting the nonlinear results.

\begin{figure}[ht!]
\begin{center}
\includegraphics[width=\columnwidth]{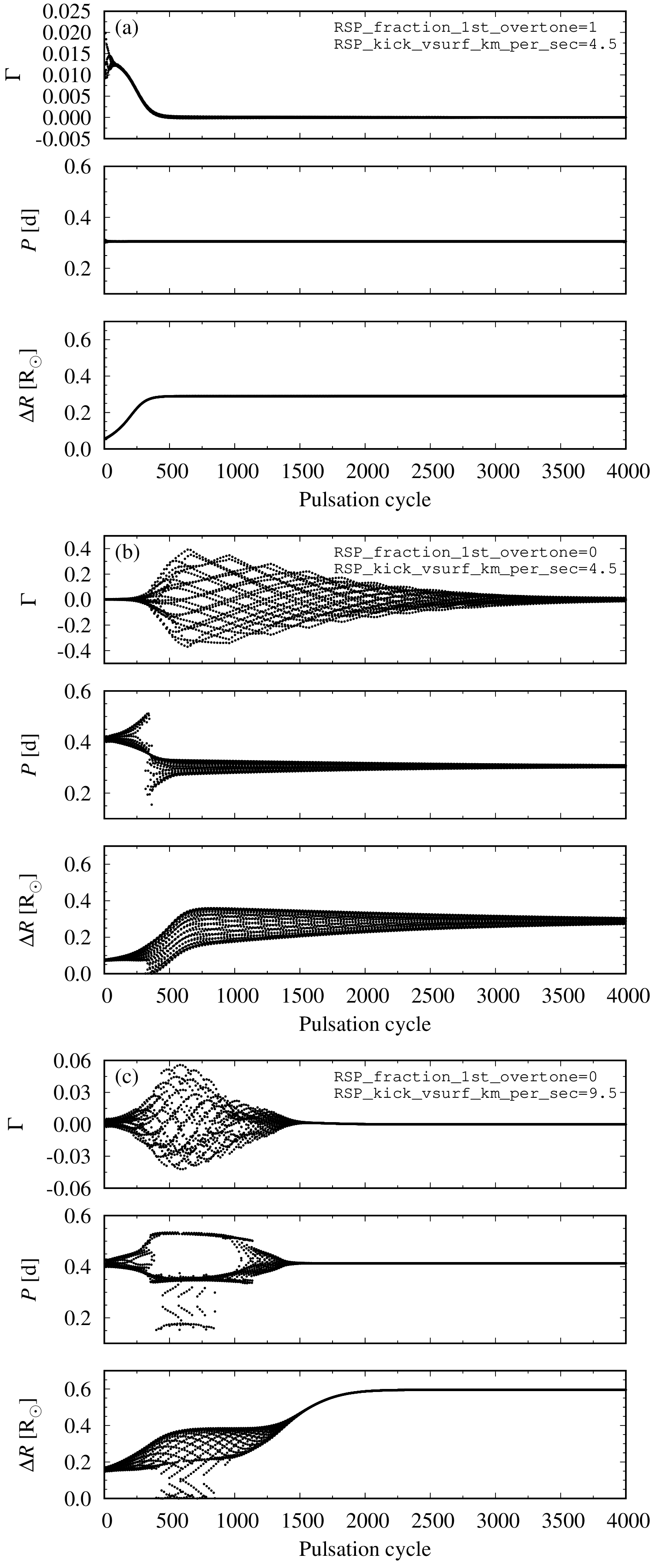}
\end{center}
\vspace{-0.15in}
\caption{Fractional growth rate $\Gamma$, period $P$,
and amplitude of radius variation $\Delta R$ during 4,000-cycle 
integrations of the same RR~Lyrae model as in Figure~\ref{fig:rsp_grtest}
with three different initial conditions labelled (a), (b), and (c).
\label{fig:rsp_integ}}
\end{figure}

\subsubsection{Different Perturbations, Different Periods}\label{s.rsp.action.whichP}

Which of the perturbed modes attains large-amplitude pulsations 
in the nonlinear regime may depend on the initial conditions \citep[e.g.,][]{rsp_s14}.
Figure~\ref{fig:rsp_integ} shows the results of longer time integrations 
for the RR~Lyrae model shown in Figure~\ref{fig:rsp_grtest}. 
The upper triplet of panels, case~(a), is for a 1O-mode initialization 
with a 4.5\,\kms\ amplitude. The middle triplet panel, case (b), is for 
a F-mode initialization with  4.5\,\kms\ amplitude.
The lower triplet, case (c), is for an F-mode initialization with a 9.5\,\kms\ amplitude. 

For case~(a), the pulsations converge towards a single, 1O-mode pulsation.
After a $\simeq$\,500 cycle transient phase the pulsation period and 
radius amplitude barely change and $\Gamma \simeq 0$.
For case~(b), the model has not converged to a single-periodic mode
after $4,000$ cycles. Despite the pure F-mode initialization, 
at $\simeq$\,300 cycles the pulsation switches toward the 1O mode.
This does not prove the model cannot pulsate in the F-mode, as case~(c) demonstrates. 
After a transient phase with beating F and 1O modes,
the 1O mode decays and the single-periodic F-mode pulsation grows to saturation. 

Figure~\ref{fig:rsp_integ} is an example of two different single-mode solutions
whose selection depends on the initial conditions.  
Two stars can have the same physical parameters but pulsate in
different modes depending on their evolutionary history.

\begin{deluxetable}{lllllllll}[!ht]
  \tablecolumns{5}
  \tablewidth{1.0\apjcolwidth}
  \tablecaption{Convective parameter sets referred to in the text as A, B, C, or D.  Note that the controls multiply base values (see Table~\ref{tab:rsp_alphas}).
  \label{tab:rsp_convectivesets}}
  \tablehead{ \colhead{Control} & \colhead{Set A} & \colhead{Set B} & \colhead{Set C} & \colhead{Set D}}
  \startdata    
\code{RSP\_alfa}  &  1.5 & 1.5 & 1.5 & 1.5 \\
\code{RSP\_alfam}  & 0.25 & 0.50 & 0.40 & 0.70 \\
\code{RSP\_alfas}  & 1.0 & 1.0 & 1.0 & 1.0 \\
\code{RSP\_alfac}  & 1.0 & 1.0 & 1.0 & 1.0 \\
\code{RSP\_alfad}  & 1.0 & 1.0 & 1.0 & 1.0 \\
\code{RSP\_alfap}  & 0.0 & 0.0 & 1.0 & 1.0 \\
\code{RSP\_alfat}  & 0.00 & 0.00 & 0.01 & 0.01 \\
\code{RSP\_gammar} & 0.0 & 1.0 & 0.0 & 1.0 \\
   \enddata
 \end{deluxetable}

 
\begin{figure}[!htb]
\begin{center}
\includegraphics[width=\columnwidth]{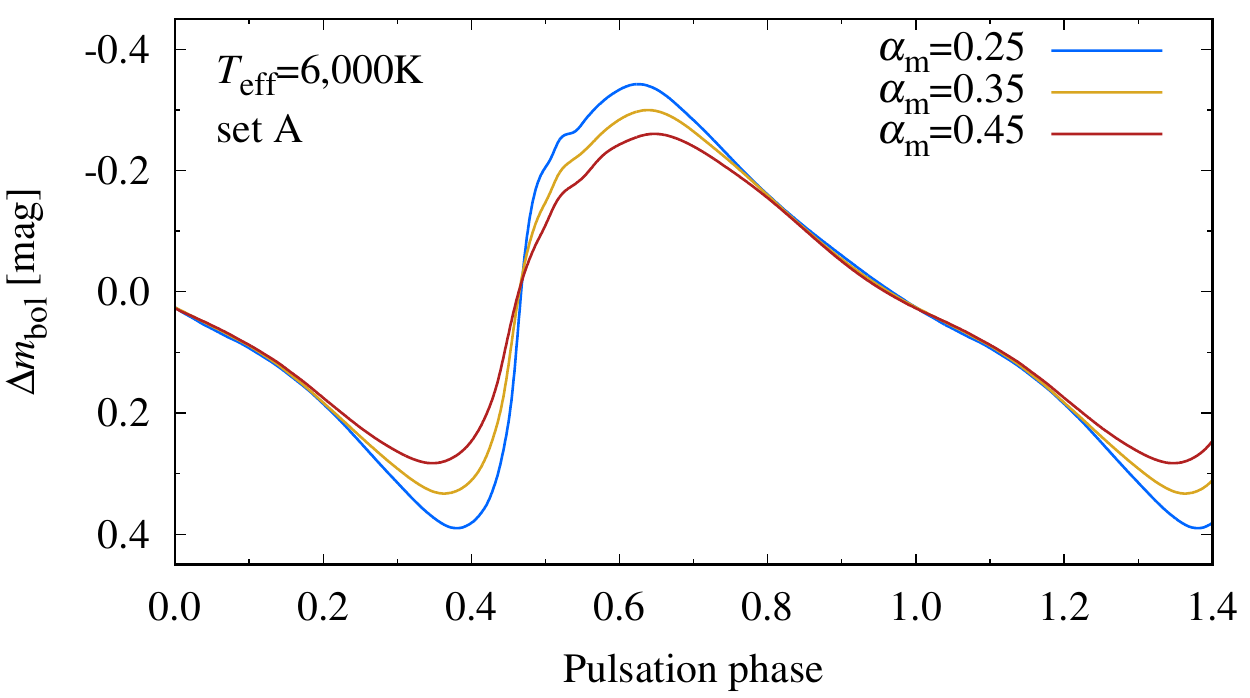}
\end{center}
\caption{Bolometric light curve of the \Teff=6,000\,\Kelvin \ Cepheid model 
of with convective parameters of set A but with varying 
eddy-viscous dissipation $\alpha_{\rm m}$.  \label{fig:rsp_ev}}
\end{figure}

\begin{figure*}[ht!]
\begin{center}
\includegraphics[width=2.1\columnwidth]{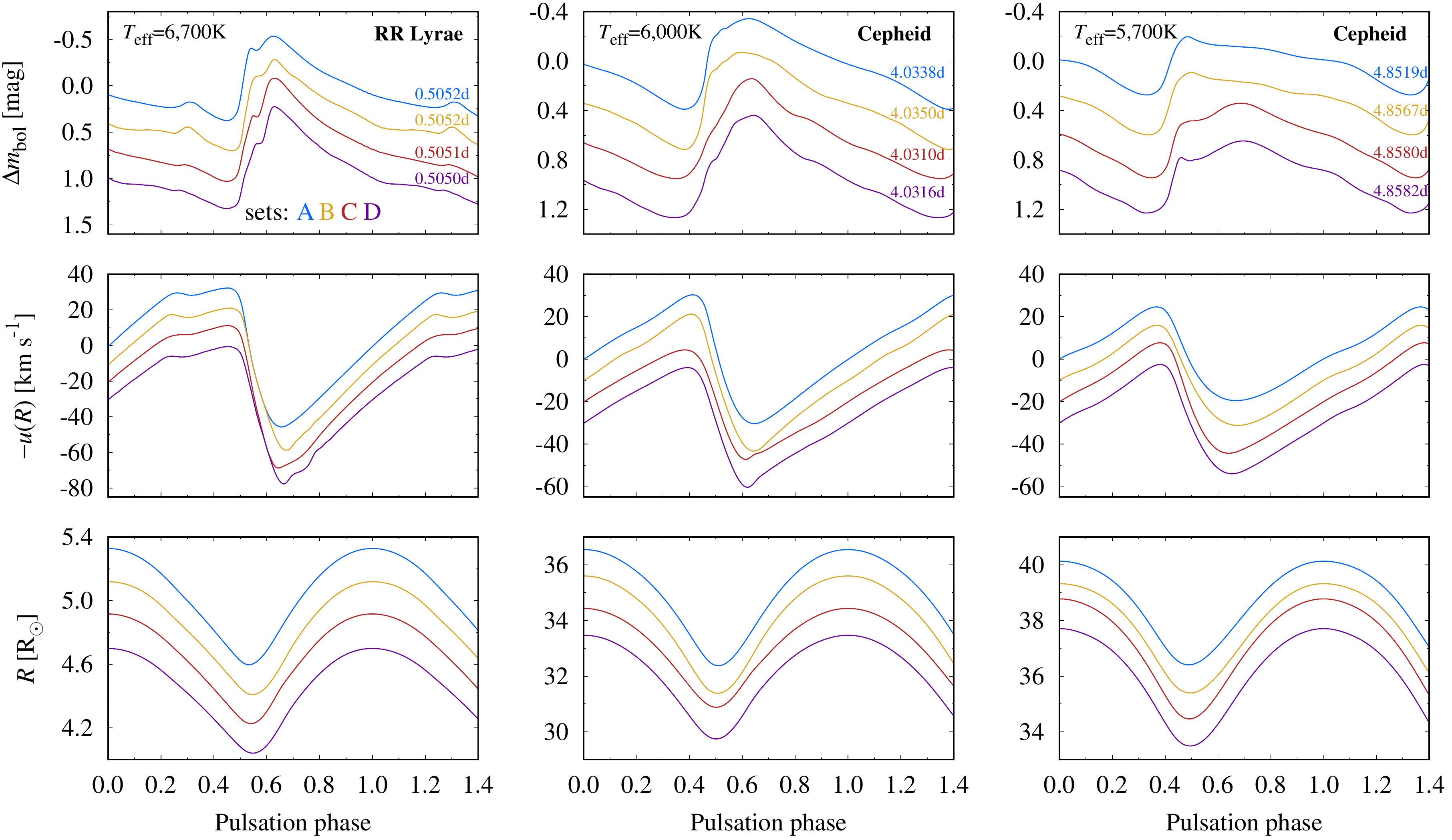}
\end{center}
\vspace{-0.15in}
\caption{Bolometric light (upper panel), radial velocity (middle panel) and radius curves (lower panel) for
an RR~Lyrae F-mode model 
(left; $M$\,=\,0.65\,\Msun, $L$\,=\,45\,\Lsun, \Teff\,=\,6,700\,\Kelvin, $X$\,=\,0.75, $Z$\,=\,0.0014) and 
two F-mode classical Cepheid models 
($M$\,=\,4.15\,\Msun, $L$\,=\,1,400\,\Lsun, $X$\,=\,0.73, $Z$\,=\,0.007) 
at \Teff=6,000\,\Kelvin \ (middle panel) and \Teff=5,700\,\Kelvin \ (right panel).
  The mass and
luminosity for the Cepheid models are close to the values derived for
OGLE-LMC-CEP-227 \citep{rsp_Pilecki2018}.  
Each curve corresponds to a set of convective parameter values listed in Table \ref{tab:rsp_convectivesets}.
The mean magnitude of the bolometric light curves is set to zero.
Light curves are vertically offset by 0.3\,mag, 
radial velocity curves by $10\,\kms$ and
radius curves by $0.2\,\Rsun$ (RR~Lyrae) or $1\,\Rsun$ (Cepheids).
\label{fig:rsp_curves}}
\end{figure*}

\begin{figure*}[ht!]
\begin{center}
\includegraphics[width=2.1\columnwidth]{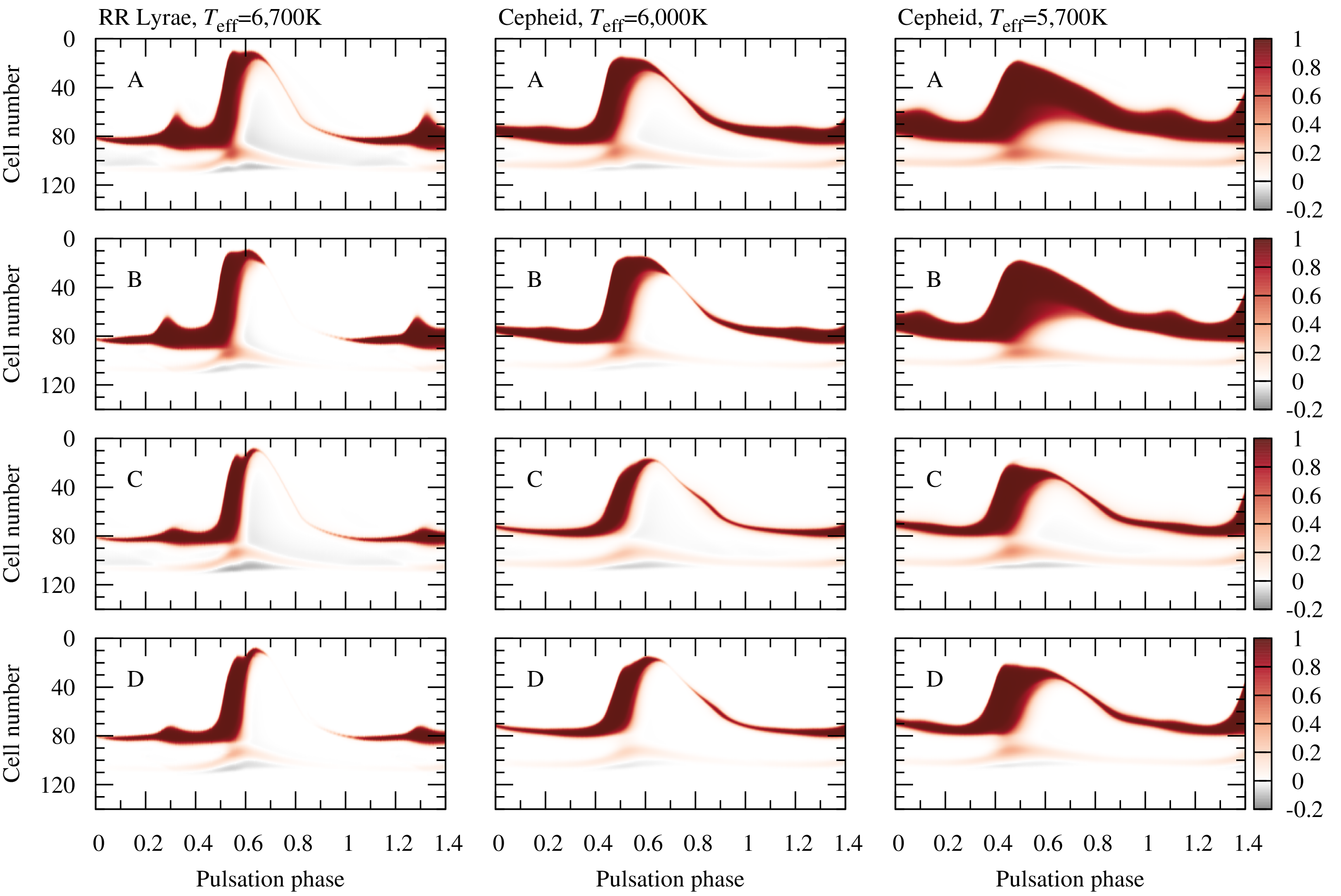}
\end{center}
\vspace{-0.15in}
\caption{Evolution of convective luminosity $L_{\rm c}/L$ for the models shown in Figure~\ref{fig:rsp_curves}. 
Cell number on the y-axis serves as a spatial coordinate, with cell 0 marking the stellar surface. 
The radiative interior of each model is not shown.
\label{fig:rsp_maps}}
\end{figure*}

\subsubsection{Convection Parameter Sensitivity}\label{s.rsp.action.convsense}

The final state in the nonlinear regime is usually a
single-periodic oscillation. The shape of the light and radial velocity curves 
may depend on the values of the eight free parameters listed in
Table~\ref{tab:rsp_alphas}. 
In Table~\ref{tab:rsp_convectivesets}
set~A corresponds to the simplest convection model. 
Set~B adds radiative cooling, 
set~C adds turbulent pressure and turbulent flux, and 
set~D includes these effects simultaneously. 
The parameter $\alpha_{\rm m}$ has little effect on the shape of the light curve but strongly affects its amplitude.
Figure~\ref{fig:rsp_ev} shows the effect of
varying $\alpha_{\rm m}$ on the \Teff=6,000\,\Kelvin \ Cepheid model.
The free parameter $\alpha_{\rm m}$ may thus be used to match
the observed amplitude.  For sets A-D, $\alpha_{\rm m}$ was adjusted so that models with different sets have similar amplitudes.

Figure~\ref{fig:rsp_curves} shows the effect of these 
parameter sets on the shapes of the bolometric light,
photosphere radial velocity, and radius variation curves for a
saturated F-mode RR~Lyrae and two saturated F-mode classical Cepheid models.
The pulsation
periods, radial velocity curves, and radius variation curves show 
only small differences.  For the RR~Lyrae models, there are differences in the
fine structure of the light curves. For example, the bump before
minimum light is weaker when turbulent pressure and turbulent flux are
included (sets C and D). The shape of the light curve near maximum
light also differs for both the RR~Lyrae and Cepheid models.

Figure~\ref{fig:rsp_maps} shows the convective luminosity profiles
for the models of Figure~\ref{fig:rsp_curves}.
Depending on pulsation phase, one convective region (darker hues)
extends from the surface cells down to cell $\simeq$\,90. 
This convective region is associated with partial ionization of H and He. 
Another convective region (lighter hues) lies deeper in the
envelope, centered at cell $\simeq$\,110, and is 
associated with the second ionization of He. In most of the models
these two convective regions merge at pulsation phase $\simeq\,$0.5 during 
maximum contraction when both
convective regions are at their strongest and most extended. In the
cooler models, the first convective region carries nearly all
of the luminosity throughout a pulsation cycle. In the hotter
models, this convective region becomes very weak 
at pulsation phase $\simeq$\,0.8 (before maximum expansion) and is barely 
resolved as it progresses deeper into the envelope. 
This behavior is pronounced in the RR~Lyrae models with
radiative cooling (sets B and D), as cooling contributes to damping the
turbulent energy and hence the near disappearance of the
convective region.

\begin{figure}[!htb]
\begin{center}
\includegraphics[width=\columnwidth]{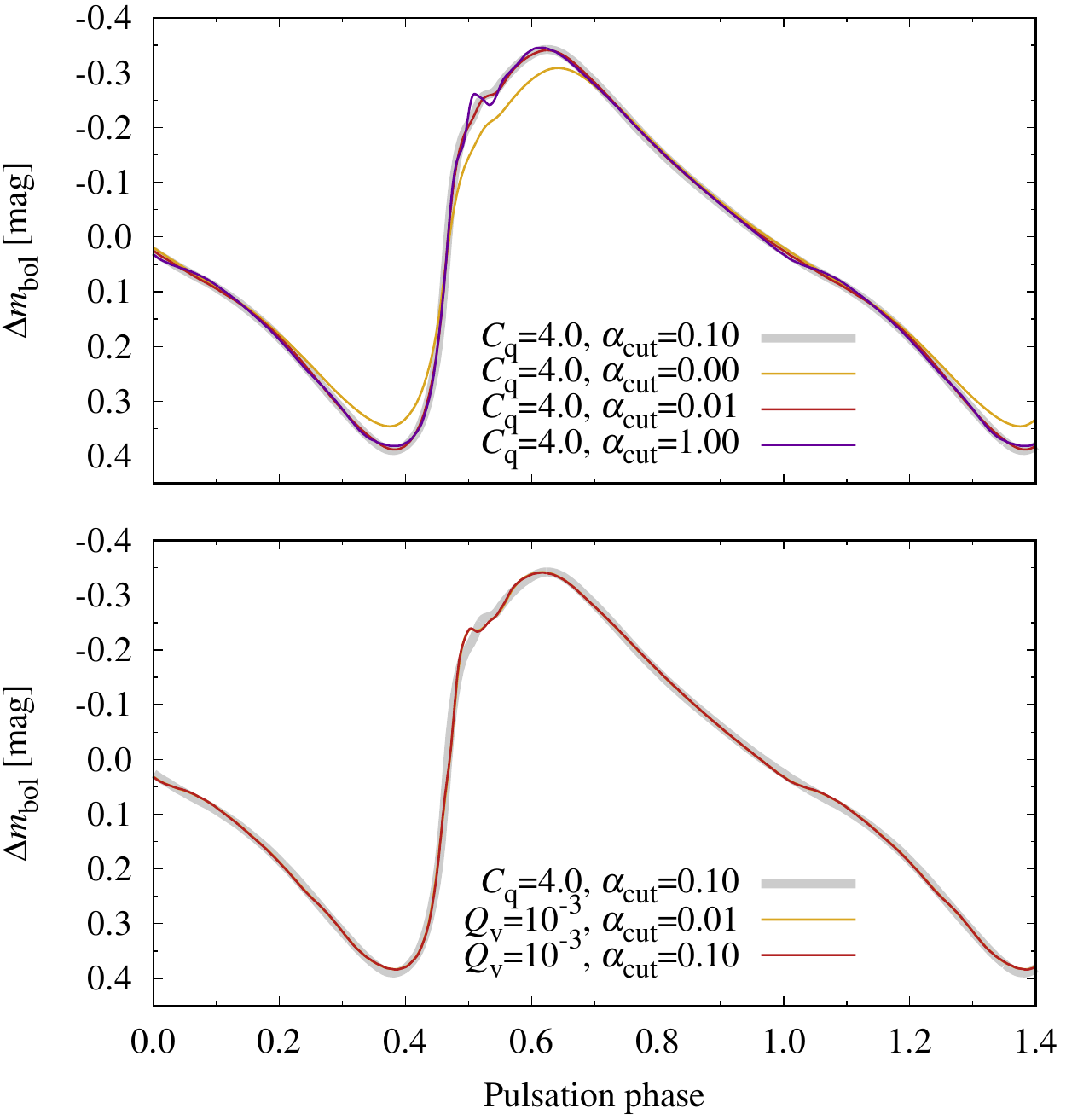}
\end{center}
\vspace{-0.15in}
\caption{Sensitivity of the bolometric light curve shape of Cepheid model on artificial
viscosity: $\alpha_{\rm cut}$ (upper panel) in the \cite{rsp_stel75} formulation,
and $\alpha_{\rm cut}$ in the \cite{rsp_TW} formulation (lower panel). 
The light curve with $C_{\rm q}$=4.0 and $\alpha_{\rm cut}$=0.1
is shown as a gray curve.
\label{fig:rsp_av}}
\end{figure}

\begin{figure}[!htb]
\begin{center}
\includegraphics[width=\columnwidth]{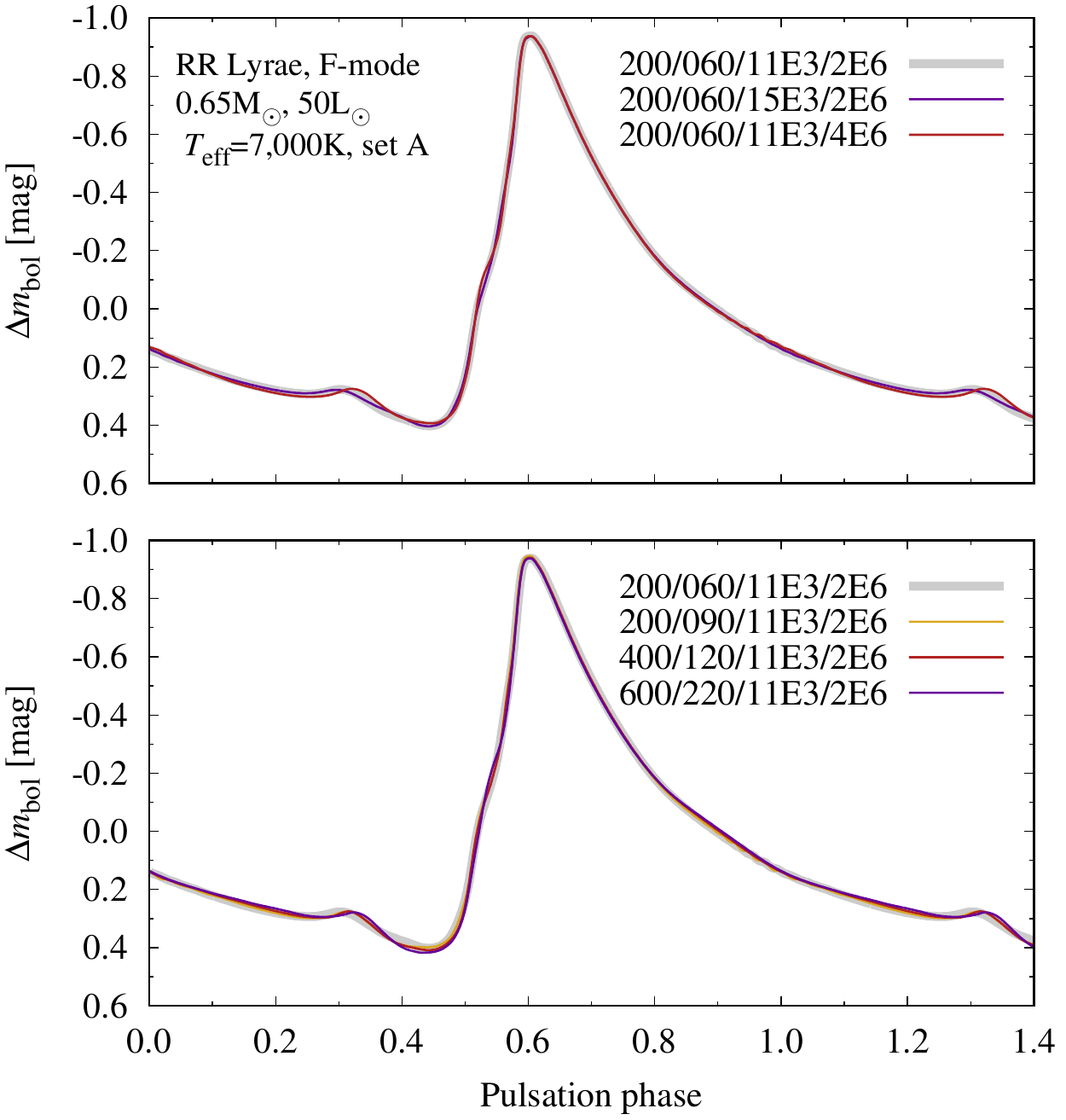}
\end{center}
\vspace{-0.15in}
\caption{Sensitivity of the bolometric light curve of an F-mode RR~Lyrae model 
to the grid, labeled as $N/N_{\rm outer}/T_{\rm anchor}/T_{\rm inner}$. 
The light curve for the default grid is shown by the gray curve. 
The upper panel shows the effects of different $T_{\rm anchor}$/$T_{\rm inner}$.
The lower panel shows the effects of different $N/N_{\rm outer}$ combinations. \label{fig:rsp_ctest}}
\end{figure}

\subsubsection{Artificial Viscosity Sensitivity}\label{s.rsp.action.avsense}

There are two parameters, $\alpha_{\rm cut}$ and $C_{\rm q}$,  that control the \cite{rsp_stel75} artificial viscosity, entering into the definition of $p_{\rm av}$ (see Table 1). 
Figure~\ref{fig:rsp_av} shows the effect of these parameters on 
light curves for the  \Teff\,=\,6,000\,\Kelvin \ Cepheid model.
The value of $C_{\rm q}$ plays a minor role. For  $C_{\rm q} = 4$, the upper panel shows a different light curve 
develops only if $\alpha_{\rm cut}$\,=\,0.0, corresponding to 
artificial viscosity acting for very small compressions, which 
leads to excessive dissipation that quenches the pulsation amplitude. 
For $\alpha_{\rm cut}$\,$\geq$\,0.01, the light curves are similar 
and roughly have the same pulsation amplitude.
When $\alpha_{\rm cut}$\,=\,0.1, artificial viscosity turns on only for strong shocks,
seen as the wiggle on the ascending branch of the light curve.
This choice ($\alpha_{\rm cut}$\,=\,0.1) numerically captures shocks without excessive dissipation,
barely affecting the light curve shape and amplitude.
\revision{While an artificial viscosity modifies the velocity structure in the envelope at each epoch,
we find that these differences are smaller than the differences for the bolometric light curves.}
The lower panel shows that the \cite{rsp_TW} form of artificial viscosity 
yields light curves with the same amplitude and qualitatively the same shape. 
Small differences are apparent at a shock-prone phase shortly before the maximum brightness.

\subsubsection{Spatial and Temporal Sensitivity}\label{s.rsp.action.gridsense}

Figure~\ref{fig:rsp_ctest} shows the sensitivity of the bolometric light curve
to the total number of cells $N$, number of cells above the anchor $N_{\rm outer}$, 
anchor location $T_{\rm anchor}$, and inner boundary location $T_{\rm inner}$
for an RR~Lyrae model
($M$\,=\,0.65\,\Msun, $L$\,=\,50\,\Lsun, \Teff\,=\,7,000\,\Kelvin, $X$\,=\,0.75, $Z$\,=\,0.0014)
with convective parameter set A. 

For classical pulsators, $T_{\rm inner}$ is typically placed at 2$\times$10$^6$\,\Kelvin,
and common choices for $T_{\rm anchor}$ are 11,000\,\Kelvin\ or
15,000\,\Kelvin. 
The upper panel shows that the light curves are weakly
sensitive to the choice of $T_{\rm inner}$ and $T_{\rm anchor}$ for this RR~Lyrae model.
Light and radial velocity curves are usually the most sensitive 
to $N$ and $N_{\rm outer}$.
The lower panel shows this effect is small for this RR~Lyrae model.
Section~\ref{sssec:rsp_BEP} shows a case with a much larger sensitivity.

The default value of 600 time steps per pulsation cycle works well
for most cases, but smaller time steps are recommended for models
that include radiative cooling, turbulent pressure, turbulent flux, 
or develop violent pulsations (e.g., the chaotic models of Section~\ref{sssec:rsp_T2CEP}).
We stress that there is no unique choice of grid or time step that will work for
all applications or guarantees convergence.  All nonlinear modeling of
variable stars should be accompanied by sensitivity and convergence tests.

\subsection{Current Limitations and Plans for the Future}\label{ssec:limitations}

\RSP\ in its present form covers most of the classical instability
strip including $\delta$~Cepheids, RR~Lyrae, High
Amplitude $\delta$~Scuti and SX~Phoenicis stars (see Figure~\ref{f.hrpulse}), 
where a single or just a few dominant radial modes are observed.
\RSP\ also has applications outside of the classical
instability strip as we show below for BLAPs.
For stars close to the main sequence, linear growth rates
are very small and thus, as we show below, long time integrations are necessary to approach 
full-amplitude nonlinear pulsations.

\RSP\ is currently of limited use for strongly non-adiabatic pulsations with large $L/M$ ratios,
including Luminous Blue Variables, Mira-type variables, and type~II Cepheids. 
For the latter, only the shortest-period BL~Her class variables
can be reliably modeled (see Section~\ref{sssec:rsp_T2CEP}). 
For the  longer-period classes of W~Vir and RV~Tau variables, either static envelopes cannot 
be constructed or nonlinear integrations break down at the onset 
due to violent relaxations of the outermost layers. 
In the extended envelopes of these variable stars the radiation-diffusion 
approximation is inadequate due to the low optical depth. 
The inclusion of pulsation-driven mass loss may also 
be necessary to study pulsations of these variable stars \citep{rsp_s16}. 

Inclusion of turbulent pressure
and flux may lead to convergence difficulties when
constructing the static initial envelope.  Cooler stellar envelopes
with higher Mach number convection are also numerically more difficult than
hotter envelope models. These difficulties are rooted in the static grid structure 
shown in Figure~\ref{f.rsp.grid}.
Future developments of \RSP\ should include a more versatile initial 
model builder, adaptive remeshing during the time integration, 
and a radiation-hydrodynamic treatment of the radiative 
energy and flux. 

\subsection{Applications of \RSP}\label{ssec:rsp_applications}

We now apply \RSP\ to variable stars.
Examples include the light curves of classical pulsators,
modeling of specific objects, and models for the dynamics of 
modulated or chaotic pulsations.

\begin{figure}
\begin{center}
\includegraphics[width=\columnwidth]{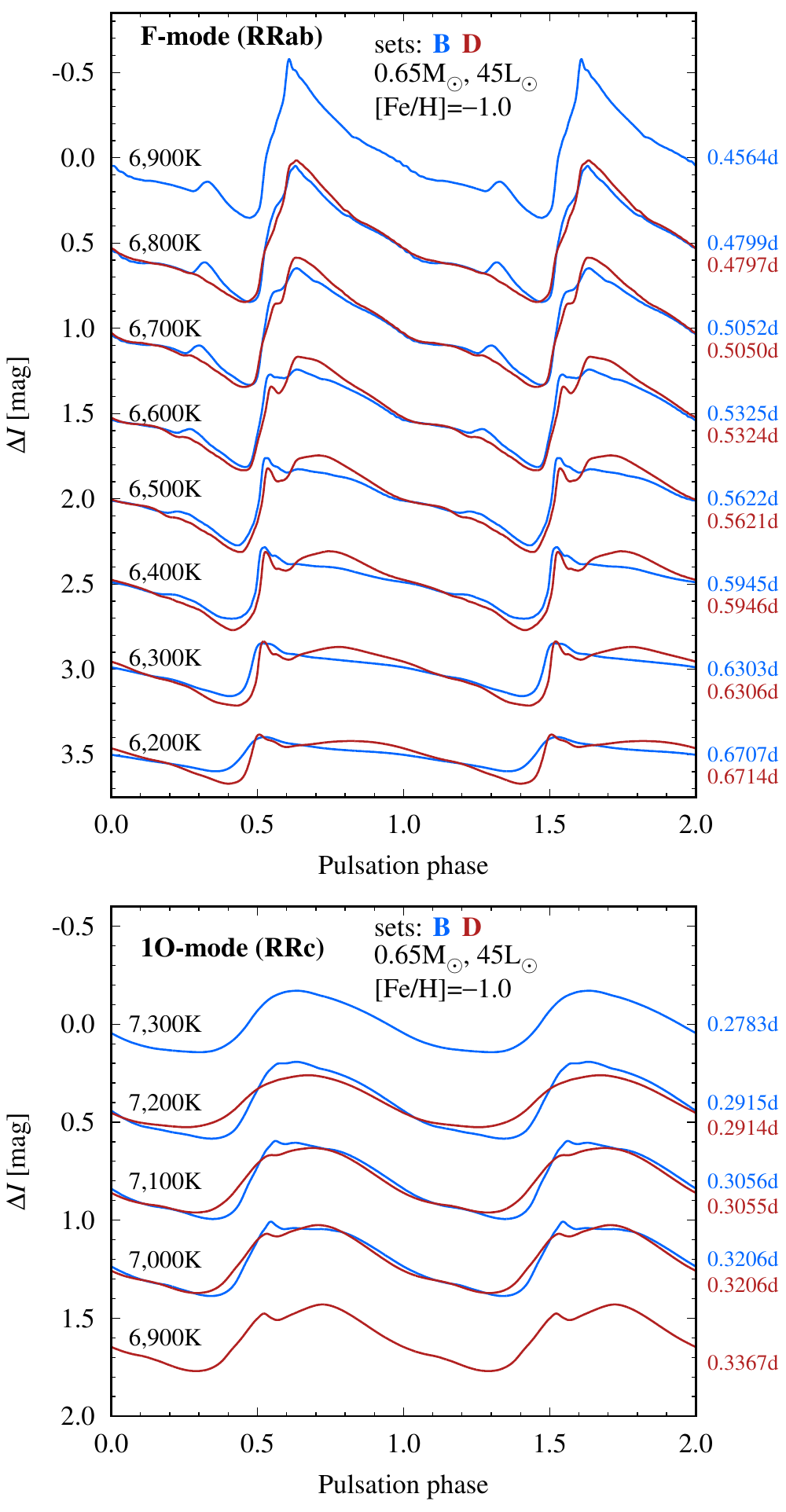}
\end{center}
\vspace{-0.18in}
\caption{
$I$-band light curves of F-mode (upper panel) and 1O-mode (lower panel) pulsators
across the instability strip for 
$M$\,=\,0.65\,\Msun, $L$\,=\,45\,\Lsun, [Fe/H]\,=\,$-$1.0, and 
convective sets B (blue) and D (red). Light curves are labelled with their \Teff \  and period,
and offset vertically to facilitate comparisons (by 0.5 in the upper panel and 0.4 in the lower panel).
\label{fig:rsp_RRL_lc1}}
\end{figure}

\begin{figure*}
\begin{center}
\includegraphics[width=\columnwidth]{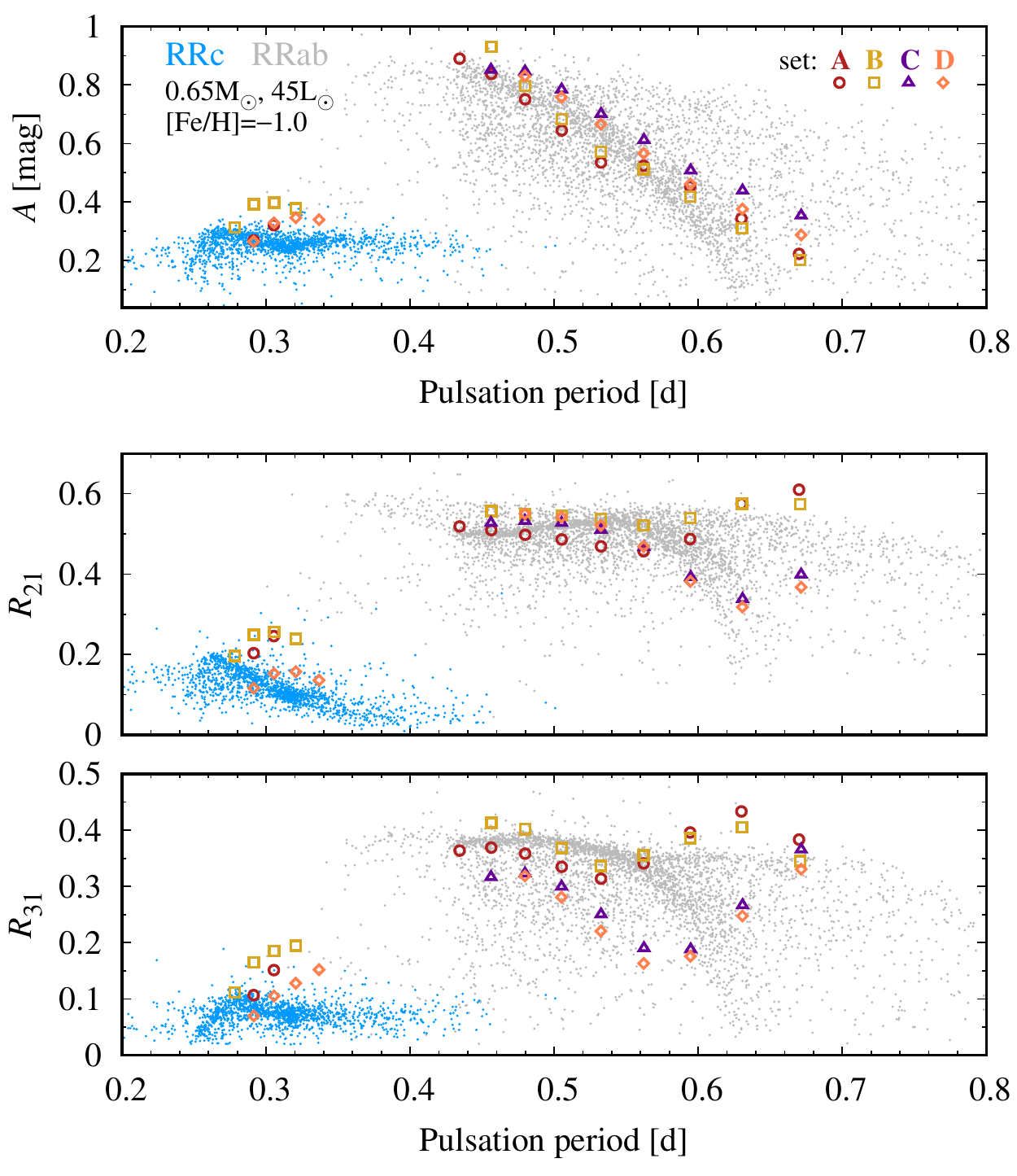}\includegraphics[width=\columnwidth]{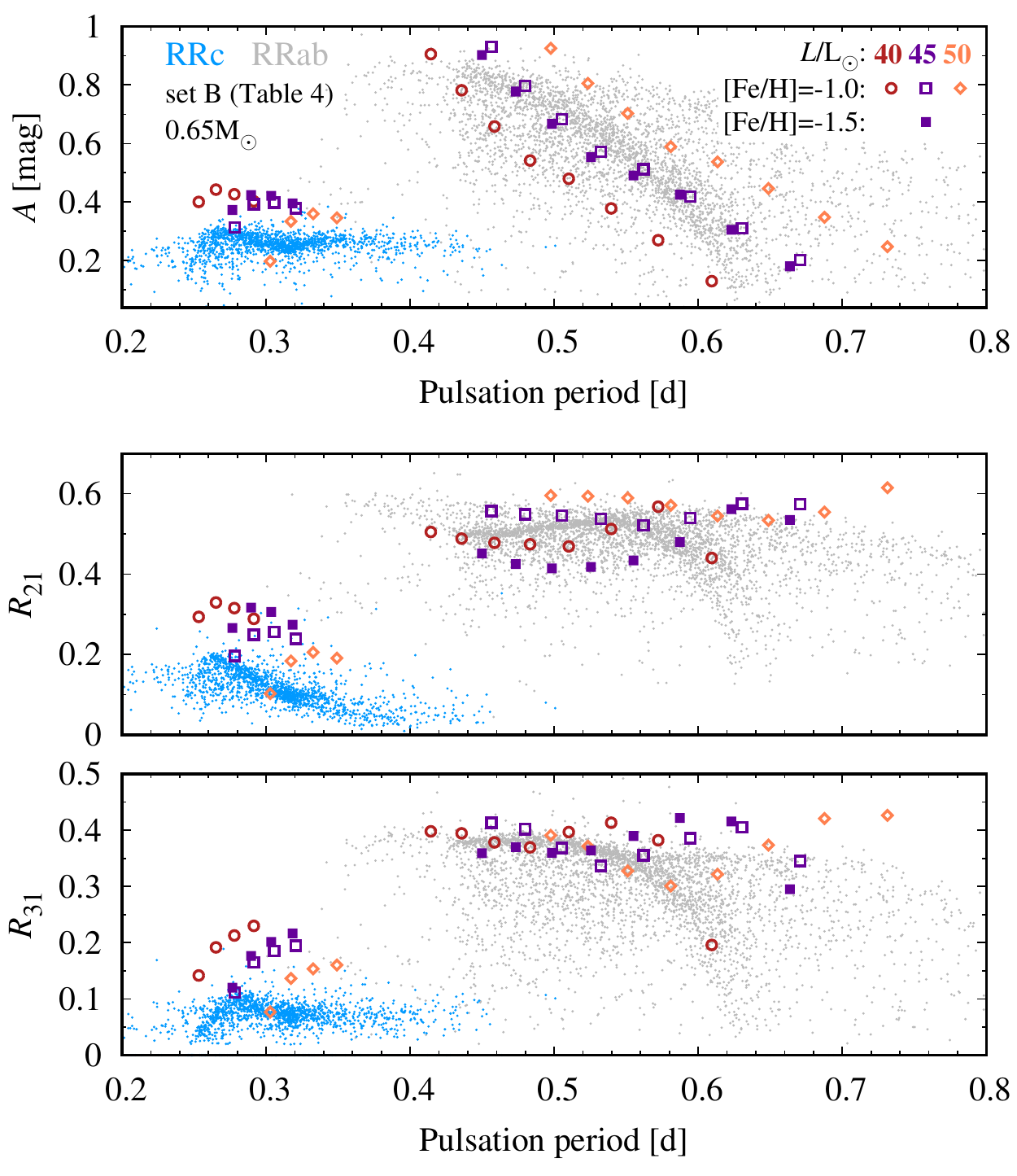}\\
\includegraphics[width=\columnwidth]{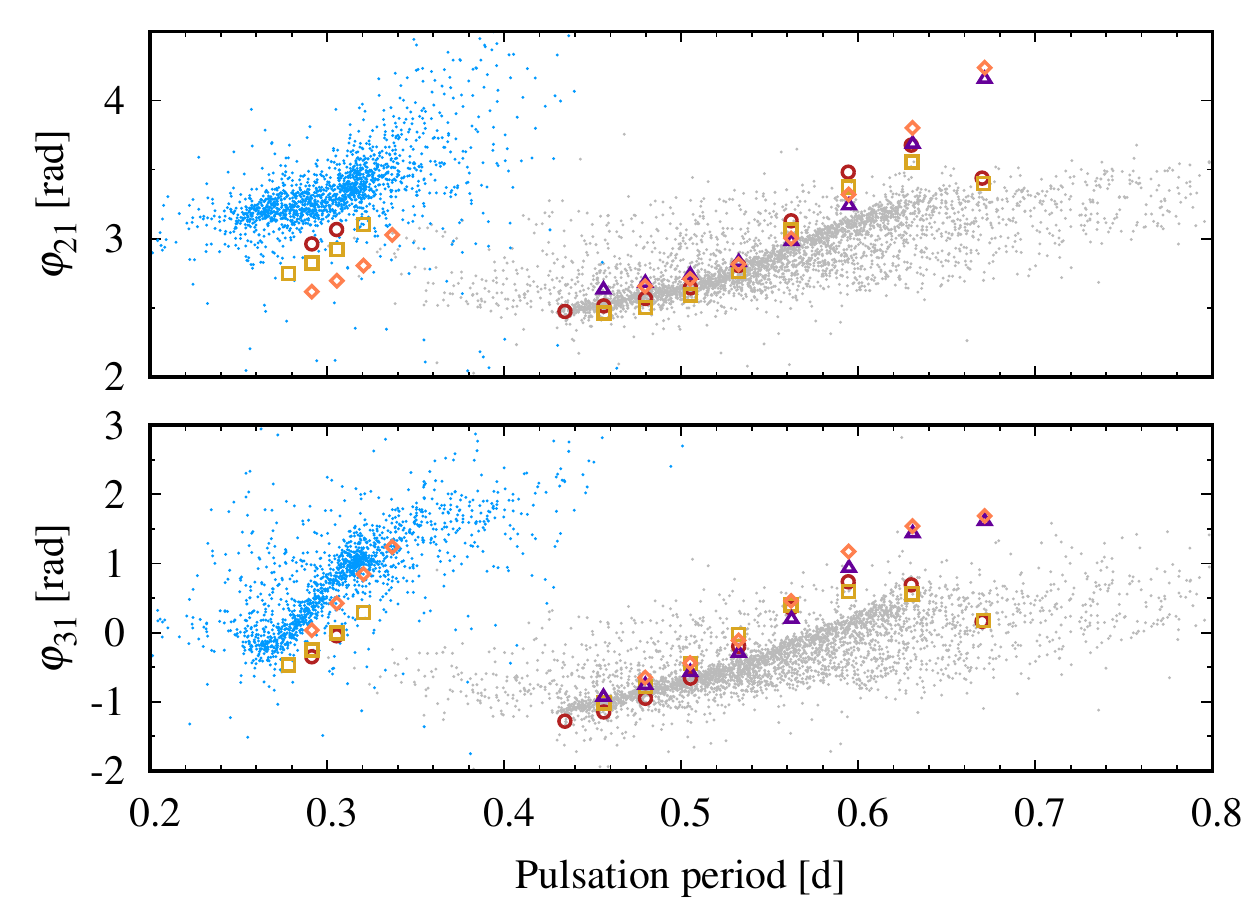}\includegraphics[width=\columnwidth]{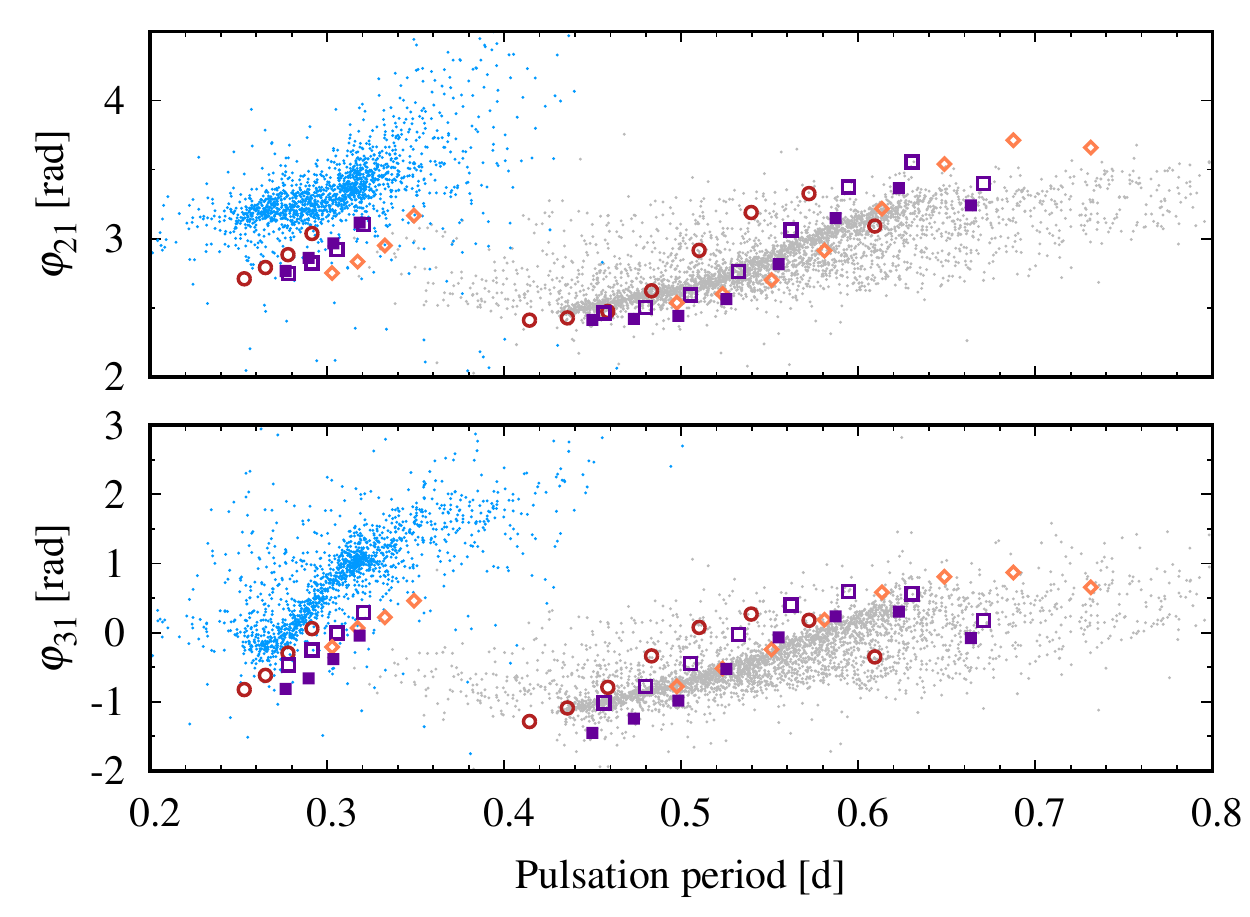}
\end{center}
\vspace{-0.15in}
\caption{
Comparison of the peak-to-peak amplitude and low-order Fourier
decomposition parameters of $I$-band light curves of Galactic bulge
RRab stars (gray dots) and RRc stars (blue dots) with
synthetic light curves (symbols). In the left panels physical parameters
are fixed, except for \Teff\ and different
convective parameter sets. In the right panels,
the convective model is fixed and physical
parameters are varied. Observational data are from
\cite{rsp_o4_rrl}.\label{fig:rsp_RRL_four}}
\end{figure*}

\subsubsection{RR~Lyrae variables}\label{sssec:rsp_RRL}

We consider two sequences of RR~Lyrae-type models.
The first sequence has $M$\,=\,0.65\,\Msun, $L$\,=\,45\,\Lsun \ and [Fe/H]\,=\,$-$1.0 ($X$\,=\,0.75,
$Z$\,=\,0.0014) with \Teff \ varying in 100 \Kelvin\ steps for convective sets A--D. 
Figure~\ref{fig:rsp_RRL_lc1} shows a gallery of $I$-band light curves from
this sequence.  The upper panel shows F-mode pulsators (commonly known as RRab stars)
and the lower panel shows  1O-mode 
pulsators (known as RRc stars).
The latter have
smaller amplitudes and are less nonlinear in shape 
(i.e., more sinusoidal) than the F-mode light curves.  
Models with different convective settings differ the most near minimum
and maximum light. For example, F-mode models with convective set B develop a bump
preceding minimum light that is absent in the light curves with
convective set D. On the other hand, F-mode models with cooler \Teff\ 
from convective set D develop broad, double-peaked light curve maxima 
that are absent in models from convective set B.

To compare the overall morphology of $I$-band light curves from these sequences
with OGLE observations, we perform a Fourier decomposition of the
synthetic light curves
\begin{equation}
I(t)=A_0+\sum_k A_k\sin(2\pi k f t +\phi_k)\,,\label{eq:fs}
\end{equation}
where $f$ is the pulsation frequency, and $A_k$ and $\phi_k$ are amplitudes
and phases, respectively. We then construct the amplitude ratios $R_{k1}$
and epoch-independent phase differences $\varphi_{k1}$ \citep{rsp_sl81}:
\begin{equation}
 R_{k1}=\frac{A_k}{A_1}\,,\quad \varphi_{k1}=\phi_k-k\phi_1\,.
\end{equation}

Observationally derived values of $R_{k1}$ and $\varphi_{k1}$ are taken from the OGLE
catalog \citep{rsp_o4_rrl} and shown in Figure~\ref{fig:rsp_RRL_four} by gray dots 
for RRab (F-mode) stars and blue dots for RRc (1O-mode) stars.
The observations show that the Fourier parameters follow progressions with pulsation period, 
traced by the highest density of data points, but with significant scatter.  
Fourier parameters from the model $I$-band light curves 
are shown with colored symbols. Left panels are for the first sequence of
models. Right panels are for the second sequence, which has
$M$\,=\,0.65\,\Msun, convective set B, \Teff \ varying in $100\,\Kelvin$ steps, 
and either $L$\,=\,(40, 45, 50)\Lsun \  at
[Fe/H]\,=\,$-$1.0  or $L$\,=\,45\,\Lsun\ at
[Fe/H]\,=\,$-$1.5. 

The left panels show that F-mode pulsators with convective sets A
and B progress similarly.  Models with convective sets C
and D also progress similarly but are
qualitatively different than those with convective sets A and
B. These differences are more pronounced for cooler, longer period,
models.  The overall match of the F-mode decompositions with the
OGLE RRab stars (gray points) is reasonable but shows
some systematic differences.  For example, the model values of
$\varphi_{31}$ are larger than the observed values.  However, the
model physical parameters except \Teff \ are fixed, while this
is not the case for the OGLE RRab stars.

The match between the 1O-mode models and the OGLE RRc stars
(blue points) is worse.  The amplitudes and the amplitude ratios
are systematically too large. The Fourier phases are
systematically too small. This may indicate different convective
parameters are needed to reproduce the observed light curve shapes of
F-mode and 1O-mode pulsators. Note the 1O-mode instability
strip with convective set C  is smaller than the F-mode instability strip 
at the luminosity considered. Models from set C, where the 1O-mode is
linearly unstable and the integration is initialized with a
1O-mode velocity perturbation, all switch to an  F-mode
pulsation.

The right panels in Figure~\ref{fig:rsp_RRL_four} show that the light curve shapes are sensitive to the
physical parameters.  By varying the luminosity and metallicity in a narrow range,
the model sequences match the OGLE values.
However, no sequence considered follows the OGLE progression, because RR~Lyrae in the Galactic
bulge are characterized by mass, luminosity and metallicity distributions that cannot 
be reproduced with a single sequence.

\begin{figure}[ht!]
\begin{center}
\includegraphics[width=\columnwidth]{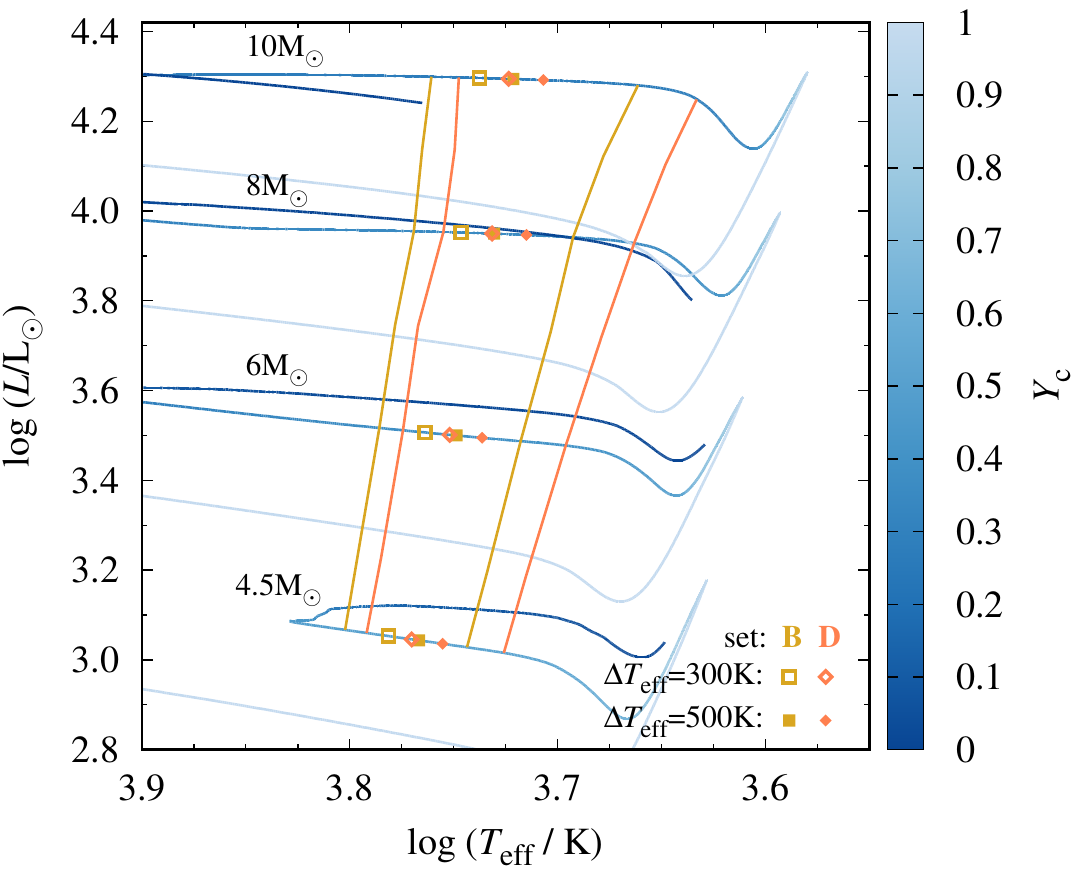}
\end{center}
\vspace{-0.15in}
\caption{
Evolutionary tracks in the HR diagram, shaded to show the core He mass fraction, $Y_{\rm c}$.
Blue and red edges of the instability strip for convective sets B and D
are shown, along with the locations 
where non-linear Cepheid models are computed (symbols). 
\label{fig:rsp_DCEP_hr}}
\end{figure}

\subsubsection{Classical F-mode Cepheids}\label{sssec:rsp_DCEP}

Cepheids display a feature known as the Hertzsprung progression
\citep{rsp_hbp}.
A secondary bump in their light and radial velocity curves appears near
minimum light on the descending branch when $P$\,$\simeq$\,5\,d. The bump
moves towards earlier phases on the descending branch as the period increases and is coincident 
with maximum light when $P$\,$\simeq$\,10 d.
The bump then moves onto the ascending branch for longer periods and disappears 
at $P$\,$\simeq$\,20\,d.
The bump is driven by a 2:1 resonance between the F-mode and a damped 2O-mode
\citep[e.g.,][]{rsp_ss76,rsp_bmk90}.
This behavior is reflected in Cepheids' Fourier parameters, which
follow more complex progressions with pulsation period than the
RR~Lyrae stars. 

\begin{figure}[ht!]
\begin{center}
\includegraphics[width=\columnwidth]{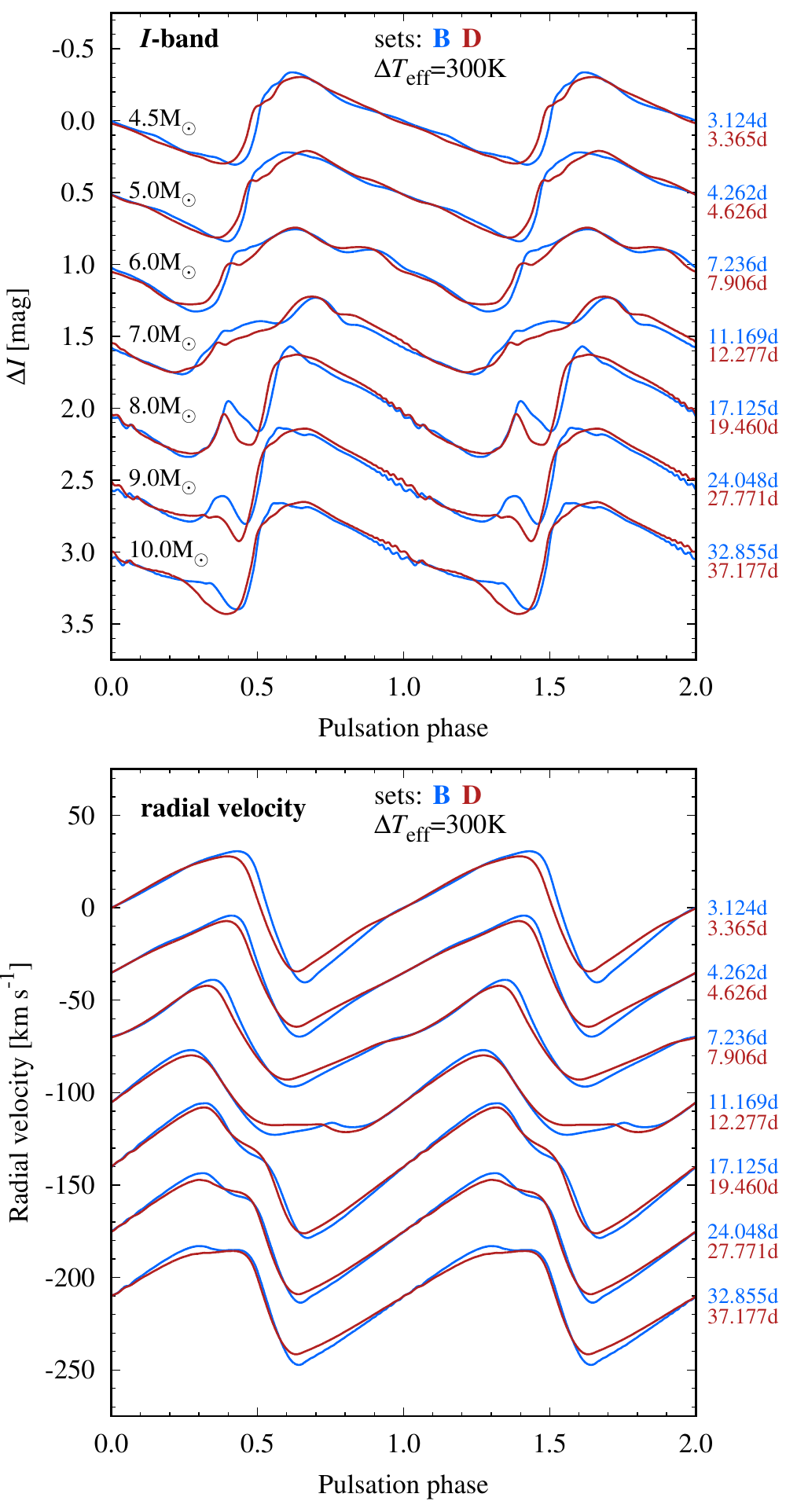}
\end{center}
\vspace{-0.15in}
\caption{
$I$-band light curves (upper panel) and radial velocity curves (lower panel) 
for the Cepheid models with a $\Delta$\Teff\,=\,300\,\Kelvin\ offset from
the blue edge and convective sets B and D. Light curves are labelled
with their pulsation periods and offset vertically by 0.5 mag or 35 \kms\
to facilitate comparison. Radial velocity curves follow the same order.
\label{fig:rsp_DCEP_lc}}
\end{figure}

\begin{figure*}
\begin{center}
\includegraphics[width=\columnwidth]{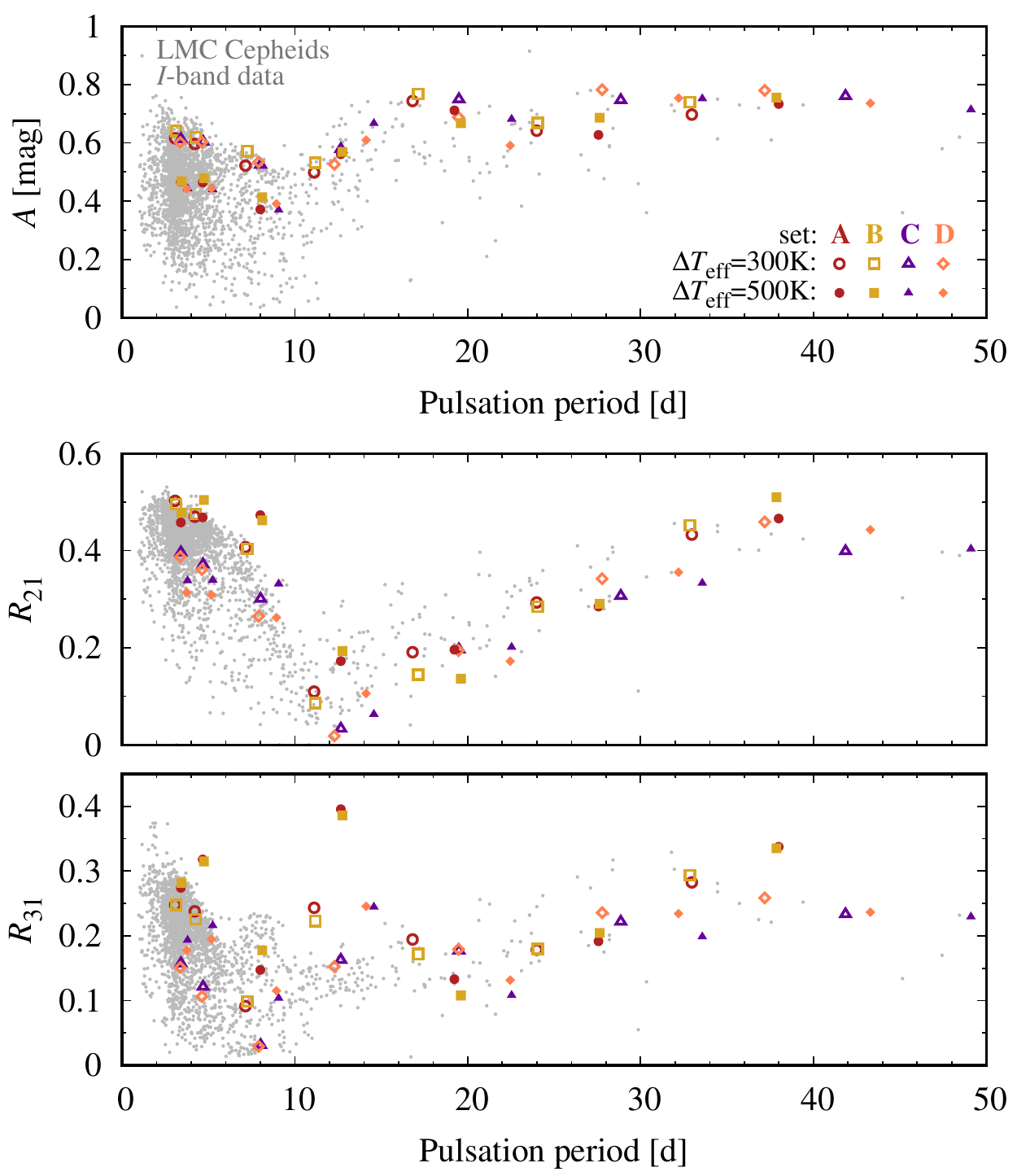}\includegraphics[width=\columnwidth]{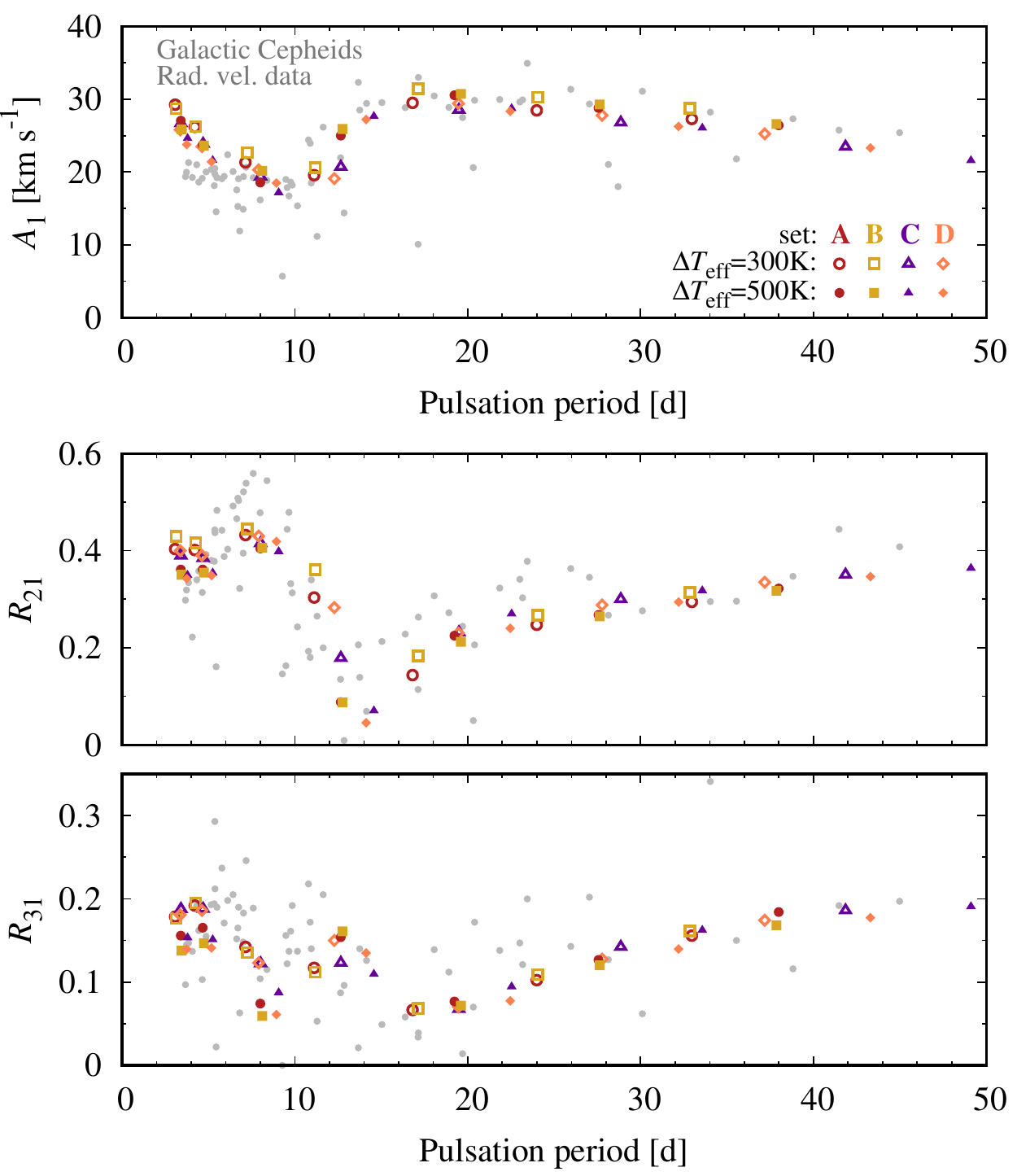}\\
\includegraphics[width=\columnwidth]{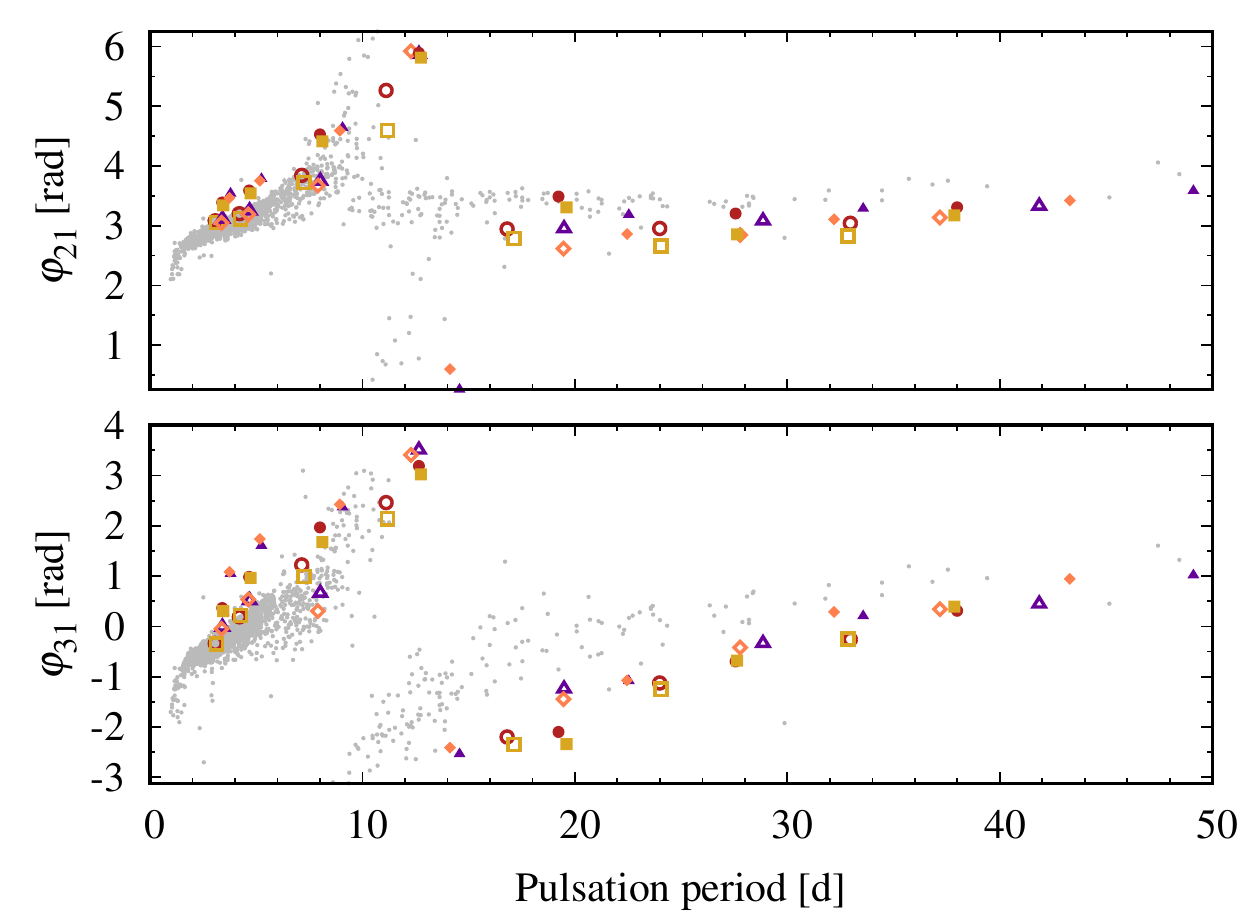}\includegraphics[width=\columnwidth]{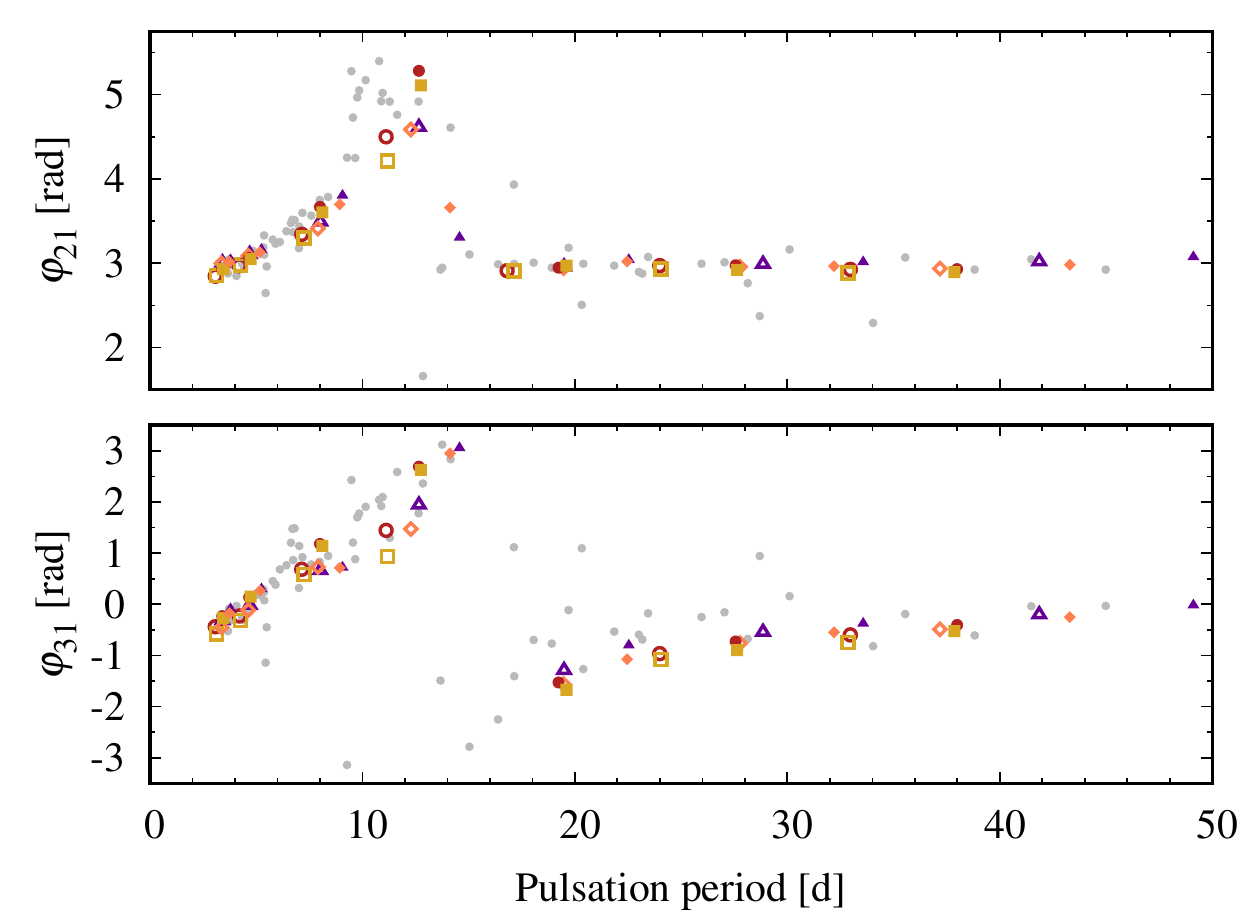}
\end{center}
\vspace{-0.15in}
\caption{
Amplitudes and Fourier parameters for 
$I$-band light curves (left panel) and radial velocity curves (right panel). 
Observations are marked with gray dots. Light curve data are from
\citet{rsp_o4_cep_mc} and \citet{rsp_shallow} and radial velocity curve data 
are from \cite{rsp_storm}.
For the light curves the peak-to-peak amplitude is shown,
while for radial velocity curves the Fourier amplitude is scaled with 
a $p$-factor of 1.3. Models are plotted with colored symbols.
\label{fig:rsp_DCEP_four}}
\end{figure*}

We consider 4.5, 5.0, 6.0, 7.0, 8.0, 9.0, and
10.0\,\Msun\ models with $X$\,=\,0.736, $Z$\,=\,0.008 and convective parameter sets A--D. 
Further details are in the test suite example \code{5M\_cepheid\_blue\_loop}. 
Figure~\ref{fig:rsp_DCEP_hr} shows the evolutionary tracks during 
the core He burning, blue-loop phase for a selection of masses.
Blue and red edges of the instability strip 
computed with \RSP\ for convective sets B and D are shown. Edges for
convective sets A and C largely overlap those for convective sets B and D, respectively. 
Nonlinear pulsation models are computed with a
$\Delta$\Teff\,=\,300\,\Kelvin\ or 500\,\Kelvin\  
offset from the blue edge.

Figure~\ref{fig:rsp_DCEP_lc} shows $I$-band light curves and radial
velocity curves for the $\Delta$\Teff\,=\,300\,\Kelvin\ models.  Due
to the shift in the location of blue edge models, models from convective sets
B and D have different \Teff\ and consequently different pulsation
periods.
Figure~\ref{fig:rsp_DCEP_four} compares Fourier parameters of the
Cepheid models with those derived from LMC Cepheid light curves 
\citep[left panel,][]{rsp_o4_cep_mc,rsp_shallow} and 
Galactic F-mode Cepheid radial velocity curves \citep[right panel,][]{rsp_storm}.  
The $I$-band light curves are more sensitive to the convective
parameters than the radial velocity curves for both
$\Delta$\Teff\ offsets.  As with the RR~Lyrae models,
the Fourier parameters for convective sets A and B are similar to each other, as are
those for sets C and D.  The radial velocity Fourier parameters
follow tighter progressions than the $I$-band light curves.

The Fourier phases in the left panels of Figure~\ref{fig:rsp_DCEP_four} are systematically larger than
the observationally-inferred values for $P\lesssim 10$\,d, 
with the difference being larger for the cooler $\Delta$\Teff\,=\,500\,\Kelvin\ models.  
For $P\gtrsim 10$\,d the model Fourier phases are systematically smaller.  
Large discrepancies for the radial velocity curves are absent
in the right panels of Figure~\ref{fig:rsp_DCEP_four}, except for
the amplitudes and $R_{21}$ ratio at the shortest periods. 
The projection (or $p$) factor is the ratio of the 
pulsation velocity to the radial velocity deduced from spectral 
line-profile observations, dependent on at least rotation and gravity darkening (see Section~\ref{s.rot}),
and plays a role in the amplitudes.
We use $p$\,=\,1.3, 
close to the average value of determinations
based on eclipsing binary Cepheid systems \citep{rsp_Pilecki2018} and
interferometric methods \citep[e.g.,][]{rsp_breitfelder}.

This brief survey is not exhaustive
as only a few masses and two model sequences are explored.
The $M-L$ relation is also important for Cepheids
as evolutionary tracks depend on overshooting, rotation, and metallicity.

\subsubsection{Type II Cepheids}\label{sssec:rsp_T2CEP}

Type~II Cepheids are more similar to RR~Lyrae stars than to classical
Cepheids due to their lower masses ($M$\,$\simeq$\,0.5\,\Msun), Pop~II chemical composition, and 
evolutionary history.  Their masses are similar to RR~Lyrae
but they cross the instability strip at larger luminosities and pulsate
with longer periods ($P$\,$>$\,1\,d).  Type II Cepheids are
F-mode pulsators except for a few double-mode stars pulsating
simultaneously in the F and 1O modes \citep{rsp_s+18,rsp_ogle_gc} and
two recently discovered 1O-mode pulsators \citep{rsp_1oblher}.
BL Her variables are a subclass of Type II Cepheids with 
$P$\,$\lesssim$\,4\,d.

Nonlinear radiative models of Type~II Cepheids revealed a variety of
complex dynamics including period-doubled and deterministic chaos
pulsations \citep[e.g.,][]{rsp_bk87,rsp_kb88,rsp_bm92}.  With convective pulsation models, 
modulated pulsations were also found \citep{rsp_sm12}.
\cite{rsp_bm92} discovered period-doubled pulsations in their survey
of BL~Her models and predicted that they should be observed in BL~Her
variables. The period-doubling is caused by a 3:2 resonance of the 
F and 1O modes and nonlinear phase synchronization \citep{rsp_mb90};
pulsations repeat only after two cycles of the F-mode.

\begin{figure}[ht!]
\begin{center}
\includegraphics[width=\columnwidth]{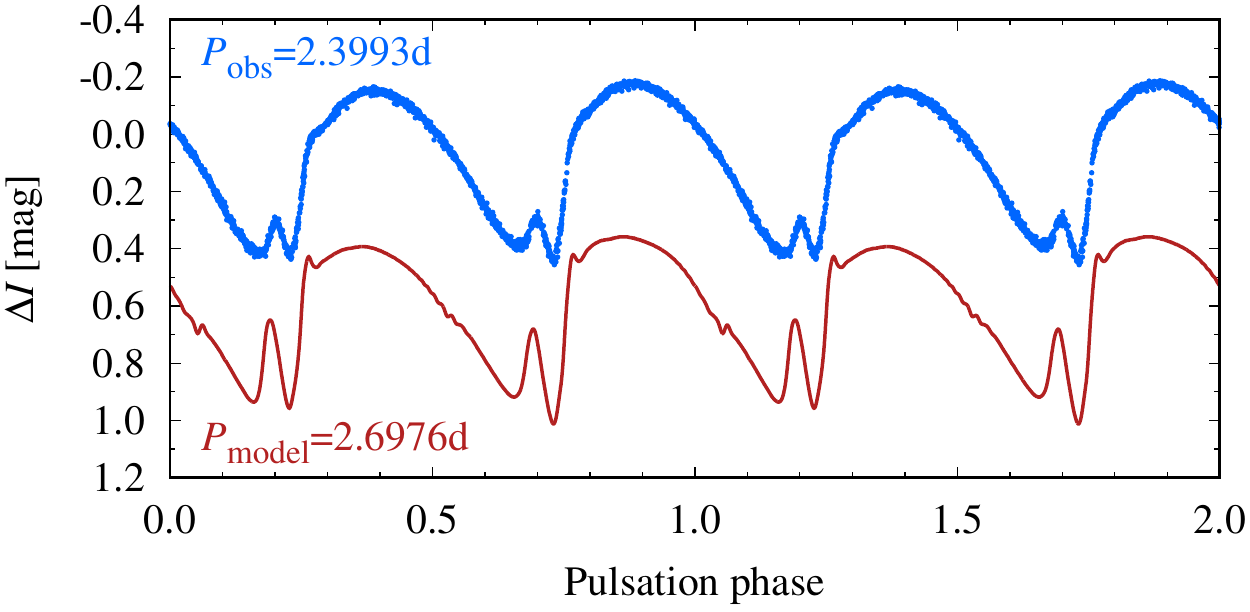}
\end{center}
\vspace{-0.15in}
\caption{
Comparison of the observed $I$-band light curve of OGLE-BLG-T2CEP-279 (blue dots) 
with the pulsation model (red curve). The period doubling effect is recognizable 
upon comparing consecutive maxima and minima.  The model is vertically offset by 0.6.
\label{fig:rsp_279}}
\end{figure}

\cite{rsp_PDogle} report a period-doubled BL~Her star, OGLE-BLG-T2CEP-279.
We adopt 
$M$\,=\,0.6\,\Msun, $L$\,=\,184\,\Lsun, \Teff\,=\,6050\,\Kelvin, $X$\,=\,0.76, $Z$=0.01,
and convective parameter set A. These are nearly the same physical
parameters \cite{rsp_s+12} chose for a T2CEP-279 model survey.
The observed and model light curves are shown in Figure~\ref{fig:rsp_279}.
The model amplitudes of the bump at minimum light of the period doubled cycle are 
larger and the shape of the light maximum is more pronounced
relative to the observed features.
The model period, $P_{\rm model}$\,=\,2.6976\,d, is longer than
the observed period, $P_{\rm obs}$\,=\,2.3993\,d. 
Still, the light curves are qualitatively similar,  
devoid of fine-tuning, and demonstrate that \RSP \ can
be used to model specific stars.

\begin{figure}[ht!]
\begin{center}
\includegraphics[width=\columnwidth]{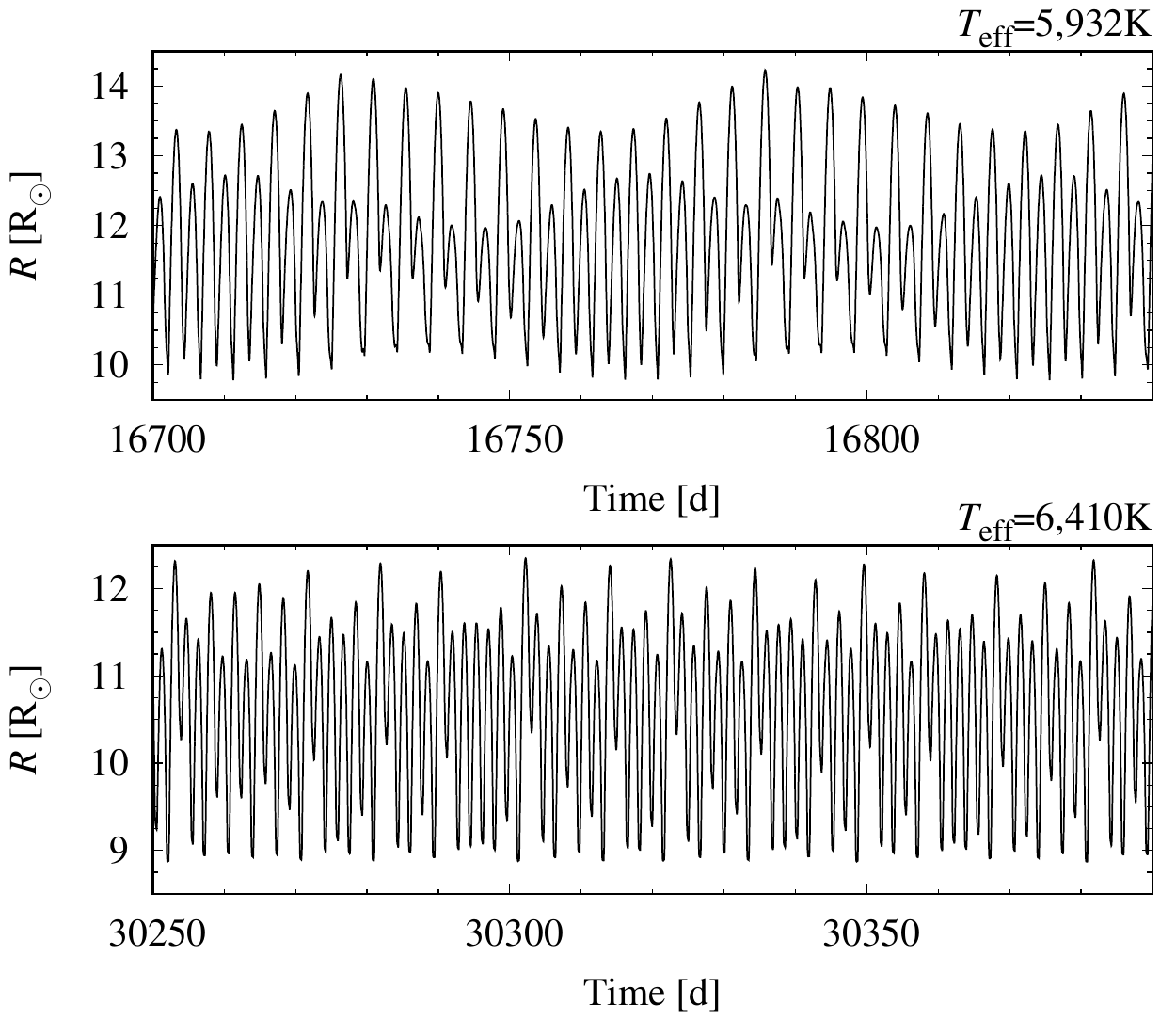}
\end{center}
\vspace{-0.15in}
\caption{
Radius variation in two models differing only in their \Teff\
that show periodic modulation (upper panel) and deterministic chaos (lower panel).
\label{fig:rsp_mcrad}}
\end{figure}

\begin{figure}[ht!]
\begin{center}
\includegraphics[width=\columnwidth]{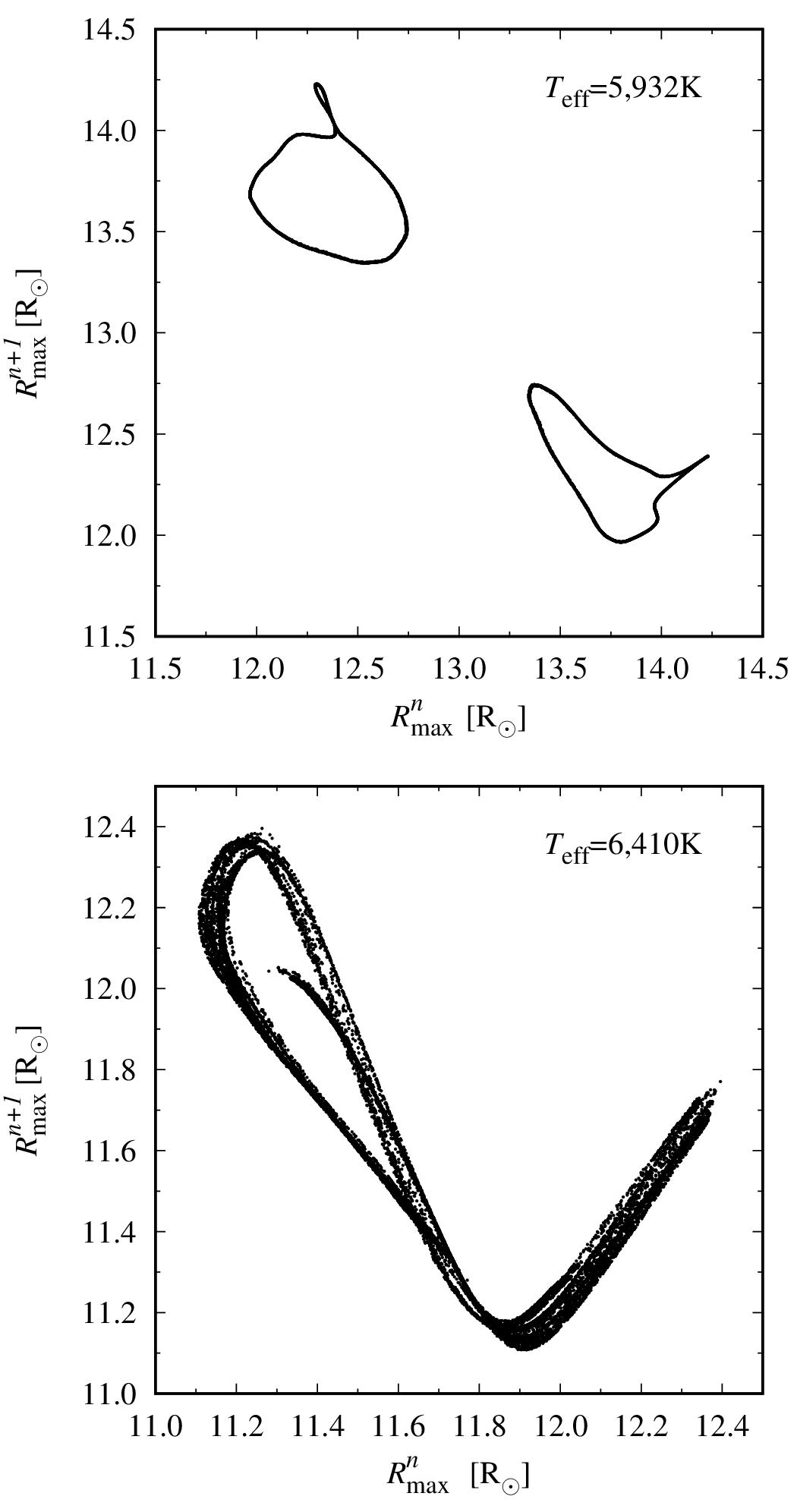}
\end{center}
\vspace{-0.15in}
\caption{Maximum radius return maps for the models in Figure~\ref{fig:rsp_mcrad}.\label{fig:rsp_retmap}}
\end{figure}

Figure~\ref{fig:rsp_mcrad} shows the radius variation for two 
models with $M$\,=\,0.55\,\Msun, $L$\,=\,136\,\Lsun, $X$\,=\,0.76, $Z$\,=\,0.0001
and convective set A but with a reduced eddy-viscosity, $\alpha_{\rm m}$\,=\,0.05 (yielding unrealistic light curves).
The two models differ only in \Teff\,=\,5,932\,\Kelvin \ (upper panel) 
and \Teff\,=\,6410\,\Kelvin \ (lower panel). 
Figure~\ref{fig:rsp_retmap} shows the return map of maximum radii 
for these two models, plotting the maximum radii for 
each pulsation cycle $R_{\rm max}^{n}$ versus the preceding $R_{\rm max}^{n-1}$.

For the cooler \Teff\,=\,5,932\,\Kelvin\ model,
the upper panel of Figure~\ref{fig:rsp_mcrad} shows a cyclic modulation in the envelope of
the period-doubled pulsation.  The modulation
period of $\simeq$\,57\,d is longer than the pulsation period of
$\simeq$\,2.3\,d.  The return map in the upper panel of Figure~\ref{fig:rsp_retmap}
is constructed from $\simeq$\,8,000 pulsation cycles and shows two loops, 
corresponding to alternating smaller and larger maximum radii.  
Since the modulation period is not commensurate with the pulsation
period, the return maps develop a locus of points that form the closed
lobes.  Light curve modulation is common in RR~Lyrae stars (Blazhko
effect) and periodic pulsation modulation was recently discovered in
BL~Her variables \citep{rsp_s+18}.

For the hotter \Teff\,=\,6,410\,\Kelvin\ model,
the radius variation appears irregular in the lower panel of Figure~\ref{fig:rsp_mcrad}.
The return map in the lower panel of Figure~\ref{fig:rsp_retmap} reveals a strange attractor,
an example of deterministic chaos in nonlinear models.
Tracing time series from models is simple, but
tracing chaotic dynamics in observations is 
difficult \citep[e.g.,][]{rsp_dfcyg}. Chaotic dynamics is
reported in a few type-II Cepheids with longer periods, 
in the RV~Tau variable star range, and in semi-regular variable stars
\citep[e.g.,][]{rsp_bksm96,rsp_kbsm98,rsp_bkc04,rsp_dfcyg}.  While the
\Teff\,=\,6,410\,\Kelvin\  model has a shorter period, in the BL~Her range,
such models may provide insight into chaotic
dynamics in pulsating stars \citep[see][]{rsp_plachy13,rsp_sm14}.

\subsubsection{Binary Evolution Pulsators}
\label{sssec:rsp_BEP}

Very-low-mass stars do not enter the classical instability strip within
a Hubble time.  However, mass loss from the more massive component 
in a close interacting binary can lead to a low-mass star that then 
evolves through the instability strip.
The Binary Evolution Pulsator (BEP, OGLE-BLG-RRLYR-02792) is the
prototype of this new class of pulsators \citep{pietrzynski_bep2012,smolec_bep2013}.
The BEP's variability is similar to an F-mode RR~Lyrae pulsator but with a dynamical mass of
$\simeq$\,0.26\,\Msun.

\begin{figure}[ht!]
\begin{center}
\includegraphics[width=\columnwidth]{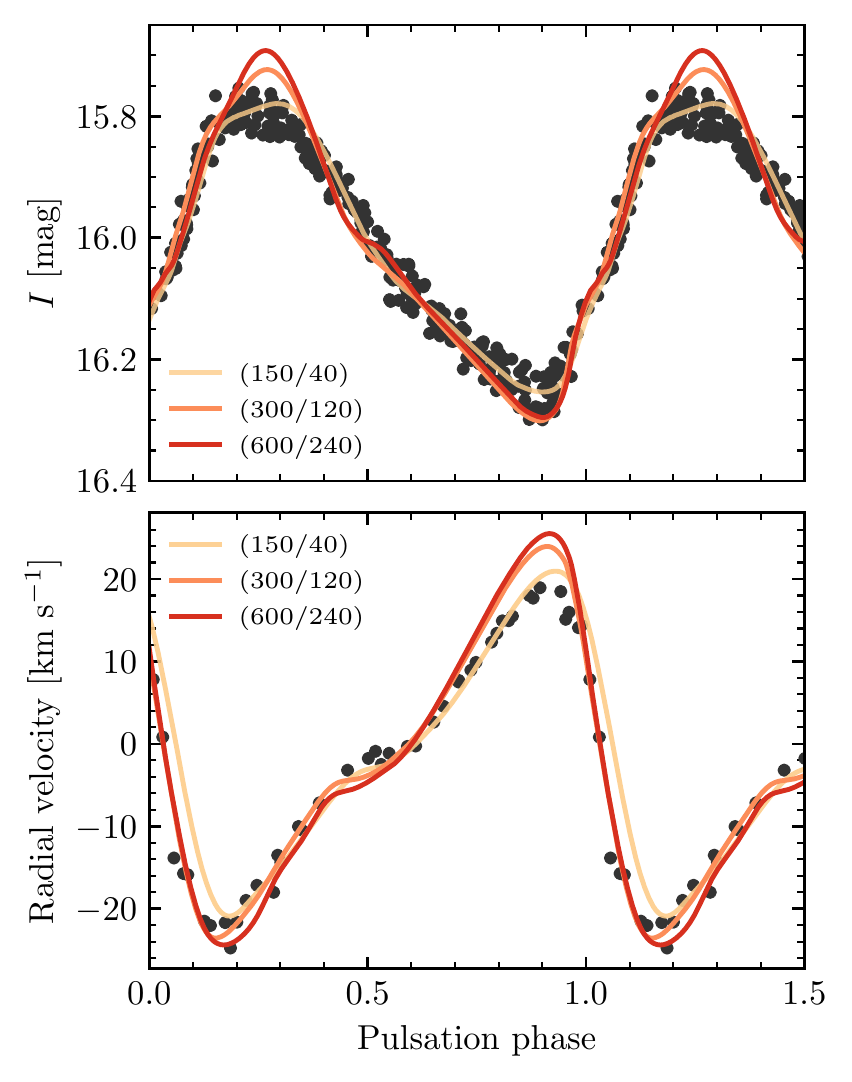}
\end{center}
\vspace{-0.20in}
\caption{$I$-band light (upper panel) and radial velocity (lower panel) curves of the BEP OGLE-BLG-RRLYR-02792
  as a function of pulsation phase. Observations are shown as circles \citep{pietrzynski_bep2012} 
  with the radial velocity multiplied by a projection factor of $p$\,=\,1.2. 
  Model curves are shown at three resolutions labeled by ($N/N_{\rm outer}$).
\label{fig:rsp_BEPtest1}}
\end{figure}

Following \cite{smolec_bep2013}, we adopt 
$M$\,=\,0.26\,\Msun, $L$\,=\,33\,\Lsun, \Teff\,=\,6,910\,\Kelvin, 
$X$\,=\,0.7, $Z$=0.01, and convective set A.
Figure~\ref{fig:rsp_BEPtest1} shows $I$-band light curves and radial
velocity curves for the BEP models.
Results for a coarse grid ($N/N_{\rm outer}$=150/40, gold curves)
qualitatively match these observations and have a period of 0.6373\,d 
close to the observed 0.6275\,d period.  We also show 
results for a medium grid (300/120, orange curves) and a fine grid (300/120, red curves).
The differences are most pronounced around maximum and minimum light.
The shape of the light and radial velocity curves approach convergence
only on grids with $\gtrsim$\, 600 cells.
The amplitude of the model curves are sensitive to
convective parameters and can be fine-tuned for a better match
with observations.

\subsubsection{Blue Large Amplitude Pulsators}
\label{sssec:rsp_BLAP}

The origin of BLAPs, introduced in Section~\ref{s.intro}, is unknown.
They have been modeled as $\simeq$\,0.3\,\Msun \ shell H-burning stars that are 
progenitors of  low mass WDs and $\simeq$\,1.0\,\Msun \ stars undergoing core He-burning
\citep{pietrukowicz_2017_aa, romero_blap2018, wu_blap2018, byrne_2018_aa}.
Though mass loss in a close interacting binary must be invoked for both
hypotheses, none of the BLAPs are known to be in a binary. 
Figure \ref{f.hrpulse} shows that BLAPs are located near sdBVs in the HR
diagram. The latter have non-canonical abundance profiles that
are strongly affected by radiative levitation (see Section~\ref{s.radlev}). Following the linear
study of \citet{romero_blap2018}, we adopt $Z$\,=\,0.05, to account for the increased envelope metallicity caused by radiative levitation.

\begin{figure}[ht!]
\begin{center}
\includegraphics[width=\columnwidth]{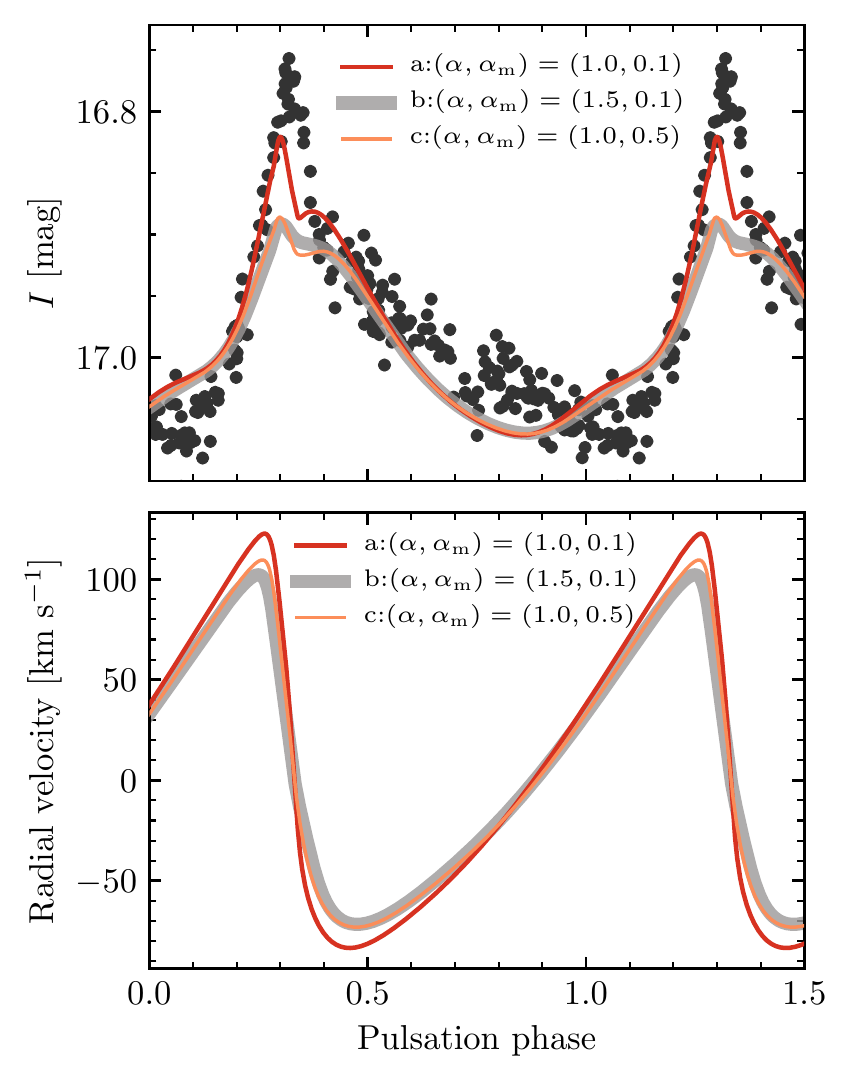}
\end{center}
\vspace{-0.15in}
\caption{$I$-band light (upper panel) and radial velocity (lower panel) curves 
  for OGLE-BLAP-011. Observations are shown as circles \citep{pietrukowicz_2017_aa}.
  \RSP \ models for three convective sets are shown under the $\simeq$\,1.0\,\Msun \ 
  core He-burning hypothesis.
  Pulsation periods of the models range between
  34.27 and 34.34~min and the observed period is 34.87~min.
\label{fig:rsp_BLAPtest1}}
\end{figure}

We explored nonlinear models with 
$M$\,=\,1.0\,\Msun, \Teff\,=30,000~\Kelvin, $L$\,=\,430\,\Lsun,
and three convection parameter sets.  
\revision{These envelope models
used $N$\,=\,280, $N_{\mathrm{outer}}$\,=\,140, $T_{\mathrm{anchor}}$\,=\,2$\times$10$^{5}$~K, and $T_{\mathrm{inner}}$\,=\,6$\times$10$^{6}$~K.}
The LNA analyses show the F and 1O modes are unstable. 
Figure~\ref{fig:rsp_BLAPtest1} compares the OGLE-BLAP-011
and 1O-mode ($P\simeq35$ min)  model $I$-band light curves.
The qualitative agreement is reasonable.
Figure~\ref{fig:rsp_BLAPtest1} shows the model radial velocity curves
have amplitudes of $\simeq$\,200~km~s$^{-1}$, a pronounced temporal
asymmetry, and light maxima that are narrower than light minima.

These BLAP models lie far off the classical instability strip so that
the initial model builder (see Section~\ref{s.rsp.action.grid}), 
which is optimized for classical pulsators, failed to relax 
the initial models  to complete hydrostatic equilibrium.  
An option is to switch off the relaxation process and 
commence the time integration with a near-hydrostatic-equilibrium initial model.
The price to pay is the LNA growth rates are only indicative, not accurate. 

\begin{figure}[ht!]
\begin{center}
\includegraphics[width=\columnwidth]{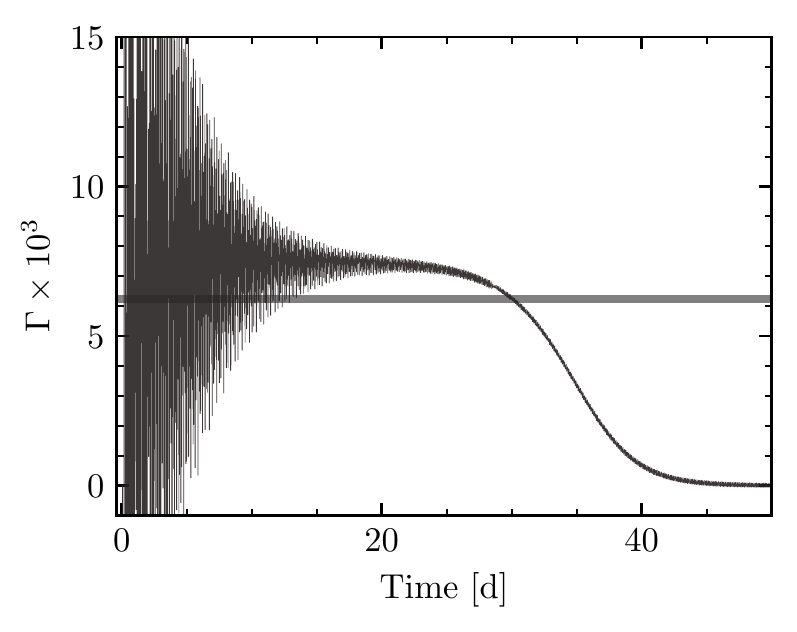}
\end{center}
\vspace{-0.15in}
\caption{Evolution of the \RSP\ growth rate $\Gamma$ starting
from the unrelaxed initial model for convection set b in
Figure~\ref{fig:rsp_BLAPtest1}.
The horizontal line
shows the LNA growth rate computed from the unrelaxed initial model.
\label{fig:rsp_BLAPgrowth}}
\end{figure}

Figure~\ref{fig:rsp_BLAPgrowth} shows the evolution of $\Gamma$ for the $(\alpha$,$\alpha_{\rm m})$=(1.5,0.1) 
model in Figure~\ref{fig:rsp_BLAPtest1}.
Before $\simeq$\,10\,d, $\Gamma$ fluctuates around a mean $\simeq$\,7.5$\times$10$^{-3}$.
The fluctuations diminish with time but $\Gamma$ remains above the 
LNA value of 6.25$\times$10$^{-3}$ up to $\simeq$\,30~d.
Between 30-50~d $\Gamma$ diminishes 
and approaches zero, the sign that the nonlinear 
pulsation saturates at its terminal amplitude, 
yielding the results of Figure~\ref{fig:rsp_BLAPtest1}.   
In contrast to $\Gamma$, the period of the 
nonlinear pulsations remain close to the LNA period.
In cases where the initial model cannot be relaxed, the initial $\Gamma$ should not be expected to match the LNA analysis.

\subsubsection{High Amplitude Delta Scuti}\label{sssec:rsp_DSCT}

High-amplitude $\delta$ Scuti (HADS) pulsators are defined to have
$V$-band light curve amplitudes greater than 0.1 mag.
HADS lie close to the main-sequence (MS; see Figure~\ref{f.hrpulse}),
where growth rates are usually much smaller than those for 
RR~Lyrae stars or classical Cepheids.
This implies long time integrations are needed to
drive nonlinear pulsations to saturation.

We consider a stellar model with 
$M$\,=\,2\,\Msun, $L$\,=\,30\,\Lsun, \Teff\,=\,6,900\,\Kelvin,
$X$\,=\,0.7, $Z$=0.01, and convective set A.
This represents a star evolving
towards the Hertzsprung gap.  
The LNA analysis of the initial model reveals that the F and 1O modes 
are linearly unstable, with growth rates of 1$\times$10$^{-6}$ and 6$\times$10$^{-5}$, respectively.

\begin{figure}[ht!]
\begin{center}
\includegraphics[width=\columnwidth]{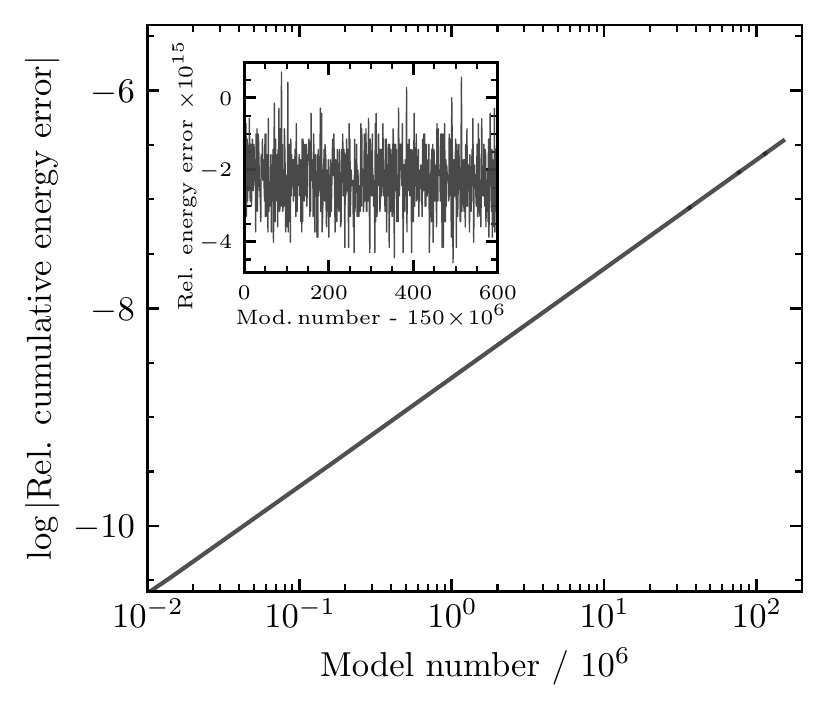}
\end{center}
\vspace{-0.15in}
\caption{Evolution of the relative cumulative energy error from
  model 10,000 ($\simeq$\,15 cycles) to $\simeq$\,150 million (at $\simeq$\,250,000 cycles).
  The inset shows the per-step relative energy error over a cycle (600 steps).
\label{fig:rsp_ErrorEvolution}}
\end{figure}

Section~\ref{s.energy} emphasizes the importance of numerical energy
conservation.  Figure~\ref{fig:rsp_ErrorEvolution} shows the evolution
of the relative cumulative error in the energy for a 250,000 cycle integration
($\simeq$\,150 million time steps, at 600 steps per cycle).  The relative cumulative error
grows from $\simeq$\,3$\times$10$^{-11}$ after about 15 cycles to
$\simeq$\,3$\times$10$^{-7}$ after 250,000 cycles.  The inset figure
shows that the per-step relative error in the energy scatters
around $-$2.5$\times$10$^{-15}$ but is
systematically different than zero.

\begin{figure}[ht!]
\begin{center}
\includegraphics[width=\columnwidth]{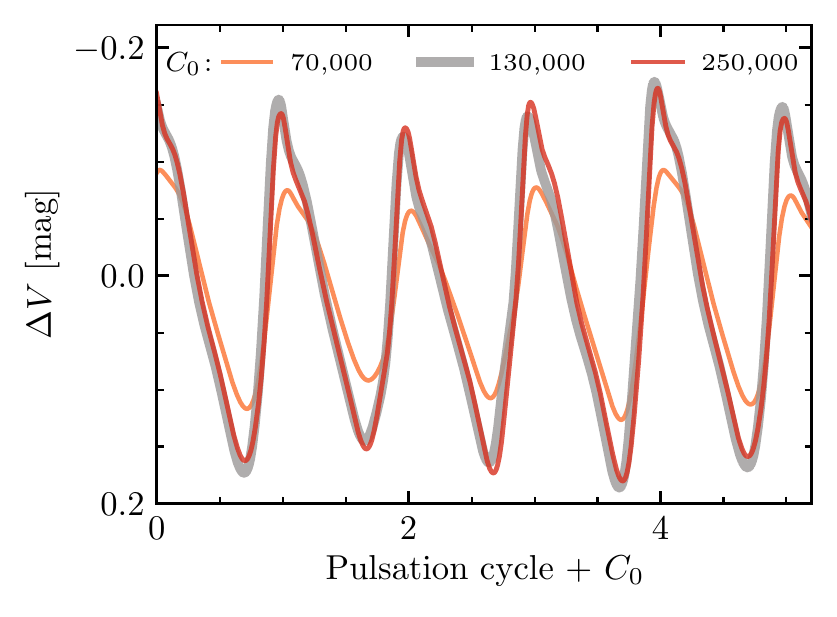}
\end{center}
\vspace{-0.15in}
\caption{$V$-band light curves of the $\delta$~Sct model at 
  three stages during a very-long ($\simeq$\,150 million time step) integration.
  For each stage, we plot the light curve for 5 pulsation cycles.
  The legend gives the cycle numbers when the snapshots were taken.
\label{fig:rsp_DSctVlight}}
\end{figure}

Figure~\ref{fig:rsp_DSctVlight} shows that after 70,000 cycles the 
asymmetric $V$-band light curves have an amplitude of $\simeq$\,0.2\,mag.
The dominant pulsation period is 0.127~d, close to the LNA period of 0.1269~d for 
the 1O-mode. The amplitude, though, varies cyclically over about four
pulsation cycles, reflecting the presence of the F-mode.
Continuing the integration to 130,000 cycles leads to a more saturated 
light curve with an amplitude of $\simeq$\,0.3~mag and an unchanged dominant pulsation
period. Extending the integration to 250,000 cycles does not lead to any 
significant changes.

\begin{figure}
\begin{center}
\includegraphics[width=\columnwidth]{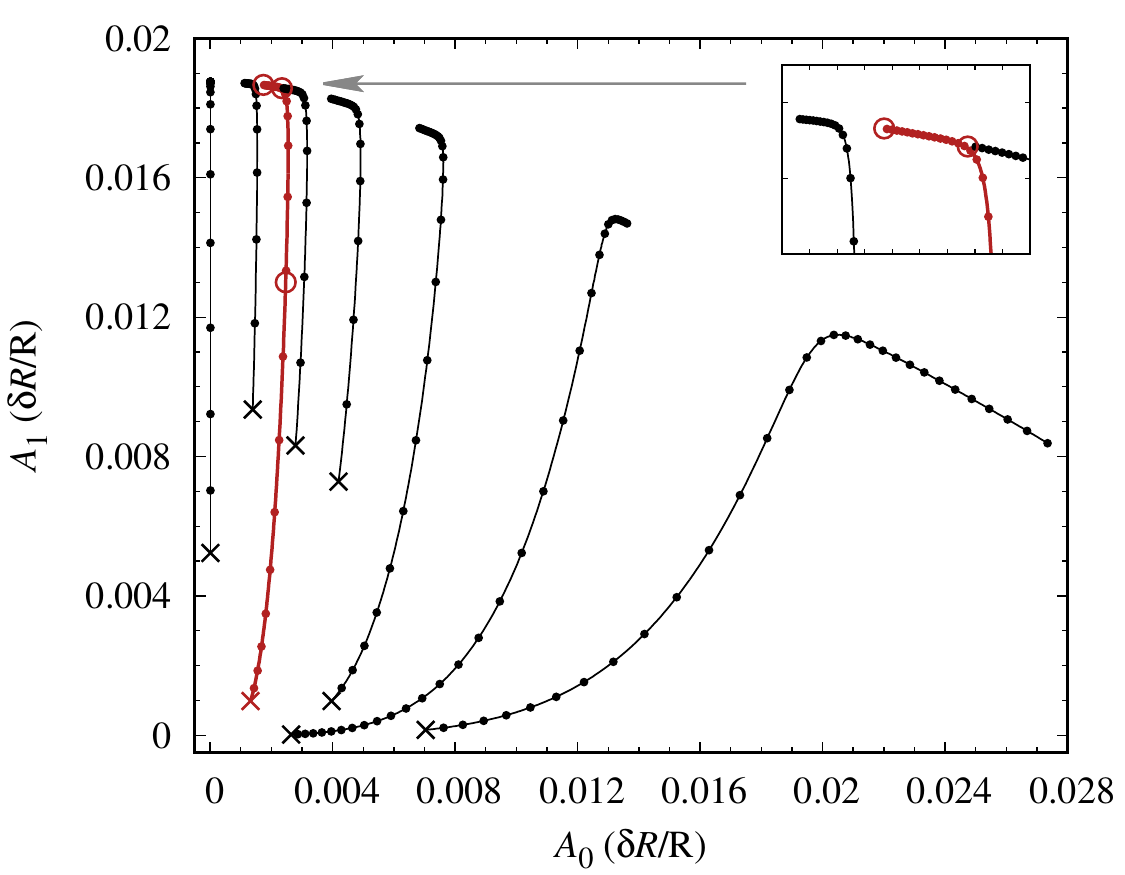}
\end{center}
\vspace{-0.15in}
\caption{
Evolution of fractional radius amplitude, $\delta R/R$, for the
1O mode, $A_1$ and F mode, $A_0$. Eight trajectories begin at locations
marked with a cross. Circles mark 1,000\,d intervals along trajectories.
The red curve corresponds to the model in Figure~\ref{fig:rsp_DSctVlight} 
with the three open circles marking the amplitudes after
70,000, 130,000, and 250,000 cycles. The inset on the upper right zooms
into the slowing amplitude evolution of the red curve late in the simulation.
\label{fig:rsp_trajectories}}
\end{figure}

The robustness in the light curves between 130,000 and 250,000 cycles 
might suggest that the long-term behavior is that of a double-mode HADS model.  
However, additional integrations with different initial perturbations,
chosen to more adequately sample the neighboring amplitude phase space,
are necessary to assess the possible mode selection.
Figure~\ref{fig:rsp_trajectories} shows the evolution of these integrations
in the amplitude-amplitude diagram using the analytical signal method \citep[e.g.,][]{rsp_kbsc02}. The amplitude behavior of the model integration
shown in Figure~\ref{fig:rsp_DSctVlight} is traced out by the red line.
After initially rapid evolution, all the amplitude trajectories
develop an arc along which evolution slows markedly. 
The model in Figure~\ref{fig:rsp_DSctVlight} and neighboring trajectories 
bend to the left toward smaller F-mode amplitudes.
Their likely final state is that of a single-periodic 1O pulsation. 
In contrast, the two right-most trajectories bend to the right
towards larger, more dominant, F-mode amplitudes.
Despite the limited sampling of phase space in Figure~\ref{fig:rsp_trajectories}, 
we cautiously conclude that the most likely 
long-term outcome of this $\delta$~Sct model is a 
single-periodic 1O-mode or F-mode pulsator depending on the initial 
conditions (see Section \ref{s.rsp.action.whichP}).

\section{Energy Conservation}\label{s.energy}

For the following discussion we define the total energy $E$ of a model to
be the sum of the internal, potential, and kinetic energies, ignoring
rotational and turbulent energy which are currently not included in the
energy accounting.  To support improved numerical energy
conservation\footnote{Note that we are discussing numerical issues in
  the code rather than questions of the physical completeness and
  validity of the equations.  We will often use the term numerical
  energy conservation to make this distinction explicit.}, $\MESAstar$
provides an option to use what we call the \code{dedt}-form of the energy equation:
\begin{equation}
  \label{eq:dedt_eqn}
  \DDt{e} =  \epsilon - \ddm{}\left(L + p \area\,u\right)
  - \DDt{}\left(\frac{1}{2}u^2 - \frac{Gm}{r}  \right)~.
\end{equation}
This form was introduced in \mesafour\ and provides an alternative  to the \code{dLdm}-form of the energy equation (Equation~(11) in \mesaone).
When the time
derivative terms are combined, the result is more easily recognizable as an equation
for the time evolution of local specific total energy (left hand side)
due to local source terms\footnote{This
includes energy from nuclear reactions ($\epsnuc$) and thermal
neutrino losses ($-\epsnu$), as well as terms associated with other
processes such as accretion (see Section~\ref{s.mdot}).  Importantly,
$\epsilon$ does not include $\epsgrav$, the specific rate of change of gravothermal energy, as that source term is not present
when using a total form of the energy equation (see \mesafour, Section 8).} 
($\epsilon$) and local fluxes between cells (the $\partial/\partial m$ term):
\begin{equation}
  \label{eq:total_conservation}
  \DDt{}\left(e + \frac{1}{2}u^2 - \frac{Gm}{r} \right)
  =   
\epsilon - \ddm{}\left(L + p \area\,u\right) 
~.
\end{equation}
The error in numerical energy conservation $ E_{\rm error}$ is
the extent to which the time- and mass-integrated Equation~(\ref{eq:total_conservation}) is not satisfied 
when solved in a discretized, finite-mass form.
This section discusses recent efforts to improve numerical energy conservation in $\MESAstar$.

Recall (from \mesaone, Section 6.3) the generalized Newton-Raphson scheme 
used by $\MESAstar$  to solve the stellar equations
\begin{equation}
0 = \vec{F}(\vec{y}) = \vec{F}(\vec{y}_i + \delta \vec{y}_i) = \vec{F}(\vec{y}_i) + \left[ \frac{d\vec{F}}{d\vec{y}} \right]_i \delta \vec{y}_i + O(\delta \vec{y}_i^{\,2})
\,,
\end{equation}
where $\vec{y}_i$ is the trial solution for the $i$-th iteration, $\vec{F}(\vec{y}_i)$ is the
residual, $\delta \vec{y}_i$ is the correction, and
$[d\vec{F}/d\vec{y}]_i$ is the Jacobian matrix.  The residual is the
left over difference between the left- and the right-hand sides of the
equation we are trying to solve, while the correction is the change in
the primary variable that is calculated by Newton's rule.
The solver generates a series of trial
solutions until it produces one that is acceptable according to given
convergence criteria.  In $\MESAstar$ the trial solution is not
accepted until the magnitudes of all corrections and residuals become
smaller than specified tolerances\footnote{The GARching STellar
  Evolution Code (GARSTEC) is the only other stellar evolution code we
  are aware of that considers residuals as well as corrections in
  deciding when to accept a trial solution~\citep{GARSTEC:2008}.
  Several other codes consider corrections but not, as far as we can  tell, 
  residuals~\citep{2008Ap&SS.316...75R,2008Ap&SS.316...13C,2008Ap&SS.316...31D,2008Ap&SS.316...83S,MONSTAR:C}.}.
If no acceptable trial solution has been found in the allowed maximum
number of iterations, the solver rejects the attempt and forces a
retry with a smaller time step.

If the numerical accuracy of the partial derivatives 
forming the Jacobian matrix is not excellent, the
reduction in magnitude of the residuals can stall after a few
iterations.  For this reason, $\MESAstar$ has provided a means to use
a tight tolerance on residuals for an initial sequence of iterations
and then switch to a much relaxed tolerance if no acceptable solution
has been found.  The benefit of this is that residuals will be driven
down when possible, but if the residuals stall at a level above the
initial tolerance, the system will still be able to take a step as
long as the corrections can be adequately reduced. The cost of
relaxing the tolerance for residuals of the total energy equation is
the creation of numerical energy conservation errors.  To
obtain good numerical energy conservation we must be able to drive down
residuals to low levels, and to do that we must have numerically
accurate partial derivatives.  This has motivated a
major effort to improve partials, and we can now require the solver to
keep iterating until it reduces the residuals to a low level that
gives good numerical energy conservation.  The most significant
changes to improve the numerical accuracy of partials were in the \eos\ 
module and are discussed in Appendix~\ref{s.eos}.

\subsection{Gold Tolerances}\label{s.goldtols}

\begin{figure*}[!htb]
\centering
\includegraphics[width=\textwidth]{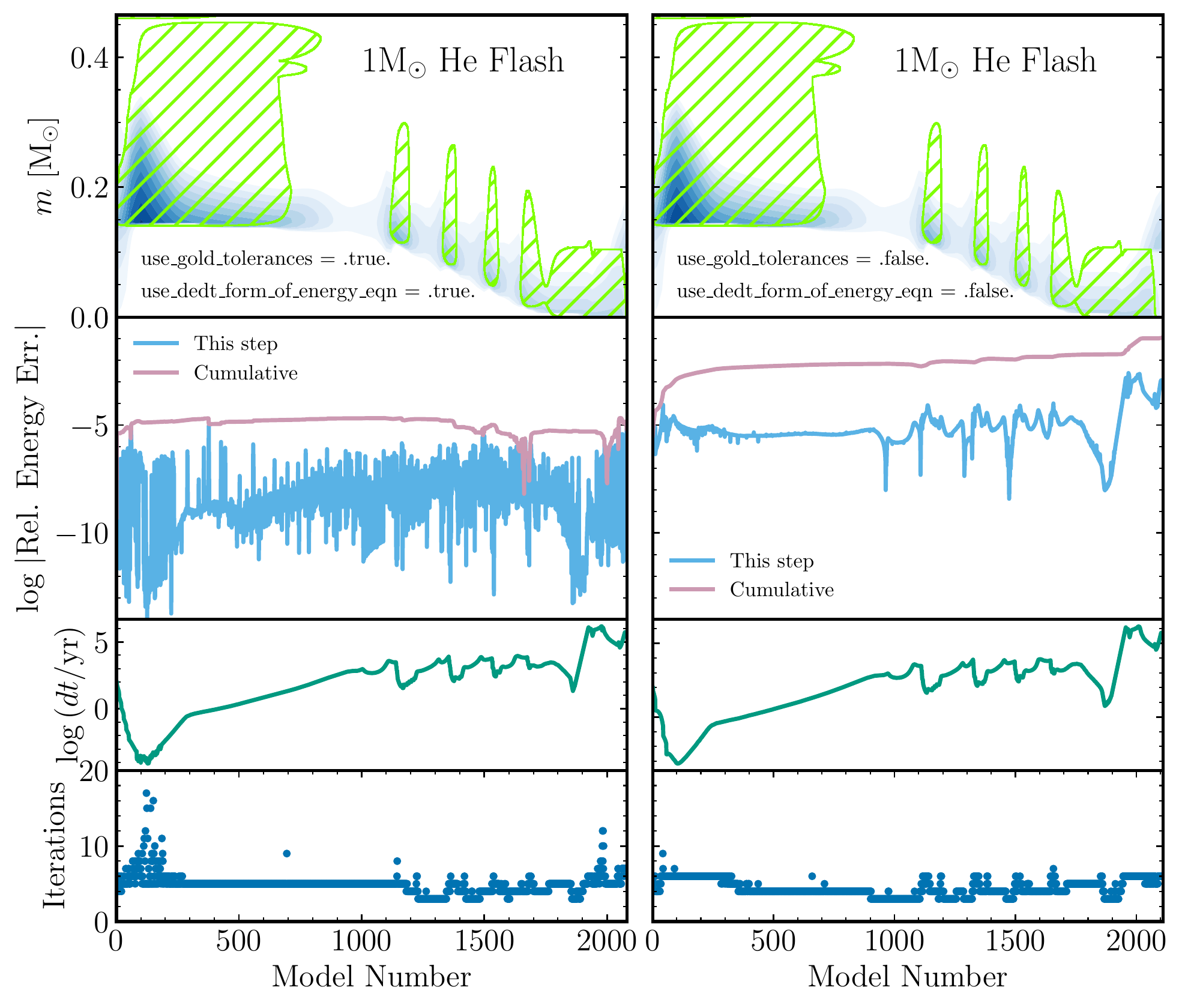}
\caption{
Evolution of a 1\,\Msun\ model during the He flash and core He burning using gold
tolerances and the \code{dedt}-form of the energy equation (left panel), and
without gold tolerances and using the \code{dLdm}-form of the energy
equation (right panel).  The first panel shows a Kippenhahn plot, where green
hatched regions are convective and blue shading shows region of
nuclear energy generation.  The second panel shows the
absolute error in energy conservation divided by the total stellar
energy, both per step and cumulative. 
The third panel shows the
time step, while in the fourth panel we report the
number of iterations required by the Newton-Raphson solver.
Gold tolerances keep the cumulative errors small during the He flash
at the price of a larger number of iterations during the first $\approx$200 time steps.
}
\label{f.kipp_err_heflash}
\end{figure*}

To improve energy conservation,
a new standard ``gold tolerances'' is now the
default in $\MESAstar$.
%
%
This uses tight 
tolerances that apply
even after an arbitrarily large number of iterations.
%
%
As a result, steps with poor residuals will be
rejected, thereby ensuring that if a run succeeds with gold tolerances
enabled while using the total energy equation given above, it will
have good energy conservation.  
To show example improvements from this
new strategy, in Table~\ref{tab:energy} we report the results of
calculations of a $1\,\Msun$ model during the main sequence and 
the He flash with gold tolerances and compare to the old
approach. The $1\,\Msun$ models are evolved from the ZAMS
until the core H mass fraction reaches a value of
$10^{-6}$. The He flash models start at off-center He
ignition and terminate when the core He mass fraction drops to 
$10^{-3}$.  The cumulative energy error is the sum of the energy
conservation errors at each of the steps. The relative cumulative
energy error is the cumulative energy error divided by the final total
energy.  This is now much less than 1\% during the He flash, a
notoriously difficult evolutionary phase from the numerical
perspective.  The evolution of the errors in energy conservation, as
well as the number of iterations required by the solver and the
adopted time step, are shown in Figure~\ref{f.kipp_err_heflash} for the
He flash runs. This figure clearly show the superiority of the model
adopting gold tolerances and the \code{dedt}-form of the equation in
terms of energy conservation.

Not all cases in the $\MESAstar$ test suite are currently able to use gold tolerances.  
This is primarily because of remaining problems with the numerical
accuracy of certain partials, especially in the  PC EOS  \citep{potekhin_2010_aa}  that is used by
WD models (see Appendix~\ref{s.eos}) and on the boundary
between the OPAL EOS \citep{rogers_2002_aa} and the SCVH EOS \citep{saumon_1995_aa}.
There are also problems with the numerical accuracy of some partials
associated with nuclear reactions at the high temperatures encountered during late
stages of evolution, such as Si burning.  To address these
situations, there are controls to allow gold tolerances to be turned
off automatically for steps that either require use of the PC EOS or
that have extremely high temperatures.  To provide feedback, the value of the relative
cumulative energy error is monitored and a warning message is written if it exceeds a
specified value (2\% is the default setting).  

\begin{deluxetable}{lll}[!ht]
  \tablecolumns{3}
 \tablewidth{1.0\apjcolwidth}
  \tablecaption{Relative cumulative energy error for $1\,\Msun$ runs. \label{tab:energy}}
  \tablehead{ \colhead{ \hfil} & \colhead{Main Sequence \hfil} & \colhead{ He Flash \hfil} }
    
  \startdata 
     \code{dedt} + Gold Tolerances     &    0.3\%   &  0.0006\%  \\
     \code{dLdm}                       &    14\%   & 12\%  
  \enddata
\end{deluxetable}

\subsection{Definition of $\epsnuc$ Source Term}\label{s.epsnuc}

Previous \MESA\ papers have not given a precise definition of $\epsnuc$.
Motivated by numerical convenience, \MESA\ formerly exploited the fact
that the $\epsilon$ source term contained the sum of $\epsgrav$ and
$\epsnuc$ and included the response of the internal energy 
to composition changes due to nuclear reactions in $\epsnuc$ instead of in $\epsgrav$.
However, since \dedt\ does not include $\epsgrav$,
this is no longer an appropriate choice.

In the current approach, $\epsnuc$ is
evaluated in the \net\ module as a sum over reactions.  Schematically,
\begin{equation}
  \epsnuc = \NA \sum_{\mathrm{reactions}} \left(Q_i - Q_{\nu, i}\right) R_i~,
  \label{eq:epsnuc-schematic}
\end{equation}
where $R_i$ is the molar rate of reaction $i$, $Q_{i}$ is the
change in rest mass energy between the products and reactants, and
$Q_{\nu, i}$ is the per-reaction average energy of the neutrino (if
present).  We note the equivalence
\begin{equation}
  \sum_{\mathrm{reactions}} Q_i R_i \equiv -\sum_{\mathrm{isotopes}} M_i c^2 \dot{Y_i} ~,
  \label{eq:epsnuc-hix}
\end{equation}
where $M_i$ is the rest mass of isotope $i$ and $\dot{Y}_i$ is the
rate of change of the molar fraction.  The \texttt{approx} family of
\MESA\ nuclear networks exploit this equivalence and do not strictly
follow Equation~\eqref{eq:epsnuc-schematic}.  The right hand side of
Equation~\eqref{eq:epsnuc-hix} is a common nuclear physics definition
of $\epsnuc$ \citep[e.g., Equation~11 in][]{Hix2006}.  The \MESA\
definition of \epsnuc\ differs in subtracting off the nuclear neutrino losses,
thus enforcing the assumption that they free-stream out of the star.

\subsection{Mass Changes}\label{s.mdot}

The methods described above perform well for numerical energy
conservation when the stellar mass is constant, but extensions are necessary
for cases where the mass changes.  For energy accounting, we must
specify the amount of energy we expect the new mass to introduce and
the amount we expect departing mass to remove. To that end we assume
that mass being added has the same specific energy as the surface of
the model at the start of the step.  For mass being removed, we assume
that it leaves with a specific energy between what it had at the start
of the step and the value at the surface at the start of the step.
The exact amount depends on the amount of energy that leaks out of the
material as it approaches the surface during the time step.  For low
rates of mass loss there will be adequate time for the material to
adjust so that it leaves with the initial surface value, but for high
rates there may not be enough time for adjustment, so that it leaves
with a specific energy closer to its starting value.  The details of
this are presented below.

In addition to providing accurate accounting for the total energy of the model, it is
important to ensure that the energy changes from mass loss or gain are
distributed properly within the model.  As a guide for
this we use the analytic calculations of~\citet{2004ApJ...600..390T}.
Our new procedure improves upon these by also 
calculating the distribution of energy in systems with long thermal times, 
allowing \MESA\ to handle the limit of rapid accretion. 
We confirm the numerical energy conservation of this method using the $>20$
cases in the $\MESAstar$ test suite that have mass changes and fully
support gold tolerances and \dedt\ of the energy equation.
Using the new scheme, each of these completes the test run
with a cumulative error in total energy $<2$\,\%.
In addition, test cases that depend on the internal distribution of accretion heating continue
to yield the expected results.

\subsubsection{Methodology}

Because \MESA\ works on a Lagrangian mesh, it handles accretion and
mass loss in a two-stage process (\mesathree, Section 7).  In stage I
the masses of certain cells are increased or decreased as needed to
give the desired end-of-step total mass, but no attempt is
made to ensure energy conservation at this stage.  In stage II the
model thus produced is evolved in time by an amount $\dt$.  This
separation of stages means that the time step is only taken for a
model of fixed mass.  However, because the energy of the model changes
in stage I, a correction must be added to stage II to make the overall
step consistent with energy conservation.  Thus, we introduce a new
source term $\epsilon_{\dot{M}}$ that accounts for the heating
associated with mass changes.

The change in the mass of cell $k$ during stage I results 
from the difference between the 
outward\footnote{For $k>1$, this flux is from cell $k$ to cell $k-1$; for $k=1$ this flux is out of the model.}  
mass flux 
$F_{m}$ through each cell face.
This flux obeys
\begin{align}
	dm_{\Mmid,k} - dm_{\Mstart,k} = \dt \left(F_{m,k+1} - F_{m,k}\right)
	\label{eq:mass_flux}
\end{align}
and
\begin{align}
	M_{\Mmid} - M_{\Mstart} = - \dt F_{m,1} = \dt \dot{M}.
\end{align}
During stage~I the temperature, density, and velocity of each cell
are held fixed but the composition is updated to track the flow of material between cells.
The subscript \Mstart\ is used for quantities at the start of stage~I.
The subscript \Mmid\ is used for quantities at end of stage I, which is the start of stage~II.
No subscript is used for quantities evaluated at the end of the time step (after stage~II).
There is no mass change during stage~II, so $dm_k = dm_{\Mmid,k}$.
In the following we write $dm_{k}$ rather than $dm_{\Mmid,k}$.

The change in energy for cell $k$ during stage I is
\begin{align}
\label{eq:mdot_cons}
 E_{\Mmid,k} - E_{\Mstart,k} =& dm_{k}  \mathcal{E}_{\Mmid,k} -  dm_{\Mstart,k}  \mathcal{E}_{\Mstart,k} \\
                             =& dm_{k} \left(\mathcal{E}_{\Mmid,k} - \mathcal{E}_{\Mstart,k}\right)\nonumber\\
                              &+(dm_{k} - dm_{\Mstart,k}) \mathcal{E}_{\Mstart,k} \nonumber,
\end{align}
where $\mathcal{E}_k$ is the total specific energy of cell $k$, given by the sum 
of specific potential, kinetic, and internal energies.
Neglecting changes in specific energy owing to changes in composition, the difference in $\mathcal{E}_k$ across stage I is
\begin{align}
\label{eq:dEdPhi}
  \mathcal{E}_{\Mmid,k} - \mathcal{E}_{\Mstart,k} =
  \left(\frac{G m_{k,\rm C}}{r_{k,\rm C}}\right)_{\Mstart} - \left(\frac{G m_{k,\rm C}}{r_{k,\rm C}}\right)_{\Mmid}~.
\end{align}
where $m_{k,\rm C}$ and $r_{k,\rm C}$ are the mass coordinate and radius at the center of mass of the cell (\mesafour).

We now introduce $\epsilon_{k,{\rm I}}$, the effective source term in cell $k$ during stage I.
This is defined by writing the change in energy 
in flux-conservative form as
\begin{align}
\label{eq:fluxI}
E_{\Mmid,k} - E_{\Mstart,k} &= \dt \left(dm_{k} \epsilon_{k,{\rm I}} + F_{e,k+1,{\rm I}} - F_{e,k,{\rm I}}\right),
\end{align}
where  $F_{e,k,{\rm I}}$ is the outward flux of energy across face $k$ owing to work and material passing through face $k$.
Inserting Equation~\eqref{eq:mdot_cons} and rearranging we obtain
\begin{align}
\label{eq:eps_mdot_0}
  \epsilon_{k,{\rm I}} \equiv& \frac{\mathcal{E}_{\Mmid,k} - \mathcal{E}_{\Mstart,k}}{\dt} + \frac{\mathcal{E}_{\Mstart,k}}{\dt} \left(1 - \frac{dm_{\Mstart,k}}{dm_k }\right) \nonumber\\
                             &- \frac{F_{e,k+1,{\rm I}} - F_{e,k,{\rm I}}}{dm_{k}}.
\end{align}
The energy flux is
\begin{align}
\label{eq:stageIflux}
  F_{e,k,{\rm I}} &= p_{{\rm face},k} \area_k \vhat_{k,{\rm I}} + \efacek{k} F_{m,k} \,,
\end{align}
where 
\begin{align}
  \vhat_{k,{\rm I}} \equiv \frac{r_{\Mmid,k}-r_{\Mstart,k}}{\dt}~,
\end{align}
and the face values $\eface$ are 
interpolated\footnote{To ensure that $\eface$ has smooth derivatives this is done in the 
same manner as $\bar{T}$ in \mesaone. At the surface $\efacek{1} = \mathcal{E}_{\Mstart,1}$.} 
from the cell values $\mathcal{E}_{\Mstart}$.
\revision{The radial coordinate after state~I is calculated using the updated cell masses, holding cell densities fixed.}
The change in total
energy of the model during stage~I is
\begin{align}
\label{eq:stageIchange}
  \sum_k \left( E_{\Mmid,k} - E_{\Mstart,k}\right) =
  &-\dt \left ( p_{{\rm face},1} \area_1 \vhat_{1,{\rm I}} + \efacek{1} F_{m,1} \right)
    \nonumber\\
  &+ \dt \sum_k  dm_k \epsilon_{k,{\rm I}}~ .
\end{align}
The term in parentheses  on the right-hand side accounts for the
energy of new material entering or leaving the model and the work done
in the process.  The additional $\epsilon_{k,{\rm I}}$ term implies
the need for a corrective source term $\epsilon_{\dot{M}}$ which must
be added during stage~II so that energy is properly accounted for.

To determine this new source we consider the ratio of the thermal time scale
\begin{align}
  \tau_{{\rm th},k} \equiv \left|\frac{x_{m}c_{{\rm p}} T (\nabla_{\rm rad} - \nabla_{\rm ad})}{L}\right|_{\Mstart, k} ,
  \label{eq:tauth}
\end{align}
to the mass-change time scale
\begin{align}
  \tau_{{\rm m},k} \equiv \left|\frac{x_{m}}{F_{m}}\right|_{\Mstart, k},
\end{align}
where $x_{m,k}$ is the mass above face $k$.
When this ratio is small the thermodynamic state of material is a function primarily of depth \citep{Sugimoto1970,Sugimoto1975}.
This means that as material moves from cell to cell it is a good approximation to the time evolution to suppose that it adopts the state of whichever cell it is in (\mesathree).
In the opposing limit ($\tau_{\rm th} \gg \tau_{\rm m}$) the entropy of material adjusts minimally as it moves from cell to cell.

We account for the effects of the thermal and mass-change time scales by tracking the heating of material as it moves from cell to cell in stage I and estimating what fraction of that heat is released as part of $L$ versus carried by the material.
Consider the path taken by an infinitesimal fluid element as it moves from face $k$ to face $k+1$ due to accretion.
Over the course of this adjustment the material changes state and releases some heat.
We take this heat to be given by 
\begin{align}
  dq_k = -\dt \epsilon_{k,{\rm I}}\frac{dm_{k} }{dm_{\pass,k}},
\end{align}
where
\begin{align}
dm_{\pass,k} = dm_{\Mstart,k} + \dt \left(\max(0,F_{m,k+1}) - \min(0,F_{m,k})\right)
\end{align}
is the total mass of material that at any point during stage I was inside cell $k$.
\revision{This evenly distributes the heating which occurs in a cell over all material that starts in, ends in, or passes through the cell.}
When the thermal time is long, however, the fluid does not have a chance to release all of this heat before it has finished crossing the cell.
We estimate the fraction of this heat it releases as the leak fraction
\begin{align}
  f_{{\rm leak},k} &\equiv \min\left(1,\frac{\tau_{{\rm m},k}}{\tau_{{\rm th},k}}\right).
\end{align}
This parameterises the extent to which material flowing through a cell follows the implied energy gradient ($f_{{\rm leak},k} \approx 1$) versus evolving adiabatically ($f_{{\rm leak},k} \ll 1$).


We define $\delta E_{k,j}$ to be the amount of energy that the material which ends up in cell $j$ had not leaked by the time it reached face $k$.
This is zero for $k=1$ and for $k > j$, and for all other faces is given by
\begin{align}
  \delta E_{k+1,j} = (1 - f_{{\rm leak},k}) \left(\delta E_{k,j} + dq_k \mathcal{M}_{k,j}\right),
  \label{eq:depo}
\end{align}
where $\mathcal{M}_{k,j}$ is the amount of material which ends in cell $j$ which passes through cell $k$ during stage I.
The heat which is actually released in cell $k$ is then given by
\begin{align}
  \epsilon_{\dot{M},k} = \frac{\delta E_{k,k} + \sum_{j\neq k} f_{{\rm leak},k} \left(\delta E_{k,j} + dq_k \mathcal{M}_{k,j}\right)}{\dt dm_k}.
\end{align}
This is our new corrective source term.
The same procedure may be used in the case of mass loss, but with $\delta E_{N+1,j} =0$ instead of $\delta E_{1,j}$ and $-1$ in the subscript on the left-hand side of Equation~\eqref{eq:depo} rather than $+1$.
In the limit of long thermal times most heat is retained and the resulting evolution is adiabatic.
In the limit of short thermal times we recover the results of \mesathree.

Along the lines of Equation~\eqref{eq:fluxI}, the energy change of cell $k$ during stage II may now be written in flux-conservative form as
\begin{align}
  \label{eq:dedt_mdot}
  E_k - E_{\Mmid,k} &=\dt \left[ dm_k (\epsilon_{\dot{M},k} + \epsilon )+ F_{e,k+1,\mathrm{II}} - F_{e,k,\mathrm{II}}\right],
\end{align}
where
\begin{equation}
  F_{e, k, {\rm II}} = L_k + p_{{\rm face},k}  \area_k \vhat_{k,{\rm II}}
\end{equation}
and
\begin{equation}
  \vhat_{k,{\rm II}} \equiv \frac{r_{k}-r_{\Mmid,k}}{\dt}~.
\end{equation}
\MESA\ solves Equation~\eqref{eq:dedt_mdot} when using \dedt\ with mass changes.
Because the time evolution is implicit all sources are evaluated at the end of stage II.
Nuclear burning is evaluated as if material spends the whole step in the cell in which it ends stage I.
While this is usually a good approximation it may break down when $\dt\dot{M}$ is large relative to the masses of cells in burning regions.

When $\dot{M} \geq 0$ the above procedure for redistributing energy is conservative, so that
\begin{align}
  \sum_k \epsilon_{k,\mathrm{I}} \dt dm_k + \sum_k \epsilon_{\dot{M},k} \dt dm_k  = 0.
\end{align}
The first sum represents the effect of stage I; the second sum represents the effect of stage II.
When $\dot{M} < 0$ the equality is instead
\begin{align}
  \sum_k \epsilon_{k,\mathrm{I}} \dt dm_k + \sum_k \epsilon_{\dot{M},k} \dt dm_k  &= -\delta E_{1,0},
\end{align}
with the additional term $\delta E_{1,0}$ being the energy carried out of the star.
This term is explicitly accounted for in the \code{MESA} energy budget, so in both cases energy is properly accounted for.

\subsubsection{Results}

\begin{figure}[!htb]\centering
\includegraphics[width=1.0\columnwidth]{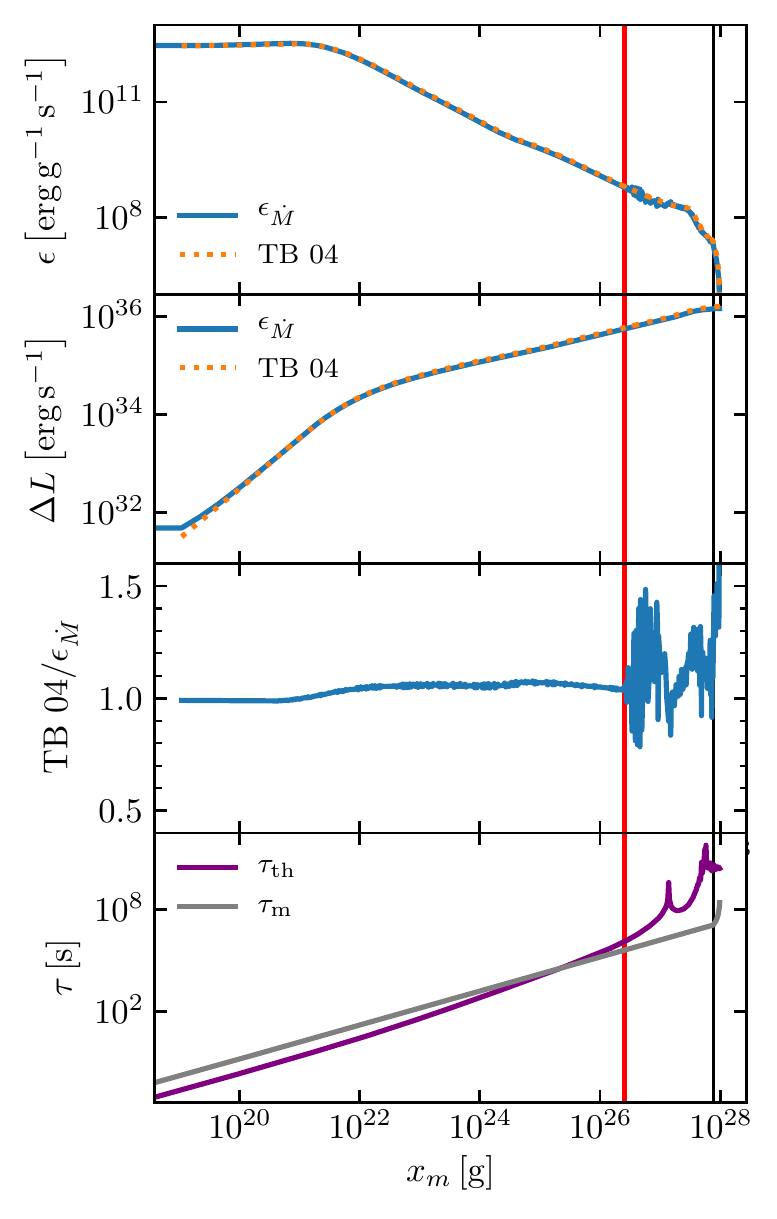}
\caption{Comparison of $\epsilon_{\dot{M}}$ and the analytic accreting heating expression from~\citet{2004ApJ...600..390T} (TB~04) for a $0.33\,\Msun$ He WD accreting He and N at a rate of $10^{-5}\Msunyr$. The first panel shows both as functions of depth $x_m$.
The second panel shows the same but integrated inward with respect to mass.
The third panel shows the ratio of the TB~04 expression to $\epsilon_{\dot{M}}$.
The final panel shows the thermal time scale $\tau_{\rm th}$ and the mass-change time scale $\tau_{\rm m}$.
All material to the left of the red vertical line is material which is new in this time step.
To the right of the vertical black line \MESA\ transitions to a Lagrangian mesh which continues over the remainder of the core.
}
\label{fig:custom_rates}
\end{figure}

To examine the behavior of $\epsilon_{\dot{M}}$ we modeled a $0.33\,\Msun$ WD accreting He and N at a rate of $10^{-5}\,\Msunyr$.
The first panel of Figure~\ref{fig:custom_rates} shows a profile of $\epsilon_{\dot{M}}$ along with the analytic accretion heating calculations of~\citet{2004ApJ...600..390T}.
For the most part the two agree closely.
The second panel shows the mass integral of the same inward from the surface while the third shows their ratio.
Around $x_m \approx 10^{26}{\rm g}$, $\tau_{\rm th}$ becomes long relative to $\tau_{\rm m}$ and the two prescriptions differ because that of~\citet{2004ApJ...600..390T} is only applicable where $\tau_{\rm th} \ll \tau_{\rm m}$.
The $\epsilon_{\dot{M}}$ term handles both limits.

\begin{figure}[!htb]\centering
\includegraphics[width=1.0\columnwidth]{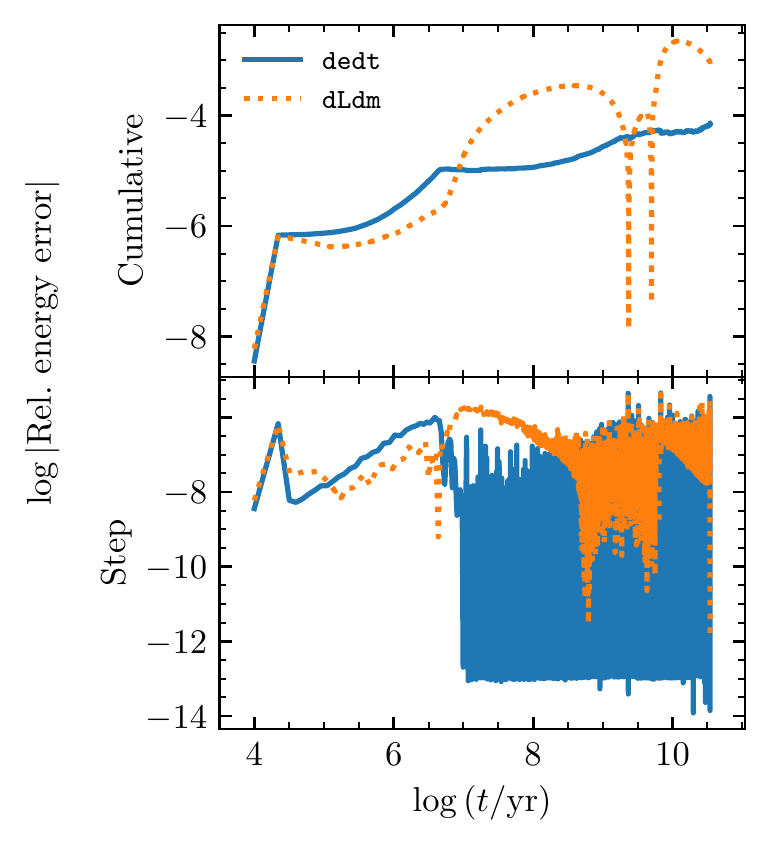}
\caption{Comparison of energy conservation for a $0.3\,\Msun$ He WD model accreting an H/He mixture at a rate of $10^{-10}\,\Msunyr$. The upper panel shows the relative cumulative error and the lower shows the error in each step.
}

\label{fig:he_wd_err}
\end{figure}

To demonstrate improved energy conservation during mass changes with \dedt\ we model accretion onto a $0.3\,\Msun$ He WD with an initial $\log(\Tc/\K) = 6.7$.
The accretion rate is fixed at $10^{-10}\,\Msunyr$.
Nuclear reactions are disabled throughout the run. 
This is repeated with \dLdm.
The relative cumulative error in energy conservation is shown in the upper panel of Figure~\ref{fig:he_wd_err}, and the relative error in each step is shown in the lower panel.
Near the beginning of the run there is a period where \dLdm\ performs better, however at those early times both forms do a good job, conserving energy to one part in $\sim 10^5$.
At later times \dedt\ produces less error, staying below one part in $10^4$ while \dLdm\ yields cumulative errors greater than one part in $10^3$.

\begin{figure}[!htb]\centering
\includegraphics[width=1.0\columnwidth]{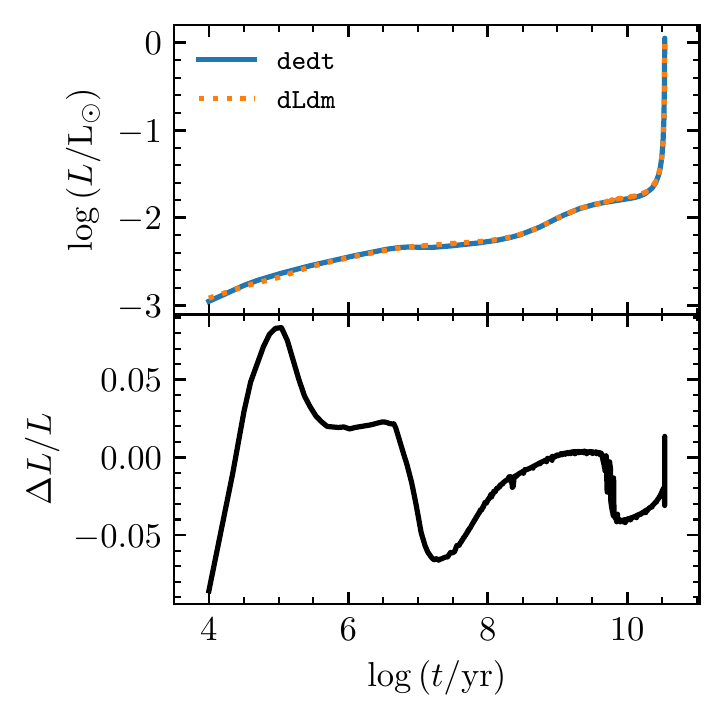}
\caption{Comparison of the surface luminosity for a $0.3\,\Msun$ He WD model accreting a H/He mixture at $10^{-10}\,\Msunyr$ (same as Figure~\ref{fig:he_wd_err}).
The upper panel shows the surface luminosity and the lower shows the relative difference between \dLdm\ and \dedt, all as functions of time since the start of accretion.
}
\label{fig:he_wd_mdot}
\end{figure}

\begin{figure}[!htb]\centering
\includegraphics[width=1.0\columnwidth]{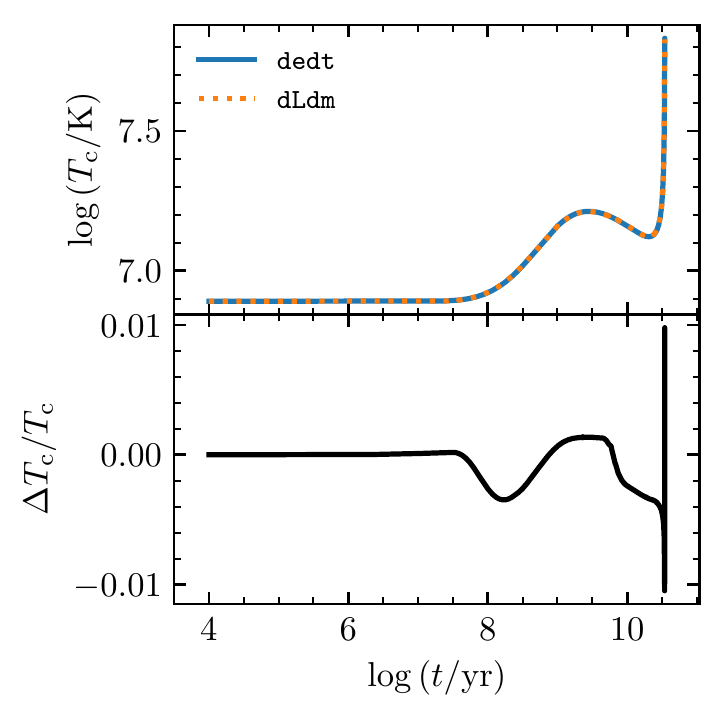}
\caption{Comparison of the central temperature for a $0.3\,\Msun$ He WD model accreting a H/He mixture at  $10^{-10}\,\Msunyr$ (same as Figure~\ref{fig:he_wd_err}).
The upper panel shows the central temperature and the lower shows the relative difference between \dLdm\ and \dedt, all as functions of time since the start of accretion.
}
\label{fig:he_wd_mdot_tc}
\end{figure}

We prefer using \dedt\  because of its improved energy conservation and handling of long thermal times.
We now explore the consequences for stellar evolution of these different prescriptions.
The upper panel of Figure~\ref{fig:he_wd_mdot} shows the luminosity for the same case as Figure~\ref{fig:he_wd_err}, again for both \dedt\ and \dLdm.
The difference is shown in the lower panel.
The largest differences are at early times as the models adjust to the accretion.
After that, both yield results similar to a few percent, and the relative difference only improves as the luminosity increases.
Figure~\ref{fig:he_wd_mdot_tc} shows \Tc\ for the same two runs as a function of time.
The differences are small.
This is because the core lies beneath the region where cell masses are adjusted significantly, so the precise handling of mass changes only matters for the core insofar as the core temperature is sensitive to the luminosity.

\begin{figure}[!htb]\centering
\includegraphics[width=1.0\columnwidth]{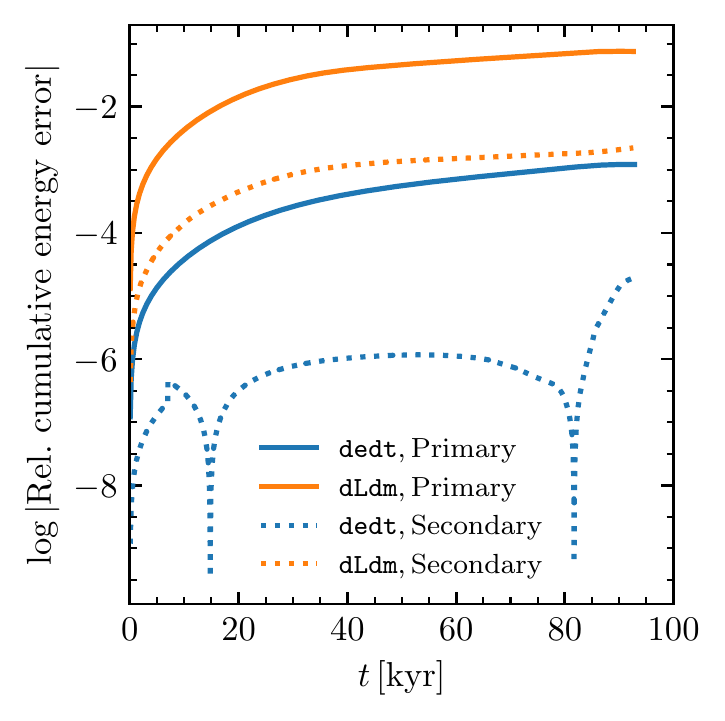}
\caption{Comparison of energy conservation for the case of Figure~4 of \mesathree: a binary system with mass transfer from the $8\,\Msun$ primary to the $6.5\,\Msun$ secondary and an initial orbital period of $3$ days.
The relative cumulative energy error computed using both the \code{dedt}- and \code{dLdm}-forms is shown as a function of time after the start of accretion.
}
\label{fig:binary_err}
\end{figure}

\begin{figure}[!htb]\centering
\includegraphics[width=1.0\columnwidth]{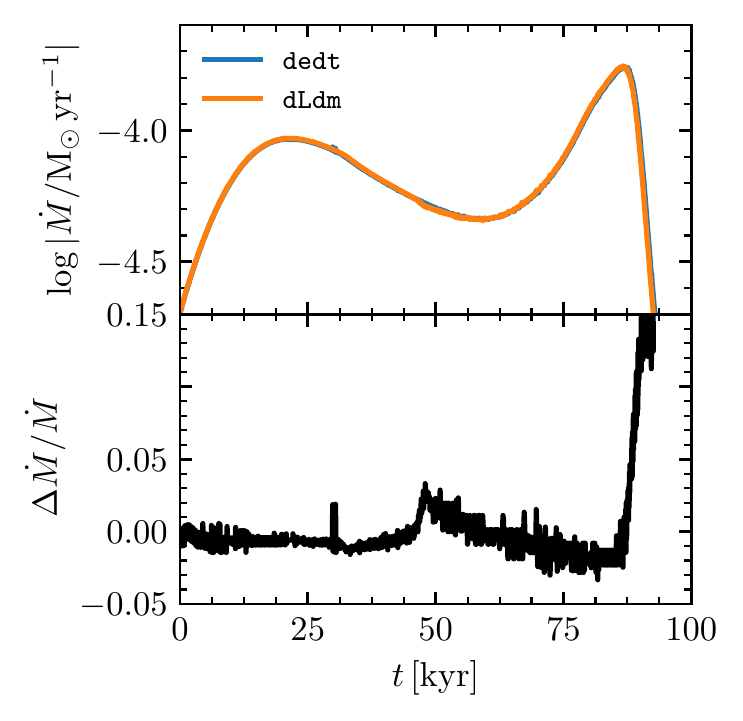}
\caption{Comparison of mass transfer rates for the case of Figure~4 of \mesathree: a binary system with mass transfer from the $8\,\Msun$ primary to the $6.5\,\Msun$ secondary and an initial orbital period of $3$ days.
The upper panel shows the mass transfer rate computed using both the \code{dedt}- and \code{dLdm}-forms as a function of time after the start of accretion.
Both are similar to the corresponding curves in Figure~4 of \mesathree.
The lower panel shows the relative difference between $\dot{M}$ computed with the \code{dedt}- and \code{dLdm}-forms.
The spike at the end is due to mass transfer terminating at slightly different times.
}
\label{fig:binary_mdot}
\end{figure}

Mass changes are ubiquitous in binary stellar evolution.
To demonstrate the effect of \dedt\ we model the evolution of a binary system with an $8\,\Msun$ primary and a $6.5\,\Msun$ secondary with an initial orbital period of $3$ days.
This is the same example as in Figure~4 of \mesathree.
The relative cumulative error in energy conservation is shown in the upper panel of Figure~\ref{fig:binary_err}.
The model is also run with \dLdm.
The errors are shown independently for the primary and the secondary.
Figure~\ref{fig:binary_mdot} compares the mass transfer rate for this system computed using each energy equation.
The differences are typically of order $1$\,\%.

\section{Rotation}\label{s.rot}

For a rigidly-rotating star in hydrostatic equilibrium, surfaces of constant pressure (isobars) coincide with the equipotential surfaces defined by 
the Roche potential $\Psi$,
\begin{eqnarray}
   \Psi(r,\theta) = \Phi(r,\theta) - \frac{r^2\Omega^2\sin^2\theta}{2},
\end{eqnarray}
where $\Phi$ is the standard Newtonian potential, $\theta$ is the polar angle,
and $\Omega$ is the angular frequency of rotation.
In one-dimensional stellar evolution calculations the effects of rotation on the stellar structure are usually captured by a simple modification of the stellar equations. These retain their regular form, but include two correction 
factors, 
\begin{eqnarray}
   f_{P}=\frac{4\pi r_{\Psi}^4}{Gm_\Psi S_\Psi}\frac{1}{\langle g^{-1}
   \rangle},\quad f_{
   T}=\left(\frac{4\pi r_{\Psi}^2}{S_\Psi}\right)^2\frac{1}{\langle
g\rangle \langle g^{-1}\rangle},\label{eq:fdef}
\end{eqnarray}
where $r_{\Psi}$ is the volume-equivalent radius of an isobar,
and $m_{\Psi}$ and $S_{\Psi}$ are the mass inside that isobar and its
surface area respectively \citep{KippenhahnThomas1970, EndalSofia1976}.
The effective gravity is $g=|\nabla \Psi|$,
while $\langle g\rangle$ and $\langle g^{-1}\rangle$ are surface averages over the
equipotential.%
\footnote{In \mesatwo\ we defined the volume equivalent
radius as $r_P$, while here we adopt the symbol $r_\Psi$ as used by
\citet{EndalSofia1976}. This change is to prevent confusion between
$r_\Psi$ and the polar radius of an isobar, which we denote as $r_{\rm p}$.}

This approach is still applicable to the case of a differentially-rotating star under the assumption of shellular rotation,
in which shells are rigidly-rotating, isobaric surfaces with
rotation frequency $\Omega_\Psi$ \citep{EndalSofia1976,Zahn1992}.
The averages in Equation~(\ref{eq:fdef}) are performed over each isobar \citep{MeynetMaeder1997}.

Until now, \MESA\ used the method of \citet{EndalSofia1976}, which considers deviations of the Roche
potential from spherical symmetry \citep{Kopal1959},
to compute the $f_P$ and $f_T$ factors.
One issue is that to ensure numerical stability,  this approach requires a floor on the correction factors ($f_P=0.75$ and $f_T=0.95$, \mesatwo),
corresponding to a maximum rotation rate of 60\% of critical rotation (the angular frequency at which the centrifugal force would match gravity at the stellar equator).

As stars are centrally condensed and rotational frequencies close to
critical are typically reached only in the outermost layers, $\Psi$ is well approximated by the potential of a point mass in rapidly
rotating layers (e.g., \citealt{Maeder2009}).
This justifies using the Newtonian potential as $\Phi=-Gm_\Psi/r$ for the calculation of $f_{P}$ and $f_{T}$, such that they are only
functions of the fraction of critical rotation  $\omega\equiv\Omega_\Psi/\Omega_{\rm crit,\Psi}=\Omega_\Psi/\sqrt{Gm_\Psi/r_{\rm e}^3}$.
Here $r_{\rm e}$ is the equatorial radius of the isobar and $\Omega_{\rm
crit,\Psi}$ is the rotational frequency at which the centrifugal force is equal
to gravity at the equator of the isobar.

We describe a new implementation of centrifugal effects
in \texttt{MESA}, which makes use of analytical fits to the Roche
potential of a point mass, improving the calculation of rotating stars to $\omega\approx 0.9$.

\subsection{The Roche Potential of a Rigidly-Rotating, Single Star}\label{sec:roche}

For a point mass $m_\Psi$, the dimensionless Roche potential $\Psi'=\Psi/(Gm_\Psi\Omega)^{2/3}$ can be
written in terms of the dimensionless radius $r'=r/(Gm_\Psi/\Omega^2)^{1/3}$ as
\begin{eqnarray}
   \Psi' = -\frac{1}{r'}-\frac{r'^2\sin^2\theta}{2}. \label{eq:phi}
\end{eqnarray}
Note that
\begin{eqnarray}
   r_{\rm e}'=\frac{r_{\rm e}}{(Gm_\Psi/\Omega^2)^{1/3}}=\omega^{2/3},
\end{eqnarray}
such that evaluating Equation~\eqref{eq:phi} for
$\theta=0$ ($r'=r_{\rm p}'$) and $\theta=\pi/2$ ($r'=r_{\rm e}'$) provides the ratio of the polar radius $r_{\rm
p}$ to $r_{\rm e}$ as a function of $\omega$,
\begin{eqnarray}
   \frac{r_{\rm e}}{r_{\rm p}}=1+\frac{1}{2}\omega^2.\label{eq:re_div_rp}
\end{eqnarray}

\begin{figure}
   \begin{center}
   \includegraphics[width=\columnwidth]{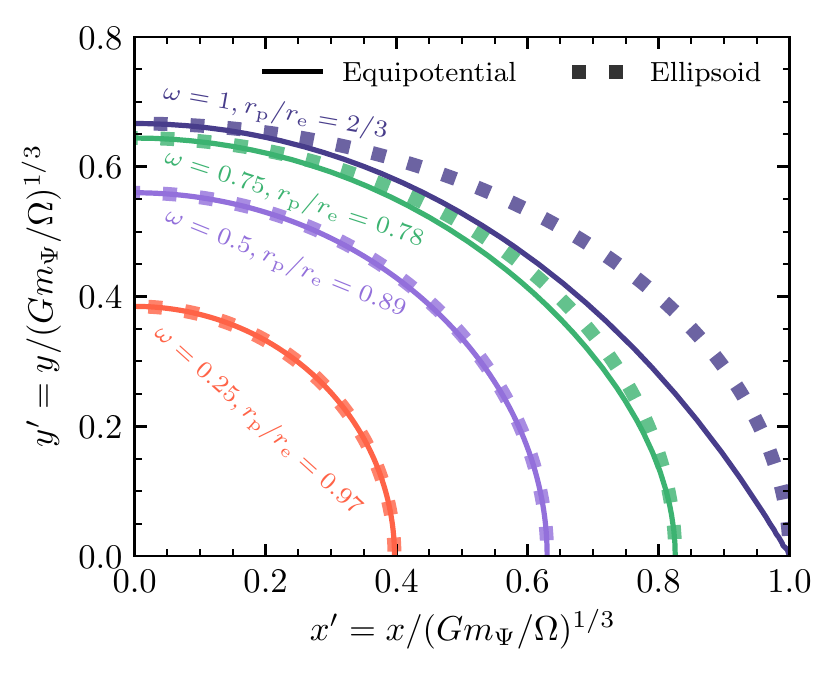}
   \end{center}
   \caption{Equipotential lines of the dimensionless Roche potential $\Psi'$
   given by Equation~\eqref{eq:phi}. Solid lines show the equipotentials for which
   $\omega$ is equal to $0.25,0.5,0.75$ and $1$. Dashed lines show ellipses with the same polar and equatorial radii as the
   actual equipotentials.
   \label{fig:equip}}
\end{figure}

Figure \ref{fig:equip} shows how the equipotential surfaces change with
increasing $\omega$. For $\omega \lesssim 0.5$ the equipotentials are
approximately given by oblate spheroids, while for $\omega$ close to unity a
cusp develops at the equator. For this critically rotating surface, the
polar radius is $2/3$ of the equatorial one. 

\begin{figure}[ht!]
   \begin{center}
   \includegraphics[width=\columnwidth]{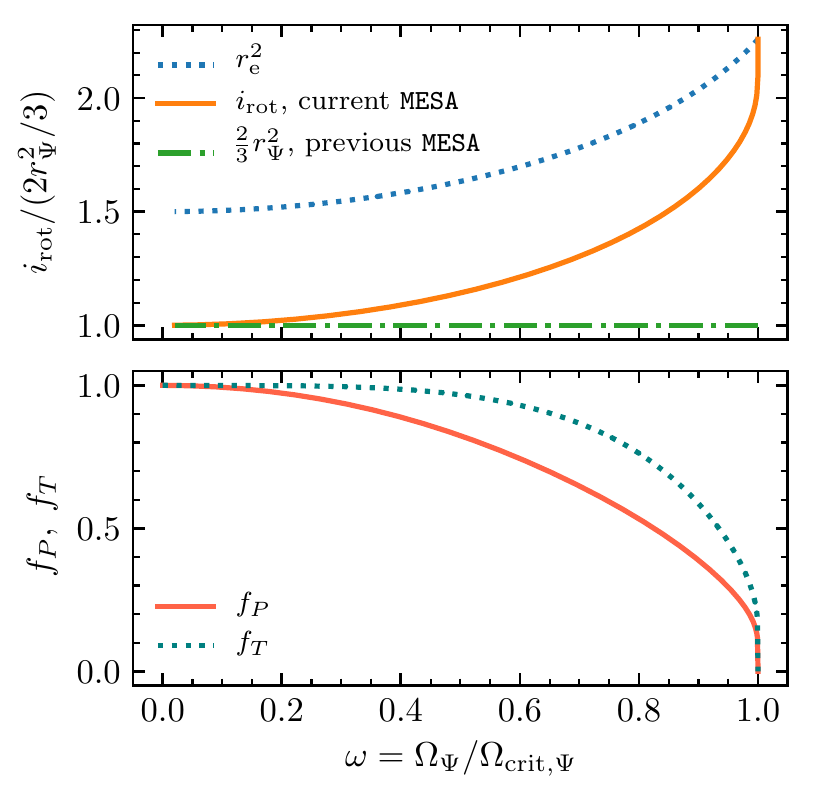}
   \end{center}
   \caption{The upper panel shows the specific moment of inertia for a shell of material at different
   values of $\omega$, normalized by $2/3r_{\Psi}^2$. The model implemented
   previously in \texttt{MESA} is shown with a dot-dashed green line, and the
   new $\omega$-dependent model for $i_{\rm rot}$ is shown with a solid orange
   line. As $\omega\rightarrow1$, the moment of inertia becomes that of a ring with radius $r_{\rm e}$.
   The lower panel shows the $f_{P}$ and $f_{T}$ factors as a function of
   $\omega=\Omega/\Omega_{\rm crit}$.
   }\label{fig:irotfpft}
\end{figure}

By determining the asymptotic behavior of the Roche potential in the
limit $\omega\rightarrow 0$, and, when possible, also in the limit
$\omega\rightarrow1$, we have constructed analytic fits to properties of
interest. These are described in
Appendix \ref{a.rot}, and include fits for the equatorial radius
$r_{\rm e}(r_\Psi,\omega)$, the centrifugal corrections $f_P(\omega)$ and
$f_T(\omega)$, and the volumes and surface areas of Roche equipotentials,
$V_\Psi(r_{\rm e},\omega)$ and $S_\Psi(r_{\rm e},\omega)$.  Previous versions of
\texttt{MESA} approximated the specific moment of inertia of isobaric surfaces as
that of a thin spherical shell with radius $r_\Psi$, $i_{\rm
rot}=2r_\Psi^2/3$, but now default to a fit of the form $i_{\rm rot}(r_{\rm e},\omega)$. Figure \ref{fig:irotfpft}
shows the resulting fits for $i_{\rm rot}$, $f_P$ and $f_T$.
The new implementation results in values of the specific moment of inertia that
are larger by a factor of two as $\omega\rightarrow 1$.

\subsection{Implementation in Stellar Evolution Instruments}\label{sec:rotimp}
To include these fits into a stellar evolution calculation that uses
the shellular approximation, a value of $\omega$ must be determined for
a given specific angular momentum $j_{\rm rot}$, $m_{\Psi}$, and $r_{\Psi}$.
From $\Omega_\Psi=\omega\Omega_{\rm crit,\Psi}=\omega\sqrt{Gm_\Psi/r_{\rm e}^3}$ and
$j_{\rm rot}=i_{\rm rot}\Omega_\Psi$, we find
\begin{eqnarray}
   \frac{j_{\rm rot}}{\sqrt{Gm_\Psi r_{\Psi}}}
   =\omega\frac{i_{\rm rot}}{r_{\rm
   e}^2}\sqrt{\frac{r_{\rm e}}{r_{\Psi}}}.\label{eq:wiii}
\end{eqnarray}
For a given $j_{\rm rot}$, $m_\Psi$ and $r_{\Psi}$, the left-hand side can be
directly evaluated, while the right-hand side is a monotonic function of $\omega$ for
$0\le\omega\le 1$.
We  compute the solution to this equation  for each cell in the stellar
model using a bisection method. Given $\omega$, we then use the computed fits to
determine the values required by the structure equations: $r_{\rm e}$, $r_{\rm
p}$, $i_{\rm rot}$, $\Omega_\Psi$,
$f_{P}$ and $f_{T}$. As in the previous implementation of rotation in
\texttt{MESA}, we evaluate these quantities explicitly at the beginning
and at the end of a step. The analytical nature of the fits allows
the possibility of a fully-coupled and implicit implementation in the future.

\begin{figure}[ht!]
   \begin{center}
   \includegraphics[width=\columnwidth]{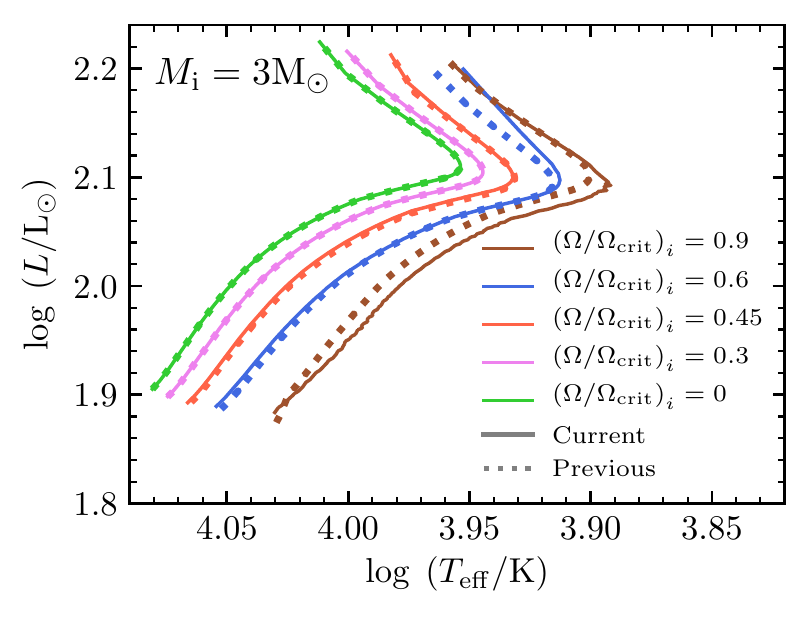}
   \end{center}
   \caption{Stellar evolution calculations with initial mass $3\,\Msun$ and solar
   metallicity, from ZAMS up to terminal age main sequence (TAMS). Different colors
   indicate different initial rotation rates, defined in terms of the ratio of the
   rotational frequency to the critical value at the surface at ZAMS. Solid
   and dotted lines indicate calculations done with the current and the
   previous implementations of rotation in \texttt{MESA}, respectively.}\label{fig:rot3m}
\end{figure}

Figure \ref{fig:rot3m} shows a $3\,\Msun$ solar metallicity model
with different initial rotational velocities, evolved using the current and previous implementations of rotation in \texttt{MESA}.
Both methods agree for rotation rates $\omega<0.5$. Differences at higher rotation rates are due to the aforementioned 
floor on $f_P$ and $f_T$.
In our current approach we also define a floor on $f_P$ and $f_T$
in terms of a maximum value $\omega_{\rm max}$, beyond which
the effects are truncated to $f_P(\omega_{\rm max})$ and $f_T(\omega_{\rm max})$.
We find that calculations using the new strategy are numerically stable near critical rotation, with the simulations
shown in Figure \ref{fig:rot3m} being performed with $\omega_{\rm
  max}=0.9$.  This is in comparison to the previous method, that for rapidly-rotating models set a floor on  $f_P$ and $f_T$ corresponding to their values at $\omega \approx 0.6$.
Therefore, \MESA\ can now consistently calculate shellular rotation models closer to critical rotation.

\subsection{Gravity Darkening Corrections}\label{s.gdark}

Rotating stars are subject to gravity darkening
\citep{von-zeipel_1924_aa,von-zeipel_1924_ab}.  The variation of flux
over the surface and the distorted stellar shape imply that the
observed properties of the star vary with the angle between the
rotation axis of the star and the line of sight (LOS). Here we
describe our approach to calculating geometric factors that allow the
intrinsic surface quantities $L$ and $\Teff$ to be corrected for
projection effects along a given LOS. By \emph{intrinsic} we mean the
total $L$ emitted by the star and the $\Teff$ associated with this $L$
given the total surface area of the star and the Stefan-Boltzmann law.
This is done in two steps.  First, we solve the gravity darkening
problem for an arbitrary surface element of a rotating star, and 
second, calculate the projection of the gravity-darkened surface along
the LOS.

\subsubsection{The Gravity Darkening Model}

\begin{figure}[htb!]
  \centering
  \includegraphics[width=\columnwidth]{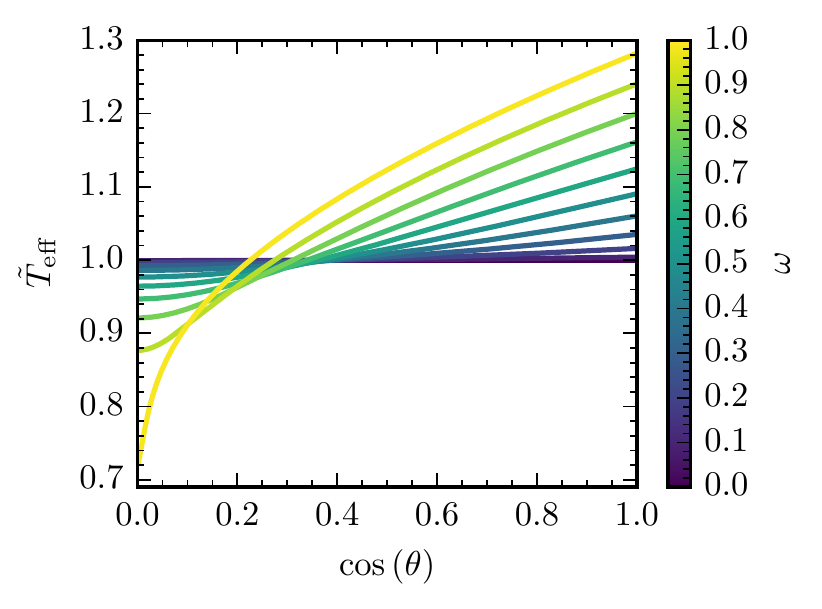}
  \caption{Solution of the ELR model for $0 \leq \omega \leq 1$; each curve plots the variation of $\rm{\tilde{T}_{eff}}$ as a function of $\cos(\theta)$ for a different value of $\omega$. Recall that $\cos(\theta)=1$ corresponds to the pole and $\cos(\theta)=0$ to the equator. 
   \label{fig:ELR}}
\end{figure}

\begin{figure*}[htb!]
  \centering
  \includegraphics[width=0.9\textwidth]{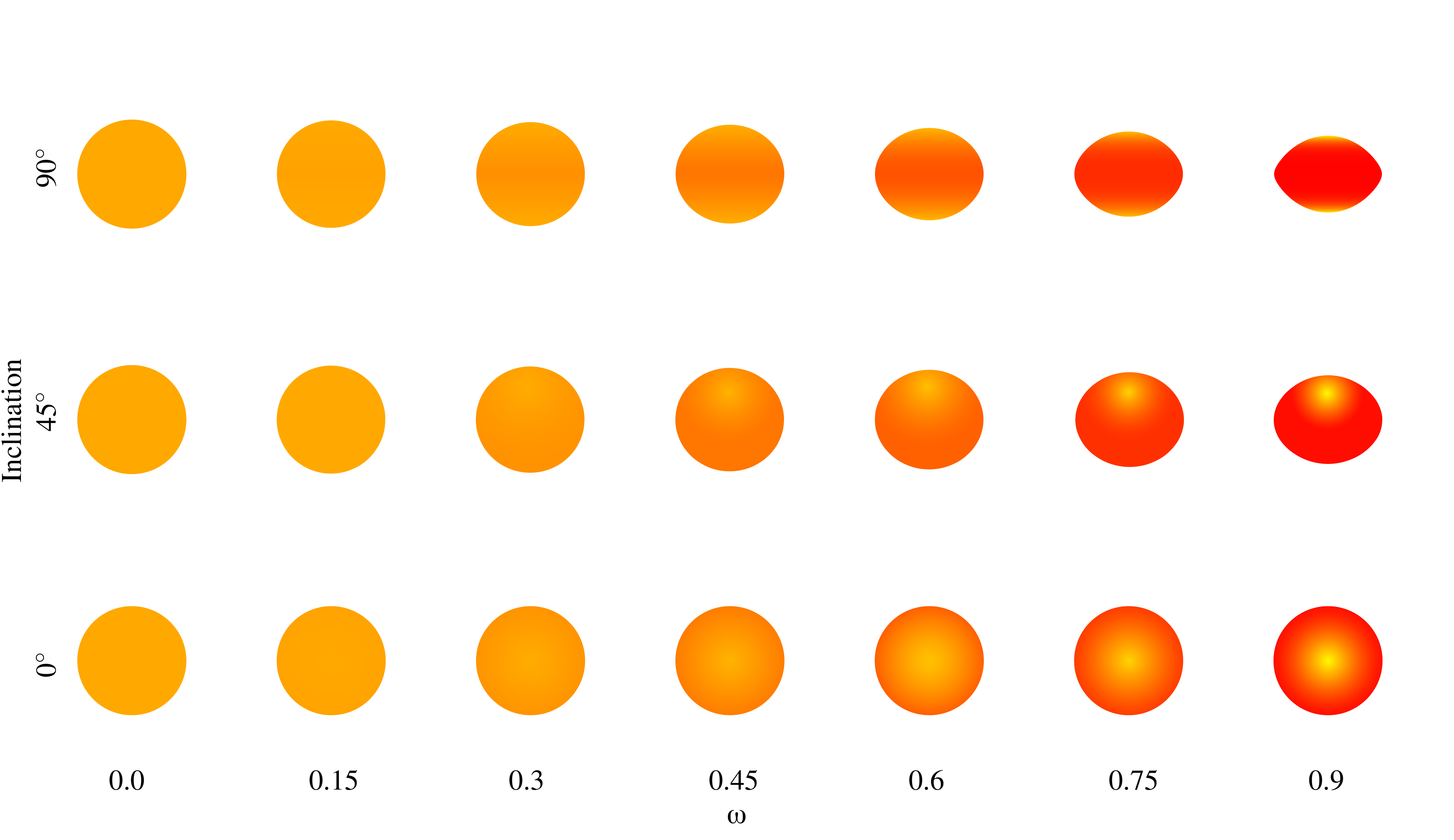}
  \caption{A grid of Roche equipotential surfaces for a range of rotation rates ($\omega$) and
    inclination angles ($i$). The color of the surface corresponds to the variation in $\tilde{T}_{\rm eff}$ with red corresponding
    to 0.87 and pale yellow to 1.25; the stars with $\omega=0$ have ${\tilde{T}_{\rm eff}}=1$.\label{fig:Roche}}
\end{figure*}

We use the gravity darkening model of \citet[][hereafter ELR]{ELR2011}, where it is assumed that the radiative flux is directed antiparallel to the effective surface
gravity.  At a point on the stellar surface with polar angle $\theta$, we find the value of the scaled photosphere radius $\tilde{r}=R/R_{\rm e}$  by solving
\begin{equation}
  \frac{1}{\tilde{r}} + \frac{\omega^2}{2} \tilde{r}^2 \sin^2 \theta = 1 + \frac{\omega^2}{2}.
\end{equation}
We then solve 
\begin{equation}
  \cos \vartheta + \ln \tan(\vartheta/2) = \frac{1}{3} \omega^2 \tilde{r}^3 \cos^3 \theta + \cos \theta + \ln \tan(\theta/2),
  \label{eq:blurgh}
\end{equation}
for the modified angular variable $\vartheta$.  Using ELR Equation\ (31), we use this value to obtain the local \Teff.

Figure \ref{fig:ELR} shows the variation of $\tilde{T}_{\rm eff} = \Teff [L/ (4 \pi \sigma R_{\rm e}^2)]^{-1/4}$ over a range of $\theta$ for a series of curves with different values of $\omega$.
When $\omega=0$, 
$\tilde{T}_{\rm eff} = 1$ for all $\theta$. When $\omega=1$, $\tilde{T}_{\rm eff}$ varies by nearly a factor of 2 between the pole and the equator.

\subsubsection{Projection Effects and Correction Factors}

We are interested in the \emph{projected}---the directional average over the surface along the LOS---$\Teff$ and $L$. The two parameters governing the problem are $\omega$ and the inclination angle, $i$, of the LOS with
respect to the rotation axis of the star: $i = 90^\circ$ when the LOS is in the plane of the equator. We denote the LOS unit vector $\hat{l}(i)$ and the projected surface area
$\Sigma_{\rm proj}$. Figure~\ref{fig:Roche} shows a grid of Roche equipotential surfaces for different values of $\omega$
and $i$. The color describes the variation of $\tilde{T}_{\rm eff}$ over the surface.

To calculate the luminosity projected along the LOS, $L_{\rm proj}$, requires the surface integral
\begin{equation}\label{eq:lproj}
  L_{\rm proj} = 4 \times \iint \limits_{d\vec{\Sigma} \cdot \hat{l} > 0} F d\vec{\Sigma} \cdot \hat{l},
\end{equation}
where only the flux projected toward the observer, i.e., $d\vec{\Sigma} \cdot \hat{l} > 0$, is kept.
The emergent specific intensity from each surface element is assumed to be isotropic.
Once $L_{\rm proj}$ and $\Sigma_{\rm proj}$ are known, the projected $\Teff$ can be obtained from the Stefan-Boltzmann law
\begin{equation}\label{eq:tproj}
T_{\rm eff,proj} = \left( \frac{L_{\rm proj}}{\sigma \Sigma_{\rm proj}} \right)^{1/4}.
\end{equation}
As noted by \citet{georgy2014}, the ratio of $L_{\rm proj}/L$, and likewise the ratio
of $T_{\rm eff,proj}/\Teff$, are geometric factors that depend only on $\omega$ and $i$.
We  define gravity darkening coefficients
\begin{equation}
  C_L(\omega,i) \equiv L_{\rm proj}(\omega,i) / L
\end{equation}
and
\begin{equation}
  C_T(\omega,i) \equiv T_{\rm eff,proj}(\omega,i) / \Teff.
\end{equation}

\begin{figure*}[htb!]
  \centering
  \includegraphics[width=\columnwidth]{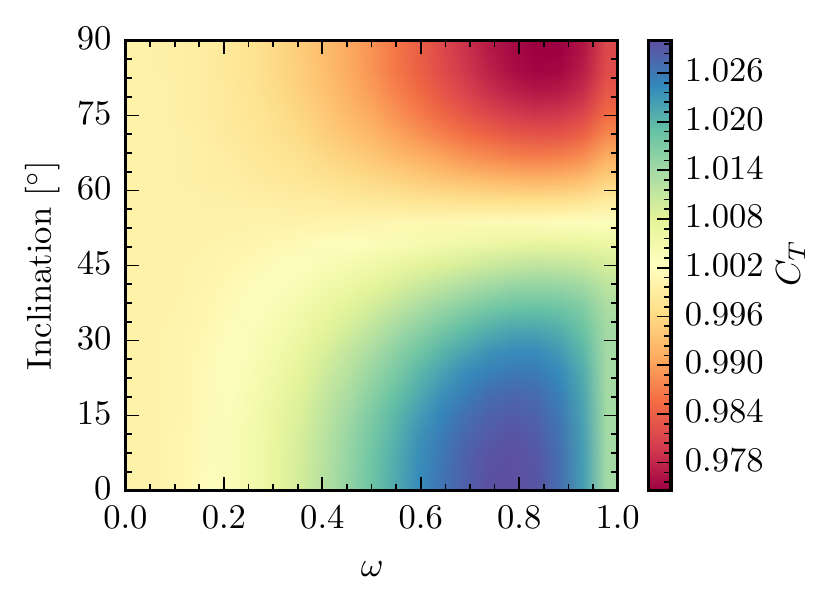}
  \includegraphics[width=\columnwidth]{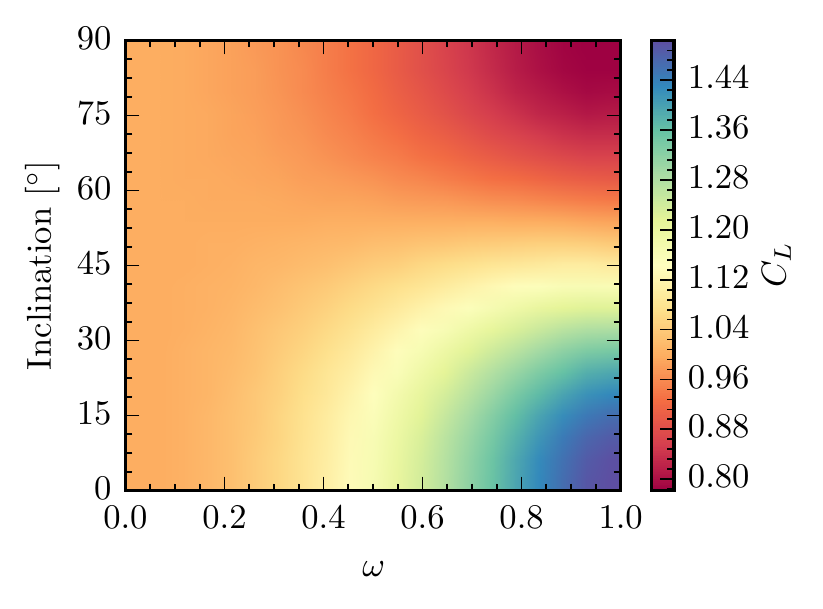} 
  \caption{The variation of $C_T$ (left panel) and $C_L$ (right panel) in the ($\omega,i$)-plane. Note the
    color scale is different in the two panels. \label{fig:coeff}}
\end{figure*}

\begin{figure*}[htb!]
  \centering
  \includegraphics[width=\textwidth]{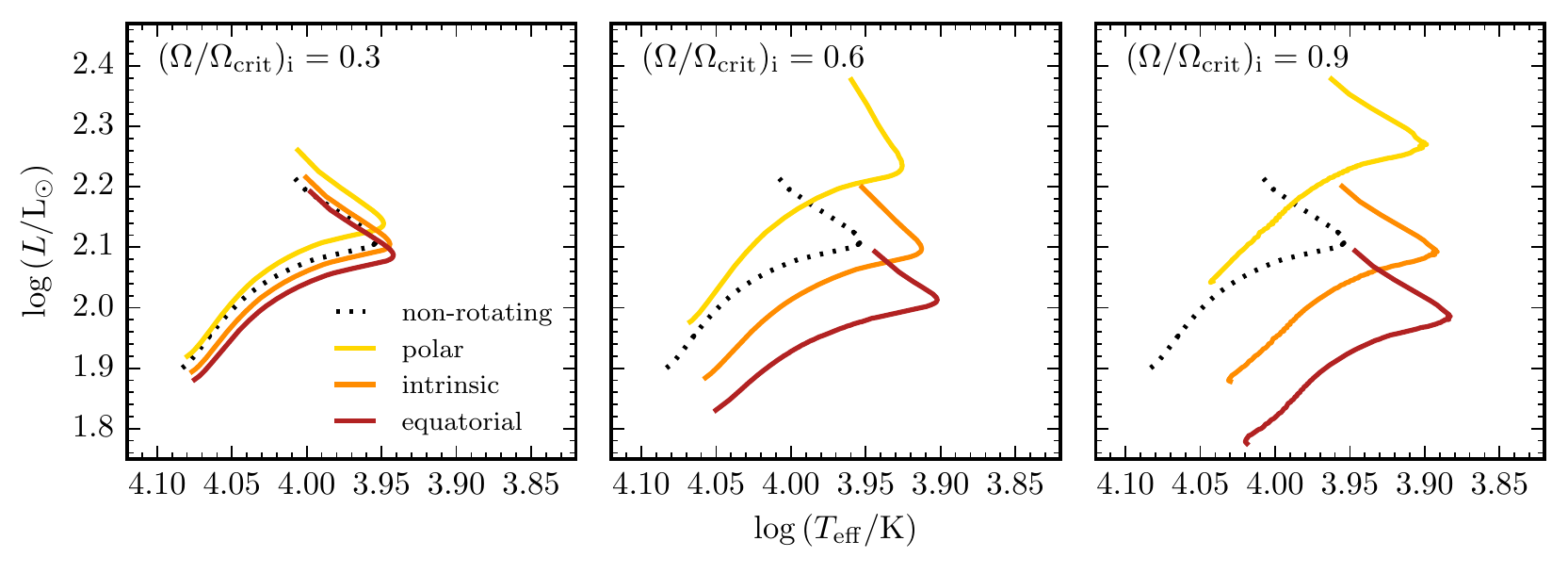} 
  \caption{ Three rotating tracks from Figure~\ref{fig:rot3m} showing the effect of gravity darkening in the HR diagram of a $3\,\Msun$ model from ZAMS to TAMS. The orange line plots the intrinsic values; the yellow line plots the polar projection; the red line plots the equatorial projection. The dotted line shows the evolution of the non-rotating track.\label{fig:HRD} }
\end{figure*}

$C_T$ and $C_L$ are tabulated in \MESA\ for the valid domain of $\omega$ and $i$; values
are readily obtained via bicubic interpolation. The projected $L$ and $\Teff$ as
viewed from the pole and the equator can be output in the \MESA\ history file; other inclination
angles can be accessed via \code{run\_star\_extras}.

Figure \ref{fig:coeff} shows how $C_T$ and $C_L$ vary over the
($\omega$,$i$)-plane. These can be compared with the upper panels of Figure 2
in \citet{georgy2014}. The variation of $C_L$ is greater than that of $C_T$
due to the $\frac{1}{4}$-power relationship between $L$ and $\Teff$. At $\omega=1$, where
the geometric factors are the largest, $C_L$ varies from $-20\%$ to
$+50\%$ while $C_T$ varies by only about $\pm2.5\%$. 

Figure \ref{fig:coeff} shows a slight, but noticeable, decrease in $C_T$
for $\omega \simeq 1$ whereas the comparable figures from \citet{georgy2014} do not.
When we
calculate the coefficients using oblate spheroids instead of Roche equipotential surfaces, we see no
such decrease in $C_T$ for near-critical rotation.

Figure~\ref{fig:HRD} demonstrates the effect of gravity darkening on 3 of the $3\,\Msun$ tracks from Figure~\ref{fig:rot3m} in the HR diagram. In each panel of Figure~\ref{fig:HRD} the non-rotating track is shown for reference as the dotted line. The magnitude of the gravity darkening effect increases with $\omega$ and is more substantial for $L$ than for $\Teff$. Relative to the intrinsic track, the polar projection is brighter and hotter, and the equatorial projection is cooler and fainter.

\section{Convective Boundaries and Semiconvection Regions} \label{s.cpm}

The correct treatment of convective boundaries continues to be a challenging problem.
In this section, we discuss three approaches: the ``sign-change'' algorithm (\mesaone, \mesatwo),
an improved approach called ``predictive mixing'' (\mesafour),
and a new ``convective premixing'' scheme that addresses several remaining issues.

Early versions of \MESA\ located convective boundaries by searching
for sign changes in the discriminant $y$ 
\revision{, defined by $y = y_{\rm S} \equiv \gradr - \grada$ when
the Schwarzschild criterion is used to assess convective stability, or by $y = y_{\rm L}
\equiv \gradr - \gradL$ when the Ledoux criterion is used}; here $\gradr$,
  $\grada$ and $\gradL$ are the radiative, adiabatic and Ledoux
  temperature gradients, respectively. As demonstrated in \mesafour,
  this sign-change algorithm can fail at convective boundaries that
  exhibit composition discontinuities. Typically, the failing cases
  exhibit $\gradr$ appreciably larger than $\grada$ on the convective
  side of the boundary.  Instead, as argued by
  \citet{gabriel_2014_aa}, physical consistency within local mixing
  length theory (MLT) dictates that $\gradr = \grada$ should hold on
  the convective side.

In \mesafour\ we introduced the predictive mixing scheme for
treatment of convective boundaries. 
This scheme improves on the sign-change algorithm  by allowing each
convection region to expand during a time step until its boundaries
satisfy $\gradr = \grada$ on their convective side. The expansion is
achieved by modifying convective diffusivities in the cells on the
radiative side of a boundary. In \mesafour, we applied predictive mixing in five scenarios: a growing
convective core in a $1.5\,\Msun$ star on the MS; a
retreating convective core in a $16\,\Msun$ star on the MS; growing
convective cores in $1\,\Msun$ and $3\,\Msun$ stars during the core
Helium burning (CHeB) phase; and an evolving convective envelope in a
$1\,\Msun$ star on the MS. Predictive mixing is able to achieve the
desired $\gradr=\grada$ outcome in most of these cases.

However, two cases shown in \mesafour\ 
 continue to exhibit
$\gradr > \grada$ on the convective side of a
boundary of a convection region,
highlighting the need for further work and motivating us to develop a new scheme for treating
convective boundaries. This scheme, which we dub ``convective
premixing'', draws inspiration from earlier work by
\citet{castellani_1985_aa} and \citet{mowlavi_1994_aa}. In the
following, we first discuss why predictive mixing sometimes fails. Then, we describe
the new convective premixing scheme in detail
(Section~\ref{s.cpm.scheme}), and demonstrate its application in
various evolutionary scenarios
(Sections~\ref{s.cpm.M16}--\ref{s.cpm.M1.5}).



\subsection{The Failure of Predictive Mixing} \label{s.cpm.pred}

The $16\,\Msun$ MS scenario presented in Figure~4 of \mesafour\ exhibits
a convective shell above the abundance-gradient region with
$\gradr > \grada$ at its lower boundary.
If predictive mixing is applied to the shell, the lower boundary advances downward
to merge with the core while the upper boundary remains fixed in position.
The end result is that the entire
abundance-gradient region mixes into the core, delivering
significant quantities of fresh H. Such behavior is unphysical,
and in \mesafour\ we made the pragmatic choice to avoid this outcome
by disabling predictive mixing for the convective shell.


Additionally, the $1\,\Msun$ CHeB scenario presented in Figure 6 of
\mesafour\ exhibits $\gradr > \grada$ at the upper boundary of its
convective core.  This issue cannot be resolved by predictive mixing,
as discussed in \mesafour.

In both of these scenarios, the problem encountered is an unavoidable
consequence of the design of the predictive mixing scheme, which
manipulates the diffusion coefficients at cell faces and then relies
on \mesa's abundance solver (\mesaone, Section 6.2) to update the
model.  In contrast, the new convective premixing scheme directly
updates the abundances, as we now describe.

\subsection{The Convective Premixing Scheme} \label{s.cpm.scheme}

\begin{figure*}
\begin{center}
\includegraphics{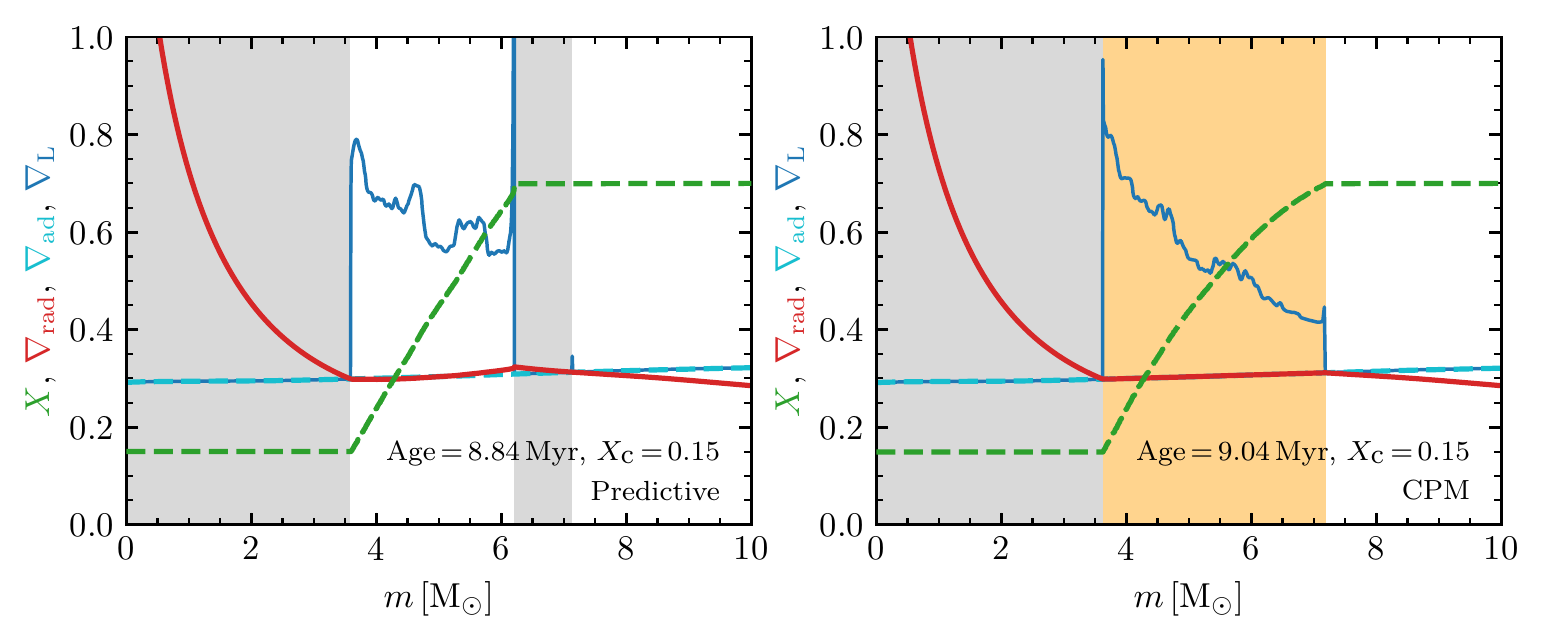}
\caption{Profiles of $\gradr$, $\grada$, $\gradL$, and $\Xhyd$ as a
  function of mass coordinate, in the inner part of the $16\,\Msun$ MS
  star at $\Xcore=0.15$. The panels show the separate runs described
  in the text. Gray (gold) shading indicates convection
  (semiconvection) regions.}
\label{fig:cpm-M16-profiles}
\end{center}
\end{figure*}

The convective premixing (CPM) scheme is applied at the start of each
time step, before any structural or compositional changes that arise
due to the evolution of the star. It proceeds by finding the cells
where $y > 0$ on one face (convective) and $y < 0$ on the other face
(radiative). For each of these initial boundary cells, the algorithm
considers whether $y$ on the radiative face would change if the
adjacent cell outside the convection region is mixed completely with
the rest of the convection region. This putative mixing is performed
at constant cell pressure and temperature, and involves recalculating
abundances, densities, opacities and the various temperature gradients
(\gradr, \grada, \gradL) throughout the convection region plus the
adjacent cell.

If the radiative face of the boundary cell becomes convective during
this putative mixing, then the mixing is committed to the model,
overwriting the composition profile throughout the entire
(newly extended) convection region. Then, the next adjacent cell
outside the convection region is considered for incorporation. This
process continues iteratively until the radiative face of the current
convective boundary remains radiative during the putative mixing.

At a given point in its evolution, a star typically exhibits multiple
convective boundaries (for instance, the left panel of
Figure~\ref{fig:cpm-M16-profiles} shows three). The order in which CPM
processes these boundaries is determined by evaluating a
characteristic mixing timescale $\dr{} / \vconv$ for the initial
boundary cells; here, $\vconv$ is the convective velocity on the $y >
0$ face of the boundary cell, and $\dr{}$ is its radial extent. The
boundary with the smallest time scale is processed first, then the
boundary with the next-smallest time scale, and so on.

In CPM the mixing is treated as instantaneous.  Given that the convective diffusivity $\Dconv$ is well in
excess of $10^{10}\,{\rm cm^{2} s^{-1}}$ even near convective
boundaries (see, e.g., Figures~11 and~29 of \mesatwo), the
characteristic mixing timescale is $\tau \sim \Delta r^{2}/\Dconv
\lesssim 10^{4}\,{\rm yr} $ for a typical region with extent $\Delta r
\lesssim \Rsun$. This is small compared to typical nuclear timescales,
and so the assumption of instantaneous mixing seems warranted for all
but the most rapid evolutionary phases.

During its iterations, CPM naturally handles the transition of cell
faces inside the convection region from convective to radiative. This
typically happens either at the boundary opposite to the advancing one
(causing that boundary to retreat), or at a point inside the
convection region (causing the region to split). Because mixing through a face ceases when it transitions from
convective to radiative, newly-transitioned faces should be very close
to convective neutrality. To improve how closely neutrality is
achieved, our CPM implementation divides the cell outside the
advancing boundary into a number of virtual sub-cells. Each of these
sub-cells is mixed into the convective region in turn, until all have
been incorporated. The number of sub-cells is automatically adjusted to
ensure that, during each sub-cell mix, at most a single face within
the current convective region transitions to radiative.

\subsection{Evolution of a Retreating Convective Core on the Main Sequence} \label{s.cpm.M16}

\begin{figure*}
\begin{center}
\includegraphics{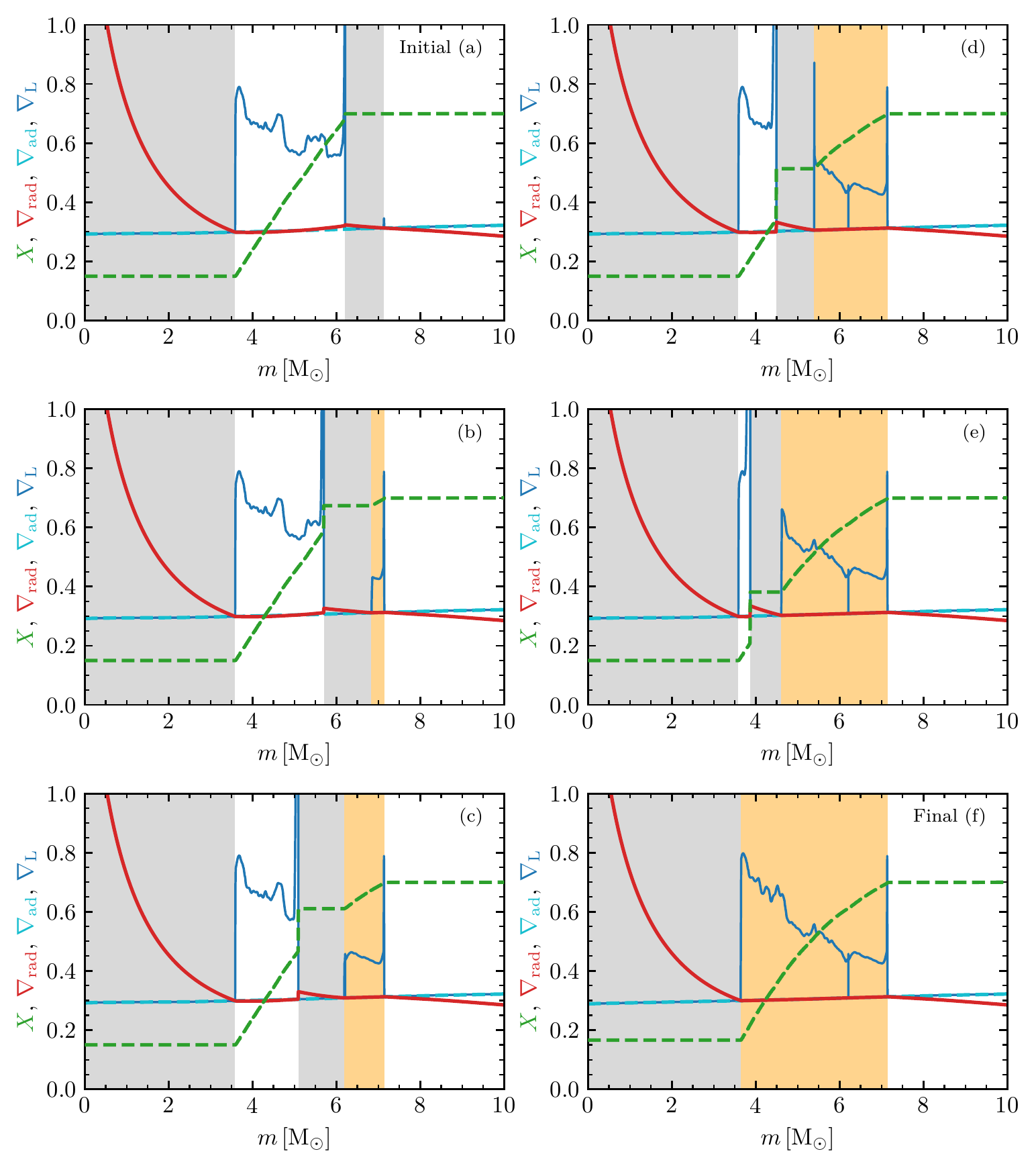}
\caption{Profiles of $\gradr$, $\grada$, $\gradL$, and $\Xhyd$ as a
  function of mass coordinate, in the inner part of the $16\,\Msun$ MS
  star at $\Xcore=0.15$. The panels show the outcome of an artificial
  switch from the predictive mixing scheme to the CPM scheme. The
  initial state (a) is shown in the upper left panel, and the
  subsequent panels, running in alphabetical order (b),\ldots,(f),
  show how this state evolves during the CPM iterations.
  Gray (gold) shading indicates convection
  (semiconvection) regions.}
\label{fig:cpm-M16-movie}
\end{center}
\end{figure*}

We evolve a $16\,\Msun$ star from the ZAMS to the TAMS.  This is the
scenario considered in Section~2.3 of \mesafour, and illustrates the
behavior of MS stars with retreating convective cores. Here, and for
the scenarios presented in the following sections, we assume an
initial He mass fraction $Y=0.28$ and an initial metal mass fraction
$Z=0.02$, and we ignore rotation and mass loss.
Figure~\ref{fig:cpm-M16-profiles} plots the profiles of \gradr,
\grada, \gradL, and \Xhyd\ in the inner parts of the star, at a point
nearing the TAMS ($\Xcore=0.15$). The left panel illustrates a run
using the predictive mixing scheme applied at the boundary of the
convective core, while the right panel shows a run using the CPM
scheme; in both \revision{calculations}, the Ledoux  \revision{criterion is used to assess convective stability,
  so that $y = y_{\rm L}$.}

In the left panel, the convective shell discussed in
Section~\ref{s.cpm.pred} can clearly be seen at the top of the
abundance-gradient region, spanning mass coordinates $6.2 \lesssim
m/\Msun \lesssim 7.1$. In the right panel, the shell is absent and the
abundance-gradient region is wider and shallower, extending all the
way out to $m/\Msun \approx 7.2$. Moreover, the region is very close
to adiabatic ($\gradr = \grada$) , in contrast
to the super-adiabatic stratification ($\gradr > \grada$) seen in the
left panel.

\begin{figure}
\begin{center}
\includegraphics{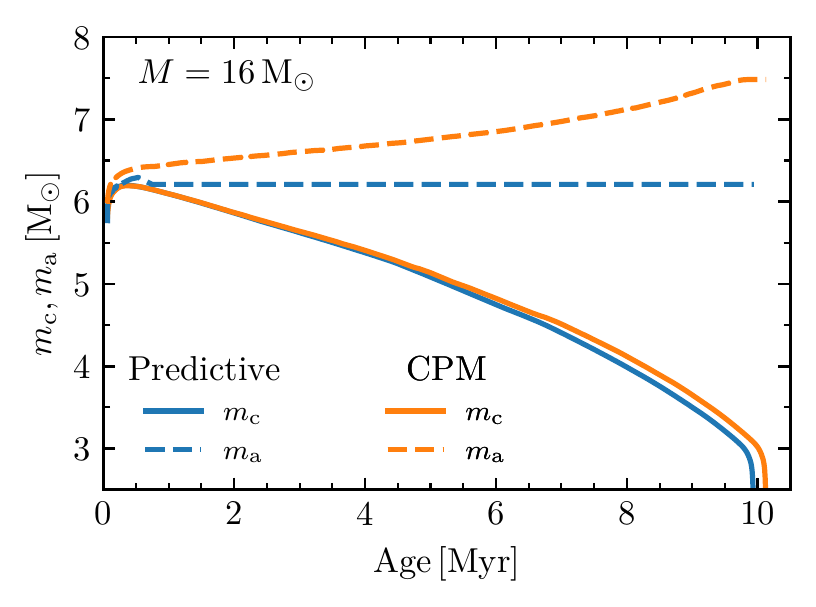}
\caption{Mass coordinates of the convective core boundary (\mcore) and
  the top of the abundance-gradient region (\mabund) as a function of
  MS age, for the 16\,\Msun\ star. Different line colors
  show the separate runs discussed in the text.}
\label{fig:cpm-M16-bounds}
\end{center}
\end{figure}

\citet{schwarzschild_1958_aa} were the first to predict the appearance
of 
\revision{adiabatically stratified} abundance-gradient regions outside
the convective cores of massive MS stars, labeling them
`semiconvection regions'. The key feature of a semiconvection region
is that  \revision{the adiabatic
  stratification} is continuously maintained by a gradual adjustment
of opacity due to the changing abundance profile.

To illustrate how this adjustment naturally arises within the CPM scheme,
we simulate a single evolutionary time step of the $16\,\Msun$ star
where we artificially switch from the predictive scheme to CPM. The
starting configuration is the profile shown in the left panel of
Figure~\ref{fig:cpm-M16-profiles}. This panel is reproduced as the
upper-left panel of Figure~\ref{fig:cpm-M16-movie}; the subsequent
panels then show the evolving profiles at selective intermediate
stages during the CPM iterations. A clear
narrative emerges from these panels: as the lower boundary of the
convective shell advances inward, the cell faces near the upper
boundary of the shell transition from convective to radiative, causing
that boundary to retreat inward. Because there are no abundance
gradients within the shell itself, the Ledoux and Schwarzschild
 criteria give the same condition for this transition to
occur: $\gradr = \grada$. The overall effect is that the shell
propagates inward as a whole, leaving behind it a `wake' with an
adiabatic stratification. Eventually, the propagating shell merges
with the core, leading to a final state (seen in the lower-right panel
of Figure~\ref{fig:cpm-M16-movie}) that closely resembles the right
panel of Figure~\ref{fig:cpm-M16-profiles} (the small differences are
because the former has an evolutionary history determined by the
predictive mixing scheme, while the latter has a history determined by
CPM).

The seminal paper by \citet{schwarzschild_1958_aa} triggered
significant interest in semiconvection regions, with a particular
focus on 
\revision{their final stratification}.  \citet{sakashita_1961_aa}
argued that 
\revision{$y_{\rm L} = 0$} should apply in semiconvection regions,
rather than  \revision{$y_{\rm S} = 0$} as
originally proposed. However, \citet{kato_1966_aa} reasoned that
because  \revision{the former
  stratification is} super-adiabatic ($\gradr > \grada$), slow mixing
by overstable g-mode oscillations will 
\revision{drive it} toward  \revision{the same
  $y_{\rm S} = 0$ outcome}. Subsequently, \citet{gabriel_1970_aa}
suggested that Kato's mechanism is 
\revision{superfluous}, due to the appearance of propagating
convective shells that continually adjust the abundance profile to
achieve  \revision{$y_{\rm S} = 0$}. Gabriel's narrative closely mirrors the one we give above,
and the correspondence between his Figure~1 and our
Figure~\ref{fig:cpm-M16-movie} is striking.

A possible source of confusion in this discussion is the relationship
between the semiconvection region shown in the right panel of
Figure~\ref{fig:cpm-M16-profiles}, and the semiconvective mixing
discussed in Section 4.1 of \mesatwo.    \revision{The latter implements the mixing envisaged by
  \citet{kato_1966_aa}; while it will ultimately yield a $y_{\rm S} =
  0$ stratification, it is a fundamentally different mechanism than
  the propagating convective shells shown in
  Figure~\ref{fig:cpm-M16-movie}.}  A critical distinction lies in the
role played by the convective stability criterion. While our
calculations adopt the Ledoux criterion \revision{($y=y_{\rm L}$)},
repeating them with the Schwarzschild criterion \revision{($y = y_{\rm
    S}$)} leads to results very similar to the ones already shown
(although the abundance profiles are rather more jagged). In contrast,
Kato's mechanism \emph{requires} the Ledoux criterion to establish an
initially super-adiabatic stratification, \revision{which the mechanism
  then drives toward adiabaticity}.  In hindsight, it seems prudent to
reserve the label `semiconvection' for the adiabatically
 \revision{stratified} regions envisaged by
\citet{schwarzschild_1958_aa}, and avoid using it as in \mesatwo\ to
describe a mechanism that can generate these regions \citep[for
  additional examples of this conflation, see
  e.g.,][]{silva_aguirre_2010_aa,noels_2010_aa,ding_2014_aa,moore_2016_aa}.

To summarize the core and near-core evolution of the $16\,\Msun$ star,
Figure~\ref{fig:cpm-M16-bounds} plots the mass coordinate \mcore\ of the
convective core boundary as a function of MS age for the separate
runs using the predictive mixing and CPM schemes. We also show the
mass coordinate $\mabund$ of the top of the abundance-gradient region
outside the core; this coincides with the maximal extent of the core
in the predictive mixing case, and to the top of the semiconvection
region in the CPM case. During the star's evolution, the opacity change due to progressive
H depletion at the center lowers $\gradr$ throughout the core,
causing its boundary to retreat inward. In the CPM case, this retreat
is mirrored by an outward growth of the semiconvection region, a
behavior first noted by \citet{schwarzschild_1958_aa}. The growth
slowly feeds fresh H from the envelope down into the core,
which causes the core to shrink slightly less rapidly than in the
predictive mixing case, thereby marginally prolonging the star's MS
lifetime.

\begin{figure}
\begin{center}
\includegraphics{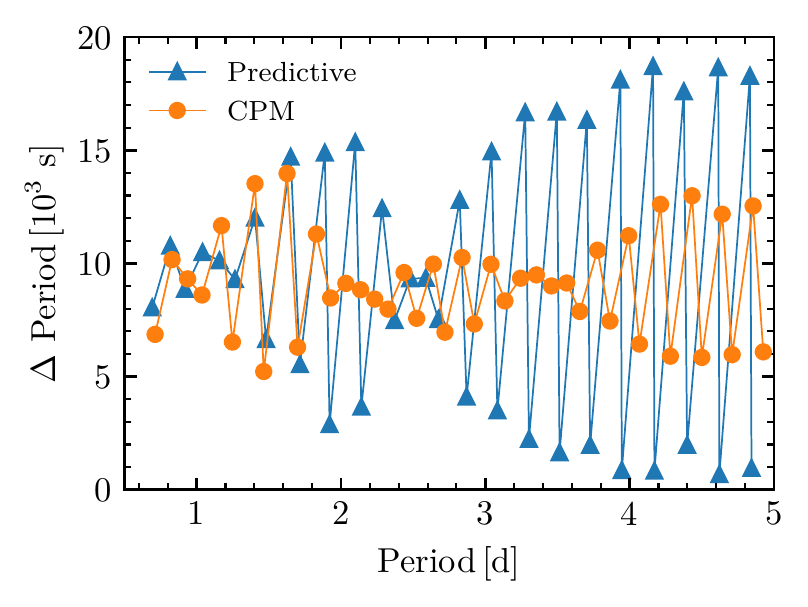}
\caption{Period spacings $\Delta P$ plotted as a function of period
  $P$, for $\ell=2$ g-modes of the 16\,\Msun\ MS star at
  $\Xcore=0.15$. Different symbols show the separate runs
  discussed in the text.}
\label{fig:cpm-M16-periods}
\end{center}
\end{figure}

As the star nears the TAMS, the mass $\mabund-\mcore$ of the
abundance-gradient region is $\approx$30\% larger in the CPM case than
with predictive mixing. This region plays a key role in determining
the shape of the \BV\ frequency near the core, and the two cases
should therefore exhibit differences in their g-mode oscillation
spectra. We explore this by using release 5.2 of
\GYRE\ \citep{townsend_2013_aa,townsend_2018_aa} to
evaluate the star's normal-mode frequencies at
$\Xcore=0.15$. Figure~\ref{fig:cpm-M16-periods} plots the resulting
period echelle diagram (period $P$ versus period-spacing $\Delta P$)
for $\ell=2$ g modes in the period range 0.5--5\,d. Both cases show
significant departures from the uniform period spacing $\Delta P
\approx 10^{4}\,{\rm s}$ predicted by asymptotic theory; these
departures are caused by the abundance-gradient region, and therefore
are an asteroseismic diagnostic of the star's age
\citep[e.g.,][]{miglio_2008_aa}. As the figure shows, the CPM scheme
yields significantly smaller non-uniformity in $\Delta P$ for $P
\gtrsim 2\,{\rm d}$ than the predictive mixing scheme. These
different outcomes are potentially testable by the \emph{TESS} mission
\citep{ricker_2016_aa}, which will observe many massive stars, some
with the long time baselines necessary to detect g-modes with
multi-day periods. It will first be necessary to explore
whether the differences persist when the additional effects of core
overshoot and rotation are included.

\subsection{Evolution of the Convective Core During Core He Burning} \label{s.cpm.M1}

\begin{figure*}
\begin{center}
\includegraphics{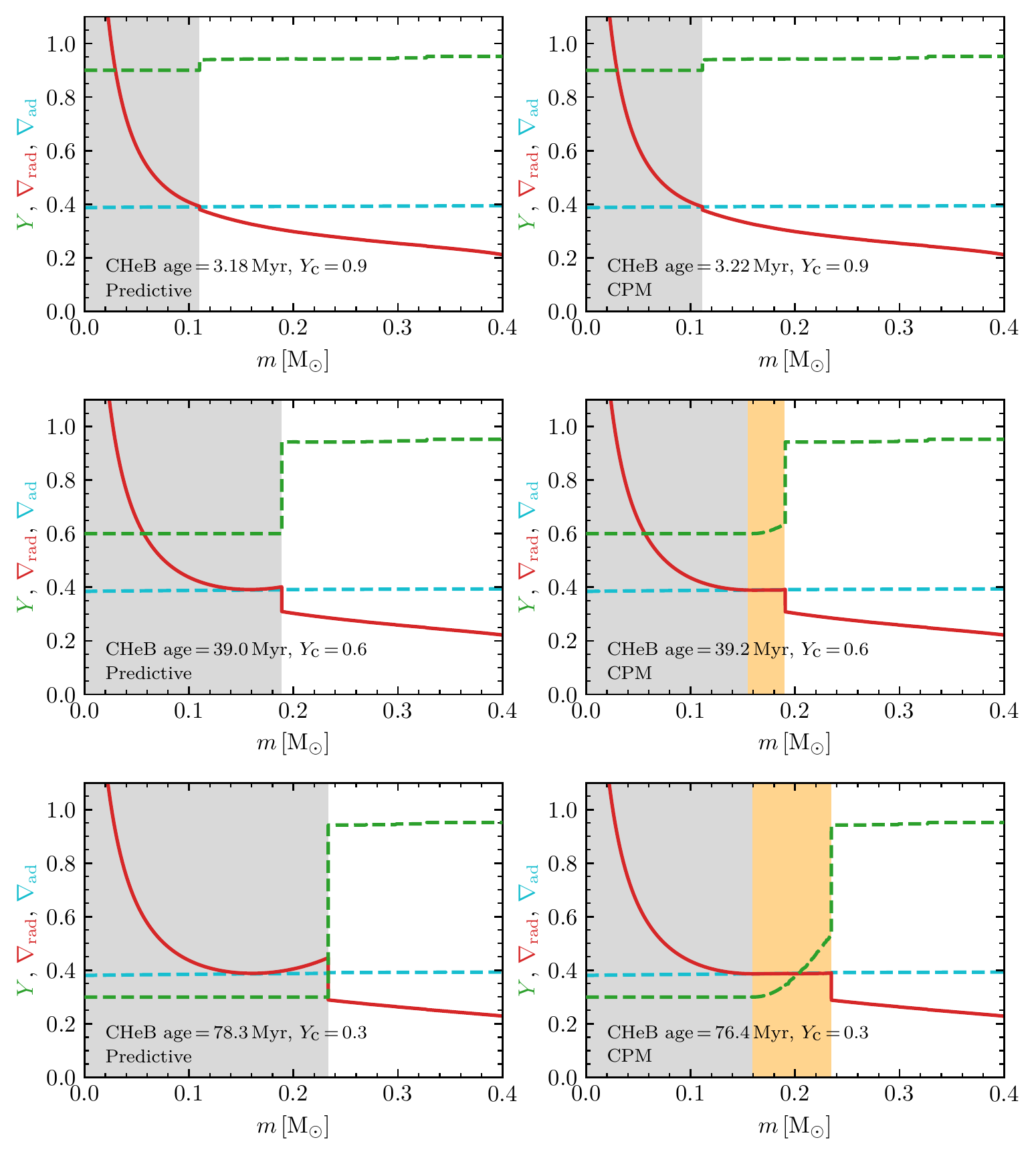}
\caption{Profiles of $\gradr$, $\grada$, and $\Yhel$  as a function of mass coordinate, in
  the inner part of the $1\,\Msun$ star. The panels correspond to
  different stages during the CHeB phase: $\Ycore=0.9$ (upper panel),
  $\Ycore = 0.6$ (middle panel), and $\Ycore = 0.3$ (lower panel). The left panels
  show the run using the predictive mixing scheme, and the right
  panels the run using CPM. Gray (gold) shading indicates convection
  (semiconvection) regions.}
\label{fig:cpm-M1-profiles}
\end{center}
\end{figure*}

\begin{figure}
\begin{center}
\includegraphics{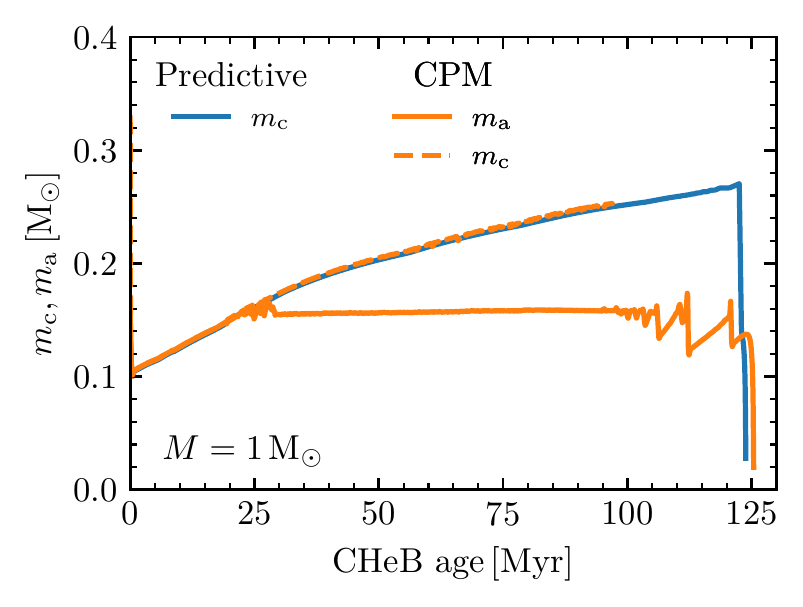}
\caption{Mass coordinates of the convective core boundary (\mcore) and
  the top of the abundance-gradient region (\mabund) as a function of
  CHeB age, for the 1\,\Msun\ star. Different line colors
  show the separate runs discussed in the text.}
\label{fig:cpm-M1-bounds}
\end{center}
\end{figure}

We now evolve a $1\,\Msun$ star through the CHeB phase; this is the
scenario considered in Section~2.4 of \mesafour, and illustrates the
behavior of He-burning stars with growing convective
cores. Figure~\ref{fig:cpm-M1-profiles} plots the profiles of \gradr,
\grada, and \Yhel\ in the inner parts of the star, at three points
during its evolution corresponding to $\Ycore=0.9$, 0.6 and 0.3. The
left panels show a run using the predictive mixing scheme, while the
right panels show a run using the CPM scheme; in both 
\revision{calculations}, the Ledoux  \revision{criterion is used to assess convective
  stability}.

The left panels, which reprise Figure~6 of \mesafour, show how a local
minimum in \gradr\ has developed by $\Ycore = 0.6$. This minimum is
held just above the $\grada$ threshold using the
\texttt{predictive\_superad\_thresh} control, with the result that the
core continues to grow slowly without splitting. With the CPM scheme, a growing semiconvection region
develops above a smaller convective core.


The emergence of semiconvection regions during the CHeB phase may have
first been proposed by \citet{schwarzschild_1969_aa}, but it was
\citet{castellani_1971_ab} and \citet{eggleton_1972_aa} who considered
the possibility in detail. Later, \citet{castellani_1985_aa} described
a `concatenated convective mixings' scheme for simulating the
formation of the semiconvection region; their Figure 3, which can be
regarded as a CHeB analog to our Figure~\ref{fig:cpm-M16-movie}, reveals
how the core splits to form a convective shell, which then propagates
outward leaving a wake with an adiabatic
stratification. \citet{mowlavi_1994_aa} demonstrated how the
\citet{castellani_1985_aa} scheme can be generalized to work in other
evolutionary phases, and together these two papers provided the
original inspiration for the CPM scheme.

To summarize the core and near-core evolution of the $1\,\Msun$ star,
Figure~\ref{fig:cpm-M1-bounds} plots the mass coordinate \mcore\ of the
convective core boundary as a function of CHeB age, for the separate
runs using the predictive mixing and CPM schemes. For the CPM run, we
also show the mass coordinate $\mabund$ of the top of the
abundance-gradient region outside the core; there is no abundance-gradient region
in the predictive mixing run. The figure shows that the semiconvection
region forms in the CPM run at an age $\approx 25\,\Myr$. The top of the
semiconvection region then closely tracks the outward growth of the
core boundary from the predictive mixing run, through to an age $\approx 95\,\Myr$. At this juncture, oscillations start to occur in
\mcore\ for the CPM run. These oscillations
are due to breathing pulses, which disrupt the semiconvection region
and introduce discontinuities in the abundance profile. Because
$\mabund$ becomes ill-defined when the discontinuities appear, we do
not plot it beyond this point. After two final, large-amplitude
pulses, the star reaches the end of the CHeB at an age $\approx
125\,\Myr$, slightly later than the final age for the predictive
mixing run.

Debate continues as to whether core breathing pulses during the CHeB phase
are physical or numerical \citep[see, e.g.,][and references
  therein]{salaris_2017_aa}. In the present context, we note that the
instantaneous mixing assumed in the CPM scheme will tend to exacerbate
the pulses, because it does not account for the finite time required
for He ingested at the top of the semiconvection region to be
transported down to the core. We are currently considering
improvements that address this shortcoming, but in the meantime we
recommend that the CPM scheme be used with caution in the late stages
($\Ycore \lesssim 0.15$) of the CHeB phase.

\subsection{Evolution of a Growing Convective Core on the Main Sequence} \label{s.cpm.M1.5}

\begin{figure*}
\begin{center}
\includegraphics{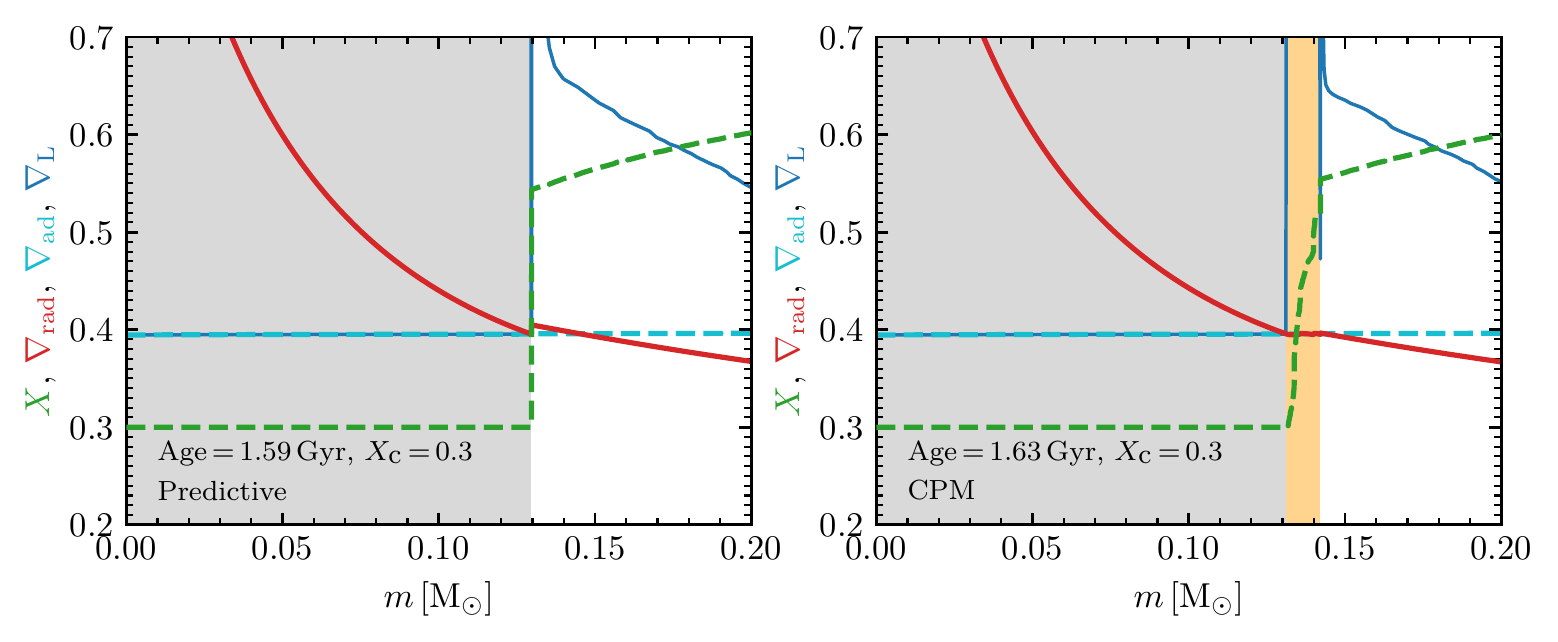}
\caption{Profiles of $\gradr$, $\grada$, $\gradL$, and $\Xhyd$ as a
  function of mass coordinate, in the inner part of the $1.5\,\Msun$
  MS star at $\Xcore=0.30$. The panels show the separate runs
  described in the text. Gray (gold) shading indicates convection
  (semiconvection) regions.}
\label{fig:cpm-M1.5-profiles}
\end{center}
\end{figure*}

\begin{figure}
\begin{center}
\includegraphics{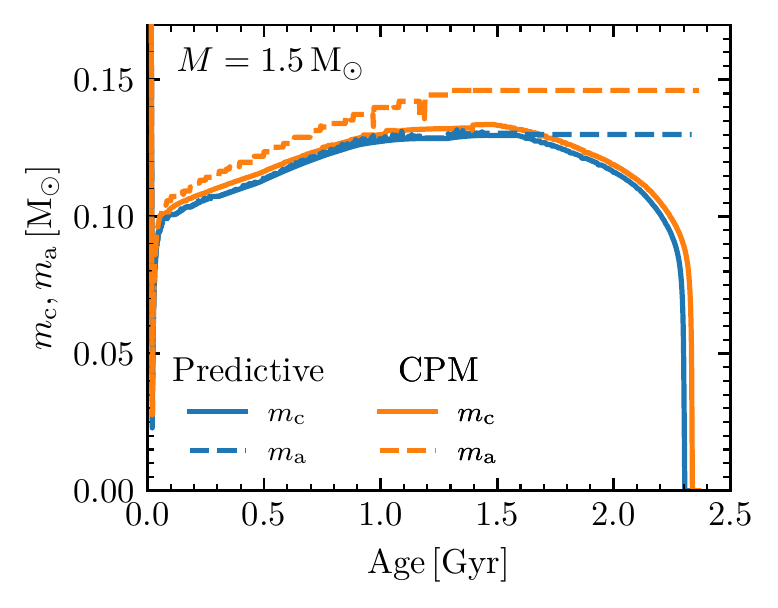}
\caption{Mass coordinates of the convective core boundary (\mcore) and
  the top of the abundance-gradient region (\mabund) as a function of
  MS age, for the 1.5\,\Msun\ star. Different line colors
  show the separate runs discussed in the text.}
\label{fig:cpm-M1.5-bounds}
\end{center}
\end{figure}

We now evolve a $1.5\,\Msun$ star from the ZAMS to the TAMS; this is
the scenario considered in Section~2.4 of \mesafour, and illustrates
the behavior of MS stars with initially growing convective
cores. Figure~\ref{fig:cpm-M1.5-profiles} plots the profiles of
\gradr, \grada, \gradL\ and \Xhyd\ in the inner parts of the star when
$\Xcore=0.30$.  The left panel illustrates a run using the predictive
mixing scheme, while the right panel shows a run using the CPM scheme;
in both  \revision{calculations}, the Ledoux
 \revision{criterion is used to
  assess convective stability}.

The left panel shows a small super-adiabatic region just outside the
core that is stabilized against convection by the abundance gradient
($\grada < \gradr < \gradL$). If slow mixing via the
\citet{kato_1966_aa} mechanism were allowed to proceed, this region
would eventually transform into a semiconvection region. The right
panel shows that the CPM scheme naturally reproduces this
semiconvection region.
\citet{ledoux_1947_aa} suggests the existence of this semiconvection region
(though not labeled as such). The mean molecular weight profile
$\mu \propto m\,p^{7/5}$ he obtains for the abundance-gradient region outside the core
yields an adiabatic stratification ($\gradr=\grada$).

To bring the present analysis to a close,
Figure~\ref{fig:cpm-M1.5-bounds} plots the mass coordinate \mcore\ of
the convective core boundary as a function of MS age, for the separate
$1.5\,\Msun$ runs using the predictive mixing and CPM schemes. We also
show the mass coordinate $\mabund$ of the top of the
abundance-gradient region outside the core. In the predictive mixing
case, this region first appears at an age $\approx 1.5\,\Gyr$, when
the core boundary reverses direction and begins to retreat inward. In
the CPM case, the abundance-gradient region is present from the start,
and coincides with the semiconvection region up until an age $\approx
1.75\,\Gyr$. At this point, $\gradr$ outside the core drops below
$\grada$, and the semiconvection region becomes radiative; then,
without any further ingestion of H from the envelope, $\mabund$ remains fixed.

\section{Parallel Performance}\label{s.parallel}
Here we provide updates on the parallel performance of \mesastar.  For each test, simulations
with different numbers of parallel threads were performed on the same computer with
no other CPU-intensive tasks taking place. The tests were performed on one Intel Xeon E5-2699V4 processor
with 22 physical cores.  Although this processor allows hyperthreading
(i.e., two threads running on one physical core), we restrict these tests to one thread per physical core
because tests found that enabling hyperthreading results in a performance penalty rather
than a benefit.

\subsection{Parallel Scaling of \MESAstar}
\mesa\ uses the OpenMP application programming interface to parallelize certain operations.
Among these we distinguish three categories. The first category is operations that are parallel per cell
in the stellar model, including the EOS,
opacity, and nuclear network. The next category concerns parts of the problem that are a
mixture of parallel and serial execution.  These include matrix manipulations, the Newton-Raphson solver,
and atomic diffusion calculations. The final category is those operations that are serial,
such as the main evolve loop and the adjustment of the total mass of the star via mass loss
or accretion.

Three \MESA\ \code{test\_suite} cases are considered:
\texttt{black\_hole}, \texttt{7M\_prems\_to\_agb}, and \code{1M\_pre\_ms\_to\_wd}.\footnote{The test
  suite cases are generally split into distinct, sequential parts. For the timing tests we used the
  longest part of each test.}
The \texttt{black\_hole} test uses 6 structure variables and a nuclear network with 22 isotopes.
The \texttt{7M\_prems\_to\_AGB} test uses 4 structure variables and a nuclear network with 10 isotopes.
The \texttt{1M\_pre\_ms\_to\_wd} test uses 4 structure variables and a nuclear network with 8 isotopes and
includes rotation.

\begin{figure}[htb!]
\centering
\includegraphics[width=1.0\columnwidth]{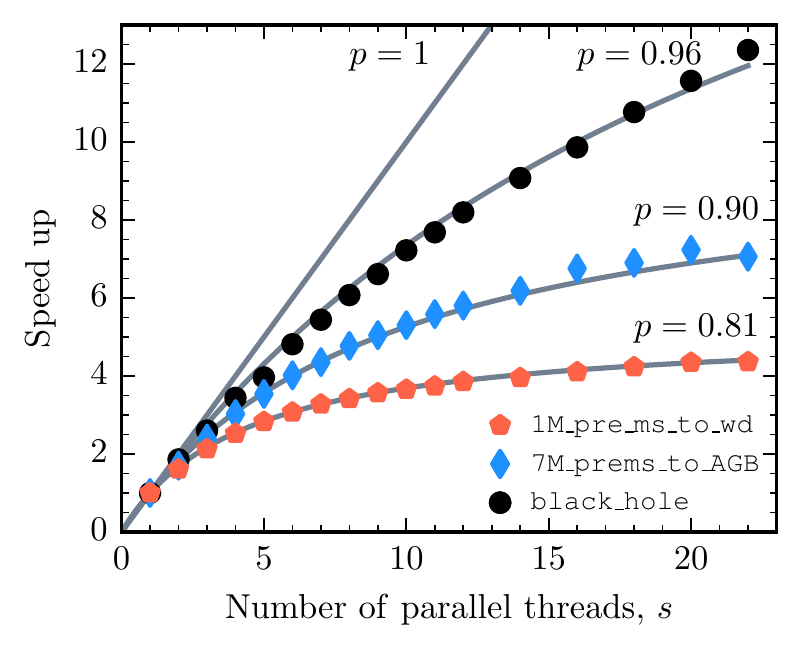}
\caption{ Speed up for the three test cases described in the text when run on different
  numbers of parallel threads from 1 to 22. 
  Lines are illustrations of Equation~\eqref{eq:spdup} for different values of $p$.}
\label{f.speedup}
\end{figure}

As a metric, we define `speed up'
to be the ratio of the measured run-time on 1 thread to that on N threads.  
The theoretical speed up,
\begin{equation}\label{eq:spdup}
\frac{1}{(1-p)+(p/s)},
\end{equation}
is predicted by \citet{Amdahl1967}, 
where, for our purposes, $s$ is the number of parallel threads 
and $p$ is the fraction of the code that will benefit from parallel execution for a given test case.

Figure \ref{f.speedup} shows the speed up for each of the three cases.
In general, we expect cases with more variables (both structure and network) to benefit
more from parallel execution. Indeed, the \code{black\_hole} case shows the best
scalability with $p \approx 0.96$, followed by \code{7M\_prems\_to\_AGB} with 
$p \approx 0.90$, and \code{1M\_pre\_ms\_to\_wd} with $p \approx 0.81$.
These results are consistent with the numbers of variables included in the respective tests.

\begin{figure*}[htb!]
\centering
\includegraphics[width=0.95\textwidth]{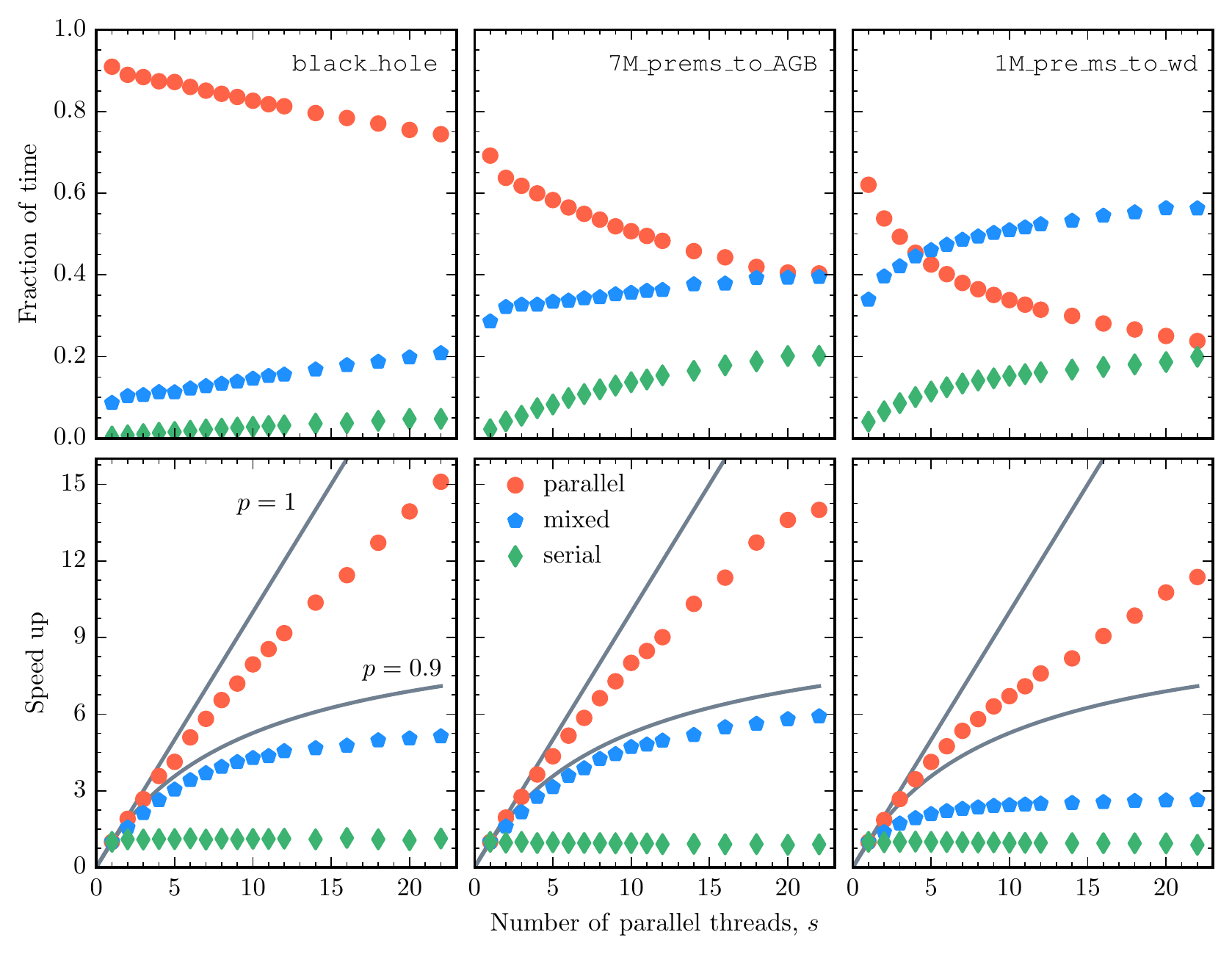}
\caption{
  Breakdown of the fraction of the execution time (upper row)
  and the speed up (lower row) of three different categories listed in the text.
  Each column is a single test case. The line in the lower panels is Equation~\eqref{eq:spdup} for $p=0.9$ and $p=1$.}
\label{f.par_combined}
\end{figure*}

For each case, Figure \ref{f.par_combined} shows the relative fraction for the three categories described above 
and the speed up. The parts of \mesastar\ that are parallel-per-cell scale nearly linearly
with the number of threads while the mixed category scales weakly
and the serial component is essentially flat. The serial portion becomes an increasing fraction of
the total runtime as the other categories decrease with an increasing number of parallel threads. This may
become a greater issue with the move towards many-core processors in the future.

\subsection{OP Mono Opacities and Radiative Levitation}\label{s.radlev}

\begin{figure}
\centering
\includegraphics[width=1.0\columnwidth]{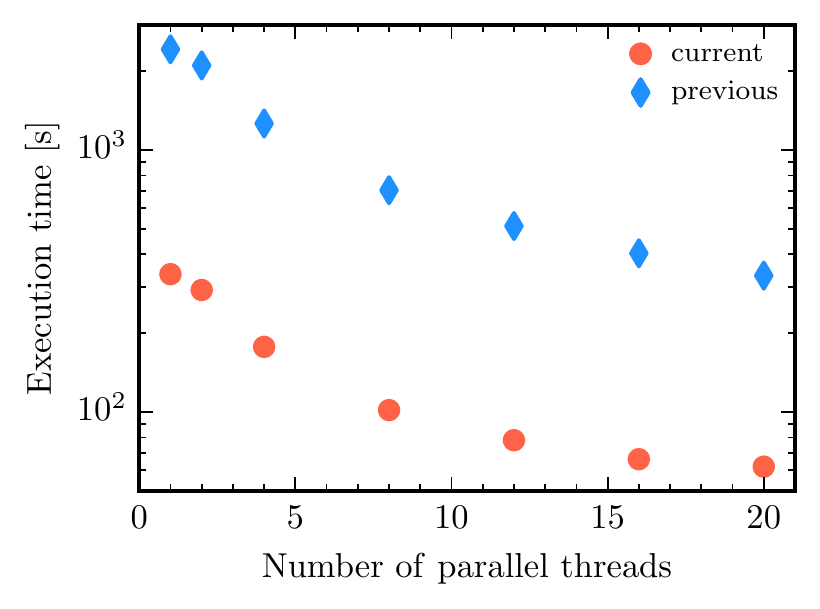}
\caption{Execution time of the \code{radiative\_levitation} test suite case. The same
  test was run using \MESA\ versions before and after the changes described herein to demonstrate the difference
  in performance obtained by these improvements.}
\label{fig:OP}
\end{figure}

\mesathree\ (Section 9) describes the inclusion of radiative
levitation in \MESA\ via the work of \citet{Hu2011}.  These
capabilities were originally developed as part of \code{STARS} \citep{eggleton_1971_aa,Pols_1995_aa} and evaluate the opacity and radiative
acceleration using the OP monochromatic opacity tables \citep{Seaton2005}.
Due to the differing approaches of \code{STARS} and \MESA, a number of
derivatives were being calculated but not used in the \MESA\
implementation of the opacity routines.  By eliminating the evaluation
of these unused quantities, and by pre-computing some frequently re-used
stimulated emission factors, we achieved at least a factor of 5 reduction
in the time required to evaluate opacities and radiative accelerations.
These optimizations translate into an improvement in total runtime relative to
previous versions of \MESA\ when making use of the
OP monochromatic opacities or radiative levitation capabilities
without any compromise to the
numerical results.

We demonstrate in Figure~\ref{fig:OP} the difference in execution time for runs of the
\code{radiative\_levitation} test suite case using \MESA\ versions before and after the changes described herein.
The computational expense of calculating the radiative accelerations
continues to dominate the runtime
of such models, accounting for more than 65\% of run-time even when using 20 parallel threads, meaning these capabilities can benefit from progress toward many-core architectures.

\section{Summary}\label{s.conclusions}

We explain significant new capabilities and improvements 
implemented in \MESA\ since the publication of \mesaone, \mesatwo, \mesathree\ and \mesafour.
The addition of the \RSP\ radial pulsation functionality in \MESAstar\
(Section~\ref{s.rsp}) provides a new capability to model radially-pulsating variable stars.
Advances to \MESA\ in 
numerical energy conservation (Section~\ref{s.energy}),
rotation factors and gravity darkening (Section~\ref{s.rot} and Appendix~\ref{a.rot}), and
convective boundaries (Section \ref{s.cpm}), 
will open opportunities for future investigations in stellar evolution.
Improvements in the computational efficiency of \MESA\ on 
current generation multicore x86 instruction set architectures
(Section~\ref{s.parallel})
will inform future development directions.
Upgrades to the EOS and nuclear reaction physics (Appendix \ref{s.updates})
will increase the robustness of stellar evolution models.
Discussion of the current treatment of fallback and 
comparisons of the thermodynamic evolution of supernova models from different software instruments (Appendix~\ref{s.snec})
will enhance the study of massive star explosions. 
Introduction of the \MESA\ Testhub software infrastructure  (Appendix~\ref{s.testhub}) for web-based, automated, 
daily examination of the \mesastar{} and \mesabinary{} test suites will lead 
to more efficient source code development.
Input files and related materials for all the figures are available at \url{http://mesastar.org}
and \revision{\url{https://doi.org/10.5281/zenodo.2582656}}.

\acknowledgements


We thank Warrick Ball and Evan Bauer for their sustained engagment with the \MESA\ project. 
Both Richa Kundu and Susmita Das graciously shared their variable star calculations. 
It is a pleasure to thank 
Conny Aerts, 
Sean Couch,
Franck Delahaye,
Luc Dessart,
Ebraheem Farag, 
Carl Fields, 
Chris Fontes,
Chris Fryer,
Jim Fuller, 
Falk Herwig, 
Thomas Janka,
Sam Jones, 
Sanjib Gupta,
Joyce Guzik,
Max Katz,
John Lattanzio,
Abhijit Majumder,
Wendell Misch,
\revision{Joey Mombarg,}
Viktoriya Morozova,
Sterl Phinney,
Eliot Quataert, 
Rene Reifarth,
Toshio Suzuki,
Katie Mussack Tamashiro,
Dean Townsley,
Todd Thompson,
Suzannah Wood and
Mike Zingale
for discussions.
We also thank the participants of the 2018 MESA Summer
School for their willingness to experiment with new capabilities.
The improvements discussed in Section 5 were in large parts discussed during RHDT and AT stays at the Munger residence.


The \MESA\ project is supported by the National Science Foundation (NSF)
under the Software Infrastructure for Sustained Innovation program grants 
(ACI-1663684, ACI-1663688, ACI-1663696).
This research benefited from interactions that were funded in part by the Gordon and Betty Moore Foundation through Grant GBMF5076 and was also supported at UCSB by the NSF under grant 17-48958.
This research was also supported at ASU by 
the NSF under grant PHY-1430152 for the Physics Frontier Center ``Joint Institute
for Nuclear Astrophysics - Center for the Evolution of the Elements'' (JINA-CEE).
Support for this work was provided by NASA through Hubble Fellowship
grant \# HST-HF2-51382.001-A awarded by the Space Telescope Science
Institute, which is operated by the Association of Universities for
Research in Astronomy, Inc., for NASA, under contract NAS5-26555. The Center for Computational Astrophysics 
at the Flatiron Institute is supported by the Simons Foundation.

\revision{P.M. acknowledges support from NSF grant AST-1517753 and the Senior Fellow of the Canadian Institute for Advanced Research (CIFAR) program in Gravity and Extreme Universe, both granted to Vassiliki Kalogera at Northwestern University.}
S.M.K thanks the Indo-US Science and Technology Forum for financial support.
A.T. is a Research Associate at the Belgian Scientific Research Fund (F.R.S-FNRS).
R.F is supported by the Netherlands Organisation for Scientific Research (NWO) through a top module 2 grant with project number 614.001.501 (PI de Mink).
R.S. acknowledges support from the IdP II 2015 0002 64 grant of the Polish Ministry of Science and Higher Education \revision{and from SONATA BIS grant, 2018/30/E/ST9/00598, from the National Science Center, Poland}.
R.H.D.T. acknowledges support from the NSF under grant AST-1716436.
M.Z. was supported by the Heising-Simons Foundation through grant \#2017-274. 
J.A.G. is supported by the National Science Foundation Graduate Research Fellowship under grant number 1650114.
A.S.J. acknowledges support from the Gordon and Betty Moore Foundation under Grant GBMF7392. 
This work was in part carried out on the Dutch national e-infrastructure with the support of SURF Cooperative.
This paper is based upon work supported by the National Aeronautics and Space Administration (NASA) under Contract No. NNG16PJ26C issued through the WFIRST Science Investigation Teams Program.
This research made extensive use of the SAO/NASA Astrophysics Data System (ADS).

\software{
\texttt{gnuplot} \citep{williams_2015_aa},
\texttt{ipython/jupyter} \citep{perez_2007_aa,kluyver_2016_aa},
\texttt{matplotlib} \citep{hunter_2007_aa},
\texttt{NumPy} \citep{der_walt_2011_aa}, and
\texttt{Python} from \href{https://www.python.org}{python.org}.
         }

\appendix
\section{Updates To Physics Modules}\label{s.updates}

\subsection{Equation of State}\label{s.eos}

\begin{figure}[!htb]
\centering
\includegraphics[width=0.5\columnwidth]{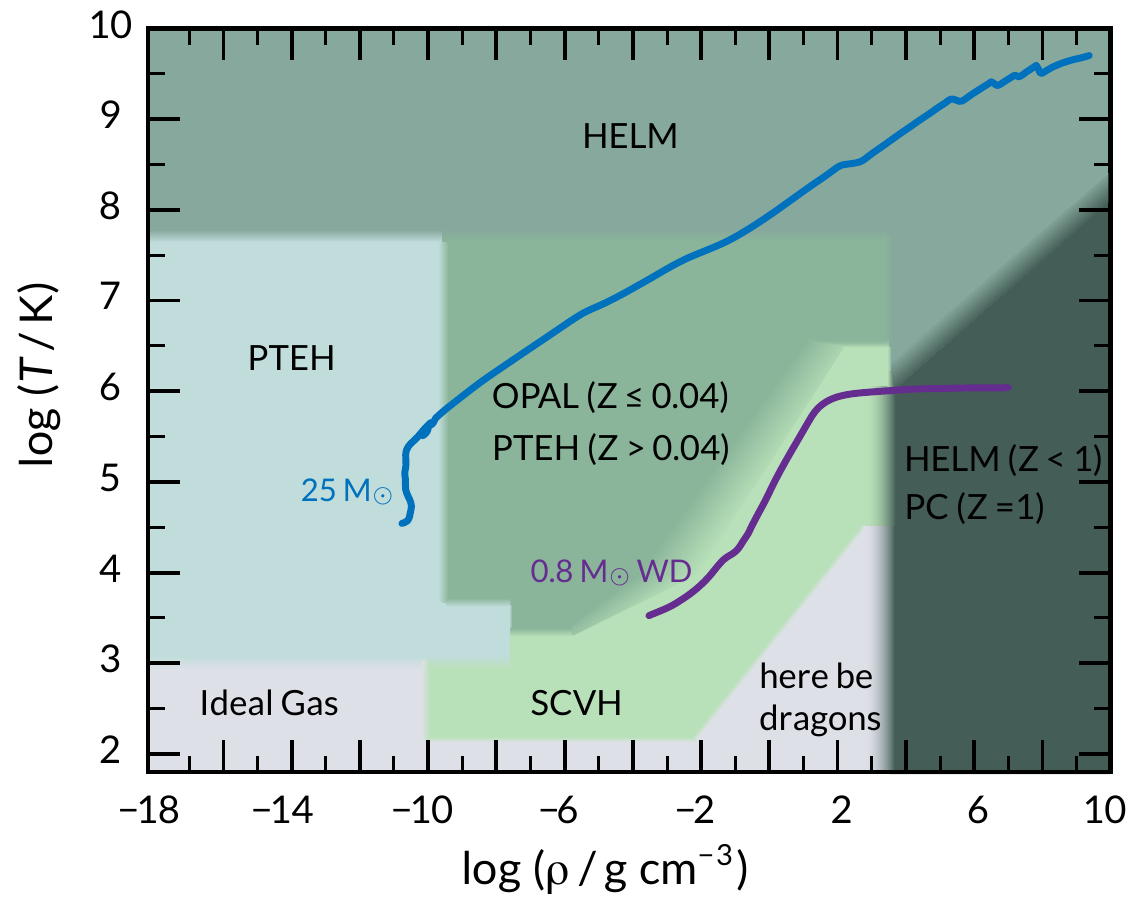}
\caption{
The $\rho-T$ coverage of the EOS used by the \eos \ module.
PTEH is from \citet{Pols_1995_aa},
HELM is from  \citet{timmes_2000_ab},
PC is from \citet{potekhin_2010_aa},
OPAL is from \citet{rogers_2002_aa},
SCVH is from \citet{saumon_1995_aa},
and the low-density cold region in the lower left 
is treated as an ideal neutral gas. 
The region between SCVH and PC is currently problematic
from input physics and numerical perspectives
and treated as an ideal gas
(see \citealt{chabrier_2019_aa} for a recent treatment that is not yet in \MESA).
The blue curve shows the profile of a 
25 \Msun\ star that has reached an iron core infall speed of 1,000 km s$^{-1}$
and the purple curve shows the profile of a 0.8 \Msun\ WD.
}
\label{f.eos}
\end{figure}

The EOS is evaluated by the \eos \ module.  Figure \ref{f.eos}
shows the default coverage in the $\rho-T$ plane.  
The new PTEH option extends the \eos \ coverage to lower densities (to
10$^{-18}$~g~cm$^{-3}$) and higher metallicities ($Z$ up to 1.0) than
allowed by OPAL ($\rho \gtrsim 10^{-10}$~g~cm$^{-3}$ and  $Z$\,$\le$\,0.04 only).
Previous versions of \eos \ use HELM to provide approximate results for the 
low density or $Z > 0.04$ cases now covered by PTEH.  The PTEH tables are created
using the approach of \citet{Pols_1995_aa} as implemented by
\citet{paxton_2004_aa} in a program derived from \citet{eggleton_1971_aa}.  PTEH includes a
solution to the Saha equations to obtain the dissociation and
ionization stages H$^+$, H, H$_2$, He, He$^+$, and He$^{++}$ and assumes full ionization of C, N, O, Ne, Mg, Si, and Fe.


\begin{figure}[!htb]
\centering
\includegraphics[width=0.5\columnwidth]{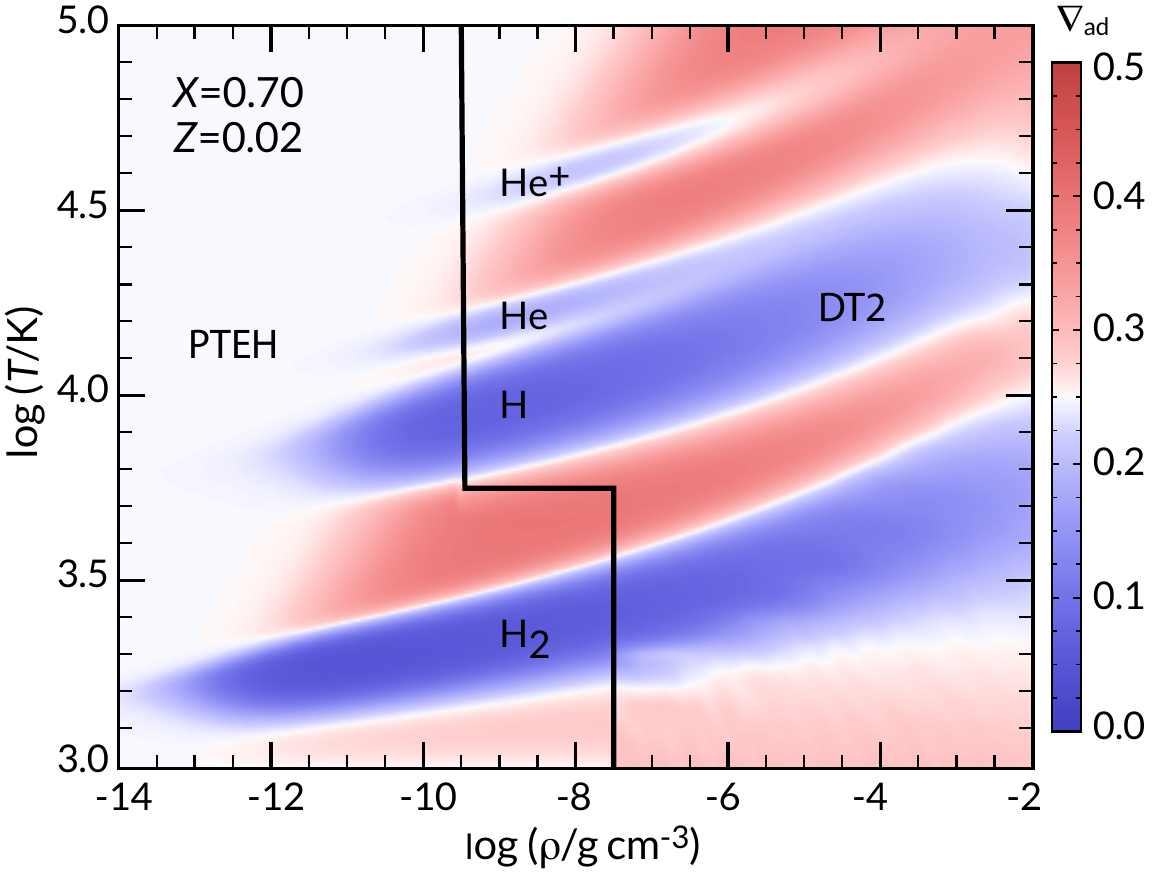}
\caption{
Adiabatic gradient in the $\rho-T$ plane for $X\,=\,0.70$ and
$Z\,=\,0.02$.  Regions undergoing H$_2$, H, He, or He$^+$
dissociation/ionization are colored blue and labelled.  Previous
versions of \eos \ truncated the H$_2$ dissociation region at
$\log$($\rho$/g\,cm$^{-3}$)=$-$10, the limit of the SCVH data in \eos. 
Other ionization bands occur at a high enough temperature to be mainly
covered by OPAL.}
\label{f.gradad}
\end{figure}

In addition to PTEH, there are two other new \eos\ options, DT2 and
ELM\footnote{DT2 is a second way to access OPAL/SCVH data using Density
  and Temperature.  ELM is a subset of HELM.}.  These are
motivated by the need for more numerically accurate partials as
discussed in Section \ref{s.energy}.  In this context, the desired numerical accuracy
of the partials is achieved by evaluating analytic partials of the
interpolating polynomials rather than by interpolating values of tabulated
partials (as is done with OPAL and SCVH data in \MESA).  As a result,
the partials correspond to how the interpolated \eos\ values will
actually change in
response to small changes of the parameters, whereas interpolated
values of partials will be less accurate predictors of the response to such variations.
All three new options use bicubic spline interpolation in high resolution tables of 
$\log(p_{\rm gas}$/erg\,cm$^{-3}$), 
$\log(e$/erg\,g$^{-1}$), and
$\log(s$/erg\,g$^{-1}\,\K^{-1}$) 
to obtain first and
second partial derivatives of these quantities with respect to
$\log$($\rho$/g\,cm$^{-3}$) and 
$\log$($T$/\K). 
The options DT2 and ELM use tables holding values derived from
OPAL/SCVH and HELM respectively.
Figure \ref{f.gradad} shows \grada\ in a region of the $\rho-T$ plane covered by DT2
and PTEH.

\begin{figure}[!htb]
\centering
\includegraphics[width=0.49\columnwidth]{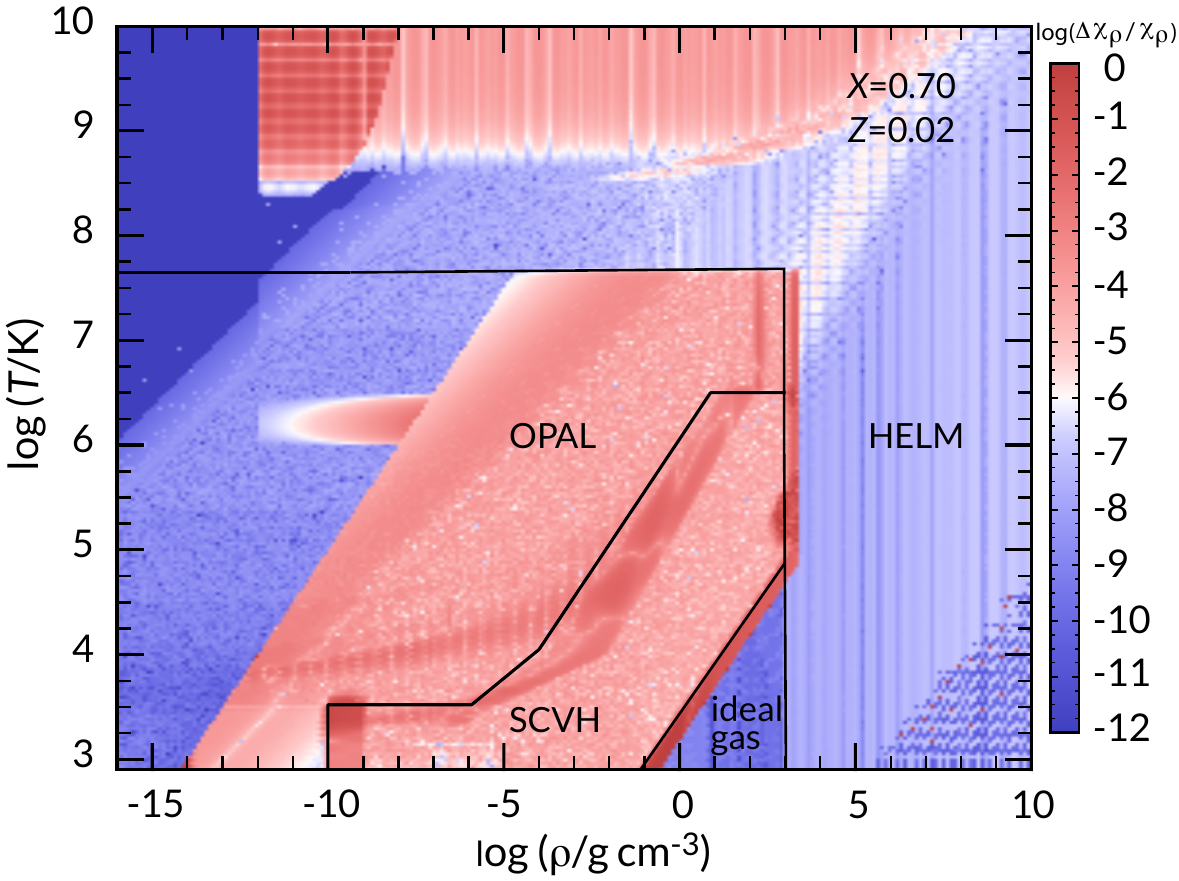}
\includegraphics[width=0.49\columnwidth]{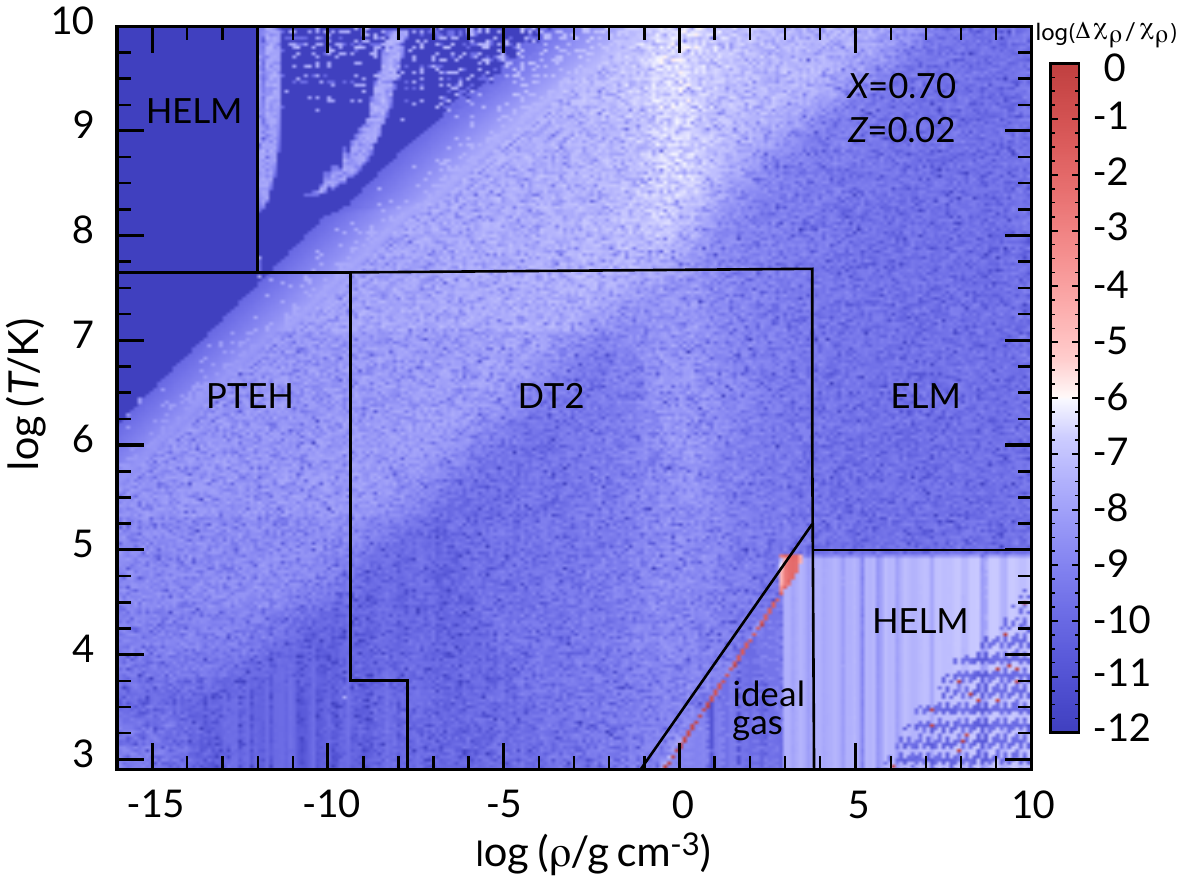}
\caption{
Relative difference in $\chi_{\rho, {\rm gas}}$
between the \eos \  derivative and a Richardson limit based numerical 
derivative across the $\rho-T$ plane. 
The left panel shows the results for the OPAL/SCVH and HELM options,
and the right panel shows the results for the PTEH, DT2, and ELM options.
Relative differences for the new \eos \ options are $\simeq$\,10$^{-7}$,
except for a region between SCVH and PC.
Note the large difference in accuracy between the old OPAL/SCVH options
using interpolated partials and the new DT2 option using partials of
interpolating polynomials.}
\label{f.chirho_dfridr}
\end{figure}

\begin{figure}[!htb]\centering
\includegraphics[width=0.49\columnwidth]{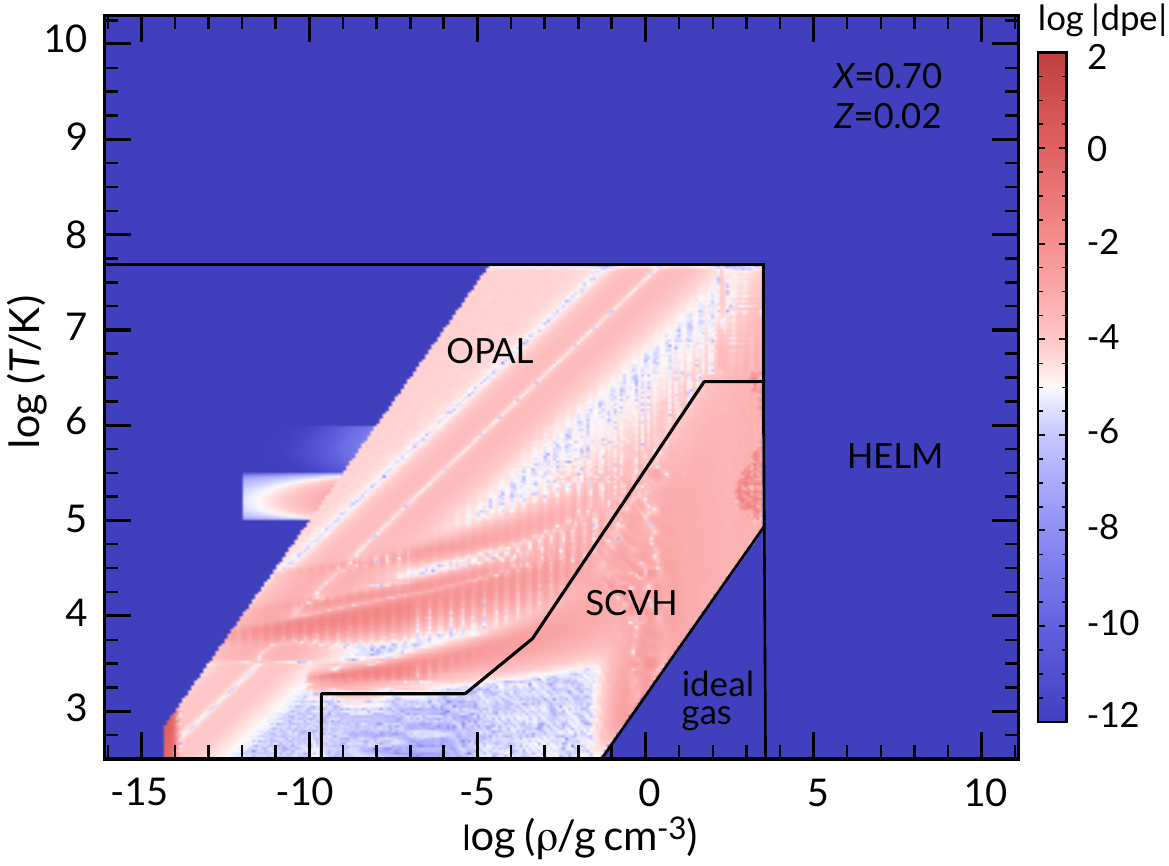}
\includegraphics[width=0.49\columnwidth]{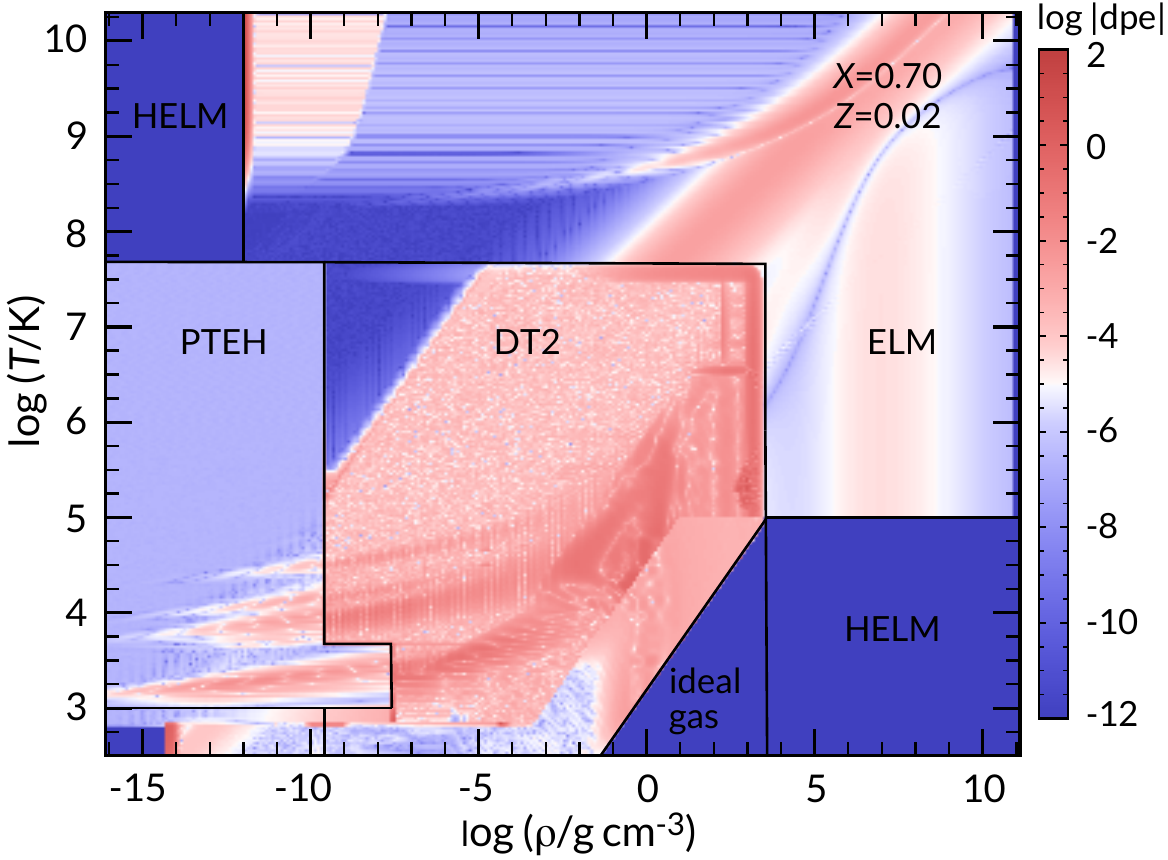}
\caption{
Thermodynamic consistency metric \code{dpe}; see Equation~\eqref{eq:dpe} for
the definition.  The left panel shows the results for OPAL/SCVH
and HELM, and the right panel shows the results for PTEH, DT2, and
ELM.  As expected, HELM gives machine precision consistency and so is
superior in this to ELM.  OPAL and SCVH have an advantage over DT2 because
they use interpolated values of thermodynamically consistent partials $-$ 
which degrades their numerical accuracy as shown in Figure~\ref{f.chirho_dfridr}.}
\label{f.dpe}
\end{figure}

\subsubsection{Blending}\label{s.eos_blend}

The control structure for blending the various EOS sources in  Figure \ref{f.eos}
considers a particular source EOS (PTEH, PC, HELM, etc.) to determine what fraction $f$
of the final result comes from that source. A recursive calling structure is used:
\begin{description}
\item[level 1] 
If $f_{\rm PTEH}$=0, then return result of level 2.
If $f_{\rm PTEH}$=1, then evaluate PTEH and return.
Otherwise, call level 2, blend result with that of PTEH, and return.

\item[level 2]
If $f_{\rm PC}$=0, then return result of level 3.
If $f_{\rm PC}$=1, then evaluate PC and return result to level 1.
Otherwise, call level 3, blend result with that of PC, and return the blended result to level 1.

\item[level 3]
\revision{%
If $f_{\rm PTEH, Z}  = 0$ (i.e., $Z \la$~0.04), then call level 4 with DT2 and return the result to level 2. If $f_{\rm PTEH, Z}  = 1$ (i.e., $Z \ga$~0.04), then call level 4 with PTEH and return the result to level 2.
Otherwise (i.e., $Z$ near $0.04$), call level 4 with each of DT2 and PTEH, blend the result in Z and return the blended result to level 2.}

\item[level 4]

\revision{%
If $f_{\rm PTEH/DT2}$=0, then return result of level 5.
If $f_{\rm PTEH/DT2}$=1, then evaluate PTEH/DT2\footnote{Read as PTEH or DT2.  There are separate level 4 routines for PTEH and DT2.} and return result to level 3.
Otherwise, call level 5, blend the result with that of PTEH/DT2 and return the blended result to level 3.}

\item[level 5]

\revision{%
If $f_{\rm ELM}$=0, then return result of level 6.
If $f_{\rm ELM}$=1, then evaluate ELM and return result to level 4.
Otherwise, call level 6, blend the result with that of ELM, and return the blended result to level 4.}

\item[level 6]
Evaluate HELM and return result to level 5.

\end{description}

The blends use a quintic polynomial with zero slope at boundaries.
The partial derivatives of the blend polynomial are included in the calculation of
the final result to maintain high numerical consistency in the blend region 
and across EOS region boundaries.

There remain challenges to providing a broad coverage EOS 
given the current need to combine multiple sources, some of which may not provide 
the necessary thermodynamic information.  For example, when the
$\log(\rho$/g\,cm$^{-3}$) blend region extended from 3.0 to 3.1, a negative
$\chi_{\rho, {\rm gas}}$ resulted because $\log(p_{\rm gas}$/erg\,cm$^{-3}$) from DT2 were
slightly greater than $\log(p_{\rm gas}$/erg\,cm$^{-3}$) from ELM. This could
give a drop in $p_{\rm gas}$ as $\log(\rho$/g\,cm$^{-3}$) transitions from DT2
to ELM.  Extending the blend region from 2.98 to 3.12 is enough to
ensure $\chi_{\rho, {\rm gas}}$ $>$ 0 in the DT2 to ELM transition.

\subsubsection{Numerical Accuracy of Partials}\label{s.eos_partials}

To check the numerical accuracy of partials using the new options, we have
compared their results with 
iteratively acquired high precision numerical derivatives
\citep{ridders_1982_aa,press_1992_aa}.
For example, Figure \ref{f.chirho_dfridr} shows 
the relative difference between the \eos\  derivative and 
a Richardson iterative numerical derivative across the $\rho-T$ plane
for $\chi_{\rho, {\rm gas}}$.  The
right panel shows that new options give relative errors of
$\simeq$\,10$^{-7}$,
while the left panel shows that the previous options have larger errors, particularly in
the OPAL/SCVH regions.   As mentioned in Section \ref{s.energy}, those
larger errors limit the ability of the $\MESAstar$\ Newton-Raphson solver to reduce
residuals and that in turn can lead to increased errors in numerical
energy conservation.

\subsubsection{Thermodynamic Consistency}\label{s.eos_consistency}

The first law of thermodynamics is an exact differential, which implies

\begin{equation}
p  = \rho^2 \ \dxdy {e} {\rho} \Biggm|_{T,Y_i} + T \ \dxdy {p} {T} \Biggm|_{\rho,Y_i}\,,  \hskip 0.6in
\dxdy {e} {T} \Biggm|_{\rho,Y_i}  = T \ \dxdy {s} {T}  \Biggm|_{\rho,Y_i}\,, \hskip 0.6in
- \dxdy {s} {\rho} \Biggm|_{T,Y_i} = \frac{1}{\rho^2} \ \dxdy {p} {T} \Biggm|_{\rho,Y_i}  
\, .
\label{eq:consist}
\end{equation}

An EOS is thermodynamically consistent if 
these relations are satisfied.
Thermodynamic inconsistency may manifest
itself as an artificial buildup or decay of the entropy 
during what should be an adiabatic flow.  Models that are
sensitive to the entropy may suffer inaccuracies if
thermodynamic consistency is systematically violated over sufficiently
long timescales. Equation~\eqref{eq:consist} may be recast in a form suitable 
for evaluating numerical inconsistencies


\begin{equation}
\begin{aligned}
\code{dpe}  = \frac{\rho^2}{p} \dxdy {e} {\rho} \Biggm|_{T,Y_i} + \frac{T}{p} \dxdy {p} {T} \Biggm|_{\rho,Y_i} & - 1 \,, \qquad \qquad \qquad \qquad \qquad
\code{dse}  = T \left( \dxdy {s} {T} \Biggm|_{\rho,Y_i} \Bigg/ \dxdy {e} {T} \Biggm|_{\rho,Y_i} \right) - 1 \,, \\
& \code{dsp}  = - \rho^2 \left( \dxdy {s} {\rho} \Biggm|_{T,Y_i}  \Bigg/ \dxdy {p} {T} \Biggm|_{\rho,Y_i}  \right) - 1  
\, .
\label{eq:dpe}
\end{aligned}
\end{equation}
Ideally, \code{dpe}, \code{dse}, and \code{dsp} are zero.
Figure \ref{f.dpe} shows the first thermodynamic consistency quantity, \code{dpe}, 
across the $\rho-T$ plane for an older (\mesa \ r8845) and current version of \eos. 
The other two thermodynamic consistency metrics, \code{dse} and \code{dsp}, show similar 
magnitudes. In general, the thermodynamic consistency with DT2 and ELM
is reduced relative to the older options directly using OPAL/SCVH  and HELM.
This is because the thermodynamic
consistency relations can only be approximated by bicubic
splines.
If possible, Hermite interpolation of the Helmholtz free energy would make use of 
partial derivatives from the EOS and guarantee thermodynamic 
consistency \citep[e.g.,][]{timmes_2000_ab}.
However, a bicubic Hermite interpolation produces discontinuities in 
second derivative quantities (e.g., $\partial{p_{\rm gas}}/\partial{\rho}|_{T,Y_i}$) 
that are problematic for \mesa's Newton-Raphson solver
and a biquintic Hermite interpolation requires partial derivatives 
that are unavailable from some constituents of the EOS patchwork.
So for now, we compromise by providing the previous options that
give better thermodynamic consistency and the new options that
provide better numerical accuracy of partials.
The impact of the thermodynamic inconsistencies of the new approach must be evaluated on a case-by-case basis.

\subsection{Nuclear Physics}\label{s.nuclear}

\subsubsection{Nuclear Reaction Rates} \label{s.rates} 

The {\it Joint Institute for Nuclear Astrophysics} (JINA) REACLIB library, 
which provides the default nuclear reaction rates for \MESA,  has 
been updated from the \textit{jina\_reaclib\_results\_v2.2} snapshot to the
\textit{default} snapshot%
\footnote{Dated 2017-10-20. Available from http://reaclib.jinaweb.org/library.php?action=viewsnapshots }.
This update includes changes to the fitting formula for a few neutron capture rates that previously returned erroneous values 
at $T$\,$\gtrsim$\,8$\times$10$^9$\,\Kelvin.
We have modified the \textit{default} snapshot to include a missing, temperature-independent $^{26}\rm{Al} \rightarrow \rm{^{26}Mg}$ weak reaction rate.
Reaction rates between the $^{26}$Al ground and meta-stable states now use \citet{gupta_2001_aa}.
Nuclear partition functions use JINA \textit{winvne\_v2.0.dat} 
table\footnote{Available from http://reaclib.jinaweb.org/associated\_files/v2.2/winvne\_v2.0.dat} 
and JINA \textit{masslib\_library\_5.data}%
\footnote{See the \MESA\ directory \code{chem/preprocessor/chem\_input\_data} }
provides atomic masses. 

A cell's temperature may exceed $\logT$\,=\,10.0 in shocks and explosive burning, 
which is beyond the range of validity for the fits to the reaction rates and the partition functions.
Previously, \MESA\ extrapolated for $\logT$\,$>$\,10.0, leading to erroneous reaction rates.
Reaction rates are now set equal to their $\logT$\,=\,10.0 values when $\logT$\,$>$\,10.0.

\MESA\ now includes the option to use the electron-capture and
$\beta$-decay rates from \citet{Suzuki2016}, which cover
\textit{sd}-shell nuclei with $A = 17-28$.  The primary application
for these tables is the evolution of high-density oxygen-neon cores
\citep[e.g.,][]{Miyaji1980, Miyaji1987, Jones2013}.
Compared to the on-the-fly weak rate approach described in \mesathree,
these tabulated rates are less computationally expensive and more accurate at high temperatures ($T\gtrsim 10^9\K$),
but less accurate at low temperatures ($T\lesssim 10^8\K$) and do not allow explicit updating of nuclear physics.

\subsubsection{Reaction Rate Screening}\label{s.screen}

The plasma coupling parameter of two reactants $\Gamma_{i,j}$ and the ion sphere radius $a_{i}$ are 
\begin{equation}\label{e:plasma_p}
\Gamma_{i,j} = \frac{Z_i Z_j e^2}{0.5\left(a_i + a_j\right) \kB T} 
\qquad ,\qquad
a_{i} = \left[ \frac{3Z_i}{4 \pi \left( Z_i n_i + Z_j n_j \right)} \right ]^{1/3}
\,,
\end{equation}
where 
$Z_i$ is the atomic charge of isotope $i$, $e$ is the electron charge, $\kB$ is the Boltzmann constant, 
and $n_i$ is the ion number density of species $i$.
\MESA\ applies screening factors to correct nuclear reaction rates for plasma interactions \citep[e.g.,][]{salpeter:54}.
Previous versions defaulted to using one set of expressions for the weak screening regime \citep[$\Gamma_{i,j}$\,$\le$\,0.3, ][]{dewitt:73, graboske:73} 
and another set of expressions for the strong screening regime \citep[$\Gamma_{i,j}$\,$\ge$\,0.8, ][]{alastuey:78, itoh:79}.
A linear blend of the weak and strong screening factors is used in the 0.3\,$<$\,$\Gamma_{i,j}$\,$<$\,0.8 intermediate regime.
Silicon burning reactions in the cores of massive stars often operate in this intermediate regime.
The numerical blending provides a smooth and
continuous function that equals the non-blended values and their derivatives at the edges.
A new default for screening, based on \citet{chugunov:07}, includes a
physical parametrization for the intermediate screening regime and reduces to the familiar weak and strong limits 
at small and large $\Gamma$ values. We extend the \citet{chugunov:07}  one-component plasma results
to a multi-component plasma following \citet{itoh:79}, where the $Z_i$ are replaced with the average charge \zbar.

\begin{figure}[!htb]
\centering
    \includegraphics[width=0.7\columnwidth]{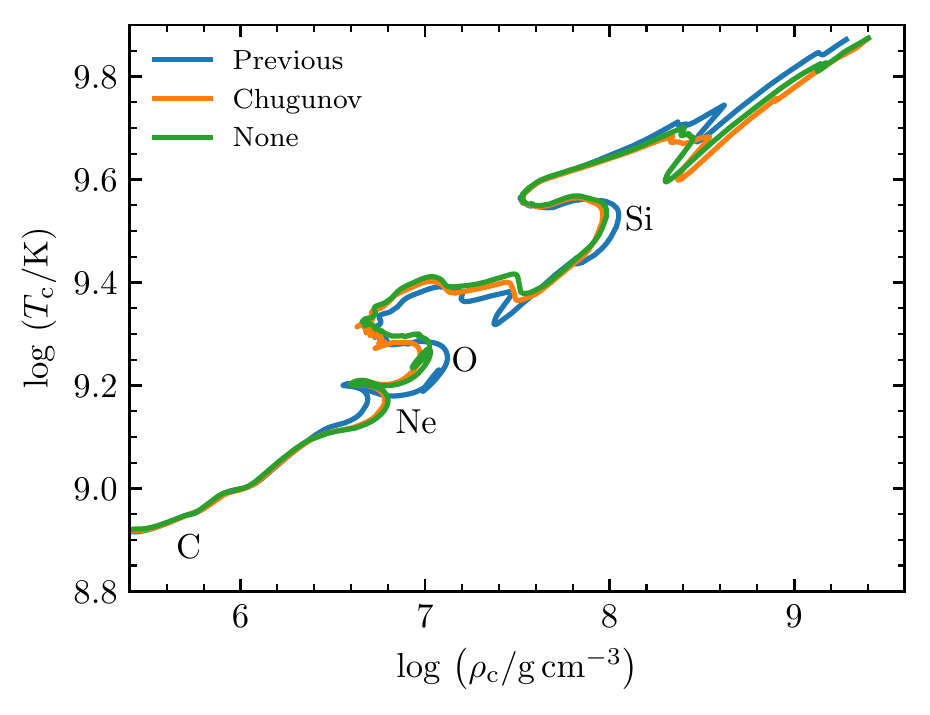}
    \caption{
    Evolution in the central ($\rho - T$) plane of a 25\,\Msun\ star until the onset of Fe core-collapse
    with three different nuclear reaction rate screening options. `Previous' denotes the previous \MESA\ default option,
    `Chugunov' is the new default option, and `None' applies no screening correction to any nuclear reaction rate.
    Locations when a fuel undergoes central ignition are labelled.
    }
    \label{f.screen_core}
\end{figure}

Figure \ref{f.screen_core} compares three \MESA\ nuclear reaction rate screening options 
on the evolution of a 25\,\Msun\ model. The differences are small from H-burning to the onset of core collapse.
However, the \citet{chugunov:07} implementation takes $\approx20\%$ fewer time steps, retries, and backups 
which indicates a numerically smoother solution.

\section{Analytical Approximations to the Roche Geometry of a Single Star}\label{a.rot}
In this appendix we compute various properties of the Roche potential of a
single star. These are used for the computation of centrifugal effects in
stellar structure, as discussed in Section \ref{s.rot}.
Through these derivations we denote dimensionless properties using an
apostrophe, with distances being normalized as
$\revision{r'=r/(Gm_\Psi/\Omega^2)^{1/3}}$
and the potential as $\Psi'=\Psi/(Gm_\Psi\Omega)^{2/3}$.

\subsection{Volume of Roche Equipotentials}

\begin{figure}
   \begin{center}
   \includegraphics[width=0.7\columnwidth]{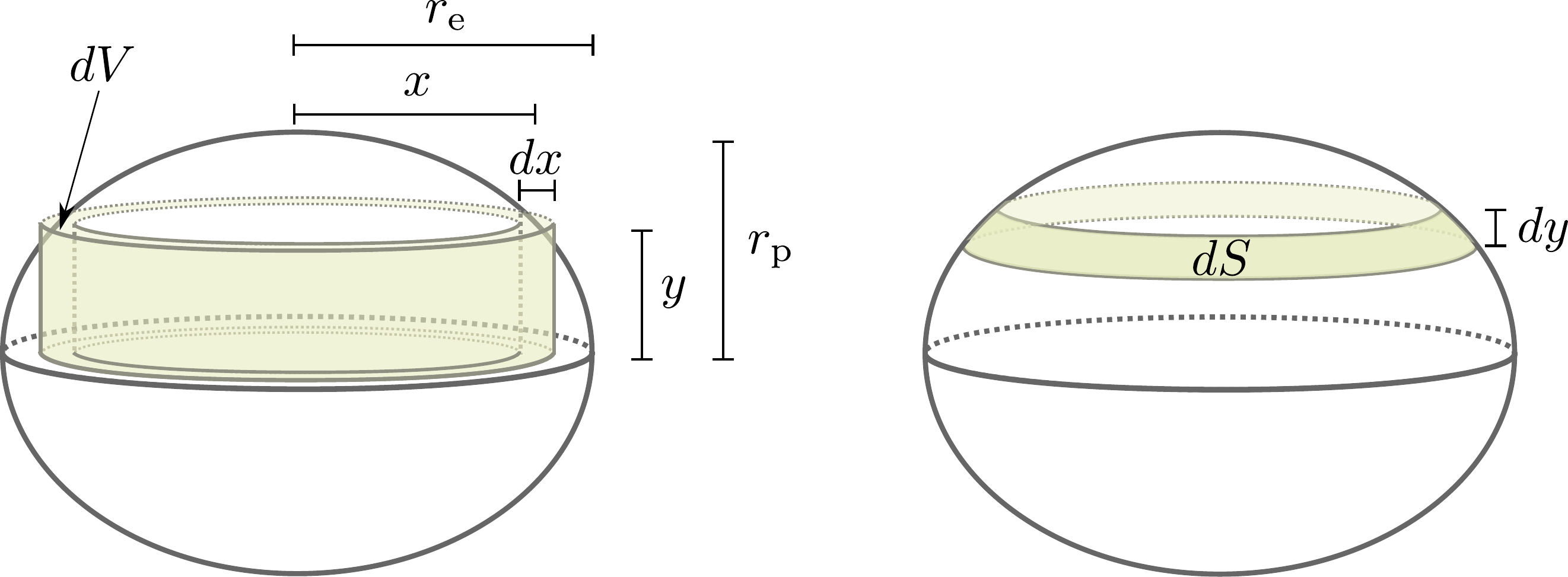}
   \end{center}
   \caption{Quantities used for the integration of the $f_{P}$ and
   $f_{T}$ factors, which require the computation of volume and surface areas of
   equipotentials. }\label{fig:drawing}
\end{figure}

Following the diagram in Figure \ref{fig:drawing}, the dimensionless volume
equivalent radius $r_{\Psi}'$ can be computed in terms of $\omega$ as
\begin{eqnarray}
   \frac{4}{3}\pi r_{\Psi}'^3&\equiv&
   V_\Psi'=2\int_{0}^{\omega^{2/3}}\frac{dV'}{dx'}dx',
   \label{eq:rphi}\\
   \frac{dV'}{dx'} &=& 2\pi x'y', \quad
   y'=\sqrt{\left(\frac{1}{\omega^{2/3}}+\frac{\omega^{4/3}}{2}-\frac{x'^2}{2}\right)^{-2}-x'^2}\nonumber,
\end{eqnarray}
where the expression for $y'(x')$ can be derived directly from the Roche
potential. The integral can be solved analytically for the case of
$\omega=1$, providing the volume of a critically rotating star in terms of its
equatorial radius \citep{Kopal1959},
\begin{eqnarray}
   V_{\Psi}(\omega=1)&=&\frac{4\pi}{3}r_{\rm
   e}^3\left[3\sqrt{3}-4+3\ln\left(\frac{3(\sqrt{3}-1)}{\sqrt{3}+1}\right)\right]\simeq\frac{4\pi}{3}\left(0.8149r_{\rm e}\right)^3.\label{eq:volcrit}
\end{eqnarray}
In the opposite case where $\omega \to 0$, the equipotential is well
approximated by an ellipsoid of revolution with
\begin{equation}
   r_{\rm p}(\omega)=r_{\rm
   e}\left(1-\frac{\omega^2}{2}+\mathcal{O}(\omega^4)\right),\label{eq:rpapprox}
\end{equation}
such that its volume is
\begin{eqnarray}
   V_\Psi(\omega)=\frac{4\pi}{3}r_{\rm
   e}^3\left(1-\frac{\omega^2}{2}+\mathcal{O}(\omega^4)\right).\label{eq:volloww}
\end{eqnarray}
By numerically integrating Equation~\eqref{eq:rphi}, a simple polynomial approximation to the volume can then be constructed, which
is consistent with the value at critical rotation given by Equation~\eqref{eq:volcrit} and the asymptotic behaviour at small $\omega$ given by
Equation~\eqref{eq:volloww}:
\begin{eqnarray}
   V_\Psi(\omega)\simeq \frac{4\pi}{3}r_{\rm
   e}^3\left(1-\frac{\omega^2}{2}+0.1149\omega^4-0.07376\omega^6\right).\label{eq:volfit}
\end{eqnarray}
This expression has an error $<0.25\%$ for $0\le\omega\le1$. In all the
asymptotic expressions considered in this work $\omega$ appears in series of even powers,
so we do not include odd terms in any fit.
Also, since the value for $\omega=1$
is fixed, Equation~\eqref{eq:volfit} is a fit with only 1 free parameter.

Stellar evolution instruments typically use the radial coordinate $r_\Psi$, so it is
useful to have polynomial fits for $r_\Psi(\omega)$ as well. In the limit of
$\omega\rightarrow0$, 
\begin{eqnarray}
   r_\Psi=r_{\rm
   e}\left(1-\frac{\omega^2}{6}+\mathcal{O}(\omega^4)\right).\label{eq:rphiapp}
\end{eqnarray}
A polynomial fit that matches this and the value at critical rotation
$0.8149r_{\rm e}$ from Equation~\eqref{eq:volcrit} is
\begin{eqnarray}
   r_\Psi = r_{\rm
   e}\left(1-\frac{\omega^2}{6}+0.01726\omega^4-0.03569\omega^6\right),\label{eq:rphifit}
\end{eqnarray}
which has an error $<0.15\%$ for $0\le\omega\le1$. Similarly, the expression
\begin{eqnarray}
   r_{\rm e} =
   r_{\Psi}\left(1+\frac{\omega^2}{6}-0.0002507\omega^4+0.06075\omega^6\right)\label{eq:refit}
\end{eqnarray}
has an error $<0.2\%$ in the same range.

\subsection{Surface Area of Roche Equipotentials}
The computation of the dimensionless surface area $S_{\Psi}'$ is given by
\begin{eqnarray}
   S_{\Psi}' = 2\int_{0}^{\omega^{2/3}}\frac{dS'}{dx'}dx',\qquad
   \frac{dS'}{dx'} = 2\pi x'
   \sqrt{\left(\frac{dy'}{dx'}\right)^2+1}.\label{eq:Surface}
\end{eqnarray}
In the limit $\omega\rightarrow 0$, $dS'/dx'$ can be approximated as
\begin{eqnarray}
   \frac{dS'}{dx'}(\omega)=\frac{2\pi x'r_{\rm e}'}{\sqrt{r_{\rm
   e}'^2-x'^2}}\left[1-\frac{1}{2}\left(\frac{x'}{r_{\rm
   e}'}\right)^2\omega^2+\mathcal{O}(\omega^4)\right]\label{eq:dSdx},
\end{eqnarray}
which upon integration provides the approximate form of the surface area of a
slowly rotating equipotential,
\begin{eqnarray}
   S_\Psi(\omega)=4\pi
   r_e^2\left(1-\frac{\omega^2}{3}+\mathcal{O}(\omega^4)\right)\label{eq:Sphiapp}.
\end{eqnarray}
This result is consistent with that for an oblate spheroid.
Using Equation~\eqref{eq:Sphiapp} and the numerically computed value
$S_\Psi(\omega=1)=8.832r_{\rm e}^2$, a fit to
$S_\Psi$ that has an error $<0.01\%$
in the range $0<\omega<1$ is
\begin{eqnarray}
   S_{\Psi}(\omega)\simeq 4\pi r_{\rm
   e}^2\left(1-\frac{\omega^2}{3}+0.08525\omega^4-0.04908\omega^6\right).
\end{eqnarray}
\subsection{Surface Averages of Gravity} \label{sec:grav}

The surface average of the dimensionless gravity $\langle g'\rangle$ is computed as
\begin{eqnarray}
   \langle g'\rangle
   &=&\frac{2}{S_\Psi'}\int_{0}^{\omega^{2/3}}g'\frac{dS'}{dx'}dx',\label{eq:gavg}
\end{eqnarray}
with an equivalent expression for $\langle g'^{-1}\rangle$. For reference, the
dimensionless gravity is given by ${g'=g/(Gm_\Psi\Omega^4)^{1/3}}$. 
In the limit
of $\omega \to 0$, it can be shown that
\begin{eqnarray}
   g'(\omega)=\frac{1}{r_{\rm
   e}'^2}\left\{1+\left[1-2\left(\frac{x'}{r_{\rm
   e}'}\right)^2\right]\omega^2\right\},\label{ref:gsmallw}
\end{eqnarray}
which combined with Equations~\eqref{eq:dSdx} and~\eqref{eq:gavg} results in
\begin{eqnarray}
   S_{\Psi}\langle g\rangle =
   4\pi
   Gm_\Psi\left(1-\frac{2}{3}\omega^2+\mathcal{O}(\omega^4)\right)\label{eq:Sgapp}.
\end{eqnarray}
By numerically computing this integral, a simple fit that matches the computed
value at $\omega=1$ and has an error $<0.35\%$ for $0\le\omega\le 1$ is
\begin{eqnarray}
   S_{\Psi}\langle g\rangle =4\pi
   Gm_\Psi\left(1-\frac{2}{3}\omega^2+0.3045\omega^4+0.001382\omega^6\right).\label{eq:fitSg}
\end{eqnarray}

Similarly the average of the inverse gravity, in the limit of $\omega \to 0$,
can be shown to be

\begin{eqnarray}
   S_{\Psi}\langle g^{-1}\rangle =
   \frac{4\pi r_{\rm
   e}^4}{Gm_\Psi}\left(1+\mathcal{O}(\omega^4)\right).\label{eq:invgapp}
\end{eqnarray}
However, this integral diverges for $\omega=1$, as for a critically rotating
star the effective gravity at its equator becomes zero. Although the expression for
$S_{\Psi}\langle g'^{-1}\rangle$ cannot be integrated in the limit
$\omega\rightarrow 1$, by comparing it to the numerical results we have verified
that it is approximately given by $S_{\Psi}\langle g^{-1}\rangle \propto
-\ln(1-\omega^4)$. Combining this information with Equation~\eqref{eq:invgapp}
we have found the following fit with an error $<0.85\%$ in the range $0\le\omega\le
0.9999$:
\begin{eqnarray}
   S_{\Psi}\langle g^{-1}\rangle =\frac{4\pi r_{\rm e}^4}{Gm_\Psi}
   A(\omega),\qquad
   A(\omega)=1-0.1076\omega^4-0.2336\omega^6-0.5583\ln(1-\omega^4).\label{eq:Sginvfit}
\end{eqnarray}

\subsection{Moments of Inertia}\label{s.rot.inertia}
The specific moment of inertia $i_{\rm rot}$
is needed to determine $\Omega_\Psi$  from the 
specific angular momentum $j_{\rm rot}$ and volume equivalent radius
$r_\Psi$.
To compute $i_{\rm rot}$, consider a shell
of material extending from $\Psi$ to $\Psi+d\Psi$. At each point in its surface, its
thickness is given by 
\begin{eqnarray}
   ds=|\nabla\Psi|^{-1}d\Psi=g^{-1}d\Psi.\label{eq:ds}
\end{eqnarray}
Assuming a constant density $\rho$ in the shell (as in the
shellular approximation), 
\begin{eqnarray}
   dm'&=&2\rho' d\Psi'\int_0^{\omega^{2/3}}g'^{-1}\frac{dS'}{dx'}dx'=\rho'd\Psi
   S_{\Psi}'\langle g'^{-1} \rangle,\label{ap:rot:equdm}
\end{eqnarray}
where $dm'=dm/m_\Psi$ and
$\rho'=\rho/\left[m_\Psi(Gm_\Psi/\Omega^2)^{-3/2}\right]$.
From Equation \eqref{ap:rot:equdm},
\begin{eqnarray}
   i_{\rm rot}' =
   \frac{\displaystyle 2\int_0^{\omega^{2/3}}x^2g'^{-1}\frac{dS'}{dx'}dx'}
   {S_{\Psi}'\langle g'^{-1} \rangle},
\end{eqnarray}
with the dimensionless specific moment of inertia defined as $i_{\rm rot}'=i_{\rm
rot}/(Gm_\Psi/\Omega^2)$.
Preserving the fit for $S_\Psi\langle g^{-1}\rangle$ given by Equation~\eqref{eq:Sginvfit} and using Equations~\eqref{eq:dSdx} and~\eqref{ref:gsmallw}, in the limit of $\omega \to 0$
\begin{eqnarray}
   i_{\rm rot}(\omega) =\frac{2}{3}r_{\rm
   e}^2\frac{1+\frac{1}{5}\omega^2+\mathcal{O}(\omega^4)}{A(\omega)}.\label{eq:irotapp}
\end{eqnarray}
Equation~\eqref{eq:ds} implies
that $i_{\rm rot}(\omega=1)=r_{\rm e}^2$, as in the extreme of critical rotation
$g=0$ at the equator and almost all mass between two close equipotentials lies
in a ring of radius $r_{\rm e}^2$. Using this
information, we construct the following fit which has an error $<0.9\%$ for $0\le
\omega \le 0.9999$:
\begin{eqnarray}
   i_{\rm rot} =\frac{2}{3}r_{\rm
   e}^2\times\frac{B(\omega)}{A(\omega)},\qquad
   B(\omega)=1+\frac{1}{5}\omega^2-0.2735\omega^4-0.4327\omega^6-\frac{3}{2}\times 0.5583\ln(1-\omega^4),\label{eq:irot}
\end{eqnarray}
where the $3/2$ factor in the last term ensures the desired result
as $\omega\to1$.


\subsection{Computation of $f_P$ And $f_T$} \label{sec:fpft}
Using all the fits constructed so far, $f_P$ and $f_T$ can be evaluated
directly. However, to provide a more compact expression, we keep only the fit for
$S_\Psi\langle g^{-1}\rangle$ and use Equations~\eqref{eq:fdef},~\eqref{eq:rphifit} and~\eqref{eq:Sgapp} to determine the behavior of the
remaining terms when $\omega\rightarrow 0$,
\begin{eqnarray}
   f_{P}(\omega)=\frac{1-\frac{2}{3}\omega^2+\mathcal{O}(\omega^4)}{A(\omega)},\qquad
   f_{T}(\omega)=\frac{1+\mathcal{O}(\omega^4)}{A(\omega)}.
\end{eqnarray}
The following fits for $f_P$ and $f_T$  are derived, which have errors $<0.8\%$ and
$<1.6\%$ in the range $0\le\omega\le0.9999$ respectively:
\begin{eqnarray}
   f_{P}(\omega)=\frac{1-\frac{2}{3}\omega^2-0.06837\omega^4-0.2495\omega^6}{A(\omega)},\qquad
   f_{T}(\omega)=\frac{1+0.2185\omega^4-0.1109\omega^6}{A(\omega)}.
\end{eqnarray}

\subsection{Determination of $\omega$}

For given values of $r_{\Psi}$, $m_\Psi$, and $j_{\rm rot}$, $\omega$ can be determined from the implicit
equation
\begin{eqnarray}
   \frac{j_{\rm rot}}{\sqrt{Gm_\Psi r_{\Psi}}}
   =\omega\frac{i_{\rm rot}}{r_{\rm
   e}^2}\sqrt{\frac{r_{\rm e}}{r_{\Psi}}}.\label{eq:wiii}
\end{eqnarray}
The left hand side can be
directly evaluated, while the right hand side is a monotonic function of $\omega$ for
$0\le\omega\le 1$.

We compute a fit to the right hand side of Equation~\eqref{eq:wiii}.
Equations~\eqref{eq:rphiapp} and~\eqref{eq:irotapp} can be used to determine the form of this
term in the limit $\omega\rightarrow 0$.
In the limit of $\omega\rightarrow 1$ all material is concentrated in a equatorial ring, such that ${j_{\rm rot}(\omega=1)=\sqrt{Gm_\Psi r_{\rm e}}}$. Using
this information, we find the following fit which has an error $<0.8\%$ in the
range $0\le\omega\le0.9999$:
\begin{eqnarray}
   \frac{j_{\rm rot}}{\sqrt{Gm_\Psi r_{\Psi}}}&=&\frac{2}{3}\frac{\omega
   C(\omega)}{A(\omega)}, \qquad C(\omega) =
   1+\frac{17}{60}\omega^2-0.3436\omega^4-0.4055\omega^6-0.9277\ln(1-\omega^4).
\end{eqnarray}
This allows the computation of partial derivatives of $\omega$ with
respect to $r_{\Psi}$ and $j_{\rm rot}$,
\begin{eqnarray}
   \frac{\partial \omega}{\partial r_{\Psi}}=-\frac{3}{4}\frac{j_{\rm
   rot}}{r^{3/2}\sqrt{Gm_\Psi}}\left[\frac{d}{d
   \omega}\left(\frac{\omega C(\omega)}{A(\omega)}\right)\right]^{-1},\qquad
   \frac{\partial \omega}{\partial j_{\rm rot}}=\frac{3}{2}\frac{1}
   {\sqrt{Gm_\Psi r_\Psi}}\left[\frac{d}{d
   \omega}\left(\frac{\omega C(\omega)}{A(\omega)}\right)\right]^{-1}.
\end{eqnarray}

\section{Core Collapse Supernova Explosions}\label{s.snec} 

\mesafour\ described modeling the evolution of core-collapse supernova
(SN) ejecta up to shock breakout with \MESA, and with \STELLA\ beyond
shock breakout.  Modifications since \mesafour\ have focused on
fallback in weak explosions of red supergiant (RSG) stars. In these
weak explosions, the total final explosion energy is positive, but
insufficient to unbind all material. Thus, some material falls
back onto the central object during the subsequent evolution.
To quantify and remove this material, we introduce two new user controls. 
First, we implement a new criterion to select which material is excised 
from the model during the ejecta evolution.\footnote{Triggered when 
\texttt{fallback\_check\_total\_energy} = \texttt{.true.} in \texttt{star\_job}.}
At each time step, \MESA\ calculates the integrated total energy from the innermost cell to cell 
$j$ above it:
\begin{equation}
E_j = \sum_{i={\rm inner}}^{j} \left(e_i - \frac{G m_i}{r_i} + \frac{1}{2} u_i^2 \right) dm_i.
\label{e.energyintegral}
\end{equation} 
If $E_j<0$, then there is a bound inner region, and \MESA\ 
continues this sum outward until it reaches a cell $k$ with local positive total energy 
($e_k - G m_k/r_k + u_k^2/2>0$). 
\MESA\ removes material inside this cell, making cell $k$ the new innermost cell.
Second, to remove any slow-moving, nearly hydrostatic material left 
near the inner boundary, a minimum innermost velocity can be specified at handoff to \STELLA. All material below the innermost cell
that has a velocity above this specified velocity is not included in \STELLA\ input files.\footnote{Controlled by the 
\texttt{star\_job} inlist parameter \texttt{stella\_skip\_inner\_v\_limit}.}
A velocity cut between 100 - 500\,km\,s$^{-1}$ has little effect on light curve properties 
and the photospheric evolution of Type IIP SNe, and can greatly reduce numerical artifacts that
may arise from an inward-propagating shock hitting the inner boundary in \STELLA. 
Such a cut can also lead to a factor of 10 or more reduction in the number of time steps required 
to produce a light curve.  While this scheme is useful in quantifying and excising fallback material,
it is not a satisfactory treatment of fallback. \revision{For a more
complete description of these fallback criteria, see Appendix A of \citet{Goldberg2019}.}

Next, we look at the evolution of material deep within the ejecta of a Type IIP supernova, 
at such large optical depths that the outcome is not sensitive 
to any particular treatment of radiation transport in different software instruments.
For this restricted regime, we compare results using \texttt{MESA+STELLA} to 
quantities derived from the open-source 1D gray radiation hydrodynamics code 
\SNEC\ \citep{Morozova2015}, both using the same \MESA\ Type IIP ejecta model at shock breakout. 
This should yield 
meaningful density and velocity comparisons nearly everywhere in the ejecta and meaningful temperature comparisons 
very deep within the ejecta. 

We explode (with $\Eexp$\,=\,10$^{51}$ erg) 
the 99em\_19 RSG progenitor model from \mesafour, 
which has a ZAMS mass of 19\,\Msun, and a mass of 17.8\,\Msun\ and a
radius of 603\,\Rsun\ at time of explosion. We excise the inner 1.5 $\Msun$  and 
explode the remaining 16.3\,\Msun\ of ejecta using a thermal bomb as described in \mesafour.
The mass of radioactive $\Ni$ is set to $\MNi$\,=\,0.03\,\Msun. 
The resulting ejecta is influenced by the inclusion of the \citet{Duffell2016} prescription for mixing 
due to the Raleigh-Taylor instability in \MESA. We then pass our \MESA\ models at shock breakout to both 
\STELLA\ and \SNEC.
To mimic the additional line opacities from metals, \SNEC\ employs an opacity floor.
We use \SNEC's default opacity 
floor recommended in \citet{Bersten2011}, which is set to
$\kappa_{\rm floor,\,core}$\,=\,0.24\,cm$^{2}$\,g$^{-1}$ for the metal-rich core material, $\kappa_{\rm floor,\,envelope}$\,=\,0.01\,cm$^{2}$\,g$^{-1}$
for the envelope, and proportional to metallicity for the intermediate region. 

\begin{figure}[!htb]
\centering
\includegraphics[width=0.5\columnwidth]{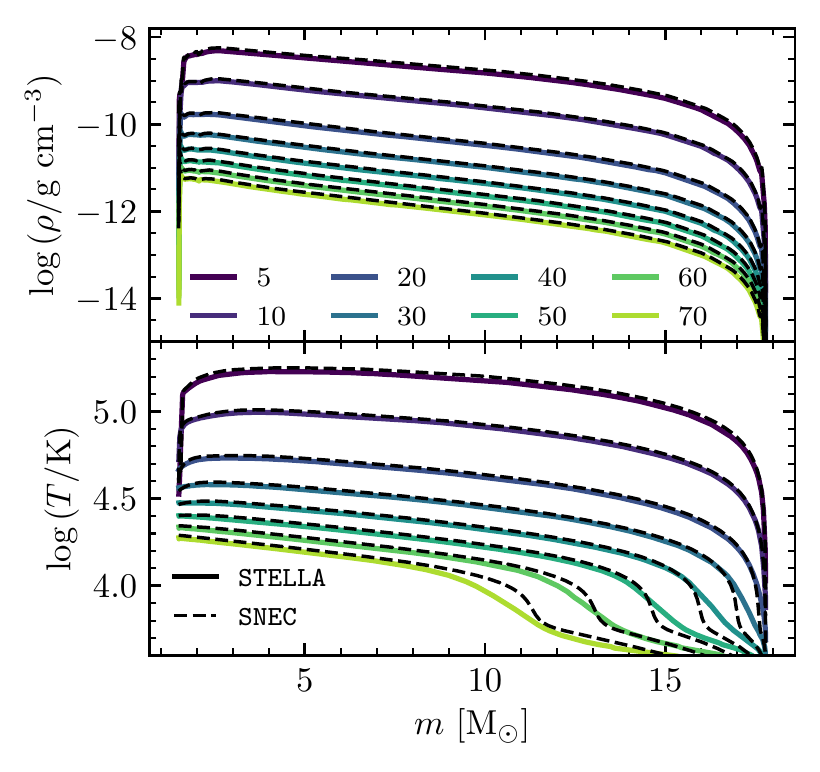} 
\caption{Density (upper panel) and temperature (lower panel) profiles in \STELLA\ (solid colored lines) 
and \SNEC\ (dashed lines) on the plateau of a Type IIP SN model with $\MNi = 0.03\ \Msun$ at
5 to 70 days after shock breakout in \MESA. Lighter colors indicate later days.}
\label{f.expansion_profiles}
\end{figure}

Figure \ref{f.expansion_profiles} shows density and 
temperature profiles as a function of mass coordinate, 
with color corresponding to the number of days after shock breakout. Both instruments agree in the density evolution of the expanding ejecta. The temperature evolution is also in agreement
in the very optically thick inner ejecta. Temperature differences at the surface 
reflect the differing treatments of opacity and radiation transfer between \SNEC\ and \STELLA. 

\begin{figure}[!htb]
\centering
\includegraphics[width=0.5\columnwidth]{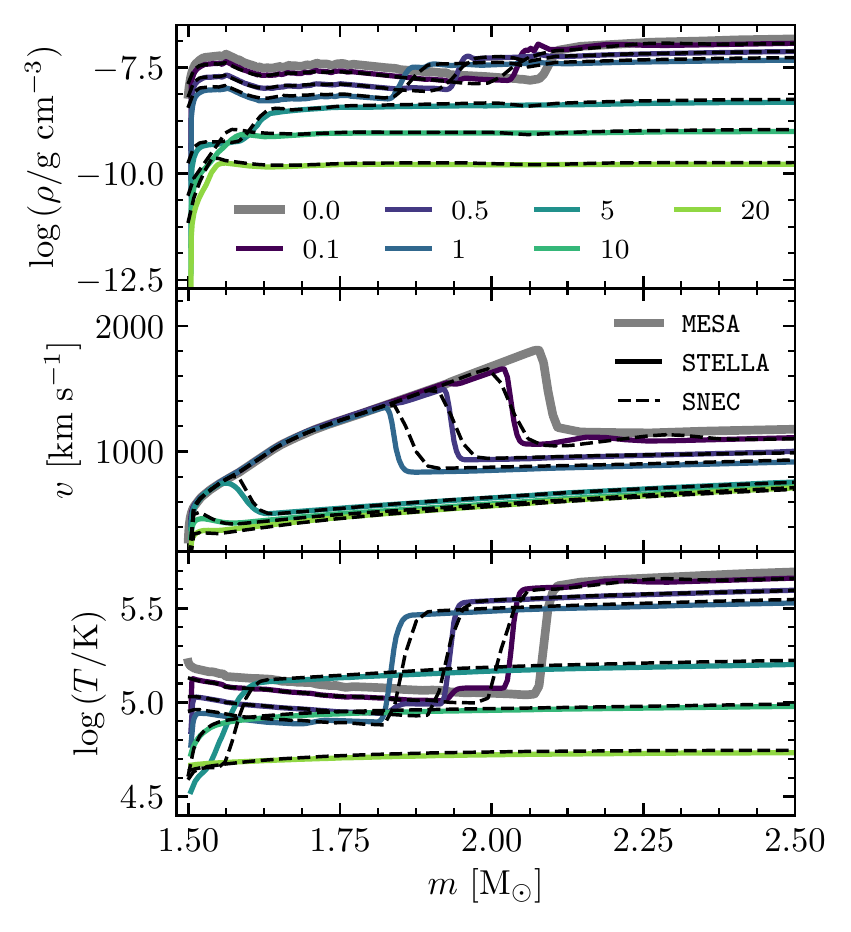} 
\caption{Density (upper panel), velocity (middle panel), and temperature (lower panel) profiles of the 
99em\_19 progenitor, exploded with $10^{51}$\,erg, up to 20 days after shock breakout. 
The reverse shock originating at  the H/He boundary makes its way back through the expanding ejecta. 
The \MESA\ profile at shock breakout (thick gray line) is used as the input for subsequent evolution in \STELLA\ (solid curves) and 
\SNEC\ (dashed curves). Numbers in the legend correspond to the day post 
shock-breakout for each profile. 
%
%
}
\label{f.RS_profiles}
\end{figure}

Figure \ref{f.RS_profiles} focuses on the deep core during the first days of the evolution
post shock-breakout at the location of the reverse shock generated at the H/He boundary 
of the initial model. Density, velocity, and temperature profiles over the first 20 days 
are plotted with the corresponding \MESA\ profile at shock breakout, which serves as the common 
input in both \STELLA\ and \SNEC.  By day 20 the reverse shock has reached
the inner boundary in both models. Although there are slight differences within the 
first day, both instruments agree on global properties of the reverse shock
and its effects on the temperature and density profiles out to day 20.

\section{\mesa\ Testhub}\label{s.testhub}

Development of the \mesa{} source code is a collaborative process with multiple
commits each day from developers working on separate parts of the
codebase.  In addition to serving as starting templates for science projects, the
test cases in \mesastar{} and \mesabinary{} exist to detect when
changes to the codebase cause unintended deviations from expected
behavior (i.e., bugs).  The number of test cases grows with time and is
currently more than 100, with a total runtime on the order of
10 hours on multicore workstations.  Since \MESA\ is
committed to supporting reproducibility by giving bit-for-bit identical
results on a variety of different hardware and software platforms (\mesathree, Section 10), the
test suite must be checked on a representative sample of host systems; just
as it takes a team to create \MESA, it takes a team to test it.

To prevent slowdowns in development that would be caused by running the test
suite on multiple hosts before every commit, we have developed the \mesa{} Testhub (\url{
testhub.mesastar.org}). The Testhub is a web application that collects and
organizes the results of test suite submissions via a companion Ruby gem 
called \code{mesa\_test}. Every day, submissions from multiple computers and
clusters with diverse hardware, operating systems, and compilers check out the
most recent revision of \mesa{}, run the test suite, and upload their results
to the Testhub. Each day, a summary e-mail is sent to developers detailing
which, if any, revisions had failing test cases submitted in the previous 24
hours. 
With a quick check of the daily email from the Testhub, developers are
able to detect cases that fail to give the expected output with
bit-for-bit identical results on different computers, 
or take more than the specified number of time steps, retries, or
backups. With daily coverage, we can promptly diagnose issues 
as soon as they arise by looking at changes from a only handful of commits while maintaining a 
brisk development pace.  The addition of the Testhub has yielded a
significant improvement in the pace and quality of \MESA\ development.

\bibliographystyle{aasjournal}

\begin{thebibliography}{}
\expandafter\ifx\csname natexlab\endcsname\relax\def\natexlab#1{#1}\fi
\providecommand{\url}[1]{\href{#1}{#1}}
\providecommand{\dodoi}[1]{doi:~\href{http://doi.org/#1}{\nolinkurl{#1}}}
\providecommand{\doeprint}[1]{\href{http://ascl.net/#1}{\nolinkurl{http://ascl.net/#1}}}
\providecommand{\doarXiv}[1]{\href{https://arxiv.org/abs/#1}{\nolinkurl{https://arxiv.org/abs/#1}}}

\bibitem[{{Abbott} {et~al.}(2017{\natexlab{a}}){Abbott}, {Abbott}, {Abbott},
  {Acernese}, {Ackley}, {Adams}, {Adams}, {Addesso}, {Adhikari}, {Adya},
  {Affeldt}, {Afrough}, {Agarwal}, {Agathos}, {Agatsuma}, {Aggarwal}, {Aguiar},
  {Aiello}, {Ain}, {Ajith}, {Allen}, {Allen}, {Allocca}, {Altin}, {Amato},
  {Ananyeva}, {Anderson}, {Anderson}, {Antier}, {Appert}, {Arai}, {Araya},
  {Areeda}, {Arnaud}, {Arun}, {Ascenzi}, {Ashton}, {Ast}, {Aston}, {Astone},
  {Aufmuth}, {Aulbert}, {AultONeal}, {Avila-Alvarez}, {Babak}, {Bacon},
  {Bader}, {Bae}, {Baker}, {Baldaccini}, {Ballardin}, {Ballmer}, {Banagiri},
  {Barayoga}, {Barclay}, {Barish}, {Barker}, {Barone}, {Barr}, {Barsotti},
  {Barsuglia}, {Barta}, {Bartlett}, {Bartos}, {Bassiri}, {Basti}, {Batch},
  {Baune}, {Bawaj}, {Bazzan}, {B{\'e}csy}, {Beer}, {Bejger}, {Belahcene},
  {Bell}, {Berger}, {Bergmann}, {Berry}, {Bersanetti}, {Bertolini},
  {Betzwieser}, {Bhagwat}, {Bhandare}, {Bilenko}, {Billingsley}, {Billman},
  {Birch}, {Birney}, {Birnholtz}, {Biscans}, {Bisht}, {Bitossi}, {Biwer},
  {Bizouard}, {Blackburn}, {Blackman}, {Blair}, {Blair}, {Blair}, {Bloemen},
  {Bock}, {Bode}, {Boer}, {Bogaert}, {Bohe}, {Bondu}, {Bonnand}, {Boom},
  {Bork}, {Boschi}, {Bose}, {Bouffanais}, {Bozzi}, {Bradaschia}, {Brady},
  {Braginsky}, {Branchesi}, {Brau}, {Briant}, {Brillet}, {Brinkmann},
  {Brisson}, {Brockill}, {Broida}, {Brooks}, {Brown}, {Brown}, {Brown},
  {Brunett}, {Buchanan}, {Buikema}, {Bulik}, {Bulten}, {Buonanno}, {Buskulic},
  {Buy}, {Byer}, {Cabero}, {Cadonati}, {Cagnoli}, {Cahillane}, {Calder{\'o}n
  Bustillo}, {Callister}, {Calloni}, {Camp}, {Canepa}, {Canizares}, {Cannon},
  {Cao}, {Cao}, {Capano}, {Capocasa}, {Carbognani}, {Caride}, {Carney},
  {Casanueva Diaz}, {Casentini}, {Caudill}, {Cavagli{\`a}}, {Cavalier},
  {Cavalieri}, {Cella}, {Cepeda}, {Cerboni Baiardi}, {Cerretani}, {Cesarini},
  {Chamberlin}, {Chan}, {Chao}, {Charlton}, {Chassande-Mottin}, {Chatterjee},
  {Chatziioannou}, {Cheeseboro}, {Chen}, {Chen}, {Cheng}, {Chincarini},
  {Chiummo}, {Chmiel}, {Cho}, {Cho}, {Chow}, {Christensen}, {Chu}, {Chua},
  {Chua}, {Chung}, {Chung}, {Ciani}, {Ciolfi}, {Cirelli}, {Cirone}, {Clara},
  {Clark}, {Cleva}, {Cocchieri}, {Coccia}, {Cohadon}, {Colla}, {Collette},
  {Cominsky}, {Constancio}, {Conti}, {Cooper}, {Corban}, {Corbitt}, {Corley},
  {Cornish}, {Corsi}, {Cortese}, {Costa}, {Coughlin}, {Coughlin}, {Coulon},
  {Countryman}, {Couvares}, {Covas}, {Cowan}, {Coward}, {Cowart}, {Coyne},
  {Coyne}, {Creighton}, {Creighton}, {Cripe}, {Crowder}, {Cullen}, {Cumming},
  {Cunningham}, {Cuoco}, {Dal Canton}, {Danilishin}, {D'Antonio}, {Danzmann},
  {Dasgupta}, {Da Silva Costa}, {Dattilo}, {Dave}, {Davier}, {Davis}, {Daw},
  {Day}, {De}, {DeBra}, {Deelman}, {Degallaix}, {De Laurentis},
  {Del{\'e}glise}, {Del Pozzo}, {Denker}, {Dent}, {Dergachev}, {De Rosa},
  {DeRosa}, {DeSalvo}, {Devenson}, {Devine}, {Dhurandhar}, {D{\'\i}az}, {Di
  Fiore}, {Di Giovanni}, {Di Girolamo}, {Di Lieto}, {Di Pace}, {Di Palma}, {Di
  Renzo}, {Doctor}, {Dolique}, {Donovan}, {Dooley}, {Doravari}, {Dorrington},
  {Douglas}, {Dovale {\'A}lvarez}, {Downes}, {Drago}, {Drever}, {Driggers},
  {Du}, {Ducrot}, {Duncan}, {Dwyer}, {Edo}, {Edwards}, {Effler}, {Eggenstein},
  {Ehrens}, {Eichholz}, {Eikenberry}, {Eisenstein}, {Essick}, {Etienne},
  {Etzel}, {Evans}, {Evans}, {Factourovich}, {Fafone}, {Fair}, {Fairhurst},
  {Fan}, {Farinon}, {Farr}, {Farr}, {Fauchon-Jones}, {Favata}, {Fays},
  {Fehrmann}, {Feicht}, {Fejer}, {Fernandez-Galiana}, {Ferrante}, {Ferreira},
  {Ferrini}, {Fidecaro}, {Fiori}, {Fiorucci}, {Fisher}, {Flaminio}, {Fletcher},
  {Fong}, {Forsyth}, {Forsyth}, {Fournier}, {Frasca}, {Frasconi}, {Frei},
  {Freise}, {Frey}, {Frey}, {Fries}, {Fritschel}, {Frolov}, {Fulda}, {Fyffe},
  {Gabbard}, {Gabel}, {Gadre}, {Gaebel}, {Gair}, {Gammaitoni}, {Ganija},
  {Gaonkar}, {Garufi}, {Gaudio}, {Gaur}, {Gayathri}, {Gehrels}, {Gemme},
  {Genin}, {Gennai}, {George}, {George}, {Gergely}, {Germain}, {Ghonge},
  {Ghosh}, {Ghosh}, {Ghosh}, {Giaime}, {Giardina}, {Giazotto}, {Gill},
  {Glover}, {Goetz}, {Goetz}, {Gomes}, {Gonz{\'a}lez}, {Gonzalez Castro},
  {Gopakumar}, {Gorodetsky}, {Gossan}, {Gosselin}, {Gouaty}, {Grado}, {Graef},
  {Granata}, {Grant}, {Gras}, {Gray}, {Greco}, {Green}, {Groot}, {Grote},
  {Grunewald}, {Gruning}, {Guidi}, {Guo}, {Gupta}, {Gupta}, {Gushwa},
  {Gustafson}, {Gustafson}, {Hall}, {Hall}, {Hammond}, {Haney}, {Hanke},
  {Hanks}, {Hanna}, {Hannam}, {Hannuksela}, {Hanson}, {Hardwick}, {Harms},
  {Harry}, {Harry}, {Hart}, {Haster}, {Haughian}, {Healy}, {Heidmann},
  {Heintze}, {Heitmann}, {Hello}, {Hemming}, {Hendry}, {Heng}, {Hennig},
  {Henry}, {Heptonstall}, {Heurs}, {Hild}, {Hoak}, {Hofman}, {Holt}, {Holz},
  {Hopkins}, {Horst}, {Hough}, {Houston}, {Howell}, {Hu}, {Huerta}, {Huet},
  {Hughey}, {Husa}, {Huttner}, {Huynh-Dinh}, {Indik}, {Ingram}, {Inta},
  {Intini}, {Isa}, {Isac}, {Isi}, {Iyer}, {Izumi}, {Jacqmin}, {Jani},
  {Jaranowski}, {Jawahar}, {Jim{\'e}nez-Forteza}, {Johnson},
  {Johnson-McDaniel}, {Jones}, {Jones}, {Jonker}, {Ju}, {Junker}, {Kalaghatgi},
  {Kalogera}, {Kandhasamy}, {Kang}, {Kanner}, {Karki}, {Karvinen}, {Kasprzack},
  {Katolik}, {Katsavounidis}, {Katzman}, {Kaufer}, {Kawabe},
  {K{\'e}f{\'e}lian}, {Keitel}, {Kemball}, {Kennedy}, {Kent}, {Key}, {Khalili},
  {Khan}, {Khan}, {Khan}, {Khazanov}, {Kijbunchoo}, {Kim}, {Kim}, {Kim}, {Kim},
  {Kim}, {Kimbrell}, {King}, {King}, {Kirchhoff}, {Kissel}, {Kleybolte},
  {Klimenko}, {Koch}, {Koehlenbeck}, {Koley}, {Kondrashov}, {Kontos},
  {Korobko}, {Korth}, {Kowalska}, {Kozak}, {Kr{\"a}mer}, {Kringel}, {Krishnan},
  {Kr{\'o}lak}, {Kuehn}, {Kumar}, {Kumar}, {Kumar}, {Kuo}, {Kutynia}, {Kwang},
  {Lackey}, {Lai}, {Landry}, {Lang}, {Lange}, {Lantz}, {Lanza},
  {Lartaux-Vollard}, {Lasky}, {Laxen}, {Lazzarini}, {Lazzaro}, {Leaci},
  {Leavey}, {Lee}, {Lee}, {Lee}, {Lee}, {Lee}, {Lehmann}, {Lenon}, {Leonardi},
  {Leroy}, {Letendre}, {Levin}, {Li}, {Libson}, {Littenberg}, {Liu}, {Lo},
  {Lockerbie}, {London}, {Lord}, {Lorenzini}, {Loriette}, {Lormand}, {Losurdo},
  {Lough}, {Lovelace}, {L{\"u}ck}, {Lumaca}, {Lundgren}, {Lynch}, {Ma},
  {Macfoy}, {Machenschalk}, {MacInnis}, {Macleod}, {Maga{\~n}a Hernandez},
  {Maga{\~n}a-Sandoval}, {Maga{\~n}a Zertuche}, {Magee}, {Majorana},
  {Maksimovic}, {Man}, {Mandic}, {Mangano}, {Mansell}, {Manske}, {Mantovani},
  {Marchesoni}, {Marion}, {M{\'a}rka}, {M{\'a}rka}, {Markakis}, {Markosyan},
  {Maros}, {Martelli}, {Martellini}, {Martin}, {Martynov}, {Marx}, {Mason},
  {Masserot}, {Massinger}, {Masso-Reid}, {Mastrogiovanni}, {Matas},
  {Matichard}, {Matone}, {Mavalvala}, {Mayani}, {Mazumder}, {McCarthy},
  {McClelland}, {McCormick}, {McCuller}, {McGuire}, {McIntyre}, {McIver},
  {McManus}, {McRae}, {McWilliams}, {Meacher}, {Meadors}, {Meidam},
  {Mejuto-Villa}, {Melatos}, {Mendell}, {Mercer}, {Merilh}, {Merzougui},
  {Meshkov}, {Messenger}, {Messick}, {Metzdorff}, {Meyers}, {Mezzani}, {Miao},
  {Michel}, {Middleton}, {Mikhailov}, {Milano}, {Miller}, {Miller}, {Miller},
  {Miller}, {Millhouse}, {Minazzoli}, {Minenkov}, {Ming}, {Mishra}, {Mitra},
  {Mitrofanov}, {Mitselmakher}, {Mittleman}, {Moggi}, {Mohan}, {Mohapatra},
  {Montani}, {Moore}, {Moore}, {Moraru}, {Moreno}, {Morriss}, {Mours},
  {Mow-Lowry}, {Mueller}, {Muir}, {Mukherjee}, {Mukherjee}, {Mukherjee},
  {Mukund}, {Mullavey}, {Munch}, {Muniz}, {Murray}, {Napier}, {Nardecchia},
  {Naticchioni}, {Nayak}, {Nelemans}, {Nelson}, {Neri}, {Nery}, {Neunzert},
  {Newport}, {Newton}, {Ng}, {Nguyen}, {Nichols}, {Nielsen}, {Nissanke},
  {Nitz}, {Noack}, {Nocera}, {Nolting}, {Normandin}, {Nuttall}, {Oberling},
  {Ochsner}, {Oelker}, {Ogin}, {Oh}, {Oh}, {Ohme}, {Oliver}, {Oppermann},
  {Oram}, {O'Reilly}, {Ormiston}, {Ortega}, {O'Shaughnessy}, {Ottaway},
  {Overmier}, {Owen}, {Pace}, {Page}, {Page}, {Pai}, {Pai}, {Palamos},
  {Palashov}, {Palomba}, {Pal-Singh}, {Pan}, {Pang}, {Pang}, {Pankow},
  {Pannarale}, {Pant}, {Paoletti}, {Paoli}, {Papa}, {Paris}, {Parker},
  {Pascucci}, {Pasqualetti}, {Passaquieti}, {Passuello}, {Patricelli},
  {Pearlstone}, {Pedraza}, {Pedurand}, {Pekowsky}, {Pele}, {Penn}, {Perez},
  {Perreca}, {Perri}, {Pfeiffer}, {Phelps}, {Piccinni}, {Pichot},
  {Piergiovanni}, {Pierro}, {Pillant}, {Pinard}, {Pinto}, {Pitkin}, {Poggiani},
  {Popolizio}, {Porter}, {Post}, {Powell}, {Prasad}, {Pratt}, {Predoi},
  {Prestegard}, {Prijatelj}, {Principe}, {Privitera}, {Prodi}, {Prokhorov},
  {Puncken}, {Punturo}, {Puppo}, {P{\"u}rrer}, {Qi}, {Qin}, {Qiu}, {Quetschke},
  {Quintero}, {Quitzow-James}, {Raab}, {Rabeling}, {Radkins}, {Raffai}, {Raja},
  {Rajan}, {Rakhmanov}, {Ramirez}, {Rapagnani}, {Raymond}, {Razzano}, {Read},
  {Regimbau}, {Rei}, {Reid}, {Reitze}, {Rew}, {Reyes}, {Ricci}, {Ricker},
  {Rieger}, {Riles}, {Rizzo}, {Robertson}, {Robie}, {Robinet}, {Rocchi},
  {Rolland}, {Rollins}, {Roma}, {Romano}, {Romano}, {Romel}, {Romie},
  {Rosi{\'n}ska}, {Ross}, {Rowan}, {R{\"u}diger}, {Ruggi}, {Ryan}, {Rynge},
  {Sachdev}, {Sadecki}, {Sadeghian}, {Sakellariadou}, {Salconi}, {Saleem},
  {Salemi}, {Samajdar}, {Sammut}, {Sampson}, {Sanchez}, {Sandberg}, {Sandeen},
  {Sanders}, {Sassolas}, {Sathyaprakash}, {Saulson}, {Sauter}, {Savage},
  {Sawadsky}, {Schale}, {Scheuer}, {Schmidt}, {Schmidt}, {Schmidt}, {Schnabel},
  {Schofield}, {Sch{\"o}nbeck}, {Schreiber}, {Schuette}, {Schulte}, {Schutz},
  {Schwalbe}, {Scott}, {Scott}, {Seidel}, {Sellers}, {Sengupta}, {Sentenac},
  {Sequino}, {Sergeev}, {Shaddock}, {Shaffer}, {Shah}, {Shahriar}, {Shao},
  {Shapiro}, {Shawhan}, {Sheperd}, {Shoemaker}, {Shoemaker}, {Siellez},
  {Siemens}, {Sieniawska}, {Sigg}, {Silva}, {Singer}, {Singer}, {Singh},
  {Singh}, {Singhal}, {Sintes}, {Slagmolen}, {Smith}, {Smith}, {Smith}, {Son},
  {Sonnenberg}, {Sorazu}, {Sorrentino}, {Souradeep}, {Spencer}, {Srivastava},
  {Staley}, {Steinke}, {Steinlechner}, {Steinlechner}, {Steinmeyer},
  {Stephens}, {Stevenson}, {Stone}, {Strain}, {Stratta}, {Strigin}, {Sturani},
  {Stuver}, {Summerscales}, {Sun}, {Sunil}, {Sutton}, {Swinkels},
  {Szczepa{\'n}czyk}, {Tacca}, {Talukder}, {Tanner}, {T{\'a}pai}, {Taracchini},
  {Taylor}, {Taylor}, {Theeg}, {Thomas}, {Thomas}, {Thomas}, {Thorne},
  {Thorne}, {Thrane}, {Tiwari}, {Tiwari}, {Tokmakov}, {Toland}, {Tonelli},
  {Tornasi}, {Torrie}, {T{\"o}yr{\"a}}, {Travasso}, {Traylor}, {Trifir{\`o}},
  {Trinastic}, {Tringali}, {Trozzo}, {Tsang}, {Tse}, {Tso}, {Tuyenbayev},
  {Ueno}, {Ugolini}, {Unnikrishnan}, {Urban}, {Usman}, {Vahi}, {Vahlbruch},
  {Vajente}, {Valdes}, {Vallisneri}, {van Bakel}, {van Beuzekom}, {van den
  Brand}, {Van Den Broeck}, {Vander-Hyde}, {van der Schaaf}, {van Heijningen},
  {van Veggel}, {Vardaro}, {Varma}, {Vass}, {Vas{\'u}th}, {Vecchio},
  {Vedovato}, {Veitch}, {Veitch}, {Venkateswara}, {Venugopalan}, {Verkindt},
  {Vetrano}, {Vicer{\'e}}, {Viets}, {Vinciguerra}, {Vine}, {Vinet}, {Vitale},
  {Vo}, {Vocca}, {Vorvick}, {Voss}, {Vousden}, {Vyatchanin}, {Wade}, {Wade},
  {Wade}, {Wald}, {Walet}, {Walker}, {Wallace}, {Walsh}, {Wang}, {Wang},
  {Wang}, {Wang}, {Wang}, {Wang}, {Ward}, {Warner}, {Was}, {Watchi}, {Weaver},
  {Wei}, {Weinert}, {Weinstein}, {Weiss}, {Wen}, {Wessel}, {We{\ss}els},
  {Westphal}, {Wette}, {Whelan}, {Whiting}, {Whittle}, {Williams}, {Williams},
  {Williamson}, {Willis}, {Willke}, {Wimmer}, {Winkler}, {Wipf}, {Wittel},
  {Woan}, {Woehler}, {Wofford}, {Wong}, {Worden}, {Wright}, {Wu}, {Wu}, {Yam},
  {Yamamoto}, {Yancey}, {Yap}, {Yu}, {Yu}, {Yvert}, {Zadro{\.Z}ny}, {Zanolin},
  {Zelenova}, {Zendri}, {Zevin}, {Zhang}, {Zhang}, {Zhang}, {Zhang}, {Zhao},
  {Zhou}, {Zhou}, {Zhu}, {Zimmerman}, {Zucker}, {Zweizig}, {LIGO Scientific},
  \& {Virgo Collaboration}}]{abbott_2017_ab}
{Abbott}, B.~P., {Abbott}, R., {Abbott}, T.~D., {et~al.} 2017{\natexlab{a}},
  \prl, 118, 221101, \dodoi{10.1103/PhysRevLett.118.221101}

\bibitem[{{Abbott} {et~al.}(2017{\natexlab{b}}){Abbott}, {Abbott}, {Abbott},
  {Acernese}, {Ackley}, {Adams}, {Adams}, {Addesso}, {Adhikari}, {Adya},
  {Affeldt}, {Afrough}, {Agarwal}, {Agathos}, {Agatsuma}, {Aggarwal}, {Aguiar},
  {Aiello}, {Ain}, {Ajith}, {Allen}, {Allen}, {Allocca}, {Altin}, {Amato},
  {Ananyeva}, {Anderson}, {Anderson}, {Angelova}, {Antier}, {Appert}, {Arai},
  {Araya}, {Areeda}, {Arnaud}, {Arun}, {Ascenzi}, {Ashton}, {Ast}, {Aston},
  {Astone}, {Atallah}, {Aufmuth}, {Aulbert}, {AultONeal}, {Austin},
  {Avila-Alvarez}, {Babak}, {Bacon}, {Bader}, {Bae}, {Baker}, {Baldaccini},
  {Ballardin}, {Ballmer}, {Banagiri}, {Barayoga}, {Barclay}, {Barish},
  {Barker}, {Barkett}, {Barone}, {Barr}, {Barsotti}, {Barsuglia}, {Barta},
  {Bartlett}, {Bartos}, {Bassiri}, {Basti}, {Batch}, {Bawaj}, {Bayley},
  {Bazzan}, {B{\'e}csy}, {Beer}, {Bejger}, {Belahcene}, {Bell}, {Berger},
  {Bergmann}, {Bero}, {Berry}, {Bersanetti}, {Bertolini}, {Betzwieser},
  {Bhagwat}, {Bhandare}, {Bilenko}, {Billingsley}, {Billman}, {Birch},
  {Birney}, {Birnholtz}, {Biscans}, {Biscoveanu}, {Bisht}, {Bitossi}, {Biwer},
  {Bizouard}, {Blackburn}, {Blackman}, {Blair}, {Blair}, {Blair}, {Bloemen},
  {Bock}, {Bode}, {Boer}, {Bogaert}, {Bohe}, {Bondu}, {Bonilla}, {Bonnand},
  {Boom}, {Bork}, {Boschi}, {Bose}, {Bossie}, {Bouffanais}, {Bozzi},
  {Bradaschia}, {Brady}, {Branchesi}, {Brau}, {Briant}, {Brillet}, {Brinkmann},
  {Brisson}, {Brockill}, {Broida}, {Brooks}, {Brown}, {Brown}, {Brunett},
  {Buchanan}, {Buikema}, {Bulik}, {Bulten}, {Buonanno}, {Buskulic}, {Buy},
  {Byer}, {Cabero}, {Cadonati}, {Cagnoli}, {Cahillane}, {Calder{\'o}n
  Bustillo}, {Callister}, {Calloni}, {Camp}, {Canepa}, {Canizares}, {Cannon},
  {Cao}, {Cao}, {Capano}, {Capocasa}, {Carbognani}, {Caride}, {Carney},
  {Casanueva Diaz}, {Casentini}, {Caudill}, {Cavagli{\`a}}, {Cavalier},
  {Cavalieri}, {Cella}, {Cepeda}, {Cerd{\'a}-Dur{\'a}n}, {Cerretani},
  {Cesarini}, {Chamberlin}, {Chan}, {Chao}, {Charlton}, {Chase}, {Chassande-
  Mottin}, {Chatterjee}, {Chatziioannou}, {Cheeseboro}, {Chen}, {Chen}, {Chen},
  {Cheng}, {Chia}, {Chincarini}, {Chiummo}, {Chmiel}, {Cho}, {Cho}, {Chow},
  {Christensen}, {Chu}, {Chua}, {Chua}, {Chung}, {Chung}, {Ciani}, {Ciolfi},
  {Cirelli}, {Cirone}, {Clara}, {Clark}, {Clearwater}, {Cleva}, {Cocchieri},
  {Coccia}, {Cohadon}, {Cohen}, {Colla}, {Collette}, {Cominsky}, {Constancio},
  {Conti}, {Cooper}, {Corban}, {Corbitt}, {Cordero-Carri{\'o}n}, {Corley},
  {Cornish}, {Corsi}, {Cortese}, {Costa}, {Coughlin}, {Coughlin}, {Coulon},
  {Countryman}, {Couvares}, {Covas}, {Cowan}, {Coward}, {Cowart}, {Coyne},
  {Coyne}, {Creighton}, {Creighton}, {Cripe}, {Crowder}, {Cullen}, {Cumming},
  {Cunningham}, {Cuoco}, {Dal Canton}, {D{\'a}lya}, {Danilishin}, {D'Antonio},
  {Danzmann}, {Dasgupta}, {Da Silva Costa}, {Dattilo}, {Dave}, {Davier},
  {Davis}, {Daw}, {Day}, {De}, {DeBra}, {Degallaix}, {De Laurentis},
  {Del{\'e}glise}, {Del Pozzo}, {Demos}, {Denker}, {Dent}, {De Pietri},
  {Dergachev}, {De Rosa}, {DeRosa}, {De Rossi}, {DeSalvo}, {de Varona},
  {Devenson}, {Dhurandhar}, {D{\'\i}az}, {Di Fiore}, {Di Giovanni}, {Di
  Girolamo}, {Di Lieto}, {Di Pace}, {Di Palma}, {Di Renzo}, {Doctor},
  {Dolique}, {Donovan}, {Dooley}, {Doravari}, {Dorrington}, {Douglas}, {Dovale
  {\'A}lvarez}, {Downes}, {Drago}, {Dreissigacker}, {Driggers}, {Du}, {Ducrot},
  {Dupej}, {Dwyer}, {Edo}, {Edwards}, {Effler}, {Eggenstein}, {Ehrens},
  {Eichholz}, {Eikenberry}, {Eisenstein}, {Essick}, {Estevez}, {Etienne},
  {Etzel}, {Evans}, {Evans}, {Factourovich}, {Fafone}, {Fair}, {Fairhurst},
  {Fan}, {Farinon}, {Farr}, {Farr}, {Fauchon-Jones}, {Favata}, {Fays}, {Fee},
  {Fehrmann}, {Feicht}, {Fejer}, {Fernandez-Galiana}, {Ferrante}, {Ferreira},
  {Ferrini}, {Fidecaro}, {Finstad}, {Fiori}, {Fiorucci}, {Fishbach}, {Fisher},
  {Fitz-Axen}, {Flaminio}, {Fletcher}, {Fong}, {Font}, {Forsyth}, {Forsyth},
  {Fournier}, {Frasca}, {Frasconi}, {Frei}, {Freise}, {Frey}, {Frey}, {Fries},
  {Fritschel}, {Frolov}, {Fulda}, {Fyffe}, {Gabbard}, {Gadre}, {Gaebel},
  {Gair}, {Gammaitoni}, {Ganija}, {Gaonkar}, {Garcia-Quiros}, {Garufi},
  {Gateley}, {Gaudio}, {Gaur}, {Gayathri}, {Gehrels}, {Gemme}, {Genin},
  {Gennai}, {George}, {George}, {Gergely}, {Germain}, {Ghonge}, {Ghosh},
  {Ghosh}, {Ghosh}, {Giaime}, {Giardina}, {Giazotto}, {Gill}, {Glover},
  {Goetz}, {Goetz}, {Gomes}, {Goncharov}, {Gonz{\'a}lez}, {Gonzalez Castro},
  {Gopakumar}, {Gorodetsky}, {Gossan}, {Gosselin}, {Gouaty}, {Grado}, {Graef},
  {Granata}, {Grant}, {Gras}, {Gray}, {Greco}, {Green}, {Gretarsson}, {Groot},
  {Grote}, {Grunewald}, {Gruning}, {Guidi}, {Guo}, {Gupta}, {Gupta}, {Gushwa},
  {Gustafson}, {Gustafson}, {Halim}, {Hall}, {Hall}, {Hamilton}, {Hammond},
  {Haney}, {Hanke}, {Hanks}, {Hanna}, {Hannam}, {Hannuksela}, {Hanson},
  {Hardwick}, {Harms}, {Harry}, {Harry}, {Hart}, {Haster}, {Haughian}, {Healy},
  {Heidmann}, {Heintze}, {Heitmann}, {Hello}, {Hemming}, {Hendry}, {Heng},
  {Hennig}, {Heptonstall}, {Heurs}, {Hild}, {Hinderer}, {Hoak}, {Hofman},
  {Holt}, {Holz}, {Hopkins}, {Horst}, {Hough}, {Houston}, {Howell}, {Hreibi},
  {Hu}, {Huerta}, {Huet}, {Hughey}, {Husa}, {Huttner}, {Huynh-Dinh}, {Indik},
  {Inta}, {Intini}, {Isa}, {Isac}, {Isi}, {Iyer}, {Izumi}, {Jacqmin}, {Jani},
  {Jaranowski}, {Jawahar}, {Jim{\'e}nez-Forteza}, {Johnson},
  {Johnson-McDaniel}, {Jones}, {Jones}, {Jonker}, {Ju}, {Junker}, {Kalaghatgi},
  {Kalogera}, {Kamai}, {Kandhasamy}, {Kang}, {Kanner}, {Kapadia}, {Karki},
  {Karvinen}, {Kasprzack}, {Katolik}, {Katsavounidis}, {Katzman}, {Kaufer},
  {Kawabe}, {K{\'e}f{\'e}lian}, {Keitel}, {Kemball}, {Kennedy}, {Kent}, {Key},
  {Khalili}, {Khan}, {Khan}, {Khan}, {Khazanov}, {Kijbunchoo}, {Kim}, {Kim},
  {Kim}, {Kim}, {Kim}, {Kim}, {Kimbrell}, {King}, {King}, {Kinley-Hanlon},
  {Kirchhoff}, {Kissel}, {Kleybolte}, {Klimenko}, {Knowles}, {Koch},
  {Koehlenbeck}, {Koley}, {Kondrashov}, {Kontos}, {Korobko}, {Korth},
  {Kowalska}, {Kozak}, {Kr{\"a}mer}, {Kringel}, {Krishnan}, {Kr{\'o}lak},
  {Kuehn}, {Kumar}, {Kumar}, {Kumar}, {Kuo}, {Kutynia}, {Kwang}, {Lackey},
  {Lai}, {Landry}, {Lang}, {Lange}, {Lantz}, {Lanza}, {Lartaux-Vollard},
  {Lasky}, {Laxen}, {Lazzarini}, {Lazzaro}, {Leaci}, {Leavey}, {Lee}, {Lee},
  {Lee}, {Lee}, {Lee}, {Lehmann}, {Lenon}, {Leonardi}, {Leroy}, {Letendre},
  {Levin}, {Li}, {Linker}, {Littenberg}, {Liu}, {Lo}, {Lockerbie}, {London},
  {Lord}, {Lorenzini}, {Loriette}, {Lormand}, {Losurdo}, {Lough}, {Lousto},
  {Lovelace}, {L{\"u}ck}, {Lumaca}, {Lundgren}, {Lynch}, {Ma}, {Macas},
  {Macfoy}, {Machenschalk}, {MacInnis}, {Macleod}, {Maga{\~n}a Hernandez},
  {Maga{\~n}a-Sandoval}, {Maga{\~n}a Zertuche}, {Magee}, {Majorana},
  {Maksimovic}, {Man}, {Mandic}, {Mangano}, {Mansell}, {Manske}, {Mantovani},
  {Marchesoni}, {Marion}, {M{\'a}rka}, {M{\'a}rka}, {Markakis}, {Markosyan},
  {Markowitz}, {Maros}, {Marquina}, {Martelli}, {Martellini}, {Martin},
  {Martin}, {Martynov}, {Mason}, {Massera}, {Masserot}, {Massinger},
  {Masso-Reid}, {Mastrogiovanni}, {Matas}, {Matichard}, {Matone}, {Mavalvala},
  {Mazumder}, {McCarthy}, {McClelland}, {McCormick}, {McCuller}, {McGuire},
  {McIntyre}, {McIver}, {McManus}, {McNeill}, {McRae}, {McWilliams}, {Meacher},
  {Meadors}, {Mehmet}, {Meidam}, {Mejuto-Villa}, {Melatos}, {Mendell},
  {Mercer}, {Merilh}, {Merzougui}, {Meshkov}, {Messenger}, {Messick},
  {Metzdorff}, {Meyers}, {Miao}, {Michel}, {Middleton}, {Mikhailov}, {Milano},
  {Miller}, {Miller}, {Miller}, {Millhouse}, {Milovich-Goff}, {Minazzoli},
  {Minenkov}, {Ming}, {Mishra}, {Mitra}, {Mitrofanov}, {Mitselmakher},
  {Mittleman}, {Moffa}, {Moggi}, {Mogushi}, {Mohan}, {Mohapatra}, {Montani},
  {Moore}, {Moraru}, {Moreno}, {Morriss}, {Mours}, {Mow-Lowry}, {Mueller},
  {Muir}, {Mukherjee}, {Mukherjee}, {Mukherjee}, {Mukund}, {Mullavey}, {Munch},
  {Mu{\~n}iz}, {Muratore}, {Murray}, {Napier}, {Nardecchia}, {Naticchioni},
  {Nayak}, {Neilson}, {Nelemans}, {Nelson}, {Nery}, {Neunzert}, {Nevin},
  {Newport}, {Newton}, {Ng}, {Nguyen}, {Nichols}, {Nielsen}, {Nissanke},
  {Nitz}, {Noack}, {Nocera}, {Nolting}, {North}, {Nuttall}, {Oberling},
  {O'Dea}, {Ogin}, {Oh}, {Oh}, {Ohme}, {Okada}, {Oliver}, {Oppermann}, {Oram},
  {O'Reilly}, {Ormiston}, {Ortega}, {O'Shaughnessy}, {Ossokine}, {Ottaway},
  {Overmier}, {Owen}, {Pace}, {Page}, {Page}, {Pai}, {Pai}, {Palamos},
  {Palashov}, {Palomba}, {Pal- Singh}, {Pan}, {Pan}, {Pang}, {Pang}, {Pankow},
  {Pannarale}, {Pant}, {Paoletti}, {Paoli}, {Papa}, {Parida}, {Parker},
  {Pascucci}, {Pasqualetti}, {Passaquieti}, {Passuello}, {Patil}, {Patricelli},
  {Pearlstone}, {Pedraza}, {Pedurand}, {Pekowsky}, {Pele}, {Penn}, {Perez},
  {Perreca}, {Perri}, {Pfeiffer}, {Phelps}, {Piccinni}, {Pichot},
  {Piergiovanni}, {Pierro}, {Pillant}, {Pinard}, {Pinto}, {Pirello}, {Pitkin},
  {Poe}, {Poggiani}, {Popolizio}, {Porter}, {Post}, {Powell}, {Prasad},
  {Pratt}, {Pratten}, {Predoi}, {Prestegard}, {Prijatelj}, {Principe},
  {Privitera}, {Prodi}, {Prokhorov}, {Puncken}, {Punturo}, {Puppo},
  {P{\"u}rrer}, {Qi}, {Quetschke}, {Quintero}, {Quitzow-James}, {Raab},
  {Rabeling}, {Radkins}, {Raffai}, {Raja}, {Rajan}, {Rajbhandari}, {Rakhmanov},
  {Ramirez}, {Ramos-Buades}, {Rapagnani}, {Raymond}, {Razzano}, {Read},
  {Regimbau}, {Rei}, {Reid}, {Reitze}, {Ren}, {Reyes}, {Ricci}, {Ricker},
  {Rieger}, {Riles}, {Rizzo}, {Robertson}, {Robie}, {Robinet}, {Rocchi},
  {Rolland}, {Rollins}, {Roma}, {Romano}, {Romel}, {Romie}, {Rosi{\'n}ska},
  {Ross}, {Rowan}, {R{\"u}diger}, {Ruggi}, {Rutins}, {Ryan}, {Sachdev},
  {Sadecki}, {Sadeghian}, {Sakellariadou}, {Salconi}, {Saleem}, {Salemi},
  {Samajdar}, {Sammut}, {Sampson}, {Sanchez}, {Sanchez}, {Sanchis-Gual},
  {Sandberg}, {Sanders}, {Sassolas}, {Sathyaprakash}, {Saulson}, {Sauter},
  {Savage}, {Sawadsky}, {Schale}, {Scheel}, {Scheuer}, {Schmidt}, {Schmidt},
  {Schnabel}, {Schofield}, {Sch{\"o}nbeck}, {Schreiber}, {Schuette}, {Schulte},
  {Schutz}, {Schwalbe}, {Scott}, {Scott}, {Seidel}, {Sellers}, {Sengupta},
  {Sentenac}, {Sequino}, {Sergeev}, {Shaddock}, {Shaffer}, {Shah}, {Shahriar},
  {Shaner}, {Shao}, {Shapiro}, {Shawhan}, {Sheperd}, {Shoemaker}, {Shoemaker},
  {Siellez}, {Siemens}, {Sieniawska}, {Sigg}, {Silva}, {Singer}, {Singh},
  {Singhal}, {Sintes}, {Slagmolen}, {Smith}, {Smith}, {Smith}, {Somala}, {Son},
  {Sonnenberg}, {Sorazu}, {Sorrentino}, {Souradeep}, {Spencer}, {Srivastava},
  {Staats}, {Staley}, {Steinke}, {Steinlechner}, {Steinlechner}, {Steinmeyer},
  {Stevenson}, {Stone}, {Stops}, {Strain}, {Stratta}, {Strigin}, {Strunk},
  {Sturani}, {Stuver}, {Summerscales}, {Sun}, {Sunil}, {Suresh}, {Sutton},
  {Swinkels}, {Szczepa{\'n}czyk}, {Tacca}, {Tait}, {Talbot}, {Talukder},
  {Tanner}, {T{\'a}pai}, {Taracchini}, {Tasson}, {Taylor}, {Taylor}, {Tewari},
  {Theeg}, {Thies}, {Thomas}, {Thomas}, {Thomas}, {Thorne}, {Thrane}, {Tiwari},
  {Tiwari}, {Tokmakov}, {Toland}, {Tonelli}, {Tornasi}, {Torres- Forn{\'e}},
  {Torrie}, {T{\"o}yr{\"a}}, {Travasso}, {Traylor}, {Trinastic}, {Tringali},
  {Trozzo}, {Tsang}, {Tse}, {Tso}, {Tsukada}, {Tsuna}, {Tuyenbayev}, {Ueno},
  {Ugolini}, {Unnikrishnan}, {Urban}, {Usman}, {Vahlbruch}, {Vajente},
  {Valdes}, {van Bakel}, {van Beuzekom}, {van den Brand}, {Van Den Broeck},
  {Vander-Hyde}, {van der Schaaf}, {van Heijningen}, {van Veggel}, {Vardaro},
  {Varma}, {Vass}, {Vas{\'u}th}, {Vecchio}, {Vedovato}, {Veitch}, {Veitch},
  {Venkateswara}, {Venugopalan}, {Verkindt}, {Vetrano}, {Vicer{\'e}}, {Viets},
  {Vinciguerra}, {Vine}, {Vinet}, {Vitale}, {Vo}, {Vocca}, {Vorvick},
  {Vyatchanin}, {Wade}, {Wade}, {Wade}, {Walet}, {Walker}, {Wallace}, {Walsh},
  {Wang}, {Wang}, {Wang}, {Wang}, {Wang}, {Ward}, {Warner}, {Was}, {Watchi},
  {Weaver}, {Wei}, {Weinert}, {Weinstein}, {Weiss}, {Wen}, {Wessel},
  {We{\ss}els}, {Westerweck}, {Westphal}, {Wette}, {Whelan}, {Whiting},
  {Whittle}, {Wilken}, {Williams}, {Williams}, {Williamson}, {Willis},
  {Willke}, {Wimmer}, {Winkler}, {Wipf}, {Wittel}, {Woan}, {Woehler},
  {Wofford}, {Wong}, {Worden}, {Wright}, {Wu}, {Wysocki}, {Xiao}, {Yamamoto},
  {Yancey}, {Yang}, {Yap}, {Yazback}, {Yu}, {Yu}, {Yvert}, {Zadro{\.z}ny},
  {Zanolin}, {Zelenova}, {Zendri}, {Zevin}, {Zhang}, {Zhang}, {Zhang}, {Zhang},
  {Zhao}, {Zhou}, {Zhou}, {Zhu}, {Zhu}, {Zimmerman}, {Zucker}, {Zweizig},
  {(LIGO Scientific Collaboration}, \& {Virgo Collaboration}}]{abbott_2017_ac}
---. 2017{\natexlab{b}}, \apj, 851, L35, \dodoi{10.3847/2041-8213/aa9f0c}

\bibitem[{{Abbott} {et~al.}(2017{\natexlab{c}}){Abbott}, {Abbott}, {Abbott},
  {Acernese}, {Ackley}, {Adams}, {Adams}, {Addesso}, {Adhikari}, {Adya},
  {Affeldt}, {Afrough}, {Agarwal}, {Agathos}, {Agatsuma}, {Aggarwal}, {Aguiar},
  {Aiello}, {Ain}, {Ajith}, {Allen}, {Allen}, {Allocca}, {Altin}, {Amato},
  {Ananyeva}, {Anderson}, {Anderson}, {Angelova}, {Antier}, {Appert}, {Arai},
  {Araya}, {Areeda}, {Arnaud}, {Arun}, {Ascenzi}, {Ashton}, {Ast}, {Aston},
  {Astone}, {Atallah}, {Aufmuth}, {Aulbert}, {AultONeal}, {Austin},
  {Avila-Alvarez}, {Babak}, {Bacon}, {Bader}, {Bae}, {Baker}, {Baldaccini},
  {Ballardin}, {Ballmer}, {Banagiri}, {Barayoga}, {Barclay}, {Barish},
  {Barker}, {Barkett}, {Barone}, {Barr}, {Barsotti}, {Barsuglia}, {Barta},
  {Barthelmy}, {Bartlett}, {Bartos}, {Bassiri}, {Basti}, {Batch}, {Bawaj},
  {Bayley}, {Bazzan}, {B{\'e}csy}, {Beer}, {Bejger}, {Belahcene}, {Bell},
  {Berger}, {Bergmann}, {Bero}, {Berry}, {Bersanetti}, {Bertolini},
  {Betzwieser}, {Bhagwat}, {Bhandare}, {Bilenko}, {Billingsley}, {Billman},
  {Birch}, {Birney}, {Birnholtz}, {Biscans}, {Biscoveanu}, {Bisht}, {Bitossi},
  {Biwer}, {Bizouard}, {Blackburn}, {Blackman}, {Blair}, {Blair}, {Blair},
  {Bloemen}, {Bock}, {Bode}, {Boer}, {Bogaert}, {Bohe}, {Bondu}, {Bonilla},
  {Bonnand}, {Boom}, {Bork}, {Boschi}, {Bose}, {Bossie}, {Bouffanais}, {Bozzi},
  {Bradaschia}, {Brady}, {Branchesi}, {Brau}, {Briant}, {Brillet}, {Brinkmann},
  {Brisson}, {Brockill}, {Broida}, {Brooks}, {Brown}, {Brown}, {Brunett},
  {Buchanan}, {Buikema}, {Bulik}, {Bulten}, {Buonanno}, {Buskulic}, {Buy},
  {Byer}, {Cabero}, {Cadonati}, {Cagnoli}, {Cahillane}, {Calder{\'o}n
  Bustillo}, {Callister}, {Calloni}, {Camp}, {Canepa}, {Canizares}, {Cannon},
  {Cao}, {Cao}, {Capano}, {Capocasa}, {Carbognani}, {Caride}, {Carney},
  {Casanueva Diaz}, {Casentini}, {Caudill}, {Cavagli{\`a}}, {Cavalier},
  {Cavalieri}, {Cella}, {Cepeda}, {Cerd{\'a}-Dur{\'a}n}, {Cerretani},
  {Cesarini}, {Chamberlin}, {Chan}, {Chao}, {Charlton}, {Chase}, {Chassande-
  Mottin}, {Chatterjee}, {Chatziioannou}, {Cheeseboro}, {Chen}, {Chen}, {Chen},
  {Cheng}, {Chia}, {Chincarini}, {Chiummo}, {Chmiel}, {Cho}, {Cho}, {Chow},
  {Christensen}, {Chu}, {Chua}, {Chua}, {Chung}, {Chung}, {Ciani}, {Ciolfi},
  {Cirelli}, {Cirone}, {Clara}, {Clark}, {Clearwater}, {Cleva}, {Cocchieri},
  {Coccia}, {Cohadon}, {Cohen}, {Colla}, {Collette}, {Cominsky}, {Constancio},
  {Conti}, {Cooper}, {Corban}, {Corbitt}, {Cordero-Carri{\'o}n}, {Corley},
  {Cornish}, {Corsi}, {Cortese}, {Costa}, {Coughlin}, {Coughlin}, {Coulon},
  {Countryman}, {Couvares}, {Covas}, {Cowan}, {Coward}, {Cowart}, {Coyne},
  {Coyne}, {Creighton}, {Creighton}, {Cripe}, {Crowder}, {Cullen}, {Cumming},
  {Cunningham}, {Cuoco}, {Dal Canton}, {D{\'a}lya}, {Danilishin}, {D'Antonio},
  {Danzmann}, {Dasgupta}, {Da Silva Costa}, {Dattilo}, {Dave}, {Davier},
  {Davis}, {Daw}, {Day}, {De}, {DeBra}, {Degallaix}, {De Laurentis},
  {Del{\'e}glise}, {Del Pozzo}, {Demos}, {Denker}, {Dent}, {De Pietri},
  {Dergachev}, {De Rosa}, {DeRosa}, {De Rossi}, {DeSalvo}, {de Varona},
  {Devenson}, {Dhurandhar}, {D{\'\i}az}, {Di Fiore}, {Di Giovanni}, {Di
  Girolamo}, {Di Lieto}, {Di Pace}, {Di Palma}, {Di Renzo}, {Doctor},
  {Dolique}, {Donovan}, {Dooley}, {Doravari}, {Dorrington}, {Douglas}, {Dovale
  {\'A}lvarez}, {Downes}, {Drago}, {Dreissigacker}, {Driggers}, {Du}, {Ducrot},
  {Dupej}, {Dwyer}, {Edo}, {Edwards}, {Effler}, {Eggenstein}, {Ehrens},
  {Eichholz}, {Eikenberry}, {Eisenstein}, {Essick}, {Estevez}, {Etienne},
  {Etzel}, {Evans}, {Evans}, {Factourovich}, {Fafone}, {Fair}, {Fairhurst},
  {Fan}, {Farinon}, {Farr}, {Farr}, {Fauchon-Jones}, {Favata}, {Fays}, {Fee},
  {Fehrmann}, {Feicht}, {Fejer}, {Fernandez-Galiana}, {Ferrante}, {Ferreira},
  {Ferrini}, {Fidecaro}, {Finstad}, {Fiori}, {Fiorucci}, {Fishbach}, {Fisher},
  {Fitz-Axen}, {Flaminio}, {Fletcher}, {Fong}, {Font}, {Forsyth}, {Forsyth},
  {Fournier}, {Frasca}, {Frasconi}, {Frei}, {Freise}, {Frey}, {Frey}, {Fries},
  {Fritschel}, {Frolov}, {Fulda}, {Fyffe}, {Gabbard}, {Gadre}, {Gaebel},
  {Gair}, {Gammaitoni}, {Ganija}, {Gaonkar}, {Garcia-Quiros}, {Garufi},
  {Gateley}, {Gaudio}, {Gaur}, {Gayathri}, {Gehrels}, {Gemme}, {Genin},
  {Gennai}, {George}, {George}, {Gergely}, {Germain}, {Ghonge}, {Ghosh},
  {Ghosh}, {Ghosh}, {Giaime}, {Giardina}, {Giazotto}, {Gill}, {Glover},
  {Goetz}, {Goetz}, {Gomes}, {Goncharov}, {Gonz{\'a}lez}, {Gonzalez Castro},
  {Gopakumar}, {Gorodetsky}, {Gossan}, {Gosselin}, {Gouaty}, {Grado}, {Graef},
  {Granata}, {Grant}, {Gras}, {Gray}, {Greco}, {Green}, {Gretarsson}, {Groot},
  {Grote}, {Grunewald}, {Gruning}, {Guidi}, {Guo}, {Gupta}, {Gupta}, {Gushwa},
  {Gustafson}, {Gustafson}, {Halim}, {Hall}, {Hall}, {Hamilton}, {Hammond},
  {Haney}, {Hanke}, {Hanks}, {Hanna}, {Hannam}, {Hannuksela}, {Hanson},
  {Hardwick}, {Harms}, {Harry}, {Harry}, {Hart}, {Haster}, {Haughian}, {Healy},
  {Heidmann}, {Heintze}, {Heitmann}, {Hello}, {Hemming}, {Hendry}, {Heng},
  {Hennig}, {Heptonstall}, {Heurs}, {Hild}, {Hinderer}, {Hoak}, {Hofman},
  {Holt}, {Holz}, {Hopkins}, {Horst}, {Hough}, {Houston}, {Howell}, {Hu},
  {Huerta}, {Huet}, {Hughey}, {Husa}, {Huttner}, {Huynh-Dinh}, {Indik}, {Inta},
  {Intini}, {Isa}, {Isac}, {Isi}, {Iyer}, {Izumi}, {Jacqmin}, {Jani},
  {Jaranowski}, {Jawahar}, {Jim{\'e}nez-Forteza}, {Johnson}, {Johnson-
  McDaniel}, {Jones}, {Jones}, {Jonker}, {Ju}, {Junker}, {Kalaghatgi},
  {Kalogera}, {Kamai}, {Kandhasamy}, {Kang}, {Kanner}, {Kapadia}, {Karki},
  {Karvinen}, {Kasprzack}, {Katolik}, {Katsavounidis}, {Katzman}, {Kaufer},
  {Kawabe}, {K{\'e}f{\'e}lian}, {Keitel}, {Kemball}, {Kennedy}, {Kent}, {Key},
  {Khalili}, {Khan}, {Khan}, {Khan}, {Khazanov}, {Kijbunchoo}, {Kim}, {Kim},
  {Kim}, {Kim}, {Kim}, {Kim}, {Kimbrell}, {King}, {King}, {Kinley-Hanlon},
  {Kirchhoff}, {Kissel}, {Kleybolte}, {Klimenko}, {Knowles}, {Koch},
  {Koehlenbeck}, {Koley}, {Kondrashov}, {Kontos}, {Korobko}, {Korth},
  {Kowalska}, {Kozak}, {Kr{\"a}mer}, {Kringel}, {Krishnan}, {Kr{\'o}lak},
  {Kuehn}, {Kumar}, {Kumar}, {Kumar}, {Kuo}, {Kutynia}, {Kwang}, {Lackey},
  {Lai}, {Landry}, {Lang}, {Lange}, {Lantz}, {Lanza}, {Lartaux-Vollard},
  {Lasky}, {Laxen}, {Lazzarini}, {Lazzaro}, {Leaci}, {Leavey}, {Lee}, {Lee},
  {Lee}, {Lee}, {Lee}, {Lehmann}, {Lenon}, {Leonardi}, {Leroy}, {Letendre},
  {Levin}, {Li}, {Linker}, {Littenberg}, {Liu}, {Lo}, {Lockerbie}, {London},
  {Lord}, {Lorenzini}, {Loriette}, {Lormand}, {Losurdo}, {Lough}, {Lousto},
  {Lovelace}, {L{\"u}ck}, {Lumaca}, {Lundgren}, {Lynch}, {Ma}, {Macas},
  {Macfoy}, {Machenschalk}, {MacInnis}, {Macleod}, {Maga{\~n}a Hernandez},
  {Maga{\~n}a-Sandoval}, {Maga{\~n}a Zertuche}, {Magee}, {Majorana},
  {Maksimovic}, {Man}, {Mandic}, {Mangano}, {Mansell}, {Manske}, {Mantovani},
  {Marchesoni}, {Marion}, {M{\'a}rka}, {M{\'a}rka}, {Markakis}, {Markosyan},
  {Markowitz}, {Maros}, {Marquina}, {Marsh}, {Martelli}, {Martellini},
  {Martin}, {Martin}, {Martynov}, {Mason}, {Massera}, {Masserot}, {Massinger},
  {Masso-Reid}, {Mastrogiovanni}, {Matas}, {Matichard}, {Matone}, {Mavalvala},
  {Mazumder}, {McCarthy}, {McClelland}, {McCormick}, {McCuller}, {McGuire},
  {McIntyre}, {McIver}, {McManus}, {McNeill}, {McRae}, {McWilliams}, {Meacher},
  {Meadors}, {Mehmet}, {Meidam}, {Mejuto-Villa}, {Melatos}, {Mendell},
  {Mercer}, {Merilh}, {Merzougui}, {Meshkov}, {Messenger}, {Messick},
  {Metzdorff}, {Meyers}, {Miao}, {Michel}, {Middleton}, {Mikhailov}, {Milano},
  {Miller}, {Miller}, {Miller}, {Millhouse}, {Milovich-Goff}, {Minazzoli},
  {Minenkov}, {Ming}, {Mishra}, {Mitra}, {Mitrofanov}, {Mitselmakher},
  {Mittleman}, {Moffa}, {Moggi}, {Mogushi}, {Mohan}, {Mohapatra}, {Montani},
  {Moore}, {Moraru}, {Moreno}, {Morisaki}, {Morriss}, {Mours}, {Mow-Lowry},
  {Mueller}, {Muir}, {Mukherjee}, {Mukherjee}, {Mukherjee}, {Mukund},
  {Mullavey}, {Munch}, {Mu{\~n}iz}, {Muratore}, {Murray}, {Napier},
  {Nardecchia}, {Naticchioni}, {Nayak}, {Neilson}, {Nelemans}, {Nelson},
  {Nery}, {Neunzert}, {Nevin}, {Newport}, {Newton}, {Ng}, {Nguyen}, {Nichols},
  {Nielsen}, {Nissanke}, {Nitz}, {Noack}, {Nocera}, {Nolting}, {North},
  {Nuttall}, {Oberling}, {O'Dea}, {Ogin}, {Oh}, {Oh}, {Ohme}, {Okada},
  {Oliver}, {Oppermann}, {Oram}, {O'Reilly}, {Ormiston}, {Ortega},
  {O'Shaughnessy}, {Ossokine}, {Ottaway}, {Overmier}, {Owen}, {Pace}, {Page},
  {Page}, {Pai}, {Pai}, {Palamos}, {Palashov}, {Palomba}, {Pal-Singh}, {Pan},
  {Pan}, {Pang}, {Pang}, {Pankow}, {Pannarale}, {Pant}, {Paoletti}, {Paoli},
  {Papa}, {Parida}, {Parker}, {Pascucci}, {Pasqualetti}, {Passaquieti},
  {Passuello}, {Patil}, {Patricelli}, {Pearlstone}, {Pedraza}, {Pedurand},
  {Pekowsky}, {Pele}, {Penn}, {Perez}, {Perreca}, {Perri}, {Pfeiffer},
  {Phelps}, {Piccinni}, {Pichot}, {Piergiovanni}, {Pierro}, {Pillant},
  {Pinard}, {Pinto}, {Pirello}, {Pitkin}, {Poe}, {Poggiani}, {Popolizio},
  {Porter}, {Post}, {Powell}, {Prasad}, {Pratt}, {Pratten}, {Predoi},
  {Prestegard}, {Prijatelj}, {Principe}, {Privitera}, {Prix}, {Prodi},
  {Prokhorov}, {Puncken}, {Punturo}, {Puppo}, {P{\"u}rrer}, {Qi}, {Quetschke},
  {Quintero}, {Quitzow-James}, {Raab}, {Rabeling}, {Radkins}, {Raffai}, {Raja},
  {Rajan}, {Rajbhandari}, {Rakhmanov}, {Ramirez}, {Ramos-Buades}, {Rapagnani},
  {Raymond}, {Razzano}, {Read}, {Regimbau}, {Rei}, {Reid}, {Reitze}, {Ren},
  {Reyes}, {Ricci}, {Ricker}, {Rieger}, {Riles}, {Rizzo}, {Robertson}, {Robie},
  {Robinet}, {Rocchi}, {Rolland}, {Rollins}, {Roma}, {Romano}, {Romano},
  {Romel}, {Romie}, {Rosi{\'n}ska}, {Ross}, {Rowan}, {R{\"u}diger}, {Ruggi},
  {Rutins}, {Ryan}, {Sachdev}, {Sadecki}, {Sadeghian}, {Sakellariadou},
  {Salconi}, {Saleem}, {Salemi}, {Samajdar}, {Sammut}, {Sampson}, {Sanchez},
  {Sanchez}, {Sanchis-Gual}, {Sandberg}, {Sanders}, {Sassolas},
  {Sathyaprakash}, {Saulson}, {Sauter}, {Savage}, {Sawadsky}, {Schale},
  {Scheel}, {Scheuer}, {Schmidt}, {Schmidt}, {Schnabel}, {Schofield},
  {Sch{\"o}nbeck}, {Schreiber}, {Schuette}, {Schulte}, {Schutz}, {Schwalbe},
  {Scott}, {Scott}, {Seidel}, {Sellers}, {Sengupta}, {Sentenac}, {Sequino},
  {Sergeev}, {Shaddock}, {Shaffer}, {Shah}, {Shahriar}, {Shaner}, {Shao},
  {Shapiro}, {Shawhan}, {Sheperd}, {Shoemaker}, {Shoemaker}, {Siellez},
  {Siemens}, {Sieniawska}, {Sigg}, {Silva}, {Singer}, {Singh}, {Singhal},
  {Sintes}, {Slagmolen}, {Smith}, {Smith}, {Smith}, {Somala}, {Son},
  {Sonnenberg}, {Sorazu}, {Sorrentino}, {Souradeep}, {Spencer}, {Srivastava},
  {Staats}, {Staley}, {Steinke}, {Steinlechner}, {Steinlechner}, {Steinmeyer},
  {Stevenson}, {Stone}, {Stops}, {Strain}, {Stratta}, {Strigin}, {Strunk},
  {Sturani}, {Stuver}, {Summerscales}, {Sun}, {Sunil}, {Suresh}, {Sutton},
  {Swinkels}, {Szczepa{\'n}czyk}, {Tacca}, {Tait}, {Talbot}, {Talukder},
  {Tanner}, {T{\'a}pai}, {Taracchini}, {Tasson}, {Taylor}, {Taylor}, {Tewari},
  {Theeg}, {Thies}, {Thomas}, {Thomas}, {Thomas}, {Thorne}, {Thrane}, {Tiwari},
  {Tiwari}, {Tokmakov}, {Toland}, {Tonelli}, {Tornasi}, {Torres-Forn{\'e}},
  {Torrie}, {T{\"o}yr{\"a}}, {Travasso}, {Traylor}, {Trinastic}, {Tringali},
  {Trozzo}, {Tsang}, {Tse}, {Tso}, {Tsukada}, {Tsuna}, {Tuyenbayev}, {Ueno},
  {Ugolini}, {Unnikrishnan}, {Urban}, {Usman}, {Vahlbruch}, {Vajente},
  {Valdes}, {Vallisneri}, {van Bakel}, {van Beuzekom}, {van den Brand}, {Van
  Den Broeck}, {Vander- Hyde}, {van der Schaaf}, {van Heijningen}, {van
  Veggel}, {Vardaro}, {Varma}, {Vass}, {Vas{\'u}th}, {Vecchio}, {Vedovato},
  {Veitch}, {Veitch}, {Venkateswara}, {Venugopalan}, {Verkindt}, {Vetrano},
  {Vicer{\'e}}, {Viets}, {Vinciguerra}, {Vine}, {Vinet}, {Vitale}, {Vo},
  {Vocca}, {Vorvick}, {Vyatchanin}, {Wade}, {Wade}, {Wade}, {Walet}, {Walker},
  {Wallace}, {Walsh}, {Wang}, {Wang}, {Wang}, {Wang}, {Wang}, {Ward}, {Warner},
  {Was}, {Watchi}, {Weaver}, {Wei}, {Weinert}, {Weinstein}, {Weiss}, {Wen},
  {Wessel}, {We{\ss}els}, {Westerweck}, {Westphal}, {Wette}, {Whelan},
  {Whitcomb}, {Whiting}, {Whittle}, {Wilken}, {Williams}, {Williams},
  {Williamson}, {Willis}, {Willke}, {Wimmer}, {Winkler}, {Wipf}, {Wittel},
  {Woan}, {Woehler}, {Wofford}, {Wong}, {Worden}, {Wright}, {Wu}, {Wysocki},
  {Xiao}, {Yamamoto}, {Yancey}, {Yang}, {Yap}, {Yazback}, {Yu}, {Yu}, {Yvert},
  {Zadro{\.Z}ny}, {Zanolin}, {Zelenova}, {Zendri}, {Zevin}, {Zhang}, {Zhang},
  {Zhang}, {Zhang}, {Zhao}, {Zhou}, {Zhou}, {Zhu}, {Zhu}, {Zimmerman},
  {Zucker}, {Zweizig}, {LIGO Scientific Collaboration}, \& {Virgo
  Collaboration}}]{abbott_2017_ad}
---. 2017{\natexlab{c}}, \prl, 119, 141101,
  \dodoi{10.1103/PhysRevLett.119.141101}

\bibitem[{{Abbott} {et~al.}(2017{\natexlab{d}}){Abbott}, {Abbott}, {Abbott},
  {Acernese}, {Ackley}, {Adams}, {Adams}, {Addesso}, {Adhikari}, {Adya},
  {Affeldt}, {Afrough}, {Agarwal}, {Agathos}, {Agatsuma}, {Aggarwal}, {Aguiar},
  {Aiello}, {Ain}, {Ajith}, {Allen}, {Allen}, {Allocca}, {Altin}, {Amato},
  {Ananyeva}, {Anderson}, {Anderson}, {Angelova}, {Antier}, {Appert}, {Arai},
  {Araya}, {Areeda}, {Arnaud}, {Arun}, {Ascenzi}, {Ashton}, {Ast}, {Aston},
  {Astone}, {Atallah}, {Aufmuth}, {Aulbert}, {AultONeal}, {Austin},
  {Avila-Alvarez}, {Babak}, {Bacon}, {Bader}, {Bae}, {Baker}, {Baldaccini},
  {Ballardin}, {Ballmer}, {Banagiri}, {Barayoga}, {Barclay}, {Barish},
  {Barker}, {Barkett}, {Barone}, {Barr}, {Barsotti}, {Barsuglia}, {Barta},
  {Barthelmy}, {Bartlett}, {Bartos}, {Bassiri}, {Basti}, {Batch}, {Bawaj},
  {Bayley}, {Bazzan}, {B{\'e}csy}, {Beer}, {Bejger}, {Belahcene}, {Bell},
  {Berger}, {Bergmann}, {Bero}, {Berry}, {Bersanetti}, {Bertolini},
  {Betzwieser}, {Bhagwat}, {Bhandare}, {Bilenko}, {Billingsley}, {Billman},
  {Birch}, {Birney}, {Birnholtz}, {Biscans}, {Biscoveanu}, {Bisht}, {Bitossi},
  {Biwer}, {Bizouard}, {Blackburn}, {Blackman}, {Blair}, {Blair}, {Blair},
  {Bloemen}, {Bock}, {Bode}, {Boer}, {Bogaert}, {Bohe}, {Bondu}, {Bonilla},
  {Bonnand}, {Boom}, {Bork}, {Boschi}, {Bose}, {Bossie}, {Bouffanais}, {Bozzi},
  {Bradaschia}, {Brady}, {Branchesi}, {Brau}, {Briant}, {Brillet}, {Brinkmann},
  {Brisson}, {Brockill}, {Broida}, {Brooks}, {Brown}, {Brown}, {Brunett},
  {Buchanan}, {Buikema}, {Bulik}, {Bulten}, {Buonanno}, {Buskulic}, {Buy},
  {Byer}, {Cabero}, {Cadonati}, {Cagnoli}, {Cahillane}, {Calder{\'o}n
  Bustillo}, {Callister}, {Calloni}, {Camp}, {Canepa}, {Canizares}, {Cannon},
  {Cao}, {Cao}, {Capano}, {Capocasa}, {Carbognani}, {Caride}, {Carney},
  {Casanueva Diaz}, {Casentini}, {Caudill}, {Cavagli{\`a}}, {Cavalier},
  {Cavalieri}, {Cella}, {Cepeda}, {Cerd{\'a}-Dur{\'a}n}, {Cerretani},
  {Cesarini}, {Chamberlin}, {Chan}, {Chao}, {Charlton}, {Chase}, {Chassande-
  Mottin}, {Chatterjee}, {Chatziioannou}, {Cheeseboro}, {Chen}, {Chen}, {Chen},
  {Cheng}, {Chia}, {Chincarini}, {Chiummo}, {Chmiel}, {Cho}, {Cho}, {Chow},
  {Christensen}, {Chu}, {Chua}, {Chua}, {Chung}, {Chung}, {Ciani}, {Ciolfi},
  {Cirelli}, {Cirone}, {Clara}, {Clark}, {Clearwater}, {Cleva}, {Cocchieri},
  {Coccia}, {Cohadon}, {Cohen}, {Colla}, {Collette}, {Cominsky}, {Constancio},
  {Conti}, {Cooper}, {Corban}, {Corbitt}, {Cordero-Carri{\'o}n}, {Corley},
  {Cornish}, {Corsi}, {Cortese}, {Costa}, {Coughlin}, {Coughlin}, {Coulon},
  {Countryman}, {Couvares}, {Covas}, {Cowan}, {Coward}, {Cowart}, {Coyne},
  {Coyne}, {Creighton}, {Creighton}, {Cripe}, {Crowder}, {Cullen}, {Cumming},
  {Cunningham}, {Cuoco}, {Dal Canton}, {D{\'a}lya}, {Danilishin}, {D'Antonio},
  {Danzmann}, {Dasgupta}, {Da Silva Costa}, {Dattilo}, {Dave}, {Davier},
  {Davis}, {Daw}, {Day}, {De}, {DeBra}, {Degallaix}, {De Laurentis},
  {Del{\'e}glise}, {Del Pozzo}, {Demos}, {Denker}, {Dent}, {De Pietri},
  {Dergachev}, {De Rosa}, {DeRosa}, {De Rossi}, {DeSalvo}, {de Varona},
  {Devenson}, {Dhurandhar}, {D{\'\i}az}, {Di Fiore}, {Di Giovanni}, {Di
  Girolamo}, {Di Lieto}, {Di Pace}, {Di Palma}, {Di Renzo}, {Doctor},
  {Dolique}, {Donovan}, {Dooley}, {Doravari}, {Dorrington}, {Douglas}, {Dovale
  {\'A}lvarez}, {Downes}, {Drago}, {Dreissigacker}, {Driggers}, {Du}, {Ducrot},
  {Dupej}, {Dwyer}, {Edo}, {Edwards}, {Effler}, {Ehrens}, {Eichholz},
  {Eikenberry}, {Eisenstein}, {Essick}, {Estevez}, {Etienne}, {Etzel}, {Evans},
  {Evans}, {Factourovich}, {Fafone}, {Fair}, {Fairhurst}, {Fan}, {Farinon},
  {Farr}, {Farr}, {Fauchon-Jones}, {Favata}, {Fays}, {Fee}, {Fehrmann},
  {Feicht}, {Fejer}, {Fernandez-Galiana}, {Ferrante}, {Ferreira}, {Ferrini},
  {Fidecaro}, {Finstad}, {Fiori}, {Fiorucci}, {Fishbach}, {Fisher},
  {Fitz-Axen}, {Flaminio}, {Fletcher}, {Fong}, {Font}, {Forsyth}, {Forsyth},
  {Fournier}, {Frasca}, {Frasconi}, {Frei}, {Freise}, {Frey}, {Frey}, {Fries},
  {Fritschel}, {Frolov}, {Fulda}, {Fyffe}, {Gabbard}, {Gadre}, {Gaebel},
  {Gair}, {Gammaitoni}, {Ganija}, {Gaonkar}, {Garcia-Quiros}, {Garufi},
  {Gateley}, {Gaudio}, {Gaur}, {Gayathri}, {Gehrels}, {Gemme}, {Genin},
  {Gennai}, {George}, {George}, {Gergely}, {Germain}, {Ghonge}, {Ghosh},
  {Ghosh}, {Ghosh}, {Giaime}, {Giardina}, {Giazotto}, {Gill}, {Glover},
  {Goetz}, {Goetz}, {Gomes}, {Goncharov}, {Gonz{\'a}lez}, {Gonzalez Castro},
  {Gopakumar}, {Gorodetsky}, {Gossan}, {Gosselin}, {Gouaty}, {Grado}, {Graef},
  {Granata}, {Grant}, {Gras}, {Gray}, {Greco}, {Green}, {Gretarsson},
  {Griswold}, {Groot}, {Grote}, {Grunewald}, {Gruning}, {Guidi}, {Guo},
  {Gupta}, {Gupta}, {Gushwa}, {Gustafson}, {Gustafson}, {Halim}, {Hall},
  {Hall}, {Hamilton}, {Hammond}, {Haney}, {Hanke}, {Hanks}, {Hanna}, {Hannam},
  {Hannuksela}, {Hanson}, {Hardwick}, {Harms}, {Harry}, {Harry}, {Hart},
  {Haster}, {Haughian}, {Healy}, {Heidmann}, {Heintze}, {Heitmann}, {Hello},
  {Hemming}, {Hendry}, {Heng}, {Hennig}, {Heptonstall}, {Heurs}, {Hild},
  {Hinderer}, {Hoak}, {Hofman}, {Holt}, {Holz}, {Hopkins}, {Horst}, {Hough},
  {Houston}, {Howell}, {Hreibi}, {Hu}, {Huerta}, {Huet}, {Hughey}, {Husa},
  {Huttner}, {Huynh- Dinh}, {Indik}, {Inta}, {Intini}, {Isa}, {Isac}, {Isi},
  {Iyer}, {Izumi}, {Jacqmin}, {Jani}, {Jaranowski}, {Jawahar},
  {Jim{\'e}nez-Forteza}, {Johnson}, {Jones}, {Jones}, {Jonker}, {Ju}, {Junker},
  {Kalaghatgi}, {Kalogera}, {Kamai}, {Kandhasamy}, {Kang}, {Kanner}, {Kapadia},
  {Karki}, {Karvinen}, {Kasprzack}, {Katolik}, {Katsavounidis}, {Katzman},
  {Kaufer}, {Kawabe}, {K{\'e}f{\'e}lian}, {Keitel}, {Kemball}, {Kennedy},
  {Kent}, {Key}, {Khalili}, {Khan}, {Khan}, {Khan}, {Khazanov}, {Kijbunchoo},
  {Kim}, {Kim}, {Kim}, {Kim}, {Kim}, {Kim}, {Kimbrell}, {King}, {King},
  {Kinley-Hanlon}, {Kirchhoff}, {Kissel}, {Kleybolte}, {Klimenko}, {Knowles},
  {Koch}, {Koehlenbeck}, {Koley}, {Kondrashov}, {Kontos}, {Korobko}, {Korth},
  {Kowalska}, {Kozak}, {Kr{\"a}mer}, {Kringel}, {Krishnan}, {Kr{\'o}lak},
  {Kuehn}, {Kumar}, {Kumar}, {Kumar}, {Kuo}, {Kutynia}, {Kwang}, {Lackey},
  {Lai}, {Landry}, {Lang}, {Lange}, {Lantz}, {Lanza}, {Larson},
  {Lartaux-Vollard}, {Lasky}, {Laxen}, {Lazzarini}, {Lazzaro}, {Leaci},
  {Leavey}, {Lee}, {Lee}, {Lee}, {Lee}, {Lee}, {Lehmann}, {Lenon}, {Leonardi},
  {Leroy}, {Letendre}, {Levin}, {Li}, {Linker}, {Littenberg}, {Liu}, {Lo},
  {Lockerbie}, {London}, {Lord}, {Lorenzini}, {Loriette}, {Lormand}, {Losurdo},
  {Lough}, {Lousto}, {Lovelace}, {L{\"u}ck}, {Lumaca}, {Lundgren}, {Lynch},
  {Ma}, {Macas}, {Macfoy}, {Machenschalk}, {MacInnis}, {Macleod}, {Maga{\~n}a
  Hernandez}, {Maga{\~n}a-Sandoval}, {Maga{\~n}a Zertuche}, {Magee},
  {Majorana}, {Maksimovic}, {Man}, {Mandic}, {Mangano}, {Mansell}, {Manske},
  {Mantovani}, {Marchesoni}, {Marion}, {M{\'a}rka}, {M{\'a}rka}, {Markakis},
  {Markosyan}, {Markowitz}, {Maros}, {Marquina}, {Marsh}, {Martelli},
  {Martellini}, {Martin}, {Martin}, {Martynov}, {Mason}, {Massera}, {Masserot},
  {Massinger}, {Masso-Reid}, {Mastrogiovanni}, {Matas}, {Matichard}, {Matone},
  {Mavalvala}, {Mazumder}, {McCarthy}, {McClelland}, {McCormick}, {McCuller},
  {McGuire}, {McIntyre}, {McIver}, {McManus}, {McNeill}, {McRae}, {McWilliams},
  {Meacher}, {Meadors}, {Mehmet}, {Meidam}, {Mejuto-Villa}, {Melatos},
  {Mendell}, {Mercer}, {Merilh}, {Merzougui}, {Meshkov}, {Messenger},
  {Messick}, {Metzdorff}, {Meyers}, {Miao}, {Michel}, {Middleton}, {Mikhailov},
  {Milano}, {Miller}, {Miller}, {Miller}, {Millhouse}, {Milovich-Goff},
  {Minazzoli}, {Minenkov}, {Ming}, {Mishra}, {Mitra}, {Mitrofanov},
  {Mitselmakher}, {Mittleman}, {Moffa}, {Moggi}, {Mogushi}, {Mohan},
  {Mohapatra}, {Montani}, {Moore}, {Moraru}, {Moreno}, {Morriss}, {Mours},
  {Mow-Lowry}, {Mueller}, {Muir}, {Mukherjee}, {Mukherjee}, {Mukherjee},
  {Mukund}, {Mullavey}, {Munch}, {Mu{\~n}iz}, {Muratore}, {Murray}, {Napier},
  {Nardecchia}, {Naticchioni}, {Nayak}, {Neilson}, {Nelemans}, {Nelson},
  {Nery}, {Neunzert}, {Nevin}, {Newport}, {Newton}, {Ng}, {Nguyen}, {Nguyen},
  {Nichols}, {Nielsen}, {Nissanke}, {Nitz}, {Noack}, {Nocera}, {Nolting},
  {North}, {Nuttall}, {Oberling}, {O'Dea}, {Ogin}, {Oh}, {Oh}, {Ohme}, {Okada},
  {Oliver}, {Oppermann}, {Oram}, {O'Reilly}, {Ormiston}, {Ortega},
  {O'Shaughnessy}, {Ossokine}, {Ottaway}, {Overmier}, {Owen}, {Pace}, {Page},
  {Page}, {Pai}, {Pai}, {Palamos}, {Palashov}, {Palomba}, {Pal- Singh}, {Pan},
  {Pan}, {Pang}, {Pang}, {Pankow}, {Pannarale}, {Pant}, {Paoletti}, {Paoli},
  {Papa}, {Parida}, {Parker}, {Pascucci}, {Pasqualetti}, {Passaquieti},
  {Passuello}, {Patil}, {Patricelli}, {Pearlstone}, {Pedraza}, {Pedurand},
  {Pekowsky}, {Pele}, {Penn}, {Perez}, {Perreca}, {Perri}, {Pfeiffer},
  {Phelps}, {Piccinni}, {Pichot}, {Piergiovanni}, {Pierro}, {Pillant},
  {Pinard}, {Pinto}, {Pirello}, {Pitkin}, {Poe}, {Poggiani}, {Popolizio},
  {Porter}, {Post}, {Powell}, {Prasad}, {Pratt}, {Pratten}, {Predoi},
  {Prestegard}, {Price}, {Prijatelj}, {Principe}, {Privitera}, {Prodi},
  {Prokhorov}, {Puncken}, {Punturo}, {Puppo}, {P{\"u}rrer}, {Qi}, {Quetschke},
  {Quintero}, {Quitzow-James}, {Raab}, {Rabeling}, {Radkins}, {Raffai}, {Raja},
  {Rajan}, {Rajbhandari}, {Rakhmanov}, {Ramirez}, {Ramos-Buades}, {Rapagnani},
  {Raymond}, {Razzano}, {Read}, {Regimbau}, {Rei}, {Reid}, {Reitze}, {Ren},
  {Reyes}, {Ricci}, {Ricker}, {Rieger}, {Riles}, {Rizzo}, {Robertson}, {Robie},
  {Robinet}, {Rocchi}, {Rolland}, {Rollins}, {Roma}, {Romano}, {Romel},
  {Romie}, {Rosi{\'n}ska}, {Ross}, {Rowan}, {R{\"u}diger}, {Ruggi}, {Rutins},
  {Ryan}, {Sachdev}, {Sadecki}, {Sadeghian}, {Sakellariadou}, {Salconi},
  {Saleem}, {Salemi}, {Samajdar}, {Sammut}, {Sampson}, {Sanchez}, {Sanchez},
  {Sanchis- Gual}, {Sandberg}, {Sanders}, {Sassolas}, {Sathyaprakash},
  {Saulson}, {Sauter}, {Savage}, {Sawadsky}, {Schale}, {Scheel}, {Scheuer},
  {Schmidt}, {Schmidt}, {Schnabel}, {Schofield}, {Sch{\"o}nbeck}, {Schreiber},
  {Schuette}, {Schulte}, {Schutz}, {Schwalbe}, {Scott}, {Scott}, {Seidel},
  {Sellers}, {Sengupta}, {Sentenac}, {Sequino}, {Sergeev}, {Shaddock},
  {Shaffer}, {Shah}, {Shahriar}, {Shaner}, {Shao}, {Shapiro}, {Shawhan},
  {Sheperd}, {Shoemaker}, {Shoemaker}, {Siellez}, {Siemens}, {Sieniawska},
  {Sigg}, {Silva}, {Singer}, {Singh}, {Singhal}, {Sintes}, {Slagmolen},
  {Smith}, {Smith}, {Smith}, {Somala}, {Son}, {Sonnenberg}, {Sorazu},
  {Sorrentino}, {Souradeep}, {Spencer}, {Srivastava}, {Staats}, {Staley},
  {Steinke}, {Steinlechner}, {Steinlechner}, {Steinmeyer}, {Stevenson},
  {Stone}, {Stops}, {Strain}, {Stratta}, {Strigin}, {Strunk}, {Sturani},
  {Stuver}, {Summerscales}, {Sun}, {Sunil}, {Suresh}, {Sutton}, {Swinkels},
  {Szczepa{\'n}czyk}, {Tacca}, {Tait}, {Talbot}, {Talukder}, {Tanner},
  {T{\'a}pai}, {Taracchini}, {Tasson}, {Taylor}, {Taylor}, {Tewari}, {Theeg},
  {Thies}, {Thomas}, {Thomas}, {Thomas}, {Thorne}, {Thorne}, {Thrane},
  {Tiwari}, {Tiwari}, {Tokmakov}, {Toland}, {Tonelli}, {Tornasi},
  {Torres-Forn{\'e}}, {Torrie}, {T{\"o}yr{\"a}}, {Travasso}, {Traylor},
  {Trinastic}, {Tringali}, {Trozzo}, {Tsang}, {Tse}, {Tso}, {Tsukada}, {Tsuna},
  {Tuyenbayev}, {Ueno}, {Ugolini}, {Unnikrishnan}, {Urban}, {Usman},
  {Vahlbruch}, {Vajente}, {Valdes}, {van Bakel}, {van Beuzekom}, {van den
  Brand}, {Van Den Broeck}, {Vander-Hyde}, {van der Schaaf}, {van Heijningen},
  {van Veggel}, {Vardaro}, {Varma}, {Vass}, {Vas{\'u}th}, {Vecchio},
  {Vedovato}, {Veitch}, {Veitch}, {Venkateswara}, {Venugopalan}, {Verkindt},
  {Vetrano}, {Vicer{\'e}}, {Viets}, {Vinciguerra}, {Vine}, {Vinet}, {Vitale},
  {Vo}, {Vocca}, {Vorvick}, {Vyatchanin}, {Wade}, {Wade}, {Wade}, {Walet},
  {Walker}, {Wallace}, {Walsh}, {Wang}, {Wang}, {Wang}, {Wang}, {Wang}, {Ward},
  {Warner}, {Was}, {Watchi}, {Weaver}, {Wei}, {Weinert}, {Weinstein}, {Weiss},
  {Wen}, {Wessel}, {Wessels}, {Westerweck}, {Westphal}, {Wette}, {Whelan},
  {Whitcomb}, {Whiting}, {Whittle}, {Wilken}, {Williams}, {Williams},
  {Williamson}, {Willis}, {Willke}, {Wimmer}, {Winkler}, {Wipf}, {Wittel},
  {Woan}, {Woehler}, {Wofford}, {Wong}, {Worden}, {Wright}, {Wu}, {Wysocki},
  {Xiao}, {Yamamoto}, {Yancey}, {Yang}, {Yap}, {Yazback}, {Yu}, {Yu}, {Yvert},
  {Zadro{\.z}ny}, {Zanolin}, {Zelenova}, {Zendri}, {Zevin}, {Zhang}, {Zhang},
  {Zhang}, {Zhang}, {Zhao}, {Zhou}, {Zhou}, {Zhu}, {Zhu}, {Zimmerman},
  {Zucker}, {Zweizig}, {LIGO Scientific Collaboration}, {Virgo Collaboration},
  {Wilson-Hodge}, {Bissaldi}, {Blackburn}, {Briggs}, {Burns}, {Cleveland},
  {Connaughton}, {Gibby}, {Giles}, {Goldstein}, {Hamburg}, {Jenke}, {Hui},
  {Kippen}, {Kocevski}, {McBreen}, {Meegan}, {Paciesas}, {Poolakkil}, {Preece},
  {Racusin}, {Roberts}, {Stanbro}, {Veres}, {von Kienlin}, {GBM}, {Savchenko},
  {Ferrigno}, {Kuulkers}, {Bazzano}, {Bozzo}, {Brandt}, {Chenevez},
  {Courvoisier}, {Diehl}, {Domingo}, {Hanlon}, {Jourdain}, {Laurent}, {Lebrun},
  {Lutovinov}, {Martin-Carrillo}, {Mereghetti}, {Natalucci}, {Rodi}, {Roques},
  {Sunyaev}, {Ubertini}, {INTEGRAL}, {Aartsen}, {Ackermann}, {Adams},
  {Aguilar}, {Ahlers}, {Ahrens}, {Samarai}, {Altmann}, {Andeen}, {Anderson},
  {Ansseau}, {Anton}, {Arg{\"u}elles}, {Auffenberg}, {Axani}, {Bagherpour},
  {Bai}, {Barron}, {Barwick}, {Baum}, {Bay}, {Beatty}, {Becker Tjus},
  {Bernardini}, {Besson}, {Binder}, {Bindig}, {Blaufuss}, {Blot}, {Bohm},
  {B{\"o}rner}, {Bos}, {Bose}, {B{\"o}ser}, {Botner}, {Bourbeau}, {Bourbeau},
  {Bradascio}, {Braun}, {Brayeur}, {Brenzke}, {Bretz}, {Bron},
  {Brostean-Kaiser}, {Burgman}, {Carver}, {Casey}, {Casier}, {Cheung},
  {Chirkin}, {Christov}, {Clark}, {Classen}, {Coenders}, {Collin}, {Conrad},
  {Cowen}, {Cross}, {Day}, {de Andr{\'e}}, {De Clercq}, {DeLaunay},
  {Dembinski}, {De Ridder}, {Desiati}, {de Vries}, {de Wasseige}, {de With},
  {DeYoung}, {D{\'\i}az-V{\'e}lez}, {di Lorenzo}, {Dujmovic}, {Dumm},
  {Dunkman}, {Dvorak}, {Eberhardt}, {Ehrhardt}, {Eichmann}, {Eller}, {Evenson},
  {Fahey}, {Fazely}, {Felde}, {Filimonov}, {Finley}, {Flis}, {Franckowiak},
  {Friedman}, {Fuchs}, {Gaisser}, {Gallagher}, {Gerhardt}, {Ghorbani}, {Giang},
  {Glauch}, {Gl{\"u}senkamp}, {Goldschmidt}, {Gonzalez}, {Grant}, {Griffith},
  {Haack}, {Hallgren}, {Halzen}, {Hanson}, {Hebecker}, {Heereman}, {Helbing},
  {Hellauer}, {Hickford}, {Hignight}, {Hill}, {Hoffman}, {Hoffmann},
  {Hokanson-Fasig}, {Hoshina}, {Huang}, {Huber}, {Hultqvist}, {H{\"u}nnefeld},
  {In}, {Ishihara}, {Jacobi}, {Japaridze}, {Jeong}, {Jero}, {Jones},
  {Kalaczynski}, {Kang}, {Kappes}, {Karg}, {Karle}, {Kauer}, {Keivani},
  {Kelley}, {Kheirandish}, {Kim}, {Kim}, {Kintscher}, {Kiryluk}, {Kittler},
  {Klein}, {Kohnen}, {Koirala}, {Kolanoski}, {K{\"o}pke}, {Kopper}, {Kopper},
  {Koschinsky}, {Koskinen}, {Kowalski}, {Krings}, {Kroll}, {Kr{\"u}ckl},
  {Kunnen}, {Kunwar}, {Kurahashi}, {Kuwabara}, {Kyriacou}, {Labare},
  {Lanfranchi}, {Larson}, {Lauber}, {Lesiak-Bzdak}, {Leuermann}, {Liu}, {Lu},
  {L{\"u}nemann}, {Luszczak}, {Madsen}, {Maggi}, {Mahn}, {Mancina}, {Maruyama},
  {Mase}, {Maunu}, {McNally}, {Meagher}, {Medici}, {Meier}, {Menne}, {Merino},
  {Meures}, {Miarecki}, {Micallef}, {Moment{\'e}}, {Montaruli}, {Moore},
  {Moulai}, {Nahnhauer}, {Nakarmi}, {Naumann}, {Neer}, {Niederhausen},
  {Nowicki}, {Nygren}, {Obertacke Pollmann}, {Olivas}, {O'Murchadha},
  {Palczewski}, {Pandya}, {Pankova}, {Peiffer}, {Pepper}, {P{\'e}rez de los
  Heros}, {Pieloth}, {Pinat}, {Price}, {Przybylski}, {Raab}, {R{\"a}del},
  {Rameez}, {Rawlins}, {Rea}, {Reimann}, {Relethford}, {Relich}, {Resconi},
  {Rhode}, {Richman}, {Robertson}, {Rongen}, {Rott}, {Ruhe}, {Ryckbosch},
  {Rysewyk}, {S{\"a}lzer}, {Sanchez Herrera}, {Sandrock}, {Sandroos},
  {Santander}, {Sarkar}, {Sarkar}, {Satalecka}, {Schlunder}, {Schmidt},
  {Schneider}, {Schoenen}, {Sch{\"o}neberg}, {Schumacher}, {Seckel},
  {Seunarine}, {Soedingrekso}, {Soldin}, {Song}, {Spiczak}, {Spiering},
  {Stachurska}, {Stamatikos}, {Stanev}, {Stasik}, {Stettner}, {Steuer},
  {Stezelberger}, {Stokstad}, {St{\"o}ssl}, {Strotjohann}, {Stuttard},
  {Sullivan}, {Sutherland}, {Taboada}, {Tatar}, {Tenholt}, {Ter- Antonyan},
  {Terliuk}, {Te{\v{s}}i{\'c}}, {Tilav}, {Toale}, {Tobin}, {Toscano}, {Tosi},
  {Tselengidou}, {Tung}, {Turcati}, {Turley}, {Ty}, {Unger}, {Usner},
  {Vandenbroucke}, {Van Driessche}, {van Eijndhoven}, {Vanheule}, {van Santen},
  {Vehring}, {Vogel}, {Vraeghe}, {Walck}, {Wallace}, {Wallraff}, {Wandler},
  {Wandkowsky}, {Waza}, {Weaver}, {Weiss}, {Wendt}, {Werthebach}, {Whelan},
  {Wiebe}, {Wiebusch}, {Wille}, {Williams}, {Wills}, {Wolf}, {Wood}, {Woolsey},
  {Woschnagg}, {Xu}, {Xu}, {Xu}, {Yanez}, {Yodh}, {Yoshida}, {Yuan}, {Zoll},
  {IceCube Collaboration}, {Balasubramanian}, {Mate}, {Bhalerao},
  {Bhattacharya}, {Vibhute}, {Dewangan}, {Rao}, {Vadawale}, {AstroSat Cadmium
  Zinc Telluride Imager Team}, {Svinkin}, {Hurley}, {Aptekar}, {Frederiks},
  {Golenetskii}, {Kozlova}, {Lysenko}, {Oleynik}, {Tsvetkova}, {Ulanov},
  {Cline}, {IPN Collaboration}, {Li}, {Xiong}, {Zhang}, {Lu}, {Song}, {Cao},
  {Chang}, {Chen}, {Chen}, {Chen}, {Chen}, {Chen}, {Chen}, {Cui}, {Cui},
  {Deng}, {Dong}, {Du}, {Fu}, {Gao}, {Gao}, {Gao}, {Ge}, {Gu}, {Guan}, {Guo},
  {Han}, {Hu}, {Huang}, {Huo}, {Jia}, {Jiang}, {Jiang}, {Jin}, {Jin}, {Li},
  {Li}, {Li}, {Li}, {Li}, {Li}, {Li}, {Li}, {Li}, {Li}, {Li}, {Liang}, {Liao},
  {Liu}, {Liu}, {Liu}, {Liu}, {Liu}, {Liu}, {Liu}, {Lu}, {Lu}, {Luo}, {Ma},
  {Meng}, {Nang}, {Nie}, {Ou}, {Qu}, {Sai}, {Sun}, {Tan}, {Tao}, {Tao}, {Tuo},
  {Wang}, {Wang}, {Wang}, {Wang}, {Wang}, {Wen}, {Wu}, {Wu}, {Xiao}, {Xu},
  {Xu}, {Yan}, {Yang}, {Yang}, {Yang}, {Zhang}, {Zhang}, {Zhang}, {Zhang},
  {Zhang}, {Zhang}, {Zhang}, {Zhang}, {Zhang}, {Zhang}, {Zhang}, {Zhang},
  {Zhang}, {Zhang}, {Zhang}, {Zhang}, {Zhang}, {Zhang}, {Zhao}, {Zhao}, {Zhao},
  {Zheng}, {Zhu}, {Zhu}, {Zou}, {Insight-HXMT Collaboration}, {Albert},
  {Andr{\'e}}, {Anghinolfi}, {Ardid}, {Aubert}, {Aublin}, {Avgitas}, {Baret},
  {Barrios- Mart{\'\i}}, {Basa}, {Belhorma}, {Bertin}, {Biagi}, {Bormuth},
  {Bourret}, {Bouwhuis}, {Br{\^a}nzaș}, {Bruijn}, {Brunner}, {Busto},
  {Capone}, {Caramete}, {Carr}, {Celli}, {Cherkaoui El Moursli}, {Chiarusi},
  {Circella}, {Coelho}, {Coleiro}, {Coniglione}, {Costantini}, {Coyle},
  {Creusot}, {D{\'\i}az}, {Deschamps}, {De Bonis}, {Distefano}, {Di Palma},
  {Domi}, {Donzaud}, {Dornic}, {Drouhin}, {Eberl}, {El Bojaddaini}, {El
  Khayati}, {Els{\"a}sser}, {Enzenh{\"o}fer}, {Ettahiri}, {Fassi}, {Felis},
  {Fusco}, {Gay}, {Giordano}, {Glotin}, {Gr{\'e}goire}, {Ruiz}, {Graf},
  {Hallmann}, {van Haren}, {Heijboer}, {Hello}, {Hern{\'a}ndez-Rey},
  {H{\"o}ssl}, {Hofest{\"a}dt}, {Hugon}, {Illuminati}, {James}, {de Jong},
  {Jongen}, {Kadler}, {Kalekin}, {Katz}, {Kiessling}, {Kouchner}, {Kreter},
  {Kreykenbohm}, {Kulikovskiy}, {Lachaud}, {Lahmann}, {Lef{\`e}vre}, {Leonora},
  {Lotze}, {Loucatos}, {Marcelin}, {Margiotta}, {Marinelli},
  {Mart{\'\i}nez-Mora}, {Mele}, {Melis}, {Michael}, {Migliozzi}, {Moussa},
  {Navas}, {Nezri}, {Organokov}, {P{\u{a}}v{\u{a}}laș}, {Pellegrino},
  {Perrina}, {Piattelli}, {Popa}, {Pradier}, {Quinn}, {Racca}, {Riccobene},
  {S{\'a}nchez-Losa}, {Salda{\~n}a}, {Salvadori}, {Samtleben}, {Sanguineti},
  {Sapienza}, {Sieger}, {Spurio}, {Stolarczyk}, {Taiuti}, {Tayalati},
  {Trovato}, {Turpin}, {T{\"o}nnis}, {Vallage}, {Van Elewyck}, {Versari},
  {Vivolo}, {Vizzoca}, {Wilms}, {Zornoza}, {Z{\'u}{\~n}iga}, {ANTARES
  Collaboration}, {Beardmore}, {Breeveld}, {Burrows}, {Cenko}, {Cusumano},
  {D'A{\`\i}}, {de Pasquale}, {Emery}, {Evans}, {Giommi}, {Gronwall}, {Kennea},
  {Krimm}, {Kuin}, {Lien}, {Marshall}, {Melandri}, {Nousek}, {Oates},
  {Osborne}, {Pagani}, {Page}, {Palmer}, {Perri}, {Siegel}, {Sbarufatti},
  {Tagliaferri}, {Tohuvavohu}, {The Swift Collaboration}, {Tavani},
  {Verrecchia}, {Bulgarelli}, {Evangelista}, {Pacciani}, {Feroci}, {Pittori},
  {Giuliani}, {Del Monte}, {Donnarumma}, {Argan}, {Trois}, {Ursi}, {Cardillo},
  {Piano}, {Longo}, {Lucarelli}, {Munar-Adrover}, {Fuschino}, {Labanti},
  {Marisaldi}, {Minervini}, {Fioretti}, {Parmiggiani}, {Gianotti}, {Trifoglio},
  {Di Persio}, {Antonelli}, {Barbiellini}, {Caraveo}, {Cattaneo}, {Costa},
  {Colafrancesco}, {D'Amico}, {Ferrari}, {Morselli}, {Paoletti}, {Picozza},
  {Pilia}, {Rappoldi}, {Soffitta}, {Vercellone}, {AGILE Team}, {Foley},
  {Coulter}, {Kilpatrick}, {Drout}, {Piro}, {Shappee}, {Siebert}, {Simon},
  {Ulloa}, {Kasen}, {Madore}, {Murguia-Berthier}, {Pan}, {Prochaska},
  {Ramirez-Ruiz}, {Rest}, {Rojas-Bravo}, {The 1M2H Team}, {Berger},
  {Soares-Santos}, {Annis}, {Alexander}, {Allam}, {Balbinot}, {Blanchard},
  {Brout}, {Butler}, {Chornock}, {Cook}, {Cowperthwaite}, {Diehl},
  {Drlica-Wagner}, {Drout}, {Durret}, {Eftekhari}, {Finley}, {Fong}, {Frieman},
  {Fryer}, {Garc{\'\i}a-Bellido}, {Gruendl}, {Hartley}, {Herner}, {Kessler},
  {Lin}, {Lopes}, {Louren{\c{c}}o}, {Margutti}, {Marshall}, {Matheson},
  {Medina}, {Metzger}, {Mu{\~n}oz}, {Muir}, {Nicholl}, {Nugent}, {Palmese},
  {Paz-Chinch{\'o}n}, {Quataert}, {Sako}, {Sauseda}, {Schlegel}, {Scolnic},
  {Secco}, {Smith}, {Sobreira}, {Villar}, {Vivas}, {Wester}, {Williams},
  {Yanny}, {Zenteno}, {Zhang}, {Abbott}, {Banerji}, {Bechtol},
  {Benoit-L{\'e}vy}, {Bertin}, {Brooks}, {Buckley-Geer}, {Burke}, {Capozzi},
  {Carnero Rosell}, {Carrasco Kind}, {Castander}, {Crocce}, {Cunha},
  {D'Andrea}, {da Costa}, {Davis}, {DePoy}, {Desai}, {Dietrich}, {Eifler},
  {Fernandez}, {Flaugher}, {Fosalba}, {Gaztanaga}, {Gerdes}, {Giannantonio},
  {Goldstein}, {Gruen}, {Gschwend}, {Gutierrez}, {Honscheid}, {James},
  {Jeltema}, {Johnson}, {Johnson}, {Kent}, {Krause}, {Kron}, {Kuehn}, {Lahav},
  {Lima}, {Maia}, {March}, {Martini}, {McMahon}, {Menanteau}, {Miller},
  {Miquel}, {Mohr}, {Nichol}, {Ogando}, {Plazas}, {Romer}, {Roodman}, {Rykoff},
  {Sanchez}, {Scarpine}, {Schindler}, {Schubnell}, {Sevilla-Noarbe}, {Sheldon},
  {Smith}, {Smith}, {Stebbins}, {Suchyta}, {Swanson}, {Tarle}, {Thomas},
  {Troxel}, {Tucker}, {Vikram}, {Walker}, {Wechsler}, {Weller}, {Carlin},
  {Gill}, {Li}, {Marriner}, {Neilsen}, {The Dark Energy Camera GW- EM
  Collaboration}, {the DES Collaboration}, {Haislip}, {Kouprianov}, {Reichart},
  {Sand}, {Tartaglia}, {Valenti}, {Yang}, {The DLT40 Collaboration}, {Benetti},
  {Brocato}, {Campana}, {Cappellaro}, {Covino}, {D'Avanzo}, {D'Elia}, {Getman},
  {Ghirlanda}, {Ghisellini}, {Limatola}, {Nicastro}, {Palazzi}, {Pian},
  {Piranomonte}, {Possenti}, {Rossi}, {Salafia}, {Tomasella}, {Amati},
  {Antonelli}, {Bernardini}, {Bufano}, {Capaccioli}, {Casella}, {Dadina}, {De
  Cesare}, {Di Paola}, {Giuffrida}, {Giunta}, {Israel}, {Lisi}, {Maiorano},
  {Mapelli}, {Masetti}, {Pescalli}, {Pulone}, {Salvaterra}, {Schipani},
  {Spera}, {Stamerra}, {Stella}, {Testa}, {Turatto}, {Vergani}, {Aresu},
  {Bachetti}, {Buffa}, {Burgay}, {Buttu}, {Caria}, {Carretti}, {Casasola},
  {Castangia}, {Carboni}, {Casu}, {Concu}, {Corongiu}, {Deiana}, {Egron},
  {Fara}, {Gaudiomonte}, {Gusai}, {Ladu}, {Loru}, {Leurini}, {Marongiu},
  {Melis}, {Melis}, {Migoni}, {Milia}, {Navarrini}, {Orlati}, {Ortu}, {Palmas},
  {Pellizzoni}, {Perrodin}, {Pisanu}, {Poppi}, {Righini}, {Saba}, {Serra},
  {Serrau}, {Stagni}, {Surcis}, {Vacca}, {Vargiu}, {Hunt}, {Jin}, {Klose},
  {Kouveliotou}, {Mazzali}, {M{\o}ller}, {Nava}, {Piran}, {Selsing}, {Vergani},
  {Wiersema}, {Toma}, {Higgins}, {Mundell}, {di Serego Alighieri}, {G{\'o}tz},
  {Gao}, {Gomboc}, {Kaper}, {Kobayashi}, {Kopac}, {Mao}, {Starling}, {Steele},
  {van der Horst}, {GRAWITA: GRAvitational Wave Inaf TeAm}, {Acero}, {Atwood},
  {Baldini}, {Barbiellini}, {Bastieri}, {Berenji}, {Bellazzini}, {Bissaldi},
  {Blandford}, {Bloom}, {Bonino}, {Bottacini}, {Bregeon}, {Buehler}, {Buson},
  {Cameron}, {Caputo}, {Caraveo}, {Cavazzuti}, {Chekhtman}, {Cheung}, {Chiang},
  {Ciprini}, {Cohen-Tanugi}, {Cominsky}, {Costantin}, {Cuoco}, {D'Ammando}, {de
  Palma}, {Digel}, {Di Lalla}, {Di Mauro}, {Di Venere}, {Dubois}, {Fegan},
  {Focke}, {Franckowiak}, {Fukazawa}, {Funk}, {Fusco}, {Gargano}, {Gasparrini},
  {Giglietto}, {Giordano}, {Giroletti}, {Glanzman}, {Green}, {Grondin},
  {Guillemot}, {Guiriec}, {Harding}, {Horan}, {J{\'o}hannesson}, {Kamae},
  {Kensei}, {Kuss}, {La Mura}, {Latronico}, {Lemoine-Goumard}, {Longo},
  {Loparco}, {Lovellette}, {Lubrano}, {Magill}, {Maldera}, {Manfreda},
  {Mazziotta}, {McEnery}, {Meyer}, {Michelson}, {Mirabal}, {Monzani},
  {Moretti}, {Morselli}, {Moskalenko}, {Negro}, {Nuss}, {Ojha}, {Omodei},
  {Orienti}, {Orlando}, {Palatiello}, {Paliya}, {Paneque}, {Pesce-Rollins},
  {Piron}, {Porter}, {Principe}, {Rain{\`o}}, {Rando}, {Razzano}, {Razzaque},
  {Reimer}, {Reimer}, {Reposeur}, {Rochester}, {Saz Parkinson}, {Sgr{\`o}},
  {Siskind}, {Spada}, {Spandre}, {Suson}, {Takahashi}, {Tanaka}, {Thayer},
  {Thayer}, {Thompson}, {Tibaldo}, {Torres}, {Torresi}, {Troja}, {Venters},
  {Vianello}, {Zaharijas}, {The Fermi Large Area Telescope Collaboration},
  {Allison}, {Bannister}, {Dobie}, {Kaplan}, {Lenc}, {Lynch}, {Murphy},
  {Sadler}, {Australia Telescope Compact Array}, {Hotan}, {James}, {Oslowski},
  {Raja}, {Shannon}, {Whiting}, {Australian SKA Pathfinder}, {Arcavi},
  {Howell}, {McCully}, {Hosseinzadeh}, {Hiramatsu}, {Poznanski}, {Barnes},
  {Zaltzman}, {Vasylyev}, {Maoz}, {Las Cumbres Observatory Group}, {Cooke},
  {Bailes}, {Wolf}, {Deller}, {Lidman}, {Wang}, {Gendre}, {Andreoni}, {Ackley},
  {Pritchard}, {Bessell}, {Chang}, {M{\"o}ller}, {Onken}, {Scalzo},
  {Ridden-Harper}, {Sharp}, {Tucker}, {Farrell}, {Elmer}, {Johnston},
  {Venkatraman Krishnan}, {Keane}, {Green}, {Jameson}, {Hu}, {Ma}, {Sun}, {Wu},
  {Wang}, {Shang}, {Hu}, {Ashley}, {Yuan}, {Li}, {Tao}, {Zhu}, {Zhang},
  {Suntzeff}, {Zhou}, {Yang}, {Orange}, {Morris}, {Cucchiara}, {Giblin},
  {Klotz}, {Staff}, {Thierry}, {Schmidt}, {OzGrav}, {(Deeper}, {Wider},
  {program}, {AST3}, {CAASTRO Collaborations}, {Tanvir}, {Levan}, {Cano}, {de
  Ugarte-Postigo}, {Gonz{\'a}lez-Fern{\'a}ndez}, {Greiner}, {Hjorth}, {Irwin},
  {Kr{\"u}hler}, {Mandel}, {Milvang-Jensen}, {O'Brien}, {Rol}, {Rosetti},
  {Rosswog}, {Rowlinson}, {Steeghs}, {Th{\"o}ne}, {Ulaczyk}, {Watson}, {Bruun},
  {Cutter}, {Figuera Jaimes}, {Fujii}, {Fruchter}, {Gompertz}, {Jakobsson},
  {Hodosan}, {J{\`e}rgensen}, {Kangas}, {Kann}, {Rabus}, {Schr{\o}der},
  {Stanway}, {Wijers}, {The VINROUGE Collaboration}, {Lipunov}, {Gorbovskoy},
  {Kornilov}, {Tyurina}, {Balanutsa}, {Kuznetsov}, {Vlasenko}, {Podesta},
  {Lopez}, {Podesta}, {Levato}, {Saffe}, {Mallamaci}, {Budnev}, {Gress},
  {Kuvshinov}, {Gorbunov}, {Vladimirov}, {Zimnukhov}, {Gabovich}, {Yurkov},
  {Sergienko}, {Rebolo}, {Serra-Ricart}, {Tlatov}, {Ishmuhametova}, {MASTER
  Collaboration}, {Abe}, {Aoki}, {Aoki}, {Asakura}, {Baar}, {Barway}, {Bond},
  {Doi}, {Finet}, {Fujiyoshi}, {Furusawa}, {Honda}, {Itoh}, {Kanda},
  {Kawabata}, {Kawabata}, {Kim}, {Koshida}, {Kuroda}, {Lee}, {Liu},
  {Matsubayashi}, {Miyazaki}, {Morihana}, {Morokuma}, {Motohara}, {Murata},
  {Nagai}, {Nagashima}, {Nagayama}, {Nakaoka}, {Nakata}, {Ohsawa}, {Ohshima},
  {Ohta}, {Okita}, {Saito}, {Saito}, {Sako}, {Sekiguchi}, {Sumi}, {Tajitsu},
  {Takahashi}, {Takayama}, {Tamura}, {Tanaka}, {Tanaka}, {Terai}, {Tominaga},
  {Tristram}, {Uemura}, {Utsumi}, {Yamaguchi}, {Yasuda}, {Yoshida}, {Zenko},
  {J-GEM}, {Adams}, {Anupama}, {Bally}, {Barway}, {Bellm}, {Blagorodnova},
  {Cannella}, {Chandra}, {Chatterjee}, {Clarke}, {Cobb}, {Cook}, {Copperwheat},
  {De}, {Emery}, {Feindt}, {Foster}, {Fox}, {Frail}, {Fremling}, {Frohmaier},
  {Garcia}, {Ghosh}, {Giacintucci}, {Goobar}, {Gottlieb}, {Grefenstette},
  {Hallinan}, {Harrison}, {Heida}, {Helou}, {Ho}, {Horesh}, {Hotokezaka}, {Ip},
  {Itoh}, {Jacobs}, {Jencson}, {Kasen}, {Kasliwal}, {Kassim}, {Kim}, {Kiran},
  {Kuin}, {Kulkarni}, {Kupfer}, {Lau}, {Madsen}, {Mazzali}, {Miller},
  {Miyasaka}, {Mooley}, {Myers}, {Nakar}, {Ngeow}, {Nugent}, {Ofek},
  {Palliyaguru}, {Pavana}, {Perley}, {Peters}, {Pike}, {Piran}, {Qi}, {Quimby},
  {Rana}, {Rosswog}, {Rusu}, {Sadler}, {Van Sistine}, {Sollerman}, {Xu}, {Yan},
  {Yatsu}, {Yu}, {Zhang}, {Zhao}, {GROWTH}, {JAGWAR}, {Caltech-NRAO},
  {TTU-NRAO}, {NuSTAR Collaborations}, {Chambers}, {Huber}, {Schultz},
  {Bulger}, {Flewelling}, {Magnier}, {Lowe}, {Wainscoat}, {Waters}, {Willman},
  {Pan-STARRS}, {Ebisawa}, {Hanyu}, {Harita}, {Hashimoto}, {Hidaka}, {Hori},
  {Ishikawa}, {Isobe}, {Iwakiri}, {Kawai}, {Kawai}, {Kawamuro}, {Kawase},
  {Kitaoka}, {Makishima}, {Matsuoka}, {Mihara}, {Morita}, {Morita}, {Nakahira},
  {Nakajima}, {Nakamura}, {Negoro}, {Oda}, {Sakamaki}, {Sasaki}, {Serino},
  {Shidatsu}, {Shimomukai}, {Sugawara}, {Sugita}, {Sugizaki}, {Tachibana},
  {Takao}, {Tanimoto}, {Tomida}, {Tsuboi}, {Tsunemi}, {Ueda}, {Ueno}, {Yamada},
  {Yamaoka}, {Yamauchi}, {Yatabe}, {Yoneyama}, {Yoshii}, {The MAXI Team},
  {Coward}, {Crisp}, {Macpherson}, {Andreoni}, {Laugier}, {Noysena}, {Klotz},
  {Gendre}, {Thierry}, {Turpin}, {Consortium}, {Im}, {Choi}, {Kim}, {Yoon},
  {Lim}, {Lee}, {Lee}, {Kim}, {Ko}, {Joe}, {Kwon}, {Kim}, {Lim}, {Choi}, {KU
  Collaboration}, {Fynbo}, {Malesani}, {Xu}, {Optical Telescope}, {Smartt},
  {Jerkstrand}, {Kankare}, {Sim}, {Fraser}, {Inserra}, {Maguire}, {Leloudas},
  {Magee}, {Shingles}, {Smith}, {Young}, {Kotak}, {Gal-Yam}, {Lyman}, {Homan},
  {Agliozzo}, {Anderson}, {Angus}, {Ashall}, {Barbarino}, {Bauer}, {Berton},
  {Botticella}, {Bulla}, {Cannizzaro}, {Cartier}, {Cikota}, {Clark}, {De Cia},
  {Della Valle}, {Dennefeld}, {Dessart}, {Dimitriadis}, {Elias-Rosa}, {Firth},
  {Fl{\"o}rs}, {Frohmaier}, {Galbany}, {Gonz{\'a}lez-Gait{\'a}n}, {Gromadzki},
  {Guti{\'e}rrez}, {Hamanowicz}, {Harmanen}, {Heintz}, {Hernandez}, {Hodgkin},
  {Hook}, {Izzo}, {James}, {Jonker}, {Kerzendorf}, {Kostrzewa-Rutkowska},
  {Kromer}, {Kuncarayakti}, {Lawrence}, {Manulis}, {Mattila}, {McBrien},
  {M{\"u}ller}, {Nordin}, {O'Neill}, {Onori}, {Palmerio}, {Pastorello},
  {Patat}, {Pignata}, {Podsiadlowski}, {Razza}, {Reynolds}, {Roy}, {Ruiter},
  {Rybicki}, {Salmon}, {Pumo}, {Prentice}, {Seitenzahl}, {Smith}, {Sollerman},
  {Sullivan}, {Szegedi}, {Taddia}, {Taubenberger}, {Terreran}, {Van Soelen},
  {Vos}, {Walton}, {Wright}, {Wyrzykowski}, {Yaron}, {pre=''(''>ePESSTO},
  {Chen}, {Kr{\"u}hler}, {Schady}, {Wiseman}, {Greiner}, {Rau}, {Schweyer},
  {Klose}, {Nicuesa Guelbenzu}, {GROND}, {Palliyaguru}, {Tech University},
  {Shara}, {Williams}, {Vaisanen}, {Potter}, {Romero Colmenero}, {Crawford},
  {Buckley}, {Mao}, {SALT Group}, {D{\'\i}az}, {Macri}, {Garc{\'\i}a Lambas},
  {Mendes de Oliveira}, {Nilo Castell{\'o}n}, {Ribeiro}, {S{\'a}nchez},
  {Schoenell}, {Abramo}, {Akras}, {Alcaniz}, {Artola}, {Beroiz}, {Bonoli},
  {Cabral}, {Camuccio}, {Chavushyan}, {Coelho}, {Colazo}, {Costa- Duarte},
  {Cuevas Larenas}, {Dom{\'\i}nguez Romero}, {Dultzin}, {Fern{\'a}ndez},
  {Garc{\'\i}a}, {Girardini}, {Gon{\c{c}}alves}, {Gon{\c{c}}alves}, {Gurovich},
  {Jim{\'e}nez-Teja}, {Kanaan}, {Lares}, {Lopes de Oliveira}, {L{\'o}pez-Cruz},
  {Melia}, {Molino}, {Padilla}, {Pe{\~n}uela}, {Placco}, {Qui{\~n}ones},
  {Ram{\'\i}rez Rivera}, {Renzi}, {Riguccini}, {R{\'\i}os-L{\'o}pez},
  {Rodriguez}, {Sampedro}, {Schneiter}, {Sodr{\'e}}, {Starck}, {Torres-Flores},
  {Tornatore}, {Zadro{\.z}ny}, {Castillo}, {TOROS: Transient Robotic
  Observatory of the South Collaboration}, {Castro-Tirado}, {Tello}, {Hu},
  {Zhang}, {Cunniffe}, {Castell{\'o}n}, {Hiriart}, {Caballero- Garc{\'\i}a},
  {Jel{\'\i}nek}, {Kub{\'a}nek}, {P{\'e}rez del Pulgar}, {Park}, {Jeong},
  {Castro Cer{\'o}n}, {Pandey}, {Yock}, {Querel}, {Fan}, {Wang}, {The BOOTES
  Collaboration}, {Beardsley}, {Brown}, {Crosse}, {Emrich}, {Franzen},
  {Gaensler}, {Horsley}, {Johnston-Hollitt}, {Kenney}, {Morales}, {Pallot},
  {Sokolowski}, {Steele}, {Tingay}, {Trott}, {Walker}, {Wayth}, {Williams},
  {Wu}, {Murchison Widefield Array}, {Yoshida}, {Sakamoto}, {Kawakubo},
  {Yamaoka}, {Takahashi}, {Asaoka}, {Ozawa}, {Torii}, {Shimizu}, {Tamura},
  {Ishizaki}, {Cherry}, {Ricciarini}, {Penacchioni}, {Marrocchesi}, {The CALET
  Collaboration}, {Pozanenko}, {Volnova}, {Mazaeva}, {Minaev}, {Krugov},
  {Kusakin}, {Reva}, {Moskvitin}, {Rumyantsev}, {Inasaridze}, {Klunko},
  {Tungalag}, {Schmalz}, {Burhonov}, {IKI-GW Follow-up Collaboration},
  {Abdalla}, {Abramowski}, {Aharonian}, {Ait Benkhali}, {Ang{\"u}ner},
  {Arakawa}, {Arrieta}, {Aubert}, {Backes}, {Balzer}, {Barnard}, {Becherini},
  {Becker Tjus}, {Berge}, {Bernhard}, {Bernl{\"o}hr}, {Blackwell},
  {B{\"o}ttcher}, {Boisson}, {Bolmont}, {Bonnefoy}, {Bordas}, {Bregeon},
  {Brun}, {Brun}, {Bryan}, {B{\"u}chele}, {Bulik}, {Capasso}, {Caroff},
  {Carosi}, {Casanova}, {Cerruti}, {Chakraborty}, {Chaves}, {Chen},
  {Chevalier}, {Colafrancesco}, {Condon}, {Conrad}, {Davids}, {Decock}, {Deil},
  {Devin}, {deWilt}, {Dirson}, {Djannati-Ata{\"\i}}, {Donath}, {O'C. Drury},
  {Dutson}, {Dyks}, {Edwards}, {Egberts}, {Emery}, {Ernenwein}, {Eschbach},
  {Farnier}, {Fegan}, {Fernandes}, {Fiasson}, {Fontaine}, {Funk},
  {F{\"u}ssling}, {Gabici}, {Gallant}, {Garrigoux}, {Gat{\'e}}, {Giavitto},
  {Giebels}, {Glawion}, {Glicenstein}, {Gottschall}, {Grondin}, {Hahn},
  {Haupt}, {Hawkes}, {Heinzelmann}, {Henri}, {Hermann}, {Hinton}, {Hofmann},
  {Hoischen}, {Holch}, {Holler}, {Horns}, {Ivascenko}, {Iwasaki},
  {Jacholkowska}, {Jamrozy}, {Jankowsky}, {Jankowsky}, {Jingo}, {Jouvin},
  {Jung-Richardt}, {Kastendieck}, {Katarzy{\'n}ski}, {Katsuragawa},
  {Kerszberg}, {Khangulyan}, {Kh{\'e}lifi}, {King}, {Klepser}, {Klochkov},
  {Klu{\'z}niak}, {Komin}, {Kosack}, {Krakau}, {Kraus}, {Kr{\"u}ger}, {Laffon},
  {Lamanna}, {Lau}, {Lees}, {Lefaucheur}, {Lemi{\`e}re}, {Lemoine-Goumard},
  {Lenain}, {Leser}, {Lohse}, {Lorentz}, {Liu}, {Lypova}, {Malyshev},
  {Marandon}, {Marcowith}, {Mariaud}, {Marx}, {Maurin}, {Maxted}, {Mayer},
  {Meintjes}, {Meyer}, {Mitchell}, {Moderski}, {Mohamed}, {Mohrmann},
  {Mor{\r{a}}}, {Moulin}, {Murach}, {Nakashima}, {de Naurois}, {Ndiyavala},
  {Niederwanger}, {Niemiec}, {Oakes}, {O'Brien}, {Odaka}, {Ohm}, {Ostrowski},
  {Oya}, {Padovani}, {Panter}, {Parsons}, {Pekeur}, {Pelletier}, {Perennes},
  {Petrucci}, {Peyaud}, {Piel}, {Pita}, {Poireau}, {Poon}, {Prokhorov},
  {Prokoph}, {P{\"u}hlhofer}, {Punch}, {Quirrenbach}, {Raab}, {Rauth},
  {Reimer}, {Reimer}, {Renaud}, {de los Reyes}, {Rieger}, {Rinchiuso},
  {Romoli}, {Rowell}, {Rudak}, {Rulten}, {Sahakian}, {Saito}, {Sanchez},
  {Santangelo}, {Sasaki}, {Schlickeiser}, {Sch{\"u}ssler}, {Schulz},
  {Schwanke}, {Schwemmer}, {Seglar-Arroyo}, {Settimo}, {Seyffert}, {Shafi},
  {Shilon}, {Shiningayamwe}, {Simoni}, {Sol}, {Spanier}, {Spir-Jacob},
  {Stawarz}, {Steenkamp}, {Stegmann}, {Steppa}, {Sushch}, {Takahashi},
  {Tavernet}, {Tavernier}, {Taylor}, {Terrier}, {Tibaldo}, {Tiziani},
  {Tluczykont}, {Trichard}, {Tsirou}, {Tsuji}, {Tuffs}, {Uchiyama}, {van der
  Walt}, {van Eldik}, {van Rensburg}, {van Soelen}, {Vasileiadis}, {Veh},
  {Venter}, {Viana}, {Vincent}, {Vink}, {Voisin}, {V{\"o}lk}, {Vuillaume},
  {Wadiasingh}, {Wagner}, {Wagner}, {Wagner}, {White}, {Wierzcholska},
  {Willmann}, {W{\"o}rnlein}, {Wouters}, {Yang}, {Zaborov}, {Zacharias},
  {Zanin}, {Zdziarski}, {Zech}, {Zefi}, {Ziegler}, {Zorn}, {{\.Z}ywucka},
  {H.~E.~S.~S. Collaboration}, {Fender}, {Broderick}, {Rowlinson}, {Wijers},
  {Stewart}, {ter Veen}, {Shulevski}, {LOFAR Collaboration}, {Kavic},
  {Simonetti}, {League}, {Tsai}, {Obenberger}, {Nathaniel}, {Taylor}, {Dowell},
  {Liebling}, {Estes}, {Lippert}, {Sharma}, {Vincent}, {Farella}, {Wavelength
  Array}, {Abeysekara}, {Albert}, {Alfaro}, {Alvarez}, {Arceo}, {Arteaga-
  Vel{\'a}zquez}, {Avila Rojas}, {Ayala Solares}, {Barber}, {Becerra Gonzalez},
  {Becerril}, {Belmont-Moreno}, {BenZvi}, {Berley}, {Bernal}, {Braun},
  {Brisbois}, {Caballero-Mora}, {Capistr{\'a}n}, {Carrami{\~n}ana}, {Casanova},
  {Castillo}, {Cotti}, {Cotzomi}, {Couti{\~n}o de Le{\'o}n}, {De Le{\'o}n}, {De
  la Fuente}, {Diaz Hernandez}, {Dichiara}, {Dingus}, {DuVernois},
  {D{\'\i}az-V{\'e}lez}, {Ellsworth}, {Engel}, {Enr{\'\i}quez-Rivera},
  {Fiorino}, {Fleischhack}, {Fraija}, {Garc{\'\i}a-Gonz{\'a}lez}, {Garfias},
  {Gerhardt}, {Gonz{\~o}lez Mu{\~n}oz}, {Gonz{\'a}lez}, {Goodman},
  {Hampel-Arias}, {Harding}, {Hernandez}, {Hernandez- Almada}, {Hona},
  {H{\"u}ntemeyer}, {Iriarte}, {Jardin-Blicq}, {Joshi}, {Kaufmann}, {Kieda},
  {Lara}, {Lauer}, {Lennarz}, {Le{\'o}n Vargas}, {Linnemann}, {Longinotti},
  {Raya}, {Luna- Garc{\'\i}a}, {L{\'o}pez-Coto}, {Malone}, {Marinelli},
  {Martinez}, {Martinez- Castellanos}, {Mart{\'\i}nez-Castro}, {Mart{\'\i
  }nez-Huerta}, {Matthews}, {Miranda-Romagnoli}, {Moreno}, {Mostaf{\'a}},
  {Nellen}, {Newbold}, {Nisa}, {Noriega-Papaqui}, {Pelayo}, {Pretz},
  {P{\'e}rez-P{\'e}rez}, {Ren}, {Rho}, {Rivi{\`e}re}, {Rosa- Gonz{\'a}lez},
  {Rosenberg}, {Ruiz-Velasco}, {Salazar}, {Salesa Greus}, {Sandoval},
  {Schneider}, {Schoorlemmer}, {Sinnis}, {Smith}, {Springer}, {Surajbali},
  {Tibolla}, {Tollefson}, {Torres}, {Ukwatta}, {Weisgarber}, {Westerhoff},
  {Wisher}, {Wood}, {Yapici}, {Yodh}, {Younk}, {Zhou}, {{\'A}lvarez}, {HAWC
  Collaboration}, {Aab}, {Abreu}, {Aglietta}, {Albuquerque}, {Albury},
  {Allekotte}, {Almela}, {Alvarez Castillo}, {Alvarez-Mu{\~n}iz}, {Anastasi},
  {Anchordoqui}, {Andrada}, {Andringa}, {Aramo}, {Arsene}, {Asorey}, {Assis},
  {Avila}, {Badescu}, {Balaceanu}, {Barbato}, {Barreira Luz}, {Becker},
  {Bellido}, {Berat}, {Bertaina}, {Bertou}, {Biermann}, {Biteau}, {Blaess},
  {Blanco}, {Blazek}, {Bleve}, {Boh{\'a}{\v{c}}ov{\'a}}, {Bonifazi}, {Borodai},
  {Botti}, {Brack}, {Brancus}, {Bretz}, {Bridgeman}, {Briechle}, {Buchholz},
  {Bueno}, {Buitink}, {Buscemi}, {Caballero-Mora}, {Caccianiga}, {Cancio},
  {Canfora}, {Caruso}, {Castellina}, {Catalani}, {Cataldi}, {Cazon}, {Chavez},
  {Chinellato}, {Chudoba}, {Clay}, {Cobos Cerutti}, {Colalillo}, {Coleman},
  {Collica}, {Coluccia}, {Concei{\c{c}}{\~a}o}, {Consolati}, {Contreras},
  {Cooper}, {Coutu}, {Covault}, {Cronin}, {D'Amico}, {Daniel}, {Dasso},
  {Daumiller}, {Dawson}, {Day}, {de Almeida}, {de Jong}, {De Mauro}, {de Mello
  Neto}, {De Mitri}, {de Oliveira}, {de Souza}, {Debatin}, {Deligny},
  {D{\'\i}az Castro}, {Diogo}, {Dobrigkeit}, {D'Olivo}, {Dorosti}, {Dos Anjos},
  {Dova}, {Dundovic}, {Ebr}, {Engel}, {Erdmann}, {Erfani}, {Escobar},
  {Espadanal}, {Etchegoyen}, {Falcke}, {Farmer}, {Farrar}, {Fauth}, {Fazzini},
  {Feldbusch}, {Fenu}, {Fick}, {Figueira}, {Filip{\v{c}}i{\v{c}}}, {Freire},
  {Fujii}, {Fuster}, {Ga{\"\i}or}, {Garc{\'\i}a}, {Gat{\'e}}, {Gemmeke},
  {Gherghel-Lascu}, {Ghia}, {Giaccari}, {Giammarchi}, {Giller}, {G{\l}as},
  {Glaser}, {Golup}, {G{\'o}mez Berisso}, {G{\'o}mez Vitale}, {Gonz{\'a}lez},
  {Gorgi}, {Gottowik}, {Grillo}, {Grubb}, {Guarino}, {Guedes}, {Halliday},
  {Hampel}, {Hansen}, {Harari}, {Harrison}, {Harvey}, {Haungs}, {Hebbeker},
  {Heck}, {Heimann}, {Herve}, {Hill}, {Hojvat}, {Holt}, {Homola},
  {H{\"o}randel}, {Horvath}, {Hrabovsk{\'y}}, {Huege}, {Hulsman}, {Insolia},
  {Isar}, {Jandt}, {Johnsen}, {Josebachuili}, {Jurysek}, {K{\"a}{\"a}p{\"a}},
  {Kampert}, {Keilhauer}, {Kemmerich}, {Kemp}, {Kieckhafer}, {Klages},
  {Kleifges}, {Kleinfeller}, {Krause}, {Krohm}, {Kuempel}, {Kukec Mezek},
  {Kunka}, {Kuotb Awad}, {Lago}, {LaHurd}, {Lang}, {Lauscher}, {Legumina},
  {Leigui de Oliveira}, {Letessier-Selvon}, {Lhenry-Yvon}, {Link}, {Lo Presti},
  {Lopes}, {L{\'o}pez}, {L{\'o}pez Casado}, {Lorek}, {Luce}, {Lucero},
  {Malacari}, {Mallamaci}, {Mandat}, {Mantsch}, {Mariazzi}, {Maris},
  {Marsella}, {Martello}, {Martinez}, {Mart{\'\i}nez Bravo}, {Mas{\'\i}as
  Meza}, {Mathes}, {Mathys}, {Matthews}, {Matthiae}, {Mayotte}, {Mazur},
  {Medina}, {Medina-Tanco}, {Melo}, {Menshikov}, {Merenda}, {Michal},
  {Micheletti}, {Middendorf}, {Miramonti}, {Mitrica}, {Mockler}, {Mollerach},
  {Montanet}, {Morello}, {Morlino}, {M{\"u}ller}, {M{\"u}ller}, {Muller},
  {M{\"u}ller}, {Mussa}, {Naranjo}, {Nguyen}, {Niculescu-Oglinzanu},
  {Niechciol}, {Niemietz}, {Niggemann}, {Nitz}, {Nosek}, {Novotny},
  {No{\v{z}}ka}, {N{\'u}{\~n}ez}, {Oikonomou}, {Olinto}, {Palatka}, {Pallotta},
  {Papenbreer}, {Parente}, {Parra}, {Paul}, {Pech}, {Pedreira}, {P{\c e}kala},
  {Pe{\~n}a-Rodriguez}, {Pereira}, {Perlin}, {Perrone}, {Peters}, {Petrera},
  {Phuntsok}, {Pierog}, {Pimenta}, {Pirronello}, {Platino}, {Plum}, {Poh},
  {Porowski}, {Prado}, {Privitera}, {Prouza}, {Quel}, {Querchfeld}, {Quinn},
  {Ramos-Pollan}, {Rautenberg}, {Ravignani}, {Ridky}, {Riehn}, {Risse},
  {Ristori}, {Rizi}, {Rodrigues de Carvalho}, {Rodriguez Fernandez}, {Rodriguez
  Rojo}, {Roncoroni}, {Roth}, {Roulet}, {Rovero}, {Ruehl}, {Saffi}, {Saftoiu},
  {Salamida}, {Salazar}, {Saleh}, {Salina}, {S{\'a}nchez}, {Sanchez-Lucas},
  {Santos}, {Santos}, {Sarazin}, {Sarmento}, {Sarmiento-Cano}, {Sato},
  {Schauer}, {Scherini}, {Schieler}, {Schimp}, {Schmidt}, {Scholten},
  {Schov{\'a}nek}, {Schr{\"o}der}, {Schr{\"o}der}, {Schulz}, {Schumacher},
  {Sciutto}, {Segreto}, {Shadkam}, {Shellard}, {Sigl}, {Silli},
  {{\v{S}}m{\'\i}da}, {Snow}, {Sommers}, {Sonntag}, {Soriano}, {Squartini},
  {Stanca}, {Stani{\v{c}}}, {Stasielak}, {Stassi}, {Stolpovskiy}, {Strafella},
  {Streich}, {Suarez}, {Suarez-Dur{\'a}n}, {Sudholz}, {Suomij{\"a}rvi},
  {Supanitsky}, {{\v{S}}up{\'\i}k}, {Swain}, {Szadkowski}, {Taboada},
  {Taborda}, {Timmermans}, {Todero Peixoto}, {Tomankova}, {Tom{\'e}}, {Torralba
  Elipe}, {Travnicek}, {Trini}, {Tueros}, {Ulrich}, {Unger}, {Urban},
  {Vald{\'e}s Galicia}, {Vali{\~n}o}, {Valore}, {van Aar}, {van Bodegom}, {van
  den Berg}, {van Vliet}, {Varela}, {Vargas C{\'a}rdenas}, {V{\'a}zquez},
  {Veberi{\v{c}}}, {Ventura}, {Vergara Quispe}, {Verzi}, {Vicha},
  {Villase{\~n}or}, {Vorobiov}, {Wahlberg}, {Wainberg}, {Walz}, {Watson},
  {Weber}, {Weindl}, {Wiede{\'n}ski}, {Wiencke}, {Wilczy{\'n}ski}, {Wirtz},
  {Wittkowski}, {Wundheiler}, {Yang}, {Yushkov}, {Zas}, {Zavrtanik},
  {Zavrtanik}, {Zepeda}, {Zimmermann}, {Ziolkowski}, {Zong}, {Zuccarello}, {The
  Pierre Auger Collaboration}, {Kim}, {Schulze}, {Bauer}, {Corral-Santana}, {de
  Gregorio- Monsalvo}, {Gonz{\'a}lez-L{\'o}pez}, {Hartmann}, {Ishwara-Chandra},
  {Mart{\'\i}n}, {Mehner}, {Misra}, {Micha{\l}owski}, {Resmi}, {ALMA
  Collaboration}, {Paragi}, {Agudo}, {An}, {Beswick}, {Casadio}, {Frey},
  {Jonker}, {Kettenis}, {Marcote}, {Moldon}, {Szomoru}, {van Langevelde},
  {Yang}, {Euro VLBI Team}, {Cwiek}, {Cwiok}, {Czyrkowski}, {Dabrowski},
  {Kasprowicz}, {Mankiewicz}, {Nawrocki}, {Opiela}, {Piotrowski}, {Wrochna},
  {Zaremba}, {{\.Z}arnecki}, {Pi of the Sky Collaboration}, {Haggard}, {Nynka},
  {Ruan}, {The Chandra Team at McGill University}, {Bland}, {Booler},
  {Devillepoix}, {de Gois}, {Hancock}, {Howie}, {Paxman}, {Sansom}, {Towner},
  {Desert Fireball Network}, {Tonry}, {Coughlin}, {Stubbs}, {Denneau},
  {Heinze}, {Stalder}, {Weiland}, {ATLAS}, {Eatough}, {Kramer}, {Kraus}, {Time
  Resolution Universe Survey}, {Troja}, {Piro}, {Becerra Gonz{\'a}lez},
  {Butler}, {Fox}, {Khandrika}, {Kutyrev}, {Lee}, {Ricci}, {Ryan},
  {S{\'a}nchez-Ram{\'\i}rez}, {Veilleux}, {Watson}, {Wieringa}, {Burgess}, {van
  Eerten}, {Fontes}, {Fryer}, {Korobkin}, {Wollaeger}, {RIMAS}, {RATIR},
  {Camilo}, {Foley}, {Goedhart}, {Makhathini}, {Oozeer}, {Smirnov}, {Fender},
  {Woudt}, \& {South Africa/MeerKAT}}]{abbott_2017_aa}
---. 2017{\natexlab{d}}, \apj, 848, L12, \dodoi{10.3847/2041-8213/aa91c9}

\bibitem[{{Abbott} {et~al.}(2017{\natexlab{e}}){Abbott}, {Abbott}, {Abbott},
  {Acernese}, {Ackley}, {Adams}, {Adams}, {Addesso}, {Adhikari}, {Adya},
  {Affeldt}, {Afrough}, {Agarwal}, {Agathos}, {Agatsuma}, {Aggarwal}, {Aguiar},
  {Aiello}, {Ain}, {Ajith}, {Allen}, {Allen}, {Allocca}, {Altin}, {Amato},
  {Ananyeva}, {Anderson}, {Anderson}, {Angelova}, {Antier}, {Appert}, {Arai},
  {Araya}, {Areeda}, {Arnaud}, {Arun}, {Ascenzi}, {Ashton}, {Ast}, {Aston},
  {Astone}, {Atallah}, {Aufmuth}, {Aulbert}, {AultONeal}, {Austin},
  {Avila-Alvarez}, {Babak}, {Bacon}, {Bader}, {Bae}, {Baker}, {Baldaccini},
  {Ballardin}, {Ballmer}, {Banagiri}, {Barayoga}, {Barclay}, {Barish},
  {Barker}, {Barkett}, {Barone}, {Barr}, {Barsotti}, {Barsuglia}, {Barta},
  {Bartlett}, {Bartos}, {Bassiri}, {Basti}, {Batch}, {Bawaj}, {Bayley},
  {Bazzan}, {B{\'e}csy}, {Beer}, {Bejger}, {Belahcene}, {Bell}, {Berger},
  {Bergmann}, {Bernuzzi}, {Bero}, {Berry}, {Bersanetti}, {Bertolini},
  {Betzwieser}, {Bhagwat}, {Bhandare}, {Bilenko}, {Billingsley}, {Billman},
  {Birch}, {Birney}, {Birnholtz}, {Biscans}, {Biscoveanu}, {Bisht}, {Bitossi},
  {Biwer}, {Bizouard}, {Blackburn}, {Blackman}, {Blair}, {Blair}, {Blair},
  {Bloemen}, {Bock}, {Bode}, {Boer}, {Bogaert}, {Bohe}, {Bondu}, {Bonilla},
  {Bonnand}, {Boom}, {Bork}, {Boschi}, {Bose}, {Bossie}, {Bouffanais}, {Bozzi},
  {Bradaschia}, {Brady}, {Branchesi}, {Brau}, {Briant}, {Brillet}, {Brinkmann},
  {Brisson}, {Brockill}, {Broida}, {Brooks}, {Brown}, {Brown}, {Brunett},
  {Buchanan}, {Buikema}, {Bulik}, {Bulten}, {Buonanno}, {Buskulic}, {Buy},
  {Byer}, {Cabero}, {Cadonati}, {Cagnoli}, {Cahillane}, {Calder{\'o}n
  Bustillo}, {Callister}, {Calloni}, {Camp}, {Canepa}, {Canizares}, {Cannon},
  {Cao}, {Cao}, {Capano}, {Capocasa}, {Carbognani}, {Caride}, {Carney},
  {Casanueva Diaz}, {Casentini}, {Caudill}, {Cavagli{\`a}}, {Cavalier},
  {Cavalieri}, {Cella}, {Cepeda}, {Cerd{\'a}-Dur{\'a}n}, {Cerretani},
  {Cesarini}, {Chamberlin}, {Chan}, {Chao}, {Charlton}, {Chase}, {Chassande-
  Mottin}, {Chatterjee}, {Cheeseboro}, {Chen}, {Chen}, {Chen}, {Cheng}, {Chia},
  {Chincarini}, {Chiummo}, {Chmiel}, {Cho}, {Cho}, {Chow}, {Christensen},
  {Chu}, {Chua}, {Chua}, {Chung}, {Chung}, {Ciani}, {Ciolfi}, {Cirelli},
  {Cirone}, {Clara}, {Clark}, {Clearwater}, {Cleva}, {Cocchieri}, {Coccia},
  {Cohadon}, {Cohen}, {Colla}, {Collette}, {Cominsky}, {Constancio}, {Conti},
  {Cooper}, {Corban}, {Corbitt}, {Cordero-Carri{\'o}n}, {Corley}, {Corsi},
  {Cortese}, {Costa}, {Coughlin}, {Coughlin}, {Coulon}, {Countryman},
  {Couvares}, {Covas}, {Cowan}, {Coward}, {Cowart}, {Coyne}, {Coyne},
  {Creighton}, {Creighton}, {Cripe}, {Crowder}, {Cullen}, {Cumming},
  {Cunningham}, {Cuoco}, {Dal Canton}, {D{\'a}lya}, {Danilishin}, {D'Antonio},
  {Danzmann}, {Dasgupta}, {Da Silva Costa}, {Dattilo}, {Dave}, {Davier},
  {Davis}, {Daw}, {Day}, {De}, {DeBra}, {Degallaix}, {De Laurentis},
  {Del{\'e}glise}, {Del Pozzo}, {Demos}, {Denker}, {Dent}, {De Pietri},
  {Dergachev}, {De Rosa}, {DeRosa}, {De Rossi}, {DeSalvo}, {de Varona},
  {Devenson}, {Dhurandhar}, {D{\'\i}az}, {Dietrich}, {Di Fiore}, {Di Giovanni},
  {Di Girolamo}, {Di Lieto}, {Di Pace}, {Di Palma}, {Di Renzo}, {Doctor},
  {Dolique}, {Donovan}, {Dooley}, {Doravari}, {Dorrington}, {Douglas}, {Dovale
  {\'A}lvarez}, {Downes}, {Drago}, {Dreissigacker}, {Driggers}, {Du}, {Ducrot},
  {Dupej}, {Dwyer}, {Edo}, {Edwards}, {Effler}, {Eggenstein}, {Ehrens},
  {Eichholz}, {Eikenberry}, {Eisenstein}, {Essick}, {Estevez}, {Etienne},
  {Etzel}, {Evans}, {Evans}, {Factourovich}, {Fafone}, {Fair}, {Fairhurst},
  {Fan}, {Farinon}, {Farr}, {Farr}, {Fauchon- Jones}, {Favata}, {Fays}, {Fee},
  {Fehrmann}, {Feicht}, {Fejer}, {Fernandez-Galiana}, {Ferrante}, {Ferreira},
  {Ferrini}, {Fidecaro}, {Finstad}, {Fiori}, {Fiorucci}, {Fishbach}, {Fisher},
  {Fitz-Axen}, {Flaminio}, {Fletcher}, {Flynn}, {Fong}, {Font}, {Forsyth},
  {Forsyth}, {Fournier}, {Frasca}, {Frasconi}, {Frei}, {Freise}, {Frey},
  {Frey}, {Fries}, {Fritschel}, {Frolov}, {Fulda}, {Fyffe}, {Gabbard}, {Gadre},
  {Gaebel}, {Gair}, {Gammaitoni}, {Ganija}, {Gaonkar}, {Garcia-Quiros},
  {Garufi}, {Gateley}, {Gaudio}, {Gaur}, {Gayathri}, {Gehrels}, {Gemme},
  {Genin}, {Gennai}, {George}, {George}, {Gergely}, {Germain}, {Ghonge},
  {Ghosh}, {Ghosh}, {Ghosh}, {Giaime}, {Giardina}, {Giazotto}, {Gill},
  {Glover}, {Goetz}, {Goetz}, {Gomes}, {Goncharov}, {Gonz{\'a}lez}, {Gonzalez
  Castro}, {Gopakumar}, {Gorodetsky}, {Gossan}, {Gosselin}, {Gouaty}, {Grado},
  {Graef}, {Granata}, {Grant}, {Gras}, {Gray}, {Greco}, {Green}, {Gretarsson},
  {Groot}, {Grote}, {Grunewald}, {Gruning}, {Guidi}, {Guo}, {Gupta}, {Gupta},
  {Gushwa}, {Gustafson}, {Gustafson}, {Halim}, {Hall}, {Hall}, {Hamilton},
  {Hammond}, {Haney}, {Hanke}, {Hanks}, {Hanna}, {Hannam}, {Hannuksela},
  {Hanson}, {Hardwick}, {Harms}, {Harry}, {Harry}, {Hart}, {Haster},
  {Haughian}, {Healy}, {Heidmann}, {Heintze}, {Heitmann}, {Hello}, {Hemming},
  {Hendry}, {Heng}, {Hennig}, {Heptonstall}, {Heurs}, {Hild}, {Hinderer},
  {Hoak}, {Hofman}, {Holt}, {Holz}, {Hopkins}, {Horst}, {Hough}, {Houston},
  {Howell}, {Hreibi}, {Hu}, {Huerta}, {Huet}, {Hughey}, {Husa}, {Huttner},
  {Huynh-Dinh}, {Indik}, {Inta}, {Intini}, {Isa}, {Isac}, {Isi}, {Iyer},
  {Izumi}, {Jacqmin}, {Jani}, {Jaranowski}, {Jawahar}, {Jim{\'e}nez-Forteza},
  {Johnson}, {Jones}, {Jones}, {Jonker}, {Ju}, {Junker}, {Kalaghatgi},
  {Kalogera}, {Kamai}, {Kandhasamy}, {Kang}, {Kanner}, {Kapadia}, {Karki},
  {Karvinen}, {Kasprzack}, {Kastaun}, {Katolik}, {Katsavounidis}, {Katzman},
  {Kaufer}, {Kawabe}, {K{\'e}f{\'e}lian}, {Keitel}, {Kemball}, {Kennedy},
  {Kent}, {Key}, {Khalili}, {Khan}, {Khan}, {Khan}, {Khazanov}, {Kijbunchoo},
  {Kim}, {Kim}, {Kim}, {Kim}, {Kim}, {Kim}, {Kimbrell}, {King}, {King},
  {Kinley-Hanlon}, {Kirchhoff}, {Kissel}, {Kleybolte}, {Klimenko}, {Knowles},
  {Koch}, {Koehlenbeck}, {Koley}, {Kondrashov}, {Kontos}, {Korobko}, {Korth},
  {Kowalska}, {Kozak}, {Kr{\"a}mer}, {Kringel}, {Krishnan}, {Kr{\'o}lak},
  {Kuehn}, {Kumar}, {Kumar}, {Kumar}, {Kuo}, {Kutynia}, {Kwang}, {Lackey},
  {Lai}, {Landry}, {Lang}, {Lange}, {Lantz}, {Lanza}, {Lartaux-Vollard},
  {Lasky}, {Laxen}, {Lazzarini}, {Lazzaro}, {Leaci}, {Leavey}, {Lee}, {Lee},
  {Lee}, {Lee}, {Lee}, {Lehmann}, {Lenon}, {Leonardi}, {Leroy}, {Letendre},
  {Levin}, {Li}, {Linker}, {Liu}, {Lo}, {Lockerbie}, {London}, {Lord},
  {Lorenzini}, {Loriette}, {Lormand}, {Losurdo}, {Lough}, {Lousto}, {Lovelace},
  {L{\"u}ck}, {Lumaca}, {Lundgren}, {Lynch}, {Ma}, {Macas}, {Macfoy},
  {Machenschalk}, {MacInnis}, {Macleod}, {Maga{\~n}a Hernandez},
  {Maga{\~n}a-Sandoval}, {Maga{\~n}a Zertuche}, {Magee}, {Majorana},
  {Maksimovic}, {Man}, {Mandic}, {Mangano}, {Mansell}, {Manske}, {Mantovani},
  {Marchesoni}, {Marion}, {M{\'a}rka}, {M{\'a}rka}, {Markakis}, {Markosyan},
  {Markowitz}, {Maros}, {Marquina}, {Martelli}, {Martellini}, {Martin},
  {Martin}, {Martynov}, {Mason}, {Massera}, {Masserot}, {Massinger},
  {Masso-Reid}, {Mastrogiovanni}, {Matas}, {Matichard}, {Matone}, {Mavalvala},
  {Mazumder}, {McCarthy}, {McClelland}, {McCormick}, {McCuller}, {McGuire},
  {McIntyre}, {McIver}, {McManus}, {McNeill}, {McRae}, {McWilliams}, {Meacher},
  {Meadors}, {Mehmet}, {Meidam}, {Mejuto-Villa}, {Melatos}, {Mendell},
  {Mercer}, {Merilh}, {Merzougui}, {Meshkov}, {Messenger}, {Messick},
  {Metzdorff}, {Meyers}, {Miao}, {Michel}, {Middleton}, {Mikhailov}, {Milano},
  {Miller}, {Miller}, {Miller}, {Milovich-Goff}, {Minazzoli}, {Minenkov},
  {Ming}, {Mishra}, {Mitra}, {Mitrofanov}, {Mitselmakher}, {Mittleman},
  {Moffa}, {Moggi}, {Mogushi}, {Mohan}, {Mohapatra}, {Montani}, {Moore},
  {Moraru}, {Moreno}, {Morriss}, {Mours}, {Mow- Lowry}, {Mueller}, {Muir},
  {Mukherjee}, {Mukherjee}, {Mukherjee}, {Mukund}, {Mullavey}, {Munch},
  {Mu{\~n}iz}, {Muratore}, {Murray}, {Napier}, {Nardecchia}, {Naticchioni},
  {Nayak}, {Neilson}, {Nelemans}, {Nelson}, {Nery}, {Neunzert}, {Nevin},
  {Newport}, {Newton}, {Ng}, {Nguyen}, {Nichols}, {Nielsen}, {Nissanke},
  {Nitz}, {Noack}, {Nocera}, {Nolting}, {North}, {Nuttall}, {Oberling},
  {O'Dea}, {Ogin}, {Oh}, {Oh}, {Ohme}, {Okada}, {Oliver}, {Oppermann}, {Oram},
  {O'Reilly}, {Ormiston}, {Ortega}, {O'Shaughnessy}, {Ossokine}, {Ottaway},
  {Overmier}, {Owen}, {Pace}, {Page}, {Page}, {Pai}, {Pai}, {Palamos},
  {Palashov}, {Palomba}, {Pal-Singh}, {Pan}, {Pan}, {Pang}, {Pang}, {Pankow},
  {Pannarale}, {Pant}, {Paoletti}, {Paoli}, {Papa}, {Parida}, {Parker},
  {Pascucci}, {Pasqualetti}, {Passaquieti}, {Passuello}, {Patil}, {Patricelli},
  {Pearlstone}, {Pedraza}, {Pedurand}, {Pekowsky}, {Pele}, {Penn}, {Perez},
  {Perreca}, {Perri}, {Pfeiffer}, {Phelps}, {Phukon}, {Piccinni}, {Pichot},
  {Piergiovanni}, {Pierro}, {Pillant}, {Pinard}, {Pinto}, {Pirello}, {Pitkin},
  {Poe}, {Poggiani}, {Popolizio}, {Porter}, {Post}, {Powell}, {Prasad},
  {Pratt}, {Pratten}, {Predoi}, {Prestegard}, {Prijatelj}, {Principe},
  {Privitera}, {Prodi}, {Prokhorov}, {Puncken}, {Punturo}, {Puppo},
  {P{\"u}rrer}, {Qi}, {Quetschke}, {Quintero}, {Quitzow-James}, {Raab},
  {Rabeling}, {Radkins}, {Raffai}, {Raja}, {Rajan}, {Rajbhandari}, {Rakhmanov},
  {Ramirez}, {Ramos-Buades}, {Rapagnani}, {Raymond}, {Razzano}, {Read},
  {Regimbau}, {Rei}, {Reid}, {Reitze}, {Ren}, {Reyes}, {Ricci}, {Ricker},
  {Rieger}, {Riles}, {Rizzo}, {Robertson}, {Robie}, {Robinet}, {Rocchi},
  {Rolland}, {Rollins}, {Roma}, {Romano}, {Romel}, {Romie}, {Rosi{\'n}ska},
  {Ross}, {Rowan}, {R{\"u}diger}, {Ruggi}, {Rutins}, {Ryan}, {Sachdev},
  {Sadecki}, {Sadeghian}, {Sakellariadou}, {Salconi}, {Saleem}, {Salemi},
  {Samajdar}, {Sammut}, {Sampson}, {Sanchez}, {Sanchez}, {Sanchis-Gual},
  {Sandberg}, {Sanders}, {Sarin}, {Sassolas}, {Sathyaprakash}, {Saulson},
  {Sauter}, {Savage}, {Sawadsky}, {Schale}, {Scheel}, {Scheuer}, {Schmidt},
  {Schmidt}, {Schnabel}, {Schofield}, {Sch{\"o}nbeck}, {Schreiber}, {Schuette},
  {Schulte}, {Schutz}, {Schwalbe}, {Scott}, {Scott}, {Seidel}, {Sellers},
  {Sengupta}, {Sentenac}, {Sequino}, {Sergeev}, {Shaddock}, {Shaffer}, {Shah},
  {Shahriar}, {Shaner}, {Shao}, {Shapiro}, {Shawhan}, {Sheperd}, {Shoemaker},
  {Shoemaker}, {Siellez}, {Siemens}, {Sieniawska}, {Sigg}, {Silva}, {Singer},
  {Singh}, {Singhal}, {Sintes}, {Rana}, {Slagmolen}, {Smith}, {Smith}, {Smith},
  {Somala}, {Son}, {Sonnenberg}, {Sorazu}, {Sorrentino}, {Souradeep}, {Sowell},
  {Spencer}, {Srivastava}, {Staats}, {Staley}, {Steinke}, {Steinlechner},
  {Steinlechner}, {Steinmeyer}, {Stevenson}, {Stone}, {Stops}, {Strain},
  {Stratta}, {Strigin}, {Strunk}, {Sturani}, {Stuver}, {Summerscales}, {Sun},
  {Sunil}, {Suresh}, {Sutton}, {Swinkels}, {Szczepa{\'n}czyk}, {Tacca}, {Tait},
  {Talbot}, {Talukder}, {Tanner}, {T{\'a}pai}, {Taracchini}, {Tasson},
  {Taylor}, {Taylor}, {Tewari}, {Theeg}, {Thies}, {Thomas}, {Thomas}, {Thomas},
  {Thorne}, {Thrane}, {Tiwari}, {Tiwari}, {Tokmakov}, {Toland}, {Tonelli},
  {Tornasi}, {Torres- Forn{\'e}}, {Torrie}, {T{\"o}yr{\"a}}, {Travasso},
  {Traylor}, {Trinastic}, {Tringali}, {Trozzo}, {Tsang}, {Tse}, {Tso},
  {Tsukada}, {Tsuna}, {Tuyenbayev}, {Ueno}, {Ugolini}, {Unnikrishnan}, {Urban},
  {Usman}, {Vahlbruch}, {Vajente}, {Valdes}, {van Bakel}, {van Beuzekom}, {van
  den Brand}, {Van Den Broeck}, {Vander-Hyde}, {van der Schaaf}, {van
  Heijningen}, {van Veggel}, {Vardaro}, {Varma}, {Vass}, {Vas{\'u}th},
  {Vecchio}, {Vedovato}, {Veitch}, {Veitch}, {Venkateswara}, {Venugopalan},
  {Verkindt}, {Vetrano}, {Vicer{\'e}}, {Viets}, {Vinciguerra}, {Vine}, {Vinet},
  {Vitale}, {Vo}, {Vocca}, {Vorvick}, {Vyatchanin}, {Wade}, {Wade}, {Wade},
  {Walet}, {Walker}, {Wallace}, {Walsh}, {Wang}, {Wang}, {Wang}, {Wang},
  {Wang}, {Ward}, {Warner}, {Was}, {Watchi}, {Weaver}, {Wei}, {Weinert},
  {Weinstein}, {Weiss}, {Wen}, {Wessel}, {We{\ss}els}, {Westerweck},
  {Westphal}, {Wette}, {Whelan}, {White}, {Whiting}, {Whittle}, {Wilken},
  {Williams}, {Williams}, {Williamson}, {Willis}, {Willke}, {Wimmer},
  {Winkler}, {Wipf}, {Wittel}, {Woan}, {Woehler}, {Wofford}, {Wong}, {Worden},
  {Wright}, {Wu}, {Wysocki}, {Xiao}, {Yamamoto}, {Yancey}, {Yang}, {Yap},
  {Yazback}, {Yu}, {Yu}, {Yvert}, {Zadro{\.z}ny}, {Zanolin}, {Zelenova},
  {Zendri}, {Zevin}, {Zhang}, {Zhang}, {Zhang}, {Zhang}, {Zhao}, {Zhou},
  {Zhou}, {Zhu}, {Zhu}, {Zimmerman}, {Zucker}, {Zweizig}, {(LIGO Scientific
  Collaboration}, \& {Virgo Collaboration}}]{abbott_2017_ae}
---. 2017{\natexlab{e}}, \apj, 851, L16, \dodoi{10.3847/2041-8213/aa9a35}

\bibitem[{{Abbott} {et~al.}(2017{\natexlab{f}}){Abbott}, {Abbott}, {Abbott},
  {Acernese}, {Ackley}, {Adams}, {Adams}, {Addesso}, {Adhikari}, {Adya},
  {Affeldt}, {Afrough}, {Agarwal}, {Agathos}, {Agatsuma}, {Aggarwal}, {Aguiar},
  {Aiello}, {Ain}, {Ajith}, {Allen}, {Allen}, {Allocca}, {Altin}, {Amato},
  {Ananyeva}, {Anderson}, {Anderson}, {Angelova}, {Antier}, {Appert}, {Arai},
  {Araya}, {Areeda}, {Arnaud}, {Arun}, {Ascenzi}, {Ashton}, {Ast}, {Aston},
  {Astone}, {Atallah}, {Aufmuth}, {Aulbert}, {AultONeal}, {Austin},
  {Avila-Alvarez}, {Babak}, {Bacon}, {Bader}, {Bae}, {Baker}, {Baldaccini},
  {Ballardin}, {Ballmer}, {Banagiri}, {Barayoga}, {Barclay}, {Barish},
  {Barker}, {Barkett}, {Barone}, {Barr}, {Barsotti}, {Barsuglia}, {Barta},
  {Bartlett}, {Bartos}, {Bassiri}, {Basti}, {Batch}, {Bawaj}, {Bayley},
  {Bazzan}, {B{\'e}csy}, {Beer}, {Bejger}, {Belahcene}, {Bell}, {Berger},
  {Bergmann}, {Bero}, {Berry}, {Bersanetti}, {Bertolini}, {Betzwieser},
  {Bhagwat}, {Bhandare}, {Bilenko}, {Billingsley}, {Billman}, {Birch},
  {Birney}, {Birnholtz}, {Biscans}, {Biscoveanu}, {Bisht}, {Bitossi}, {Biwer},
  {Bizouard}, {Blackburn}, {Blackman}, {Blair}, {Blair}, {Blair}, {Bloemen},
  {Bock}, {Bode}, {Boer}, {Bogaert}, {Bohe}, {Bondu}, {Bonilla}, {Bonnand},
  {Boom}, {Bork}, {Boschi}, {Bose}, {Bossie}, {Bouffanais}, {Bozzi},
  {Bradaschia}, {Brady}, {Branchesi}, {Brau}, {Briant}, {Brillet}, {Brinkmann},
  {Brisson}, {Brockill}, {Broida}, {Brooks}, {Brown}, {Brunett}, {Buchanan},
  {Buikema}, {Bulik}, {Bulten}, {Buonanno}, {Buskulic}, {Buy}, {Byer},
  {Cabero}, {Cadonati}, {Cagnoli}, {Cahillane}, {Calder{\'o}n Bustillo},
  {Callister}, {Calloni}, {Camp}, {Canepa}, {Canizares}, {Cannon}, {Cao},
  {Cao}, {Capano}, {Capocasa}, {Carbognani}, {Caride}, {Carney}, {Casanueva
  Diaz}, {Casentini}, {Caudill}, {Cavagli{\`a}}, {Cavalier}, {Cavalieri},
  {Cella}, {Cepeda}, {Cerd{\'a}-Dur{\'a}n}, {Cerretani}, {Cesarini},
  {Chamberlin}, {Chan}, {Chao}, {Charlton}, {Chase}, {Chassande-Mottin},
  {Chatterjee}, {Cheeseboro}, {Chen}, {Chen}, {Chen}, {Cheng}, {Chia},
  {Chincarini}, {Chiummo}, {Chmiel}, {Cho}, {Cho}, {Chow}, {Christensen},
  {Chu}, {Chua}, {Chua}, {Chung}, {Chung}, {Ciani}, {Ciolfi}, {Cirelli},
  {Cirone}, {Clara}, {Clark}, {Clearwater}, {Cleva}, {Cocchieri}, {Coccia},
  {Cohadon}, {Cohen}, {Colla}, {Collette}, {Cominsky}, {Constancio}, {Conti},
  {Cooper}, {Corban}, {Corbitt}, {Cordero-Carri{\'o}n}, {Corley}, {Corsi},
  {Cortese}, {Costa}, {Coughlin}, {Coughlin}, {Coulon}, {Countryman},
  {Couvares}, {Covas}, {Cowan}, {Coward}, {Cowart}, {Coyne}, {Coyne},
  {Creighton}, {Creighton}, {Cripe}, {Crowder}, {Cullen}, {Cumming},
  {Cunningham}, {Cuoco}, {Dal Canton}, {D{\'a}lya}, {Danilishin}, {D'Antonio},
  {Danzmann}, {Dasgupta}, {Da Silva Costa}, {Dattilo}, {Dave}, {Davier},
  {Davis}, {Daw}, {Day}, {De}, {DeBra}, {Degallaix}, {De Laurentis},
  {Del{\'e}glise}, {Del Pozzo}, {Demos}, {Denker}, {Dent}, {De Pietri},
  {Dergachev}, {De Rosa}, {DeRosa}, {De Rossi}, {DeSalvo}, {de Varona},
  {Devenson}, {Dhurandhar}, {D{\'\i}az}, {Di Fiore}, {Di Giovanni}, {Di
  Girolamo}, {Di Lieto}, {Di Pace}, {Di Palma}, {Di Renzo}, {Doctor},
  {Dolique}, {Donovan}, {Dooley}, {Doravari}, {Dorrington}, {Douglas}, {Dovale
  {\'A}lvarez}, {Downes}, {Drago}, {Dreissigacker}, {Driggers}, {Du}, {Ducrot},
  {Dupej}, {Dwyer}, {Edo}, {Edwards}, {Effler}, {Eggenstein}, {Ehrens},
  {Eichholz}, {Eikenberry}, {Eisenstein}, {Essick}, {Estevez}, {Etienne},
  {Etzel}, {Evans}, {Evans}, {Factourovich}, {Fafone}, {Fair}, {Fairhurst},
  {Fan}, {Farinon}, {Farr}, {Farr}, {Fauchon- Jones}, {Favata}, {Fays}, {Fee},
  {Fehrmann}, {Feicht}, {Fejer}, {Fernandez-Galiana}, {Ferrante}, {Ferreira},
  {Ferrini}, {Fidecaro}, {Finstad}, {Fiori}, {Fiorucci}, {Fishbach}, {Fisher},
  {Fitz-Axen}, {Flaminio}, {Fletcher}, {Fong}, {Font}, {Forsyth}, {Forsyth},
  {Fournier}, {Frasca}, {Frasconi}, {Frei}, {Freise}, {Frey}, {Frey}, {Fries},
  {Fritschel}, {Frolov}, {Fulda}, {Fyffe}, {Gabbard}, {Gadre}, {Gaebel},
  {Gair}, {Gammaitoni}, {Ganija}, {Gaonkar}, {Garcia-Quiros}, {Garufi},
  {Gateley}, {Gaudio}, {Gaur}, {Gayathri}, {Gehrels}, {Gemme}, {Genin},
  {Gennai}, {George}, {George}, {Gergely}, {Germain}, {Ghonge}, {Ghosh},
  {Ghosh}, {Ghosh}, {Giaime}, {Giardina}, {Giazotto}, {Gill}, {Glover},
  {Goetz}, {Goetz}, {Gomes}, {Goncharov}, {Gonzalez Castro}, {Gopakumar},
  {Gorodetsky}, {Gossan}, {Gosselin}, {Gouaty}, {Grado}, {Graef}, {Granata},
  {Grant}, {Gras}, {Gray}, {Greco}, {Green}, {Gretarsson}, {Groot}, {Grote},
  {Grunewald}, {Gruning}, {Guidi}, {Guo}, {Gupta}, {Gupta}, {Gushwa},
  {Gustafson}, {Gustafson}, {Halim}, {Hall}, {Hall}, {Hamilton}, {Hammond},
  {Haney}, {Hanke}, {Hanks}, {Hanna}, {Hannam}, {Hannuksela}, {Hanson},
  {Hardwick}, {Harms}, {Harry}, {Harry}, {Hart}, {Haster}, {Haughian}, {Healy},
  {Heidmann}, {Heintze}, {Heitmann}, {Hello}, {Hemming}, {Hendry}, {Heng},
  {Hennig}, {Heptonstall}, {Heurs}, {Hild}, {Hinderer}, {Hoak}, {Hofman},
  {Holgado}, {Holt}, {Holz}, {Hopkins}, {Horst}, {Hough}, {Houston}, {Howell},
  {Hreibi}, {Hu}, {Huerta}, {Huet}, {Hughey}, {Husa}, {Huttner}, {Huynh-Dinh},
  {Indik}, {Inta}, {Intini}, {Isa}, {Isac}, {Isi}, {Iyer}, {Izumi}, {Jacqmin},
  {Jani}, {Jaranowski}, {Jawahar}, {Jim{\'e}nez- Forteza}, {Johnson}, {Jones},
  {Jones}, {Jonker}, {Ju}, {Junker}, {Kalaghatgi}, {Kalogera}, {Kamai},
  {Kandhasamy}, {Kang}, {Kanner}, {Kapadia}, {Karki}, {Karvinen}, {Kasprzack},
  {Katolik}, {Katsavounidis}, {Katzman}, {Kaufer}, {Kawabe},
  {K{\'e}f{\'e}lian}, {Keitel}, {Kemball}, {Kennedy}, {Kent}, {Key}, {Khalili},
  {Khan}, {Khan}, {Khan}, {Khazanov}, {Kijbunchoo}, {Kim}, {Kim}, {Kim}, {Kim},
  {Kim}, {Kim}, {Kimball}, {Kimbrell}, {King}, {King}, {Kinley-Hanlon},
  {Kirchhoff}, {Kissel}, {Kleybolte}, {Klimenko}, {Knowles}, {Koch},
  {Koehlenbeck}, {Koley}, {Kondrashov}, {Kontos}, {Korobko}, {Korth},
  {Kowalska}, {Kozak}, {Kr{\"a}mer}, {Kringel}, {Kr{\'o}lak}, {Kuehn}, {Kumar},
  {Kumar}, {Kumar}, {Kuo}, {Kutynia}, {Kwang}, {Lackey}, {Lai}, {Landry},
  {Lang}, {Lange}, {Lantz}, {Lanza}, {Larson}, {Lartaux-Vollard}, {Lasky},
  {Laxen}, {Lazzarini}, {Lazzaro}, {Leaci}, {Leavey}, {Lee}, {Lee}, {Lee},
  {Lee}, {Lee}, {Lehmann}, {Lenon}, {Leonardi}, {Leroy}, {Letendre}, {Levin},
  {Li}, {Linker}, {Littenberg}, {Liu}, {Lo}, {Lockerbie}, {London}, {Lord},
  {Lorenzini}, {Loriette}, {Lormand}, {Losurdo}, {Lough}, {Lousto}, {Lovelace},
  {L{\"u}ck}, {Lumaca}, {Lundgren}, {Lynch}, {Ma}, {Macas}, {Macfoy},
  {Machenschalk}, {MacInnis}, {Macleod}, {Maga{\~n}a Hernandez},
  {Maga{\~n}a-Sandoval}, {Maga{\~n}a Zertuche}, {Magee}, {Majorana},
  {Maksimovic}, {Man}, {Mandic}, {Mangano}, {Mansell}, {Manske}, {Mantovani},
  {Marchesoni}, {Marion}, {M{\'a}rka}, {M{\'a}rka}, {Markakis}, {Markosyan},
  {Markowitz}, {Maros}, {Marquina}, {Martelli}, {Martellini}, {Martin},
  {Martin}, {Martynov}, {Mason}, {Massera}, {Masserot}, {Massinger},
  {Masso-Reid}, {Mastrogiovanni}, {Matas}, {Matichard}, {Matone}, {Mavalvala},
  {Mazumder}, {McCarthy}, {McClelland}, {McCormick}, {McCuller}, {McGuire},
  {McIntyre}, {McIver}, {McManus}, {McNeill}, {McRae}, {McWilliams}, {Meacher},
  {Meadors}, {Mehmet}, {Meidam}, {Mejuto-Villa}, {Melatos}, {Mendell},
  {Mercer}, {Merilh}, {Merzougui}, {Meshkov}, {Messenger}, {Messick},
  {Metzdorff}, {Meyers}, {Miao}, {Michel}, {Middleton}, {Mikhailov}, {Milano},
  {Miller}, {Miller}, {Miller}, {Millhouse}, {Milovich-Goff}, {Minazzoli},
  {Minenkov}, {Ming}, {Mishra}, {Mitra}, {Mitrofanov}, {Mitselmakher},
  {Mittleman}, {Moffa}, {Moggi}, {Mogushi}, {Mohan}, {Mohapatra}, {Montani},
  {Moore}, {Moraru}, {Moreno}, {Morriss}, {Mours}, {Mow-Lowry}, {Mueller},
  {Muir}, {Mukherjee}, {Mukherjee}, {Mukherjee}, {Mukund}, {Mullavey}, {Munch},
  {Mu{\~n}iz}, {Muratore}, {Murray}, {Napier}, {Nardecchia}, {Naticchioni},
  {Nayak}, {Neilson}, {Nelemans}, {Nelson}, {Nery}, {Neunzert}, {Nevin},
  {Newport}, {Newton}, {Ng}, {Nguyen}, {Nichols}, {Nielsen}, {Nissanke},
  {Nitz}, {Noack}, {Nocera}, {Nolting}, {North}, {Nuttall}, {Oberling},
  {O'Dea}, {Ogin}, {Oh}, {Oh}, {Ohme}, {Okada}, {Oliver}, {Oppermann}, {Oram},
  {O'Reilly}, {Ormiston}, {Ortega}, {O'Shaughnessy}, {Ossokine}, {Ottaway},
  {Overmier}, {Owen}, {Pace}, {Page}, {Page}, {Pai}, {Pai}, {Palamos},
  {Palashov}, {Palomba}, {Pal- Singh}, {Pan}, {Pan}, {Pang}, {Pang}, {Pankow},
  {Pannarale}, {Pant}, {Paoletti}, {Paoli}, {Papa}, {Parida}, {Parker},
  {Pascucci}, {Pasqualetti}, {Passaquieti}, {Passuello}, {Patil}, {Patricelli},
  {Pearlstone}, {Pedraza}, {Pedurand}, {Pekowsky}, {Pele}, {Penn}, {Perez},
  {Perreca}, {Perri}, {Pfeiffer}, {Phelps}, {Piccinni}, {Pichot},
  {Piergiovanni}, {Pierro}, {Pillant}, {Pinard}, {Pinto}, {Pirello}, {Pitkin},
  {Poe}, {Poggiani}, {Popolizio}, {Porter}, {Post}, {Powell}, {Prasad},
  {Pratt}, {Pratten}, {Predoi}, {Prestegard}, {Prijatelj}, {Principe},
  {Privitera}, {Prodi}, {Prokhorov}, {Puncken}, {Punturo}, {Puppo},
  {P{\"u}rrer}, {Qi}, {Quetschke}, {Quintero}, {Quitzow-James}, {Rabeling},
  {Radkins}, {Raffai}, {Raja}, {Rajan}, {Rajbhandari}, {Rakhmanov}, {Ramirez},
  {Ramos-Buades}, {Rapagnani}, {Raymond}, {Razzano}, {Read}, {Regimbau}, {Rei},
  {Reid}, {Reitze}, {Ren}, {Reyes}, {Ricci}, {Ricker}, {Rieger}, {Riles},
  {Rizzo}, {Robertson}, {Robie}, {Robinet}, {Rocchi}, {Rolland}, {Rollins},
  {Roma}, {Romano}, {Romel}, {Romie}, {Rosi{\'n}ska}, {Ross}, {Rowan},
  {R{\"u}diger}, {Ruggi}, {Rutins}, {Ryan}, {Sachdev}, {Sadecki}, {Sadeghian},
  {Sakellariadou}, {Salconi}, {Saleem}, {Salemi}, {Samajdar}, {Sammut},
  {Sampson}, {Sanchez}, {Sanchez}, {Sanchis- Gual}, {Sandberg}, {Sanders},
  {Sassolas}, {Sathyaprakash}, {Sauter}, {Savage}, {Sawadsky}, {Schale},
  {Scheel}, {Scheuer}, {Schmidt}, {Schmidt}, {Schnabel}, {Schofield},
  {Sch{\"o}nbeck}, {Schreiber}, {Schuette}, {Schulte}, {Schutz}, {Schwalbe},
  {Scott}, {Scott}, {Seidel}, {Sellers}, {Sengupta}, {Sentenac}, {Sequino},
  {Sergeev}, {Shaddock}, {Shaffer}, {Shah}, {Shahriar}, {Shaner}, {Shao},
  {Shapiro}, {Shawhan}, {Sheperd}, {Shoemaker}, {Shoemaker}, {Siellez},
  {Siemens}, {Sieniawska}, {Sigg}, {Silva}, {Singer}, {Singh}, {Singhal},
  {Sintes}, {Slagmolen}, {Smith}, {Smith}, {Smith}, {Somala}, {Son},
  {Sonnenberg}, {Sorazu}, {Sorrentino}, {Souradeep}, {Spencer}, {Srivastava},
  {Staats}, {Staley}, {Steinke}, {Steinlechner}, {Steinlechner}, {Steinmeyer},
  {Stevenson}, {Stone}, {Stops}, {Strain}, {Stratta}, {Strigin}, {Strunk},
  {Sturani}, {Stuver}, {Summerscales}, {Sun}, {Sunil}, {Suresh}, {Sutton},
  {Swinkels}, {Szczepa{\'n}czyk}, {Tacca}, {Tait}, {Talbot}, {Talukder},
  {Tanner}, {T{\'a}pai}, {Taracchini}, {Tasson}, {Taylor}, {Taylor}, {Tewari},
  {Theeg}, {Thies}, {Thomas}, {Thomas}, {Thomas}, {Thorne}, {Thrane}, {Tiwari},
  {Tiwari}, {Tokmakov}, {Toland}, {Tonelli}, {Tornasi}, {Torres-Forn{\'e}},
  {Torrie}, {T{\"o}yr{\"a}}, {Travasso}, {Traylor}, {Trinastic}, {Tringali},
  {Trozzo}, {Tsang}, {Tse}, {Tso}, {Tsukada}, {Tsuna}, {Tuyenbayev}, {Ueno},
  {Ugolini}, {Unnikrishnan}, {Urban}, {Usman}, {Vahlbruch}, {Vajente},
  {Valdes}, {van Bakel}, {van Beuzekom}, {van den Brand}, {Van Den Broeck},
  {Vander-Hyde}, {van der Schaaf}, {van Heijningen}, {van Veggel}, {Vardaro},
  {Varma}, {Vass}, {Vas{\'u}th}, {Vecchio}, {Vedovato}, {Veitch}, {Veitch},
  {Venkateswara}, {Venugopalan}, {Verkindt}, {Vetrano}, {Vicer{\'e}}, {Viets},
  {Vinciguerra}, {Vine}, {Vinet}, {Vitale}, {Vo}, {Vocca}, {Vorvick},
  {Vyatchanin}, {Wade}, {Wade}, {Wade}, {Walet}, {Walker}, {Wallace}, {Walsh},
  {Wang}, {Wang}, {Wang}, {Wang}, {Wang}, {Ward}, {Warner}, {Was}, {Watchi},
  {Weaver}, {Wei}, {Weinert}, {Weinstein}, {Weiss}, {Wen}, {Wessel},
  {We{\ss}els}, {Westerweck}, {Westphal}, {Wette}, {Whelan}, {Whiting},
  {Whittle}, {Wilken}, {Williams}, {Williams}, {Williamson}, {Willis},
  {Willke}, {Wimmer}, {Winkler}, {Wipf}, {Wittel}, {Woan}, {Woehler},
  {Wofford}, {Wong}, {Worden}, {Wright}, {Wu}, {Wysocki}, {Xiao}, {Yamamoto},
  {Yancey}, {Yang}, {Yap}, {Yazback}, {Yu}, {Yu}, {Yvert}, {Zadro{\.z}ny},
  {Zanolin}, {Zelenova}, {Zendri}, {Zevin}, {Zhang}, {Zhang}, {Zhang}, {Zhang},
  {Zhao}, {Zhou}, {Zhou}, {Zhu}, {Zhu}, {Zucker}, {Zweizig}, {(LIGO Scientific
  Collaboration}, \& {Virgo Collaboration}}]{abbott_2017_af}
---. 2017{\natexlab{f}}, \apj, 850, L40, \dodoi{10.3847/2041-8213/aa93fc}

\bibitem[{{Aikawa} \& {Simon}(1983)}]{rsp_TGRID}
{Aikawa}, T., \& {Simon}, N.~R. 1983, \apj, 273, 346, \dodoi{10.1086/161373}

\bibitem[{{Alastuey} \& {Jancovici}(1978)}]{alastuey:78}
{Alastuey}, A., \& {Jancovici}, B. 1978, \apj, 226, 1034,
  \dodoi{10.1086/156681}

\bibitem[{{Alcock} {et~al.}(2003){Alcock}, {Alves}, {Becker}, {Bennett},
  {Cook}, {Drake}, {Freeman}, {Geha}, {Griest}, {Kov{\'a}cs}, {Lehner},
  {Marshall}, {Minniti}, {Nelson}, {Peterson}, {Popowski}, {Pratt}, {Quinn},
  {Rodgers}, {Stubbs}, {Sutherland}, {Vandehei}, \& {Welch}}]{alcock_2003_aa}
{Alcock}, C., {Alves}, D.~R., {Becker}, A., {et~al.} 2003, \apj, 598, 597,
  \dodoi{10.1086/378689}

\bibitem[{Amdahl(1967)}]{Amdahl1967}
Amdahl, G.~M. 1967, in Proceedings of the April 18-20, 1967, Spring Joint
  Computer Conference, AFIPS '67 (Spring) (New York, NY, USA: ACM), 483--485.
\newblock \url{http://doi.acm.org/10.1145/1465482.1465560}

\bibitem[{{Artigau} {et~al.}(2014){Artigau}, {Sivaramakrishnan}, {Greenbaum},
  {Doyon}, {Goudfrooij}, {Fullerton}, {Lafreni{\`e}re}, {Volk}, {Albert},
  {Martel}, {Ford}, \& {McKernan}}]{artigau_2014_aa}
{Artigau}, {\'E}., {Sivaramakrishnan}, A., {Greenbaum}, A.~Z., {et~al.} 2014,
  in Society of Photo-Optical Instrumentation Engineers (SPIE) Conference
  Series, Vol. 9143, Space Telescopes and Instrumentation 2014: Optical,
  Infrared, and Millimeter Wave, 914340

\bibitem[{{Astropy Collaboration} {et~al.}(2018){Astropy Collaboration},
  {Price-Whelan}, {Sip{\'{o}}cz}, {G{\"u}nther}, {Lim}, {Crawford}, {Conseil},
  {Shupe}, {Craig}, {Dencheva}, {Ginsburg}, {VanderPlas}, {Bradley},
  {P{\'e}rez-Su{\'a}rez}, {de Val- Borro}, {Aldcroft}, {Cruz}, {Robitaille},
  {Tollerud}, {Ardelean}, {Babej}, {Bach}, {Bachetti}, {Bakanov}, {Bamford},
  {Barentsen}, {Barmby}, {Baumbach}, {Berry}, {Biscani}, {Boquien}, {Bostroem},
  {Bouma}, {Brammer}, {Bray}, {Breytenbach}, {Buddelmeijer}, {Burke},
  {Calderone}, {Cano Rodr{\'\i}guez}, {Cara}, {Cardoso}, {Cheedella}, {Copin},
  {Corrales}, {Crichton}, {D'Avella}, {Deil}, {Depagne}, {Dietrich}, {Donath},
  {Droettboom}, {Earl}, {Erben}, {Fabbro}, {Ferreira}, {Finethy}, {Fox},
  {Garrison}, {Gibbons}, {Goldstein}, {Gommers}, {Greco}, {Greenfield},
  {Groener}, {Grollier}, {Hagen}, {Hirst}, {Homeier}, {Horton}, {Hosseinzadeh},
  {Hu}, {Hunkeler}, {Ivezi{\'c}}, {Jain}, {Jenness}, {Kanarek}, {Kendrew},
  {Kern}, {Kerzendorf}, {Khvalko}, {King}, {Kirkby}, {Kulkarni}, {Kumar},
  {Lee}, {Lenz}, {Littlefair}, {Ma}, {Macleod}, {Mastropietro}, {McCully},
  {Montagnac}, {Morris}, {Mueller}, {Mumford}, {Muna}, {Murphy}, {Nelson},
  {Nguyen}, {Ninan}, {N{\"o}the}, {Ogaz}, {Oh}, {Parejko}, {Parley}, {Pascual},
  {Patil}, {Patil}, {Plunkett}, {Prochaska}, {Rastogi}, {Reddy Janga},
  {Sabater}, {Sakurikar}, {Seifert}, {Sherbert}, {Sherwood-Taylor}, {Shih},
  {Sick}, {Silbiger}, {Singanamalla}, {Singer}, {Sladen}, {Sooley},
  {Sornarajah}, {Streicher}, {Teuben}, {Thomas}, {Tremblay}, {Turner},
  {Terr{\'o}n}, {van Kerkwijk}, {de la Vega}, {Watkins}, {Weaver}, {Whitmore},
  {Woillez}, {Zabalza}, \& {Astropy
  Contributors}}]{astropy-collaboration_2018_aa}
{Astropy Collaboration}, {Price-Whelan}, A.~M., {Sip{\'{o}}cz}, B.~M., {et~al.}
  2018, \aj, 156, 123, \dodoi{10.3847/1538-3881/aabc4f}

\bibitem[{{Auvergne} {et~al.}(2009){Auvergne}, {Bodin}, {Boisnard}, {Buey},
  {Chaintreuil}, {Epstein}, {Jouret}, {Lam- Trong}, {Levacher}, {Magnan},
  {Perez}, {Plasson}, {Plesseria}, {Peter}, {Steller}, {Tiph{\`e}ne}, {Baglin},
  {Agogu{\'e}}, {Appourchaux}, {Barbet}, {Beaufort}, {Bellenger}, {Berlin},
  {Bernardi}, {Blouin}, {Boumier}, {Bonneau}, {Briet}, {Butler}, {Cautain},
  {Chiavassa}, {Costes}, {Cuvilho}, {Cunha- Parro}, {de Oliveira Fialho},
  {Decaudin}, {Defise}, {Djalal}, {Docclo}, {Drummond}, {Dupuis}, {Exil},
  {Faur{\'e}}, {Gaboriaud}, {Gamet}, {Gavalda}, {Grolleau}, {Gueguen},
  {Guivarc'h}, {Guterman}, {Hasiba}, {Huntzinger}, {Hustaix}, {Imbert},
  {Jeanville}, {Johlander}, {Jorda}, {Journoud}, {Karioty}, {Kerjean},
  {Lafond}, {Lapeyrere}, {Landiech}, {Larqu{\'e}}, {Laudet}, {Le Merrer},
  {Leporati}, {Leruyet}, {Levieuge}, {Llebaria}, {Martin}, {Mazy}, {Mesnager},
  {Michel}, {Moalic}, {Monjoin}, {Naudet}, {Neukirchner}, {Nguyen-Kim},
  {Ollivier}, {Orcesi}, {Ottacher}, {Oulali}, {Parisot}, {Perruchot},
  {Piacentino}, {Pinheiro da Silva}, {Platzer}, {Pontet}, {Pradines},
  {Quentin}, {Rohbeck}, {Rolland}, {Rollenhagen}, {Romagnan}, {Russ}, {Samadi},
  {Schmidt}, {Schwartz}, {Sebbag}, {Smit}, {Sunter}, {Tello}, {Toulouse},
  {Ulmer}, {Vandermarcq}, {Vergnault}, {Wallner}, {Waultier}, \&
  {Zanatta}}]{auvergne_2009_aa}
{Auvergne}, M., {Bodin}, P., {Boisnard}, L., {et~al.} 2009, \aap, 506, 411,
  \dodoi{10.1051/0004-6361/200810860}

\bibitem[{{Bailer-Jones} {et~al.}(2018){Bailer-Jones}, {Rybizki}, {Fouesneau},
  {Mantelet}, \& {Andrae}}]{bailer-jones_2018_aa}
{Bailer-Jones}, C.~A.~L., {Rybizki}, J., {Fouesneau}, M., {Mantelet}, G., \&
  {Andrae}, R. 2018, \aj, 156, 58, \dodoi{10.3847/1538-3881/aacb21}

\bibitem[{{Ball} {et~al.}(2018){Ball}, {Chaplin}, {Schofield}, {Miglio},
  {Bossini}, {Davies}, \& {Girardi}}]{ball_2018_aa}
{Ball}, W.~H., {Chaplin}, W.~J., {Schofield}, M., {et~al.} 2018, The
  Astrophysical Journal Supplement Series, 239, 34,
  \dodoi{10.3847/1538-4365/aaedbc}

\bibitem[{{Beichman} {et~al.}(2012){Beichman}, {Rieke}, {Eisenstein}, {Greene},
  {Krist}, {McCarthy}, {Meyer}, \& {Stansberry}}]{beichman_2012_aa}
{Beichman}, C.~A., {Rieke}, M., {Eisenstein}, D., {et~al.} 2012, in Society of
  Photo-Optical Instrumentation Engineers (SPIE) Conference Series, Vol. 8442,
  Space Telescopes and Instrumentation 2012: Optical, Infrared, and Millimeter
  Wave, 84422N

\bibitem[{{Benedict} {et~al.}(2002){Benedict}, {McArthur}, {Fredrick},
  {Harrison}, {Lee}, {Slesnick}, {Rhee}, {Patterson}, {Nelan}, {Jefferys}, {van
  Altena}, {Shelus}, {Franz}, {Wasserman}, {Hemenway}, {Duncombe}, {Story},
  {Whipple}, \& {Bradley}}]{benedict_2002_aa}
{Benedict}, G.~F., {McArthur}, B.~E., {Fredrick}, L.~W., {et~al.} 2002, \aj,
  123, 473, \dodoi{10.1086/338087}

\bibitem[{{Bersten} {et~al.}(2011){Bersten}, {Benvenuto}, \&
  {Hamuy}}]{Bersten2011}
{Bersten}, M.~C., {Benvenuto}, O., \& {Hamuy}, M. 2011, \apj, 729, 61,
  \dodoi{10.1088/0004-637X/729/1/61}

\bibitem[{{Bla{\v{z}}ko}(1907)}]{blazko_1907_aa}
{Bla{\v{z}}ko}, S. 1907, Astronomische Nachrichten, 175, 325,
  \dodoi{10.1002/asna.19071752002}

\bibitem[{{Bono} {et~al.}(2000){Bono}, {Marconi}, \&
  {Stellingwerf}}]{bono_2000_aa}
{Bono}, G., {Marconi}, M., \& {Stellingwerf}, R.~F. 2000, \aap, 360, 245.
\newblock \doarXiv{astro-ph/0006229}

\bibitem[{{Bono} \& {Stellingwerf}(1994)}]{rsp_IC}
{Bono}, G., \& {Stellingwerf}, R.~F. 1994, \apjs, 93, 233,
  \dodoi{10.1086/192054}

\bibitem[{{Borucki} {et~al.}(2010){Borucki}, {Koch}, {Basri}, {Batalha},
  {Brown}, {Caldwell}, {Caldwell}, {Christensen-Dalsgaard}, {Cochran},
  {DeVore}, {Dunham}, {Dupree}, {Gautier}, {Geary}, {Gilliland}, {Gould},
  {Howell}, {Jenkins}, {Kondo}, {Latham}, {Marcy}, {Meibom}, {Kjeldsen},
  {Lissauer}, {Monet}, {Morrison}, {Sasselov}, {Tarter}, {Boss}, {Brownlee},
  {Owen}, {Buzasi}, {Charbonneau}, {Doyle}, {Fortney}, {Ford}, {Holman},
  {Seager}, {Steffen}, {Welsh}, {Rowe}, {Anderson}, {Buchhave}, {Ciardi},
  {Walkowicz}, {Sherry}, {Horch}, {Isaacson}, {Everett}, {Fischer}, {Torres},
  {Johnson}, {Endl}, {MacQueen}, {Bryson}, {Dotson}, {Haas}, {Kolodziejczak},
  {Van Cleve}, {Chandrasekaran}, {Twicken}, {Quintana}, {Clarke}, {Allen},
  {Li}, {Wu}, {Tenenbaum}, {Verner}, {Bruhweiler}, {Barnes}, \&
  {Prsa}}]{borucki_2010_aa}
{Borucki}, W.~J., {Koch}, D., {Basri}, G., {et~al.} 2010, Science, 327, 977,
  \dodoi{10.1126/science.1185402}

\bibitem[{{Breitfelder} {et~al.}(2016){Breitfelder}, {M{\'e}rand}, {Kervella},
  {Gallenne}, {Szabados}, {Anderson}, \& {Le Bouquin}}]{rsp_breitfelder}
{Breitfelder}, J., {M{\'e}rand}, A., {Kervella}, P., {et~al.} 2016, \aap, 587,
  A117, \dodoi{10.1051/0004-6361/201527030}

\bibitem[{{Broeg} {et~al.}(2013){Broeg}, {Fortier}, {Ehrenreich}, {Alibert},
  {Baumjohann}, {Benz}, {Deleuil}, {Gillon}, {Ivanov}, {Liseau}, {Meyer},
  {Oloffson}, {Pagano}, {Piotto}, {Pollacco}, {Queloz}, {Ragazzoni}, {Renotte},
  {Steller}, \& {Thomas}}]{broeg_2013_aa}
{Broeg}, C., {Fortier}, A., {Ehrenreich}, D., {et~al.} 2013, in European
  Physical Journal Web of Conferences, Vol.~47, European Physical Journal Web
  of Conferences, 03005

\bibitem[{{Buchler}(2009)}]{rsp_buchlerAIP09}
{Buchler}, J.~R. 2009, in American Institute of Physics Conference Series, Vol.
  1170, American Institute of Physics Conference Series, ed. J.~A. {Guzik} \&
  P.~A. {Bradley}, 51--58

\bibitem[{{Buchler} {et~al.}(2004){Buchler}, {Koll{\'a}th}, \&
  {Cadmus}}]{rsp_bkc04}
{Buchler}, J.~R., {Koll{\'a}th}, Z., \& {Cadmus}, Jr., R.~R. 2004, \apj, 613,
  532, \dodoi{10.1086/422903}

\bibitem[{{Buchler} {et~al.}(1996){Buchler}, {Kollath}, {Serre}, \&
  {Mattei}}]{rsp_bksm96}
{Buchler}, J.~R., {Kollath}, Z., {Serre}, T., \& {Mattei}, J. 1996, \apj, 462,
  489, \dodoi{10.1086/177167}

\bibitem[{{Buchler} \& {Kovacs}(1987)}]{rsp_bk87}
{Buchler}, J.~R., \& {Kovacs}, G. 1987, \apjl, 320, L57, \dodoi{10.1086/184976}

\bibitem[{{Buchler} \& {Moskalik}(1992)}]{rsp_bm92}
{Buchler}, J.~R., \& {Moskalik}, P. 1992, \apj, 391, 736,
  \dodoi{10.1086/171384}

\bibitem[{{Buchler} {et~al.}(1990){Buchler}, {Moskalik}, \&
  {Kovacs}}]{rsp_bmk90}
{Buchler}, J.~R., {Moskalik}, P., \& {Kovacs}, G. 1990, \apj, 351, 617,
  \dodoi{10.1086/168500}

\bibitem[{{Byrne} \& {Jeffery}(2018)}]{byrne_2018_aa}
{Byrne}, C.~M., \& {Jeffery}, C.~S. 2018, \mnras, 481, 3810,
  \dodoi{10.1093/mnras/sty2545}

\bibitem[{{Castellani} {et~al.}(1985){Castellani}, {Chieffi}, {Tornambe}, \&
  {Pulone}}]{castellani_1985_aa}
{Castellani}, V., {Chieffi}, A., {Tornambe}, A., \& {Pulone}, L. 1985, \apj,
  296, 204, \dodoi{10.1086/163437}

\bibitem[{{Castellani} {et~al.}(1971){Castellani}, {Giannone}, \&
  {Renzini}}]{castellani_1971_ab}
{Castellani}, V., {Giannone}, P., \& {Renzini}, A. 1971, \apss, 10, 355

\bibitem[{{Castor} {et~al.}(1977){Castor}, {Davis}, \& {Davison}}]{rsp_DYN}
{Castor}, J., {Davis}, Jr., C.~G., \& {Davison}, D.~K. 1977, {}, Tech. Rep.
  {LA-6664, TRN: 77-008571}, {Los Alamos Scientific Lab., NM (USA)}

\bibitem[{{Chabrier} {et~al.}(2019){Chabrier}, {Mazevet}, \&
  {Soubiran}}]{chabrier_2019_aa}
{Chabrier}, G., {Mazevet}, S., \& {Soubiran}, F. 2019, \apj, 872, 51,
  \dodoi{10.3847/1538-4357/aaf99f}

\bibitem[{{Chandrasekhar}(1933)}]{chandrasekhar_1933_aa}
{Chandrasekhar}, S. 1933, \mnras, 93, 462, \dodoi{10.1093/mnras/93.6.462}

\bibitem[{{Choi} {et~al.}(2016){Choi}, {Dotter}, {Conroy}, {Cantiello},
  {Paxton}, \& {Johnson}}]{choi_2016_aa}
{Choi}, J., {Dotter}, A., {Conroy}, C., {et~al.} 2016, \apj, 823, 102,
  \dodoi{10.3847/0004-637X/823/2/102}

\bibitem[{{Christensen-Dalsgaard}(2008)}]{2008Ap&SS.316...13C}
{Christensen-Dalsgaard}, J. 2008, \apss, 316, 13,
  \dodoi{10.1007/s10509-007-9675-5}

\bibitem[{{Christy}(1964)}]{rsp_christy}
{Christy}, R.~F. 1964, Reviews of Modern Physics, 36, 555,
  \dodoi{10.1103/RevModPhys.36.555}

\bibitem[{{Chugunov} {et~al.}(2007){Chugunov}, {Dewitt}, \&
  {Yakovlev}}]{chugunov:07}
{Chugunov}, A.~I., {Dewitt}, H.~E., \& {Yakovlev}, D.~G. 2007, \prd, 76,
  025028, \dodoi{10.1103/PhysRevD.76.025028}

\bibitem[{{Clementini} {et~al.}(2001){Clementini}, {Federici}, {Corsi},
  {Cacciari}, {Bellazzini}, \& {Smith}}]{clementini_2001_aa}
{Clementini}, G., {Federici}, L., {Corsi}, C., {et~al.} 2001, \apj, 559, L109,
  \dodoi{10.1086/323973}

\bibitem[{{Conroy} {et~al.}(2018){Conroy}, {Strader}, {van Dokkum}, {Dolphin},
  {Weisz}, {Murphy}, {Dotter}, {Johnson}, \& {Cargile}}]{conroy_2018_aa}
{Conroy}, C., {Strader}, J., {van Dokkum}, P., {et~al.} 2018, \apj, 864, 111,
  \dodoi{10.3847/1538-4357/aad460}

\bibitem[{{Cranmer} \& {Winebarger}(2018)}]{cranmer_2018_aa}
{Cranmer}, S.~R., \& {Winebarger}, A.~R. 2018, arXiv e-prints,
  arXiv:1811.00461.
\newblock \doarXiv{1811.00461}

\bibitem[{{Demarque} {et~al.}(2008){Demarque}, {Guenther}, {Li}, {Mazumdar}, \&
  {Straka}}]{2008Ap&SS.316...31D}
{Demarque}, P., {Guenther}, D.~B., {Li}, L.~H., {Mazumdar}, A., \& {Straka},
  C.~W. 2008, \apss, 316, 31, \dodoi{10.1007/s10509-007-9698-y}

\bibitem[{{Deming} {et~al.}(2009){Deming}, {Seager}, {Winn}, {Miller-Ricci},
  {Clampin}, {Lindler}, {Greene}, {Charbonneau}, {Laughlin}, {Ricker},
  {Latham}, \& {Ennico}}]{deming_2009_aa}
{Deming}, D., {Seager}, S., {Winn}, J., {et~al.} 2009, Publications of the
  Astronomical Society of the Pacific, 121, 952, \dodoi{10.1086/605913}

\bibitem[{{Dewitt} {et~al.}(1973){Dewitt}, {Graboske}, \& {Cooper}}]{dewitt:73}
{Dewitt}, H.~E., {Graboske}, H.~C., \& {Cooper}, M.~S. 1973, \apj, 181, 439,
  \dodoi{10.1086/152061}

\bibitem[{{Ding} \& {Li}(2014)}]{ding_2014_aa}
{Ding}, C.~Y., \& {Li}, Y. 2014, \mnras, 438, 1137

\bibitem[{{Dorfi} \& {Drury}(1987)}]{rsp_dd}
{Dorfi}, E.~A., \& {Drury}, L.~O. 1987, Journal of Computational Physics, 69,
  175, \dodoi{10.1016/0021-9991(87)90161-6}

\bibitem[{{Dorfi} \& {Feuchtinger}(1991)}]{rsp_df91}
{Dorfi}, E.~A., \& {Feuchtinger}, M.~U. 1991, \aap, 249, 417

\bibitem[{{Dragomir} {et~al.}(2019){Dragomir}, {Teske}, {G{\"u}nther},
  {S{\'e}gransan}, {Burt}, {Huang}, {Vanderburg}, {Matthews}, {Dumusque},
  {Stassun}, {Pepper}, {Ricker}, {Vanderspek}, {Latham}, {Seager}, {Winn},
  {Jenkins}, {Beatty}, {Bouchy}, {Brown}, {Butler}, {Ciardi}, {Crane},
  {Eastman}, {Fossati}, {Francis}, {Fulton}, {Gaudi}, {Goeke}, {James},
  {Klaus}, {Kuhn}, {Lovis}, {Lund}, {McDermott}, {Paegert}, {Pepe},
  {Rodriguez}, {Sha}, {Shectman}, {Shporer}, {Siverd}, {Garcia Soto},
  {Stevens}, {Twicken}, {Udry}, {Villanueva}, {Wang}, {Wohler}, {Yao}, \&
  {Zhan}}]{dragomir_2019_aa}
{Dragomir}, D., {Teske}, J., {G{\"u}nther}, M.~N., {et~al.} 2019, \apj, 875,
  L7, \dodoi{10.3847/2041-8213/ab12ed}

\bibitem[{{Duffell}(2016)}]{Duffell2016}
{Duffell}, P.~C. 2016, \apj, 821, 76, \dodoi{10.3847/0004-637X/821/2/76}

\bibitem[{{Eggleton}(1971)}]{eggleton_1971_aa}
{Eggleton}, P.~P. 1971, \mnras, 151, 351

\bibitem[{{Eggleton}(1972)}]{eggleton_1972_aa}
---. 1972, \mnras, 156, 361

\bibitem[{{Endal} \& {Sofia}(1976)}]{EndalSofia1976}
{Endal}, A.~S., \& {Sofia}, S. 1976, \apj, 210, 184, \dodoi{10.1086/154817}

\bibitem[{{Espinosa Lara} \& {Rieutord}(2011)}]{ELR2011}
{Espinosa Lara}, F., \& {Rieutord}, M. 2011, \aap, 533, A43,
  \dodoi{10.1051/0004-6361/201117252}

\bibitem[{{Faulkner}(1968)}]{MONSTAR:C}
{Faulkner}, D.~J. 1968, \mnras, 140, 223, \dodoi{10.1093/mnras/140.2.223}

\bibitem[{{Feng} {et~al.}(2010){Feng}, {Yang}, {Xiang}, {Wu}, {Zhou}, \&
  {Zhong}}]{feng_2010_aa}
{Feng}, X., {Yang}, L., {Xiang}, C., {et~al.} 2010, \apj, 723, 300,
  \dodoi{10.1088/0004-637X/723/1/300}

\bibitem[{{Feuchtinger} {et~al.}(2000){Feuchtinger}, {Buchler}, \&
  {Koll{\'a}th}}]{rsp_fbk00}
{Feuchtinger}, M., {Buchler}, J.~R., \& {Koll{\'a}th}, Z. 2000, \apj, 544,
  1056, \dodoi{10.1086/317260}

\bibitem[{{Foreman-Mackey} {et~al.}(2013){Foreman-Mackey}, {Hogg}, {Lang}, \&
  {Goodman}}]{foreman-mackey_2013_aa}
{Foreman-Mackey}, D., {Hogg}, D.~W., {Lang}, D., \& {Goodman}, J. 2013,
  Publications of the Astronomical Society of the Pacific, 125, 306,
  \dodoi{10.1086/670067}

\bibitem[{{Fraley}(1968)}]{rsp_fraley}
{Fraley}, G.~S. 1968, \apss, 2, 96, \dodoi{10.1007/BF00651498}

\bibitem[{{Freedman} {et~al.}(2001){Freedman}, {Madore}, {Gibson}, {Ferrarese},
  {Kelson}, {Sakai}, {Mould}, {Kennicutt}, {Ford}, {Graham}, {Huchra},
  {Hughes}, {Illingworth}, {Macri}, \& {Stetson}}]{freedman_2001_aa}
{Freedman}, W.~L., {Madore}, B.~F., {Gibson}, B.~K., {et~al.} 2001, \apj, 553,
  47, \dodoi{10.1086/320638}

\bibitem[{{Gabriel}(1970)}]{gabriel_1970_aa}
{Gabriel}, M. 1970, \aap, 6, 124

\bibitem[{{Gabriel} {et~al.}(2014){Gabriel}, {Noels}, {Montalb{\'a}n}, \&
  {Miglio}}]{gabriel_2014_aa}
{Gabriel}, M., {Noels}, A., {Montalb{\'a}n}, J., \& {Miglio}, A. 2014, \aap,
  569, A63, \dodoi{10.1051/0004-6361/201423442}

\bibitem[{{Gaia Collaboration} {et~al.}(2018{\natexlab{a}}){Gaia
  Collaboration}, {Brown}, {Vallenari}, {Prusti}, {de Bruijne}, {Babusiaux},
  {Bailer-Jones}, {Biermann}, {Evans}, {Eyer}, {Jansen}, {Jordi}, {Klioner},
  {Lammers}, {Lindegren}, {Luri}, {Mignard}, {Panem}, {Pourbaix}, {Randich},
  {Sartoretti}, {Siddiqui}, {Soubiran}, {van Leeuwen}, {Walton}, {Arenou},
  {Bastian}, {Cropper}, {Drimmel}, {Katz}, {Lattanzi}, {Bakker}, {Cacciari},
  {Casta{\~n}eda}, {Chaoul}, {Cheek}, {De Angeli}, {Fabricius}, {Guerra},
  {Holl}, {Masana}, {Messineo}, {Mowlavi}, {Nienartowicz}, {Panuzzo},
  {Portell}, {Riello}, {Seabroke}, {Tanga}, {Th{\'e}venin}, {Gracia-Abril},
  {Comoretto}, {Garcia-Reinaldos}, {Teyssier}, {Altmann}, {Andrae}, {Audard},
  {Bellas-Velidis}, {Benson}, {Berthier}, {Blomme}, {Burgess}, {Busso},
  {Carry}, {Cellino}, {Clementini}, {Clotet}, {Creevey}, {Davidson}, {De
  Ridder}, {Delchambre}, {Dell'Oro}, {Ducourant}, {Fern{\'a}ndez-
  Hern{\'a}ndez}, {Fouesneau}, {Fr{\'e}mat}, {Galluccio}, {Garc{\'\i}a-Torres},
  {Gonz{\'a}lez-N{\'u}{\~n}ez}, {Gonz{\'a}lez-Vidal}, {Gosset}, {Guy},
  {Halbwachs}, {Hambly}, {Harrison}, {Hern{\'a}ndez}, {Hestroffer}, {Hodgkin},
  {Hutton}, {Jasniewicz}, {Jean-Antoine-Piccolo}, {Jordan}, {Korn},
  {Krone-Martins}, {Lanzafame}, {Lebzelter}, {L{\"o}ffler}, {Manteiga},
  {Marrese}, {Mart{\'\i}n-Fleitas}, {Moitinho}, {Mora}, {Muinonen}, {Osinde},
  {Pancino}, {Pauwels}, {Petit}, {Recio-Blanco}, {Richards}, {Rimoldini},
  {Robin}, {Sarro}, {Siopis}, {Smith}, {Sozzetti}, {S{\"u}veges}, {Torra}, {van
  Reeven}, {Abbas}, {Abreu Aramburu}, {Accart}, {Aerts}, {Altavilla},
  {{\'A}lvarez}, {Alvarez}, {Alves}, {Anderson}, {Andrei}, {Anglada Varela},
  {Antiche}, {Antoja}, {Arcay}, {Astraatmadja}, {Bach}, {Baker},
  {Balaguer-N{\'u}{\~n}ez}, {Balm}, {Barache}, {Barata}, {Barbato}, {Barblan},
  {Barklem}, {Barrado}, {Barros}, {Barstow}, {Bartholom{\'e} Mu{\~n}oz},
  {Bassilana}, {Becciani}, {Bellazzini}, {Berihuete}, {Bertone}, {Bianchi},
  {Bienaym{\'e}}, {Blanco-Cuaresma}, {Boch}, {Boeche}, {Bombrun}, {Borrachero},
  {Bossini}, {Bouquillon}, {Bourda}, {Bragaglia}, {Bramante}, {Breddels},
  {Bressan}, {Brouillet}, {Br{\"u}semeister}, {Brugaletta}, {Bucciarelli},
  {Burlacu}, {Busonero}, {Butkevich}, {Buzzi}, {Caffau}, {Cancelliere},
  {Cannizzaro}, {Cantat-Gaudin}, {Carballo}, {Carlucci}, {Carrasco},
  {Casamiquela}, {Castellani}, {Castro-Ginard}, {Charlot}, {Chemin},
  {Chiavassa}, {Cocozza}, {Costigan}, {Cowell}, {Crifo}, {Crosta}, {Crowley},
  {Cuypers}, {Dafonte}, {Damerdji}, {Dapergolas}, {David}, {David}, {de
  Laverny}, {De Luise}, {De March}, {de Martino}, {de Souza}, {de Torres},
  {Debosscher}, {del Pozo}, {Delbo}, {Delgado}, {Delgado}, {Di Matteo},
  {Diakite}, {Diener}, {Distefano}, {Dolding}, {Drazinos}, {Dur{\'a}n},
  {Edvardsson}, {Enke}, {Eriksson}, {Esquej}, {Eynard Bontemps}, {Fabre},
  {Fabrizio}, {Faigler}, {Falc{\~a}o}, {Farr{\`a}s Casas}, {Federici},
  {Fedorets}, {Fernique}, {Figueras}, {Filippi}, {Findeisen}, {Fonti},
  {Fraile}, {Fraser}, {Fr{\'e}zouls}, {Gai}, {Galleti}, {Garabato},
  {Garc{\'\i}a-Sedano}, {Garofalo}, {Garralda}, {Gavel}, {Gavras}, {Gerssen},
  {Geyer}, {Giacobbe}, {Gilmore}, {Girona}, {Giuffrida}, {Glass}, {Gomes},
  {Granvik}, {Gueguen}, {Guerrier}, {Guiraud}, {Guti{\'e}rrez-S{\'a}nchez},
  {Haigron}, {Hatzidimitriou}, {Hauser}, {Haywood}, {Heiter}, {Helmi}, {Heu},
  {Hilger}, {Hobbs}, {Hofmann}, {Holland}, {Huckle}, {Hypki}, {Icardi},
  {Jan{\ss}en}, {Jevardat de Fombelle}, {Jonker}, {Juh{\'a}sz}, {Julbe},
  {Karampelas}, {Kewley}, {Klar}, {Kochoska}, {Kohley}, {Kolenberg},
  {Kontizas}, {Kontizas}, {Koposov}, {Kordopatis}, {Kostrzewa-Rutkowska},
  {Koubsky}, {Lambert}, {Lanza}, {Lasne}, {Lavigne}, {Le Fustec}, {Le
  Poncin-Lafitte}, {Lebreton}, {Leccia}, {Leclerc}, {Lecoeur-Taibi},
  {Lenhardt}, {Leroux}, {Liao}, {Licata}, {Lindstr{\o}m}, {Lister}, {Livanou},
  {Lobel}, {L{\'o}pez}, {Managau}, {Mann}, {Mantelet}, {Marchal}, {Marchant},
  {Marconi}, {Marinoni}, {Marschalk{\'o}}, {Marshall}, {Martino}, {Marton},
  {Mary}, {Massari}, {Matijevi{\v{c}}}, {Mazeh}, {McMillan}, {Messina},
  {Michalik}, {Millar}, {Molina}, {Molinaro}, {Moln{\'a}r}, {Montegriffo},
  {Mor}, {Morbidelli}, {Morel}, {Morris}, {Mulone}, {Muraveva}, {Musella},
  {Nelemans}, {Nicastro}, {Noval}, {O'Mullane}, {Ord{\'e}novic},
  {Ord{\'o}{\~n}ez-Blanco}, {Osborne}, {Pagani}, {Pagano}, {Pailler},
  {Palacin}, {Palaversa}, {Panahi}, {Pawlak}, {Piersimoni}, {Pineau}, {Plachy},
  {Plum}, {Poggio}, {Poujoulet}, {Pr{\v{s}}a}, {Pulone}, {Racero}, {Ragaini},
  {Rambaux}, {Ramos-Lerate}, {Regibo}, {Reyl{\'e}}, {Riclet}, {Ripepi}, {Riva},
  {Rivard}, {Rixon}, {Roegiers}, {Roelens}, {Romero-G{\'o}mez}, {Rowell},
  {Royer}, {Ruiz-Dern}, {Sadowski}, {Sagrist{\`a} Sell{\'e}s}, {Sahlmann},
  {Salgado}, {Salguero}, {Sanna}, {Santana- Ros}, {Sarasso}, {Savietto},
  {Schultheis}, {Sciacca}, {Segol}, {Segovia}, {S{\'e}gransan}, {Shih},
  {Siltala}, {Silva}, {Smart}, {Smith}, {Solano}, {Solitro}, {Sordo}, {Soria
  Nieto}, {Souchay}, {Spagna}, {Spoto}, {Stampa}, {Steele},
  {Steidelm{\"u}ller}, {Stephenson}, {Stoev}, {Suess}, {Surdej}, {Szabados},
  {Szegedi-Elek}, {Tapiador}, {Taris}, {Tauran}, {Taylor}, {Teixeira},
  {Terrett}, {Teyssandier}, {Thuillot}, {Titarenko}, {Torra Clotet}, {Turon},
  {Ulla}, {Utrilla}, {Uzzi}, {Vaillant}, {Valentini}, {Valette}, {van Elteren},
  {Van Hemelryck}, {van Leeuwen}, {Vaschetto}, {Vecchiato}, {Veljanoski},
  {Viala}, {Vicente}, {Vogt}, {von Essen}, {Voss}, {Votruba}, {Voutsinas},
  {Walmsley}, {Weiler}, {Wertz}, {Wevers}, {Wyrzykowski}, {Yoldas},
  {{\v{Z}}erjal}, {Ziaeepour}, {Zorec}, {Zschocke}, {Zucker}, {Zurbach}, \&
  {Zwitter}}]{gaia-collaboration_2018_aa}
{Gaia Collaboration}, {Brown}, A.~G.~A., {Vallenari}, A., {et~al.}
  2018{\natexlab{a}}, \aap, 616, A1, \dodoi{10.1051/0004-6361/201833051}

\bibitem[{{Gaia Collaboration} {et~al.}(2018{\natexlab{b}}){Gaia
  Collaboration}, {Babusiaux}, {van Leeuwen}, {Barstow}, {Jordi}, {Vallenari},
  {Bossini}, {Bressan}, {Cantat-Gaudin}, {van Leeuwen}, {Brown}, {Prusti}, {de
  Bruijne}, {Bailer-Jones}, {Biermann}, {Evans}, {Eyer}, {Jansen}, {Klioner},
  {Lammers}, {Lindegren}, {Luri}, {Mignard}, {Panem}, {Pourbaix}, {Randich},
  {Sartoretti}, {Siddiqui}, {Soubiran}, {Walton}, {Arenou}, {Bastian},
  {Cropper}, {Drimmel}, {Katz}, {Lattanzi}, {Bakker}, {Cacciari},
  {Casta{\~n}eda}, {Chaoul}, {Cheek}, {De Angeli}, {Fabricius}, {Guerra},
  {Holl}, {Masana}, {Messineo}, {Mowlavi}, {Nienartowicz}, {Panuzzo},
  {Portell}, {Riello}, {Seabroke}, {Tanga}, {Th{\'e}venin}, {Gracia-Abril},
  {Comoretto}, {Garcia-Reinaldos}, {Teyssier}, {Altmann}, {Andrae}, {Audard},
  {Bellas-Velidis}, {Benson}, {Berthier}, {Blomme}, {Burgess}, {Busso},
  {Carry}, {Cellino}, {Clementini}, {Clotet}, {Creevey}, {Davidson}, {De
  Ridder}, {Delchambre}, {Dell'Oro}, {Ducourant}, {Fern{\'a}ndez-
  Hern{\'a}ndez}, {Fouesneau}, {Fr{\'e}mat}, {Galluccio}, {Garc{\'\i}a-Torres},
  {Gonz{\'a}lez-N{\'u}{\~n}ez}, {Gonz{\'a}lez-Vidal}, {Gosset}, {Guy},
  {Halbwachs}, {Hambly}, {Harrison}, {Hern{\'a}ndez}, {Hestroffer}, {Hodgkin},
  {Hutton}, {Jasniewicz}, {Jean-Antoine-Piccolo}, {Jordan}, {Korn},
  {Krone-Martins}, {Lanzafame}, {Lebzelter}, {L{\"o}ffler}, {Manteiga},
  {Marrese}, {Mart{\'\i}n-Fleitas}, {Moitinho}, {Mora}, {Muinonen}, {Osinde},
  {Pancino}, {Pauwels}, {Petit}, {Recio-Blanco}, {Richards}, {Rimoldini},
  {Robin}, {Sarro}, {Siopis}, {Smith}, {Sozzetti}, {S{\"u}veges}, {Torra}, {van
  Reeven}, {Abbas}, {Abreu Aramburu}, {Accart}, {Aerts}, {Altavilla},
  {{\'A}lvarez}, {Alvarez}, {Alves}, {Anderson}, {Andrei}, {Anglada Varela},
  {Antiche}, {Antoja}, {Arcay}, {Astraatmadja}, {Bach}, {Baker},
  {Balaguer-N{\'u}{\~n}ez}, {Balm}, {Barache}, {Barata}, {Barbato}, {Barblan},
  {Barklem}, {Barrado}, {Barros}, {Bartholom{\'e} Mu{\~n}oz}, {Bassilana},
  {Becciani}, {Bellazzini}, {Berihuete}, {Bertone}, {Bianchi}, {Bienaym{\'e}},
  {Blanco-Cuaresma}, {Boch}, {Boeche}, {Bombrun}, {Borrachero}, {Bouquillon},
  {Bourda}, {Bragaglia}, {Bramante}, {Breddels}, {Brouillet},
  {Br{\"u}semeister}, {Brugaletta}, {Bucciarelli}, {Burlacu}, {Busonero},
  {Butkevich}, {Buzzi}, {Caffau}, {Cancelliere}, {Cannizzaro}, {Carballo},
  {Carlucci}, {Carrasco}, {Casamiquela}, {Castellani}, {Castro-Ginard},
  {Charlot}, {Chemin}, {Chiavassa}, {Cocozza}, {Costigan}, {Cowell}, {Crifo},
  {Crosta}, {Crowley}, {Cuypers}, {Dafonte}, {Damerdji}, {Dapergolas}, {David},
  {David}, {de Laverny}, {De Luise}, {De March}, {de Martino}, {de Souza}, {de
  Torres}, {Debosscher}, {del Pozo}, {Delbo}, {Delgado}, {Delgado}, {Diakite},
  {Diener}, {Distefano}, {Dolding}, {Drazinos}, {Dur{\'a}n}, {Edvardsson},
  {Enke}, {Eriksson}, {Esquej}, {Eynard Bontemps}, {Fabre}, {Fabrizio},
  {Faigler}, {Falc{\~a}o}, {Farr{\`a}s Casas}, {Federici}, {Fedorets},
  {Fernique}, {Figueras}, {Filippi}, {Findeisen}, {Fonti}, {Fraile}, {Fraser},
  {Fr{\'e}zouls}, {Gai}, {Galleti}, {Garabato}, {Garc{\'\i}a-Sedano},
  {Garofalo}, {Garralda}, {Gavel}, {Gavras}, {Gerssen}, {Geyer}, {Giacobbe},
  {Gilmore}, {Girona}, {Giuffrida}, {Glass}, {Gomes}, {Granvik}, {Gueguen},
  {Guerrier}, {Guiraud}, {Guti{\'e}}, {Haigron}, {Hatzidimitriou}, {Hauser},
  {Haywood}, {Heiter}, {Helmi}, {Heu}, {Hilger}, {Hobbs}, {Hofmann}, {Holland},
  {Huckle}, {Hypki}, {Icardi}, {Jan{\ss}en}, {Jevardat de Fombelle}, {Jonker},
  {Juh{\'a}sz}, {Julbe}, {Karampelas}, {Kewley}, {Klar}, {Kochoska}, {Kohley},
  {Kolenberg}, {Kontizas}, {Kontizas}, {Koposov}, {Kordopatis},
  {Kostrzewa-Rutkowska}, {Koubsky}, {Lambert}, {Lanza}, {Lasne}, {Lavigne}, {Le
  Fustec}, {Le Poncin-Lafitte}, {Lebreton}, {Leccia}, {Leclerc},
  {Lecoeur-Taibi}, {Lenhardt}, {Leroux}, {Liao}, {Licata}, {Lindstr{\o}m},
  {Lister}, {Livanou}, {Lobel}, {L{\'o}pez}, {Managau}, {Mann}, {Mantelet},
  {Marchal}, {Marchant}, {Marconi}, {Marinoni}, {Marschalk{\'o}}, {Marshall},
  {Martino}, {Marton}, {Mary}, {Massari}, {Matijevi{\v{c}}}, {Mazeh},
  {McMillan}, {Messina}, {Michalik}, {Millar}, {Molina}, {Molinaro},
  {Moln{\'a}r}, {Montegriffo}, {Mor}, {Morbidelli}, {Morel}, {Morris},
  {Mulone}, {Muraveva}, {Musella}, {Nelemans}, {Nicastro}, {Noval},
  {O'Mullane}, {Ord{\'e}novic}, {Ord{\'o}{\~n}ez-Blanco}, {Osborne}, {Pagani},
  {Pagano}, {Pailler}, {Palacin}, {Palaversa}, {Panahi}, {Pawlak},
  {Piersimoni}, {Pineau}, {Plachy}, {Plum}, {Poggio}, {Poujoulet},
  {Pr{\v{s}}a}, {Pulone}, {Racero}, {Ragaini}, {Rambaux}, {Ramos-Lerate},
  {Regibo}, {Reyl{\'e}}, {Riclet}, {Ripepi}, {Riva}, {Rivard}, {Rixon},
  {Roegiers}, {Roelens}, {Romero-G{\'o}mez}, {Rowell}, {Royer}, {Ruiz-Dern},
  {Sadowski}, {Sagrist{\`a} Sell{\'e}s}, {Sahlmann}, {Salgado}, {Salguero},
  {Sanna}, {Santana- Ros}, {Sarasso}, {Savietto}, {Schultheis}, {Sciacca},
  {Segol}, {Segovia}, {S{\'e}gransan}, {Shih}, {Siltala}, {Silva}, {Smart},
  {Smith}, {Solano}, {Solitro}, {Sordo}, {Soria Nieto}, {Souchay}, {Spagna},
  {Spoto}, {Stampa}, {Steele}, {Steidelm{\"u}ller}, {Stephenson}, {Stoev},
  {Suess}, {Surdej}, {Szabados}, {Szegedi-Elek}, {Tapiador}, {Taris}, {Tauran},
  {Taylor}, {Teixeira}, {Terrett}, {Teyssandier}, {Thuillot}, {Titarenko},
  {Torra Clotet}, {Turon}, {Ulla}, {Utrilla}, {Uzzi}, {Vaillant}, {Valentini},
  {Valette}, {van Elteren}, {Van Hemelryck}, {Vaschetto}, {Vecchiato},
  {Veljanoski}, {Viala}, {Vicente}, {Vogt}, {von Essen}, {Voss}, {Votruba},
  {Voutsinas}, {Walmsley}, {Weiler}, {Wertz}, {Wevers}, {Wyrzykowski},
  {Yoldas}, {{\v{Z}}erjal}, {Ziaeepour}, {Zorec}, {Zschocke}, {Zucker},
  {Zurbach}, \& {Zwitter}}]{gaia-collaboration_2018_ab}
{Gaia Collaboration}, {Babusiaux}, C., {van Leeuwen}, F., {et~al.}
  2018{\natexlab{b}}, \aap, 616, A10, \dodoi{10.1051/0004-6361/201832843}

\bibitem[{{Gaia Collaboration} {et~al.}(2018{\natexlab{c}}){Gaia
  Collaboration}, {Katz}, {Antoja}, {Romero-G{\'o}mez}, {Drimmel}, {Reyl{\'e}},
  {Seabroke}, {Soubiran}, {Babusiaux}, {Di Matteo}, {Figueras}, {Poggio},
  {Robin}, {Evans}, {Brown}, {Vallenari}, {Prusti}, {de Bruijne}, {Bailer-
  Jones}, {Biermann}, {Eyer}, {Jansen}, {Jordi}, {Klioner}, {Lammers},
  {Lindegren}, {Luri}, {Mignard}, {Panem}, {Pourbaix}, {Randich}, {Sartoretti},
  {Siddiqui}, {van Leeuwen}, {Walton}, {Arenou}, {Bastian}, {Cropper},
  {Lattanzi}, {Bakker}, {Cacciari}, {Casta n}, {Chaoul}, {Cheek}, {De Angeli},
  {Fabricius}, {Guerra}, {Holl}, {Masana}, {Messineo}, {Mowlavi},
  {Nienartowicz}, {Panuzzo}, {Portell}, {Riello}, {Tanga}, {Th{\'e}venin},
  {Gracia-Abril}, {Comoretto}, {Garcia-Reinaldos}, {Teyssier}, {Altmann},
  {Andrae}, {Audard}, {Bellas-Velidis}, {Benson}, {Berthier}, {Blomme},
  {Burgess}, {Busso}, {Carry}, {Cellino}, {Clementini}, {Clotet}, {Creevey},
  {Davidson}, {De Ridder}, {Delchambre}, {Dell'Oro}, {Ducourant},
  {Fern{\'a}ndez- Hern{\'a}ndez}, {Fouesneau}, {Fr{\'e}mat}, {Galluccio},
  {Garc{\'\i}a-Torres}, {Gonz{\'a}lez-N{\'u}{\~n}ez}, {Gonz{\'a}lez-Vidal},
  {Gosset}, {Guy}, {Halbwachs}, {Hambly}, {Harrison}, {Hern{\'a}ndez},
  {Hestroffer}, {Hodgkin}, {Hutton}, {Jasniewicz}, {Jean-Antoine-Piccolo},
  {Jordan}, {Korn}, {Krone-Martins}, {Lanzafame}, {Lebzelter}, {L{\"o}ffler},
  {Manteiga}, {Marrese}, {Mart{\'\i}n-Fleitas}, {Moitinho}, {Mora}, {Muinonen},
  {Osinde}, {Pancino}, {Pauwels}, {Petit}, {Recio-Blanco}, {Richards},
  {Rimoldini}, {Sarro}, {Siopis}, {Smith}, {Sozzetti}, {S{\"u}veges}, {Torra},
  {van Reeven}, {Abbas}, {Abreu Aramburu}, {Accart}, {Aerts}, {Altavilla},
  {{\'A}lvarez}, {Alvarez}, {Alves}, {Anderson}, {Andrei}, {Anglada Varela},
  {Antiche}, {Arcay}, {Astraatmadja}, {Bach}, {Baker},
  {Balaguer-N{\'u}{\~n}ez}, {Balm}, {Barache}, {Barata}, {Barbato}, {Barblan},
  {Barklem}, {Barrado}, {Barros}, {Barstow}, {Bartholom{\'e} Mu{\~n}oz},
  {Bassilana}, {Becciani}, {Bellazzini}, {Berihuete}, {Bertone}, {Bianchi},
  {Bienaym{\'e}}, {Blanco-Cuaresma}, {Boch}, {Boeche}, {Bombrun}, {Borrachero},
  {Bossini}, {Bouquillon}, {Bourda}, {Bragaglia}, {Bramante}, {Breddels},
  {Bressan}, {Brouillet}, {Br{\"u}semeister}, {Brugaletta}, {Bucciarelli},
  {Burlacu}, {Busonero}, {Butkevich}, {Buzzi}, {Caffau}, {Cancelliere},
  {Cannizzaro}, {Cantat-Gaudin}, {Carballo}, {Carlucci}, {Carrasco},
  {Casamiquela}, {Castellani}, {Castro-Ginard}, {Charlot}, {Chemin},
  {Chiavassa}, {Cocozza}, {Costigan}, {Cowell}, {Crifo}, {Crosta}, {Crowley},
  {Cuypers}, {Dafonte}, {Damerdji}, {Dapergolas}, {David}, {David}, {de
  Laverny}, {De Luise}, {De March}, {de Souza}, {de Torres}, {Debosscher}, {del
  Pozo}, {Delbo}, {Delgado}, {Delgado}, {Diakite}, {Diener}, {Distefano},
  {Dolding}, {Drazinos}, {Dur{\'a}n}, {Edvardsson}, {Enke}, {Eriksson},
  {Esquej}, {Eynard Bontemps}, {Fabre}, {Fabrizio}, {Faigler}, {Falc a},
  {Farr{\`a}s Casas}, {Federici}, {Fedorets}, {Fernique}, {Filippi},
  {Findeisen}, {Fonti}, {Fraile}, {Fraser}, {Fr{\'e}zouls}, {Gai}, {Galleti},
  {Garabato}, {Garc{\'\i}a-Sedano}, {Garofalo}, {Garralda}, {Gavel}, {Gavras},
  {Gerssen}, {Geyer}, {Giacobbe}, {Gilmore}, {Girona}, {Giuffrida}, {Glass},
  {Gomes}, {Granvik}, {Gueguen}, {Guerrier}, {Guiraud}, {Guti{\'e}}, {Haigron},
  {Hatzidimitriou}, {Hauser}, {Haywood}, {Heiter}, {Helmi}, {Heu}, {Hilger},
  {Hobbs}, {Hofmann}, {Holland}, {Huckle}, {Hypki}, {Icardi}, {Jan{\ss}en},
  {Jevardat de Fombelle}, {Jonker}, {Juh{\'a}sz}, {Julbe}, {Karampelas},
  {Kewley}, {Klar}, {Kochoska}, {Kohley}, {Kolenberg}, {Kontizas}, {Kontizas},
  {Koposov}, {Kordopatis}, {Kostrzewa-Rutkowska}, {Koubsky}, {Lambert},
  {Lanza}, {Lasne}, {Lavigne}, {Le Fustec}, {Le Poncin-Lafitte}, {Lebreton},
  {Leccia}, {Leclerc}, {Lecoeur-Taibi}, {Lenhardt}, {Leroux}, {Liao}, {Licata},
  {Lindstr{\o}m}, {Lister}, {Livanou}, {Lobel}, {L{\'o}pez}, {Managau}, {Mann},
  {Mantelet}, {Marchal}, {Marchant}, {Marconi}, {Marinoni}, {Marschalk{\'o}},
  {Marshall}, {Martino}, {Marton}, {Mary}, {Massari}, {Matijevi{\v{c}}},
  {Mazeh}, {McMillan}, {Messina}, {Michalik}, {Millar}, {Molina}, {Molinaro},
  {Moln{\'a}r}, {Montegriffo}, {Mor}, {Morbidelli}, {Morel}, {Morris},
  {Mulone}, {Muraveva}, {Musella}, {Nelemans}, {Nicastro}, {Noval},
  {O'Mullane}, {Ord{\'e}novic}, {Ord{\'o}{\~n}ez-Blanco}, {Osborne}, {Pagani},
  {Pagano}, {Pailler}, {Palacin}, {Palaversa}, {Panahi}, {Pawlak},
  {Piersimoni}, {Pineau}, {Plachy}, {Plum}, {Poujoulet}, {Pr{\v{s}}a},
  {Pulone}, {Racero}, {Ragaini}, {Rambaux}, {Ramos-Lerate}, {Regibo}, {Riclet},
  {Ripepi}, {Riva}, {Rivard}, {Rixon}, {Roegiers}, {Roelens}, {Rowell},
  {Royer}, {Ruiz-Dern}, {Sadowski}, {Sagrist{\`a} Sell{\'e}s}, {Sahlmann},
  {Salgado}, {Salguero}, {Sanna}, {Santana-Ros}, {Sarasso}, {Savietto},
  {Schultheis}, {Sciacca}, {Segol}, {Segovia}, {S{\'e}gransan}, {Shih},
  {Siltala}, {Silva}, {Smart}, {Smith}, {Solano}, {Solitro}, {Sordo}, {Soria
  Nieto}, {Souchay}, {Spagna}, {Spoto}, {Stampa}, {Steele},
  {Steidelm{\"u}ller}, {Stephenson}, {Stoev}, {Suess}, {Surdej}, {Szabados},
  {Szegedi-Elek}, {Tapiador}, {Taris}, {Tauran}, {Taylor}, {Teixeira},
  {Terrett}, {Teyssandier}, {Thuillot}, {Titarenko}, {Torra Clotet}, {Turon},
  {Ulla}, {Utrilla}, {Uzzi}, {Vaillant}, {Valentini}, {Valette}, {van Elteren},
  {Van Hemelryck}, {van Leeuwen}, {Vaschetto}, {Vecchiato}, {Veljanoski},
  {Viala}, {Vicente}, {Vogt}, {von Essen}, {Voss}, {Votruba}, {Voutsinas},
  {Walmsley}, {Weiler}, {Wertz}, {Wevers}, {Wyrzykowski}, {Yoldas},
  {{\v{Z}}erjal}, {Ziaeepour}, {Zorec}, {Zschocke}, {Zucker}, {Zurbach}, \&
  {Zwitter}}]{gaia-collaboration_2018_ac}
{Gaia Collaboration}, {Katz}, D., {Antoja}, T., {et~al.} 2018{\natexlab{c}},
  \aap, 616, A11, \dodoi{10.1051/0004-6361/201832865}

\bibitem[{{Gaidos} {et~al.}(2017){Gaidos}, {Kitzmann}, \&
  {Heng}}]{gaidos_2017_aa}
{Gaidos}, E., {Kitzmann}, D., \& {Heng}, K. 2017, \mnras, 468, 3418,
  \dodoi{10.1093/mnras/stx615}

\bibitem[{{Gardner} {et~al.}(2006){Gardner}, {Mather}, {Clampin}, {Doyon},
  {Greenhouse}, {Hammel}, {Hutchings}, {Jakobsen}, {Lilly}, {Long}, {Lunine},
  {McCaughrean}, {Mountain}, {Nella}, {Rieke}, {Rieke}, {Rix}, {Smith},
  {Sonneborn}, {Stiavelli}, {Stockman}, {Windhorst}, \&
  {Wright}}]{gardner_2006_aa}
{Gardner}, J.~P., {Mather}, J.~C., {Clampin}, M., {et~al.} 2006, \ssr, 123,
  485, \dodoi{10.1007/s11214-006-8315-7}

\bibitem[{{Gautschy} \& {Saio}(1995)}]{gautschy_1995_aa}
{Gautschy}, A., \& {Saio}, H. 1995, Annual Review of Astronomy and
  Astrophysics, 33, 75, \dodoi{10.1146/annurev.aa.33.090195.000451}

\bibitem[{{Gautschy} \& {Saio}(1996)}]{gautschy_1996_aa}
---. 1996, Annual Review of Astronomy and Astrophysics, 34, 551,
  \dodoi{10.1146/annurev.astro.34.1.551}

\bibitem[{{Georgy} {et~al.}(2014){Georgy}, {Granada}, {Ekstr{\"o}m}, {Meynet},
  {Anderson}, {Wyttenbach}, {Eggenberger}, \& {Maeder}}]{georgy2014}
{Georgy}, C., {Granada}, A., {Ekstr{\"o}m}, S., {et~al.} 2014, \aap, 566, A21,
  \dodoi{10.1051/0004-6361/201423881}

\bibitem[{{Goldberg} {et~al.}(2019){Goldberg}, {Bildsten}, \&
  {Paxton}}]{Goldberg2019}
{Goldberg}, J.~A., {Bildsten}, L., \& {Paxton}, B. 2019, arXiv e-prints,
  arXiv:1903.09114.
\newblock \doarXiv{1903.09114}

\bibitem[{{Gombosi} {et~al.}(2018){Gombosi}, {van der Holst}, {Manchester}, \&
  {Sokolov}}]{gombosi_2018_aa}
{Gombosi}, T.~I., {van der Holst}, B., {Manchester}, W.~B., \& {Sokolov}, I.~V.
  2018, Living Reviews in Solar Physics, 15, 4,
  \dodoi{10.1007/s41116-018-0014-4}

\bibitem[{{Graboske} {et~al.}(1973){Graboske}, {Dewitt}, {Grossman}, \&
  {Cooper}}]{graboske:73}
{Graboske}, H.~C., {Dewitt}, H.~E., {Grossman}, A.~S., \& {Cooper}, M.~S. 1973,
  \apj, 181, 457, \dodoi{10.1086/152062}

\bibitem[{{Gupta} \& {Meyer}(2001)}]{gupta_2001_aa}
{Gupta}, S.~S., \& {Meyer}, B.~S. 2001, \prc, 64, 025805,
  \dodoi{10.1103/PhysRevC.64.025805}

\bibitem[{{Hertzsprung}(1926{\natexlab{a}})}]{hertzsprung_1926_aa}
{Hertzsprung}, E. 1926{\natexlab{a}}, Bulletin of the Astronomical Institutes
  of the Netherlands, 3, 115

\bibitem[{{Hertzsprung}(1926{\natexlab{b}})}]{rsp_hbp}
---. 1926{\natexlab{b}}, \bain, 3, 115

\bibitem[{{Hix} \& {Meyer}(2006)}]{Hix2006}
{Hix}, W.~R., \& {Meyer}, B.~S. 2006, Nuclear Physics A, 777, 188,
  \dodoi{10.1016/j.nuclphysa.2004.10.009}

\bibitem[{{Howell} {et~al.}(2014){Howell}, {Sobeck}, {Haas}, {Still},
  {Barclay}, {Mullally}, {Troeltzsch}, {Aigrain}, {Bryson}, {Caldwell},
  {Chaplin}, {Cochran}, {Huber}, {Marcy}, {Miglio}, {Najita}, {Smith},
  {Twicken}, \& {Fortney}}]{howell_2014_aa}
{Howell}, S.~B., {Sobeck}, C., {Haas}, M., {et~al.} 2014, Publications of the
  Astronomical Society of the Pacific, 126, 398, \dodoi{10.1086/676406}

\bibitem[{{Hu} {et~al.}(2011){Hu}, {Tout}, {Glebbeek}, \& {Dupret}}]{Hu2011}
{Hu}, H., {Tout}, C.~A., {Glebbeek}, E., \& {Dupret}, M.-A. 2011, \mnras, 418,
  195, \dodoi{10.1111/j.1365-2966.2011.19482.x}

\bibitem[{{Huang} {et~al.}(2018){Huang}, {Burt}, {Vanderburg}, {G{\"u}nther},
  {Shporer}, {Dittmann}, {Winn}, {Wittenmyer}, {Sha}, {Kane}, {Ricker},
  {Vanderspek}, {Latham}, {Seager}, {Jenkins}, {Caldwell}, {Collins},
  {Guerrero}, {Smith}, {Quinn}, {Udry}, {Pepe}, {Bouchy}, {S{\'e}gransan},
  {Lovis}, {Ehrenreich}, {Marmier}, {Mayor}, {Wohler}, {Haworth}, {Morgan},
  {Fausnaugh}, {Ciardi}, {Christiansen}, {Charbonneau}, {Dragomir}, {Deming},
  {Glidden}, {Levine}, {McCullough}, {Yu}, {Narita}, {Nguyen}, {Morton},
  {Pepper}, {P{\'a}l}, {Rodriguez}, {Stassun}, {Torres}, {Sozzetti}, {Doty},
  {Christensen-Dalsgaard}, {Laughlin}, {Clampin}, {Bean}, {Buchhave}, {Bakos},
  {Sato}, {Ida}, {Kaltenegger}, {Palle}, {Sasselov}, {Butler}, {Lissauer},
  {Ge}, \& {Rinehart}}]{huang_2018_aa}
{Huang}, C.~X., {Burt}, J., {Vanderburg}, A., {et~al.} 2018, \apj, 868, L39,
  \dodoi{10.3847/2041-8213/aaef91}

\bibitem[{Hunter(2007)}]{hunter_2007_aa}
Hunter, J.~D. 2007, Computing In Science \&amp; Engineering, 9, 90

\bibitem[{{Itoh} {et~al.}(1979){Itoh}, {Totsuji}, {Ichimaru}, \&
  {Dewitt}}]{itoh:79}
{Itoh}, N., {Totsuji}, H., {Ichimaru}, S., \& {Dewitt}, H.~E. 1979, \apj, 234,
  1079, \dodoi{10.1086/157590}

\bibitem[{{Jones} {et~al.}(2013){Jones}, {Hirschi}, {Nomoto}, {Fischer},
  {Timmes}, {Herwig}, {Paxton}, {Toki}, {Suzuki}, {Mart{\'{\i}}nez-Pinedo},
  {Lam}, \& {Bertolli}}]{Jones2013}
{Jones}, S., {Hirschi}, R., {Nomoto}, K., {et~al.} 2013, \apj, 772, 150,
  \dodoi{10.1088/0004-637X/772/2/150}

\bibitem[{{Kato}(1966)}]{kato_1966_aa}
{Kato}, S. 1966, \pasj, 18, 374

\bibitem[{{Keller} \& {Wood}(2006)}]{rsp_kw06}
{Keller}, S.~C., \& {Wood}, P.~R. 2006, \apj, 642, 834, \dodoi{10.1086/501115}

\bibitem[{{Kelly} {et~al.}(2018){Kelly}, {Diego}, {Rodney}, {Kaiser},
  {Broadhurst}, {Zitrin}, {Treu}, {P{\'e}rez-Gonz{\'a}lez}, {Morishita},
  {Jauzac}, {Selsing}, {Oguri}, {Pueyo}, {Ross}, {Filippenko}, {Smith},
  {Hjorth}, {Cenko}, {Wang}, {Howell}, {Richard}, {Frye}, {Jha}, {Foley},
  {Norman}, {Bradac}, {Zheng}, {Brammer}, {Benito}, {Cava}, {Christensen}, {de
  Mink}, {Graur}, {Grillo}, {Kawamata}, {Kneib}, {Matheson}, {McCully},
  {Nonino}, {P{\'e}rez-Fournon}, {Riess}, {Rosati}, {Schmidt}, {Sharon}, \&
  {Weiner}}]{kelly_2018_aa}
{Kelly}, P.~L., {Diego}, J.~M., {Rodney}, S., {et~al.} 2018, Nature Astronomy,
  2, 334, \dodoi{10.1038/s41550-018-0430-3}

\bibitem[{{Kippenhahn} \& {Thomas}(1970)}]{KippenhahnThomas1970}
{Kippenhahn}, R., \& {Thomas}, H.-C. 1970, in IAU Colloq. 4: Stellar Rotation,
  ed. A.~{Slettebak}, 20

\bibitem[{{Klein} {et~al.}(2014){Klein}, {Richards}, {Butler}, \&
  {Bloom}}]{klein_2014_aa}
{Klein}, C.~R., {Richards}, J.~W., {Butler}, N.~R., \& {Bloom}, J.~S. 2014,
  \mnras, 440, L96, \dodoi{10.1093/mnrasl/slu031}

\bibitem[{Kluyver {et~al.}(2016)Kluyver, Ragan-Kelley, P{\'e}rez, Granger,
  Bussonnier, Frederic, Kelley, Hamrick, Grout, Corlay,
  {et~al.}}]{kluyver_2016_aa}
Kluyver, T., Ragan-Kelley, B., P{\'e}rez, F., {et~al.} 2016, in Positioning and
  Power in Academic Publishing: Players, Agents and Agendas: Proceedings of the
  20th International Conference on Electronic Publishing, IOS Press, 87

\bibitem[{{Kollath} {et~al.}(1998){Kollath}, {Buchler}, {Serre}, \&
  {Mattei}}]{rsp_kbsm98}
{Kollath}, Z., {Buchler}, J.~R., {Serre}, T., \& {Mattei}, J. 1998, \aap, 329,
  147

\bibitem[{{Koll{\'a}th} {et~al.}(2002){Koll{\'a}th}, {Buchler}, {Szab{\'o}}, \&
  {Csubry}}]{rsp_kbsc02}
{Koll{\'a}th}, Z., {Buchler}, J.~R., {Szab{\'o}}, R., \& {Csubry}, Z. 2002,
  \aap, 385, 932, \dodoi{10.1051/0004-6361:20020182}

\bibitem[{{Kopal}(1959)}]{Kopal1959}
{Kopal}, Z. 1959, {Close binary systems} ({Springer US})

\bibitem[{{Kovacs} \& {Buchler}(1988)}]{rsp_kb88}
{Kovacs}, G., \& {Buchler}, J.~R. 1988, \apj, 334, 971, \dodoi{10.1086/166890}

\bibitem[{{Kuhfu{\ss}}(1986)}]{rsp_kuhfuss}
{Kuhfu{\ss}}, R. 1986, \aap, 160, 116

\bibitem[{{Leavitt}(1908)}]{leavitt_1908_aa}
{Leavitt}, H.~S. 1908, Annals of Harvard College Observatory, 60, 87

\bibitem[{{Ledoux}(1947)}]{ledoux_1947_aa}
{Ledoux}, P. 1947, \apj, 105, 305

\bibitem[{{Lindegren} {et~al.}(2018){Lindegren}, {Hern{\'a}ndez}, {Bombrun},
  {Klioner}, {Bastian}, {Ramos-Lerate}, {de Torres}, {Steidelm{\"u}ller},
  {Stephenson}, {Hobbs}, {Lammers}, {Biermann}, {Geyer}, {Hilger}, {Michalik},
  {Stampa}, {McMillan}, {Casta{\~n}eda}, {Clotet}, {Comoretto}, {Davidson},
  {Fabricius}, {Gracia}, {Hambly}, {Hutton}, {Mora}, {Portell}, {van Leeuwen},
  {Abbas}, {Abreu}, {Altmann}, {Andrei}, {Anglada}, {Balaguer-N{\'u}{\~n}ez},
  {Barache}, {Becciani}, {Bertone}, {Bianchi}, {Bouquillon}, {Bourda},
  {Br{\"u}semeister}, {Bucciarelli}, {Busonero}, {Buzzi}, {Cancelliere},
  {Carlucci}, {Charlot}, {Cheek}, {Crosta}, {Crowley}, {de Bruijne}, {de
  Felice}, {Drimmel}, {Esquej}, {Fienga}, {Fraile}, {Gai}, {Garralda},
  {Gonz{\'a}lez- Vidal}, {Guerra}, {Hauser}, {Hofmann}, {Holl}, {Jordan},
  {Lattanzi}, {Lenhardt}, {Liao}, {Licata}, {Lister}, {L{\"o}ffler},
  {Marchant}, {Martin-Fleitas}, {Messineo}, {Mignard}, {Morbidelli}, {Poggio},
  {Riva}, {Rowell}, {Salguero}, {Sarasso}, {Sciacca}, {Siddiqui}, {Smart},
  {Spagna}, {Steele}, {Taris}, {Torra}, {van Elteren}, {van Reeven}, \&
  {Vecchiato}}]{lindegren_2018_aa}
{Lindegren}, L., {Hern{\'a}ndez}, J., {Bombrun}, A., {et~al.} 2018, \aap, 616,
  A2, \dodoi{10.1051/0004-6361/201832727}

\bibitem[{{Luri} {et~al.}(2018){Luri}, {Brown}, {Sarro}, {Arenou},
  {Bailer-Jones}, {Castro-Ginard}, {de Bruijne}, {Prusti}, {Babusiaux}, \&
  {Delgado}}]{luri_2018_aa}
{Luri}, X., {Brown}, A.~G.~A., {Sarro}, L.~M., {et~al.} 2018, \aap, 616, A9,
  \dodoi{10.1051/0004-6361/201832964}

\bibitem[{{Maeder}(2009)}]{Maeder2009}
{Maeder}, A. 2009, {Physics, Formation and Evolution of Rotating Stars}
  ({Springer Berlin Heidelberg}), \dodoi{10.1007/978-3-540-76949-1}

\bibitem[{{Maeder} \& {Meynet}(2000)}]{maeder_2000_aa}
{Maeder}, A., \& {Meynet}, G. 2000, Annual Review of Astronomy and
  Astrophysics, 38, 143, \dodoi{10.1146/annurev.astro.38.1.143}

\bibitem[{{Majaess} {et~al.}(2009){Majaess}, {Turner}, \&
  {Lane}}]{majaess_2009_aa}
{Majaess}, D.~J., {Turner}, D.~G., \& {Lane}, D.~J. 2009, \mnras, 398, 263,
  \dodoi{10.1111/j.1365-2966.2009.15096.x}

\bibitem[{{Marconi}(2017)}]{rsp_MarcellaRev}
{Marconi}, M. 2017, in European Physical Journal Web of Conferences, Vol. 152,
  European Physical Journal Web of Conferences, 06001

\bibitem[{{Marconi} {et~al.}(2013){Marconi}, {Molinaro}, {Bono},
  {Pietrzy{\'n}ski}, {Gieren}, {Pilecki}, {Stellingwerf}, {Graczyk}, {Smolec},
  {Konorski}, {Suchomska}, {G{\'o}rski}, \& {Karczmarek}}]{rsp_Marconi13}
{Marconi}, M., {Molinaro}, R., {Bono}, G., {et~al.} 2013, \apjl, 768, L6,
  \dodoi{10.1088/2041-8205/768/1/L6}

\bibitem[{{Marconi} {et~al.}(2015){Marconi}, {Coppola}, {Bono}, {Braga},
  {Pietrinferni}, {Buonanno}, {Castellani}, {Musella}, {Ripepi}, \&
  {Stellingwerf}}]{rsp_Marconi15}
{Marconi}, M., {Coppola}, G., {Bono}, G., {et~al.} 2015, \apj, 808, 50,
  \dodoi{10.1088/0004-637X/808/1/50}

\bibitem[{{McComas} {et~al.}(2018){McComas}, {Christian}, {Schwadron}, {Fox},
  {Westlake}, {Allegrini}, {Baker}, {Biesecker}, {Bzowski}, {Clark}, {Cohen},
  {Cohen}, {Dayeh}, {Decker}, {de Nolfo}, {Desai}, {Ebert}, {Elliott}, {Fahr},
  {Frisch}, {Funsten}, {Fuselier}, {Galli}, {Galvin}, {Giacalone},
  {Gkioulidou}, {Guo}, {Horanyi}, {Isenberg}, {Janzen}, {Kistler}, {Korreck},
  {Kubiak}, {Kucharek}, {Larsen}, {Leske}, {Lugaz}, {Luhmann}, {Matthaeus},
  {Mitchell}, {Moebius}, {Ogasawara}, {Reisenfeld}, {Richardson}, {Russell},
  {Sok{\'o}{\l}}, {Spence}, {Skoug}, {Sternovsky}, {Swaczyna}, {Szalay},
  {Tokumaru}, {Wiedenbeck}, {Wurz}, {Zank}, \& {Zirnstein}}]{mccomas_2018_aa}
{McComas}, D.~J., {Christian}, E.~R., {Schwadron}, N.~A., {et~al.} 2018, \ssr,
  214, 116, \dodoi{10.1007/s11214-018-0550-1}

\bibitem[{{Meynet} \& {Maeder}(1997)}]{MeynetMaeder1997}
{Meynet}, G., \& {Maeder}, A. 1997, \aap, 321, 465

\bibitem[{{Miglio} {et~al.}(2008){Miglio}, {Montalb{\'a}n}, {Noels}, \&
  {Eggenberger}}]{miglio_2008_aa}
{Miglio}, A., {Montalb{\'a}n}, J., {Noels}, A., \& {Eggenberger}, P. 2008,
  \mnras, 386, 1487

\bibitem[{{Miyaji} \& {Nomoto}(1987)}]{Miyaji1987}
{Miyaji}, S., \& {Nomoto}, K. 1987, \apj, 318, 307, \dodoi{10.1086/165368}

\bibitem[{{Miyaji} {et~al.}(1980){Miyaji}, {Nomoto}, {Yokoi}, \&
  {Sugimoto}}]{Miyaji1980}
{Miyaji}, S., {Nomoto}, K., {Yokoi}, K., \& {Sugimoto}, D. 1980, \pasj, 32, 303

\bibitem[{{Moore} \& {Garaud}(2016)}]{moore_2016_aa}
{Moore}, K., \& {Garaud}, P. 2016, \apj, 817, 54

\bibitem[{{Morozova} {et~al.}(2015){Morozova}, {Piro}, {Renzo}, {Ott},
  {Clausen}, {Couch}, {Ellis}, \& {Roberts}}]{Morozova2015}
{Morozova}, V., {Piro}, A.~L., {Renzo}, M., {et~al.} 2015, \apj, 814, 63,
  \dodoi{10.1088/0004-637X/814/1/63}

\bibitem[{{Moskalik} \& {Buchler}(1990)}]{rsp_mb90}
{Moskalik}, P., \& {Buchler}, J.~R. 1990, \apj, 355, 590,
  \dodoi{10.1086/168792}

\bibitem[{{Mowlavi} \& {Forestini}(1994)}]{mowlavi_1994_aa}
{Mowlavi}, N., \& {Forestini}, M. 1994, \aap, 282, 843

\bibitem[{{Muraveva} {et~al.}(2018{\natexlab{a}}){Muraveva}, {Delgado},
  {Clementini}, {Sarro}, \& {Garofalo}}]{muraveva_2018_aa}
{Muraveva}, T., {Delgado}, H.~E., {Clementini}, G., {Sarro}, L.~M., \&
  {Garofalo}, A. 2018{\natexlab{a}}, \mnras, 481, 1195,
  \dodoi{10.1093/mnras/sty2241}

\bibitem[{{Muraveva} {et~al.}(2018{\natexlab{b}}){Muraveva}, {Garofalo},
  {Scowcroft}, {Clementini}, {Freedman}, {Madore}, \&
  {Monson}}]{muraveva_2018_ab}
{Muraveva}, T., {Garofalo}, A., {Scowcroft}, V., {et~al.} 2018{\natexlab{b}},
  \mnras, 480, 4138, \dodoi{10.1093/mnras/sty1959}

\bibitem[{{Ness} {et~al.}(2015){Ness}, {Hogg}, {Rix}, {Ho}, \&
  {Zasowski}}]{ness_2015_aa}
{Ness}, M., {Hogg}, D.~W., {Rix}, H.-W., {Ho}, A.~Y.~Q., \& {Zasowski}, G.
  2015, \apj, 808, 16, \dodoi{10.1088/0004-637X/808/1/16}

\bibitem[{{Noels} {et~al.}(2010){Noels}, {Montalban}, {Miglio}, {Godart}, \&
  {Ventura}}]{noels_2010_aa}
{Noels}, A., {Montalban}, J., {Miglio}, A., {Godart}, M., \& {Ventura}, P.
  2010, \apss, 328, 227

\bibitem[{{Papics}(2013)}]{papics_2013_aa}
{Papics}, P.~I. 2013, PhD thesis, Instituut voor Sterrenkunde, KU Leuven,
  Celestijnenlaan 200D, B-3001 Leuven, Belgium

\bibitem[{{Parker}(1958{\natexlab{a}})}]{parker_1958_aa}
{Parker}, E.~N. 1958{\natexlab{a}}, \apj, 128, 664, \dodoi{10.1086/146579}

\bibitem[{{Parker}(1958{\natexlab{b}})}]{parker_1958_ab}
---. 1958{\natexlab{b}}, Physics of Fluids, 1, 171, \dodoi{10.1063/1.1724339}

\bibitem[{{Paxton}(2004)}]{paxton_2004_aa}
{Paxton}, B. 2004, \pasp, 116, 699, \dodoi{10.1086/422345}

\bibitem[{{Paxton} {et~al.}(2011){Paxton}, {Bildsten}, {Dotter}, {Herwig},
  {Lesaffre}, \& {Timmes}}]{paxton_2011_aa}
{Paxton}, B., {Bildsten}, L., {Dotter}, A., {et~al.} 2011, \apjs, 192, 3,
  \dodoi{10.1088/0067-0049/192/1/3}

\bibitem[{{Paxton} {et~al.}(2013){Paxton}, {Cantiello}, {Arras}, {Bildsten},
  {Brown}, {Dotter}, {Mankovich}, {Montgomery}, {Stello}, {Timmes}, \&
  {Townsend}}]{paxton_2013_aa}
{Paxton}, B., {Cantiello}, M., {Arras}, P., {et~al.} 2013, \apjs, 208

\bibitem[{{Paxton} {et~al.}(2015){Paxton}, {Marchant}, {Schwab}, {Bauer},
  {Bildsten}, {Cantiello}, {Dessart}, {Farmer}, {Hu}, {Langer}, {Townsend},
  {Townsley}, \& {Timmes}}]{paxton_2015_aa}
{Paxton}, B., {Marchant}, P., {Schwab}, J., {et~al.} 2015, \apjs, 220, 15,
  \dodoi{10.1088/0067-0049/220/1/15}

\bibitem[{{Paxton} {et~al.}(2018){Paxton}, {Schwab}, {Bauer}, {Bildsten},
  {Blinnikov}, {Duffell}, {Farmer}, {Goldberg}, {Marchant}, {Sorokina},
  {Thoul}, {Townsend}, \& {Timmes}}]{paxton_2018_aa}
{Paxton}, B., {Schwab}, J., {Bauer}, E.~B., {et~al.} 2018, \apjs, 234, 34,
  \dodoi{10.3847/1538-4365/aaa5a8}

\bibitem[{P{\'e}rez \& Granger(2007)}]{perez_2007_aa}
P{\'e}rez, F., \& Granger, B.~E. 2007, Computing in Science \& Engineering, 9,
  21

\bibitem[{{Pietrukowicz} {et~al.}(2013){Pietrukowicz}, {Dziembowski},
  {Mr{\'o}z}, {Soszy{\'n}ski}, {Udalski}, {Poleski}, {Szyma{\'n}ski}, {Kubiak},
  {Pietrzy{\'n}ski}, {Wyrzykowski}, {Ulaczyk}, {Koz{\l}owski}, \&
  {Skowron}}]{pietrukowicz_2013_aa}
{Pietrukowicz}, P., {Dziembowski}, W.~A., {Mr{\'o}z}, P., {et~al.} 2013,
  \actaa, 63, 379.
\newblock \doarXiv{1311.5894}

\bibitem[{{Pietrukowicz} {et~al.}(2017){Pietrukowicz}, {Dziembowski}, {Latour},
  {Angeloni}, {Poleski}, {di Mille}, {Soszy{\'n}ski}, {Udalski},
  {Szyma{\'n}ski}, {Wyrzykowski}, {Koz{\l}owski}, {Skowron}, {Skowron},
  {Mr{\'o}z}, {Pawlak}, \& {Ulaczyk}}]{pietrukowicz_2017_aa}
{Pietrukowicz}, P., {Dziembowski}, W.~A., {Latour}, M., {et~al.} 2017, Nature
  Astronomy, 1, 0166, \dodoi{10.1038/s41550-017-0166}

\bibitem[{Pietrzynski {et~al.}(2012)Pietrzynski, Thompson, Gieren, Graczyk,
  Stepien, Bono, Moroni, Pilecki, Udalski, Soszynski, Preston, Nardetto,
  McWilliam, Roederer, Gorski, Konorski, \& Storm}]{pietrzynski_bep2012}
Pietrzynski, G., Thompson, I.~B., Gieren, W., {et~al.} 2012, Nature, 484, 75,
  \dodoi{10.1038/nature10966}

\bibitem[{{Pilecki} {et~al.}(2018){Pilecki}, {Gieren}, {Pietrzy{\'n}ski},
  {Thompson}, {Smolec}, {Graczyk}, {Taormina}, {Udalski}, {Storm}, {Nardetto},
  {Gallenne}, {Kervella}, {Soszy{\'n}ski}, {G{\'o}rski}, {Wielg{\'o}rski},
  {Suchomska}, {Karczmarek}, \& {Zgirski}}]{rsp_Pilecki2018}
{Pilecki}, B., {Gieren}, W., {Pietrzy{\'n}ski}, G., {et~al.} 2018, \apj, 862,
  43, \dodoi{10.3847/1538-4357/aacb32}

\bibitem[{{Plachy} {et~al.}(2018){Plachy}, {B{\'o}di}, \&
  {Koll{\'a}th}}]{rsp_dfcyg}
{Plachy}, E., {B{\'o}di}, A., \& {Koll{\'a}th}, Z. 2018, \mnras, 481, 2986,
  \dodoi{10.1093/mnras/sty2511}

\bibitem[{{Plachy} {et~al.}(2013){Plachy}, {Koll{\'a}th}, \&
  {Moln{\'a}r}}]{rsp_plachy13}
{Plachy}, E., {Koll{\'a}th}, Z., \& {Moln{\'a}r}, L. 2013, \mnras, 433, 3590,
  \dodoi{10.1093/mnras/stt1000}

\bibitem[{{Pols} {et~al.}(1995){Pols}, {Tout}, {Eggleton}, \&
  {Han}}]{Pols_1995_aa}
{Pols}, O.~R., {Tout}, C.~A., {Eggleton}, P.~P., \& {Han}, Z. 1995, \mnras,
  274, 964, \dodoi{10.1093/mnras/274.3.964}

\bibitem[{{Potekhin} \& {Chabrier}(2010)}]{potekhin_2010_aa}
{Potekhin}, A.~Y., \& {Chabrier}, G. 2010, Contributions to Plasma Physics, 50,
  82, \dodoi{10.1002/ctpp.201010017}

\bibitem[{{Press} {et~al.}(1992){Press}, {Teukolsky}, {Vetterling}, \&
  {Flannery}}]{press_1992_aa}
{Press}, W.~H., {Teukolsky}, S.~A., {Vetterling}, W.~T., \& {Flannery}, B.~P.
  1992, {Numerical recipes in FORTRAN. The art of scientific computing}, Vol.
  2nd ed. ({Cambridge: University Press})

\bibitem[{{Reiter} {et~al.}(1995){Reiter}, {Walsh}, \&
  {Weiss}}]{reiter_1995_aa}
{Reiter}, J., {Walsh}, L., \& {Weiss}, A. 1995, \mnras, 274, 899,
  \dodoi{10.1093/mnras/274.3.899}

\bibitem[{Richtmyer(1957)}]{rsp_richtmyer}
Richtmyer, R. 1957, Difference Methods for Initial Value Problems (New York:
  Interscience)

\bibitem[{{Ricker} {et~al.}(2016){Ricker}, {Vanderspek}, {Winn}, {Seager},
  {Berta-Thompson}, {Levine}, {Villasenor}, {Latham}, {Charbonneau}, {Holman},
  {Johnson}, {Sasselov}, {Szentgyorgyi}, {Torres}, {Bakos}, {Brown},
  {Christensen-Dalsgaard}, {Kjeldsen}, {Clampin}, {Rinehart}, {Deming}, {Doty},
  {Dunham}, {Ida}, {Kawai}, {Sato}, {Jenkins}, {Lissauer}, {Jernigan},
  {Kaltenegger}, {Laughlin}, {Lin}, {McCullough}, {Narita}, {Pepper},
  {Stassun}, \& {Udry}}]{ricker_2016_aa}
{Ricker}, G.~R., {Vanderspek}, R., {Winn}, J., {et~al.} 2016, in \procspie,
  Vol. 9904, Space Telescopes and Instrumentation 2016: Optical, Infrared, and
  Millimeter Wave, 99042B

\bibitem[{Ridders(1982)}]{ridders_1982_aa}
Ridders, C. 1982, Advances in Engineering Software (1978), 4, 75 ,
  \dodoi{https://doi.org/10.1016/S0141-1195(82)80057-0}

\bibitem[{{Rieke} {et~al.}(2015){Rieke}, {Wright}, {B{\"o}ker}, {Bouwman},
  {Colina}, {Glasse}, {Gordon}, {Greene}, {G{\"u}del}, {Henning}, {Justtanont},
  {Lagage}, {Meixner}, {N{\o}rgaard-Nielsen}, {Ray}, {Ressler}, {van Dishoeck},
  \& {Waelkens}}]{rieke_2015_aa}
{Rieke}, G.~H., {Wright}, G.~S., {B{\"o}ker}, T., {et~al.} 2015, Publications
  of the Astronomical Society of the Pacific, 127, 584, \dodoi{10.1086/682252}

\bibitem[{{Riess} {et~al.}(2016){Riess}, {Macri}, {Hoffmann}, {Scolnic},
  {Casertano}, {Filippenko}, {Tucker}, {Reid}, {Jones}, {Silverman},
  {Chornock}, {Challis}, {Yuan}, {Brown}, \& {Foley}}]{riess_2016_aa}
{Riess}, A.~G., {Macri}, L.~M., {Hoffmann}, S.~L., {et~al.} 2016, \apj, 826,
  56, \dodoi{10.3847/0004-637X/826/1/56}

\bibitem[{{Riess} {et~al.}(2018){Riess}, {Casertano}, {Yuan}, {Macri},
  {Bucciarelli}, {Lattanzi}, {MacKenty}, {Bowers}, {Zheng}, {Filippenko},
  {Huang}, \& {Anderson}}]{riess_2018_aa}
{Riess}, A.~G., {Casertano}, S., {Yuan}, W., {et~al.} 2018, \apj, 861, 126,
  \dodoi{10.3847/1538-4357/aac82e}

\bibitem[{{Rogers} \& {Nayfonov}(2002)}]{rogers_2002_aa}
{Rogers}, F.~J., \& {Nayfonov}, A. 2002, \apj, 576, 1064

\bibitem[{Romero {et~al.}(2018)Romero, Córsico, Althaus, Pelisoli, \&
  Kepler}]{romero_blap2018}
Romero, A.~D., Córsico, A.~H., Althaus, L.~G., Pelisoli, I., \& Kepler, S.~O.
  2018, MNRAS, 477, L30, \dodoi{10.1093/mnrasl/sly051}

\bibitem[{{Roxburgh}(2008)}]{2008Ap&SS.316...75R}
{Roxburgh}, I.~W. 2008, \apss, 316, 75, \dodoi{10.1007/s10509-007-9673-7}

\bibitem[{{Rydberg} {et~al.}(2013){Rydberg}, {Zackrisson}, {Lundqvist}, \&
  {Scott}}]{rydberg_2013_aa}
{Rydberg}, C.-E., {Zackrisson}, E., {Lundqvist}, P., \& {Scott}, P. 2013,
  \mnras, 429, 3658, \dodoi{10.1093/mnras/sts653}

\bibitem[{{Sakashita} \& {Hayashi}(1961)}]{sakashita_1961_aa}
{Sakashita}, S., \& {Hayashi}, C. 1961, Progress of Theoretical Physics, 26,
  942

\bibitem[{{Salaris} \& {Cassisi}(2017)}]{salaris_2017_aa}
{Salaris}, M., \& {Cassisi}, S. 2017, Royal Society Open Science, 4, 170192,
  \dodoi{10.1098/rsos.170192}

\bibitem[{{Salpeter}(1954)}]{salpeter:54}
{Salpeter}, E.~E. 1954, Australian Journal of Physics, 7, 373,
  \dodoi{10.1071/PH540373}

\bibitem[{{Saumon} {et~al.}(1995){Saumon}, {Chabrier}, \& {van
  Horn}}]{saumon_1995_aa}
{Saumon}, D., {Chabrier}, G., \& {van Horn}, H.~M. 1995, \apjs, 99, 713,
  \dodoi{10.1086/192204}

\bibitem[{{Schwarzschild} \& {H{\"a}rm}(1958)}]{schwarzschild_1958_aa}
{Schwarzschild}, M., \& {H{\"a}rm}, R. 1958, \apj, 128, 348

\bibitem[{{Schwarzschild} \& {H{\"a}rm}(1969)}]{schwarzschild_1969_aa}
---. 1969, \baas, 1, 99

\bibitem[{{Scuflaire} {et~al.}(2008){Scuflaire}, {Th{\'e}ado}, {Montalb{\'a}n},
  {Miglio}, {Bourge}, {Godart}, {Thoul}, \& {Noels}}]{2008Ap&SS.316...83S}
{Scuflaire}, R., {Th{\'e}ado}, S., {Montalb{\'a}n}, J., {et~al.} 2008, \apss,
  316, 83, \dodoi{10.1007/s10509-007-9650-1}

\bibitem[{{Seaton}(2005)}]{Seaton2005}
{Seaton}, M.~J. 2005, \mnras, 362, L1, \dodoi{10.1111/j.1365-2966.2005.00019.x}

\bibitem[{{Senarath} {et~al.}(2018){Senarath}, {Brown}, {Cluver}, {Moustakas},
  {Armus}, \& {Jarrett}}]{senarath_2018_aa}
{Senarath}, M.~R., {Brown}, M. J.~I., {Cluver}, M.~E., {et~al.} 2018, \apj,
  869, L26, \dodoi{10.3847/2041-8213/aaf4ff}

\bibitem[{{Shen} {et~al.}(2018){Shen}, {Boubert}, {G{\"a}nsicke}, {Jha},
  {Andrews}, {Chomiuk}, {Foley}, {Fraser}, {Gromadzki}, {Guillochon}, {Kotze},
  {Maguire}, {Siebert}, {Smith}, {Strader}, {Badenes}, {Kerzendorf}, {Koester},
  {Kromer}, {Miles}, {Pakmor}, {Schwab}, {Toloza}, {Toonen}, {Townsley}, \&
  {Williams}}]{shen_2018_aa}
{Shen}, K.~J., {Boubert}, D., {G{\"a}nsicke}, B.~T., {et~al.} 2018, \apj, 865,
  15, \dodoi{10.3847/1538-4357/aad55b}

\bibitem[{{Silva Aguirre} {et~al.}(2010){Silva Aguirre}, {Ballot}, {Serenelli},
  \& {Weiss}}]{silva_aguirre_2010_aa}
{Silva Aguirre}, V., {Ballot}, J., {Serenelli}, A., \& {Weiss}, A. 2010, \apss,
  328, 129

\bibitem[{{Simon} \& {Lee}(1981)}]{rsp_sl81}
{Simon}, N.~R., \& {Lee}, A.~S. 1981, \apj, 248, 291, \dodoi{10.1086/159153}

\bibitem[{{Simon} \& {Schmidt}(1976)}]{rsp_ss76}
{Simon}, N.~R., \& {Schmidt}, E.~G. 1976, \apj, 205, 162,
  \dodoi{10.1086/154259}

\bibitem[{{Smolec}(2009)}]{rsp_phd_smolec}
{Smolec}, R. 2009, PhD thesis, Nicolaus Copernicus Astronomical Center, Warsaw,
  Poland

\bibitem[{{Smolec}(2014)}]{rsp_s14}
{Smolec}, R. 2014, in IAU Symposium, Vol. 301, Precision Asteroseismology, ed.
  J.~A. {Guzik}, W.~J. {Chaplin}, G.~{Handler}, \& A.~{Pigulski}, 265--272

\bibitem[{{Smolec}(2016)}]{rsp_s16}
---. 2016, \mnras, 456, 3475, \dodoi{10.1093/mnras/stv2868}

\bibitem[{{Smolec} \& {Moskalik}(2008)}]{rsp_sm08a}
{Smolec}, R., \& {Moskalik}, P. 2008, \actaa, 58, 193.
\newblock \doarXiv{0809.1979}

\bibitem[{{Smolec} \& {Moskalik}(2010)}]{rsp_sm10}
---. 2010, \aap, 524, A40, \dodoi{10.1051/0004-6361/201014494}

\bibitem[{{Smolec} \& {Moskalik}(2012)}]{rsp_sm12}
---. 2012, \mnras, 426, 108, \dodoi{10.1111/j.1365-2966.2012.21678.x}

\bibitem[{{Smolec} \& {Moskalik}(2014)}]{rsp_sm14}
---. 2014, \mnras, 441, 101, \dodoi{10.1093/mnras/stu574}

\bibitem[{{Smolec} {et~al.}(2018){Smolec}, {Moskalik}, {Plachy},
  {Soszy{\'n}ski}, \& {Udalski}}]{rsp_s+18}
{Smolec}, R., {Moskalik}, P., {Plachy}, E., {Soszy{\'n}ski}, I., \& {Udalski},
  A. 2018, \mnras, 481, 3724, \dodoi{10.1093/mnras/sty2452}

\bibitem[{{Smolec} {et~al.}(2012){Smolec}, {Soszy{\'n}ski}, {Moskalik},
  {Udalski}, {Szyma{\'n}ski}, {Kubiak}, {Pietrzy{\'n}ski}, {Wyrzykowski},
  {Ulaczyk}, {Poleski}, {Koz{\l}owski}, \& {Pietrukowicz}}]{rsp_s+12}
{Smolec}, R., {Soszy{\'n}ski}, I., {Moskalik}, P., {et~al.} 2012, \mnras, 419,
  2407, \dodoi{10.1111/j.1365-2966.2011.19891.x}

\bibitem[{Smolec {et~al.}(2013)Smolec, Pietrzyński, Graczyk, Pilecki, Gieren,
  Thompson, Stępień, Karczmarek, Konorski, Górski, Suchomska, Bono, Prada,
  \& Nardetto}]{smolec_bep2013}
Smolec, R., Pietrzyński, G., Graczyk, D., {et~al.} 2013, MNRAS, 428, 3034,
  \dodoi{10.1093/mnras/sts258}

\bibitem[{{Snow} {et~al.}(2018){Snow}, {Botha}, {Scullion}, {McLaughlin},
  {Young}, \& {Jaeggli}}]{snow_2018_aa}
{Snow}, B., {Botha}, G.~J.~J., {Scullion}, E., {et~al.} 2018, \apj, 863, 172,
  \dodoi{10.3847/1538-4357/aad3bc}

\bibitem[{{Soraisam} {et~al.}(2018){Soraisam}, {Bildsten}, {Drout}, {Bauer},
  {Gilfanov}, {Kupfer}, {Laher}, {Masci}, {Prince}, {Kulkarni}, {Matheson}, \&
  {Saha}}]{soraisam_2018_aa}
{Soraisam}, M.~D., {Bildsten}, L., {Drout}, M.~R., {et~al.} 2018, \apj, 859,
  73, \dodoi{10.3847/1538-4357/aabc59}

\bibitem[{{Soszy{\'n}ski} {et~al.}(2019){Soszy{\'n}ski}, {Smolec}, {Udalski},
  \& {Pietrukowicz}}]{rsp_1oblher}
{Soszy{\'n}ski}, I., {Smolec}, R., {Udalski}, A., \& {Pietrukowicz}, P. 2019,
  \apj, 873, 43, \dodoi{10.3847/1538-4357/ab04ab}

\bibitem[{{Soszy{\'n}ski} {et~al.}(2011){Soszy{\'n}ski}, {Udalski},
  {Pietrukowicz}, {Szyma{\'n}ski}, {Kubiak}, {Pietrzy{\'n}ski}, {Wyrzykowski},
  {Ulaczyk}, {Poleski}, \& {Koz{\l}owski}}]{rsp_PDogle}
{Soszy{\'n}ski}, I., {Udalski}, A., {Pietrukowicz}, P., {et~al.} 2011, \actaa,
  61, 285.
\newblock \doarXiv{1112.1406}

\bibitem[{{Soszy{\'n}ski} {et~al.}(2014){Soszy{\'n}ski}, {Udalski},
  {Szyma{\'n}ski}, {Pietrukowicz}, {Mr{\'o}z}, {Skowron}, {Koz{\l}owski},
  {Poleski}, {Skowron}, {Pietrzy{\'n}ski}, {Wyrzykowski}, {Ulaczyk}, \&
  {Kubiak}}]{rsp_o4_rrl}
{Soszy{\'n}ski}, I., {Udalski}, A., {Szyma{\'n}ski}, M.~K., {et~al.} 2014,
  \actaa, 64, 177.
\newblock \doarXiv{1410.1542}

\bibitem[{{Soszy{\'n}ski} {et~al.}(2015){Soszy{\'n}ski}, {Udalski},
  {Szyma{\'n}ski}, {Skowron}, {Pietrzy{\'n}ski}, {Poleski}, {Pietrukowicz},
  {Skowron}, {Mr{\'o}z}, {Koz{\l}owski}, {Wyrzykowski}, {Ulaczyk}, \&
  {Pawlak}}]{rsp_o4_cep_mc}
---. 2015, \actaa, 65, 297.
\newblock \doarXiv{1601.01318}

\bibitem[{{Stellingwerf}(1975)}]{rsp_stel75}
{Stellingwerf}, R.~F. 1975, \apj, 195, 441, \dodoi{10.1086/153343}

\bibitem[{{Stellingwerf}(1982)}]{rsp_stel82}
---. 1982, \apj, 262, 330, \dodoi{10.1086/160425}

\bibitem[{{Storm} {et~al.}(2011){Storm}, {Gieren}, {Fouqu{\'e}}, {Barnes},
  {Pietrzy{\'n}ski}, {Nardetto}, {Weber}, {Granzer}, \&
  {Strassmeier}}]{rsp_storm}
{Storm}, J., {Gieren}, W., {Fouqu{\'e}}, P., {et~al.} 2011, \aap, 534, A94,
  \dodoi{10.1051/0004-6361/201117155}

\bibitem[{{Sugimoto}(1970)}]{Sugimoto1970}
{Sugimoto}, D. 1970, \apj, 159, 619, \dodoi{10.1086/150336}

\bibitem[{{Sugimoto} \& {Nomoto}(1975)}]{Sugimoto1975}
{Sugimoto}, D., \& {Nomoto}, K. 1975, \pasj, 27, 197

\bibitem[{{Suzuki} {et~al.}(2016){Suzuki}, {Toki}, \& {Nomoto}}]{Suzuki2016}
{Suzuki}, T., {Toki}, H., \& {Nomoto}, K. 2016, \apj, 817, 163,
  \dodoi{10.3847/0004-637X/817/2/163}

\bibitem[{{Szab{\'o}} {et~al.}(2010){Szab{\'o}}, {Koll{\'a}th}, {Moln{\'a}r},
  {Kolenberg}, {Kurtz}, {Bryson}, {Benk{\'{o}}}, {Christensen-Dalsgaard},
  {Kjeldsen}, {Borucki}, {Koch}, {Twicken}, {Chadid}, {di Criscienzo}, {Jeon},
  {Moskalik}, {Nemec}, \& {Nuspl}}]{szabo_2010_aa}
{Szab{\'o}}, R., {Koll{\'a}th}, Z., {Moln{\'a}r}, L., {et~al.} 2010, \mnras,
  409, 1244, \dodoi{10.1111/j.1365-2966.2010.17386.x}

\bibitem[{{Tassoul}(2000)}]{tassoul_2000_aa}
{Tassoul}, J.-L. 2000, {Stellar Rotation} (Cambridge: New York)

\bibitem[{{Timmes} \& {Swesty}(2000)}]{timmes_2000_ab}
{Timmes}, F.~X., \& {Swesty}, F.~D. 2000, \apjs, 126, 501,
  \dodoi{10.1086/313304}

\bibitem[{{Townsend} {et~al.}(2018){Townsend}, {Goldstein}, \&
  {Zweibel}}]{townsend_2018_aa}
{Townsend}, R.~H.~D., {Goldstein}, J., \& {Zweibel}, E.~G. 2018, \mnras, 475,
  879, \dodoi{10.1093/mnras/stx3142}

\bibitem[{{Townsend} \& {Teitler}(2013)}]{townsend_2013_aa}
{Townsend}, R.~H.~D., \& {Teitler}, S.~A. 2013, \mnras, 435, 3406,
  \dodoi{10.1093/mnras/stt1533}

\bibitem[{{Townsley} \& {Bildsten}(2004)}]{2004ApJ...600..390T}
{Townsley}, D.~M., \& {Bildsten}, L. 2004, \apj, 600, 390,
  \dodoi{10.1086/379647}

\bibitem[{{Tscharnuter} \& {Winkler}(1979)}]{rsp_TW}
{Tscharnuter}, W.~M., \& {Winkler}, K.-H.~A. 1979, Computer Physics
  Communications, 18, 171, \dodoi{10.1016/0010-4655(79)90111-5}

\bibitem[{{Turk} {et~al.}(2011){Turk}, {Smith}, {Oishi}, {Skory}, {Skillman},
  {Abel}, \& {Norman}}]{turk_2011_aa}
{Turk}, M.~J., {Smith}, B.~D., {Oishi}, J.~S., {et~al.} 2011, \apjs, 192, 9,
  \dodoi{10.1088/0067-0049/192/1/9}

\bibitem[{{Udalski} {et~al.}(2015){Udalski}, {Szyma{\'n}ski}, \&
  {Szyma{\'n}ski}}]{rsp_ogle4}
{Udalski}, A., {Szyma{\'n}ski}, M.~K., \& {Szyma{\'n}ski}, G. 2015, \actaa, 65,
  1.
\newblock \doarXiv{1504.05966}

\bibitem[{{Udalski} {et~al.}(2018){Udalski}, {Soszy{\'n}ski}, {Pietrukowicz},
  {Szyma{\'n}ski}, {Skowron}, {Skowron}, {Mr{\'o}z}, {Poleski}, {Koz{\l}owski},
  {Ulaczyk}, {Rybicki}, {Iwanek}, \& {Wrona}}]{rsp_ogle_gc}
{Udalski}, A., {Soszy{\'n}ski}, I., {Pietrukowicz}, P., {et~al.} 2018, \actaa,
  68, 315, \dodoi{10.32023/0001-5237/68.4.1}

\bibitem[{{Ulaczyk} {et~al.}(2013){Ulaczyk}, {Szyma{\'n}ski}, {Udalski},
  {Kubiak}, {Pietrzy{\'n}ski}, {Soszy{\'n}ski}, {Wyrzykowski}, {Poleski},
  {Gieren}, {Walker}, \& {Garcia-Varela}}]{rsp_shallow}
{Ulaczyk}, K., {Szyma{\'n}ski}, M.~K., {Udalski}, A., {et~al.} 2013, \actaa,
  63, 159.
\newblock \doarXiv{1306.4802}

\bibitem[{van~der Walt {et~al.}(2011)van~der Walt, Colbert, \&
  Varoquaux}]{der_walt_2011_aa}
van~der Walt, S., Colbert, S.~C., \& Varoquaux, G. 2011, Computing in Science
  Engineering, 13, 22, \dodoi{10.1109/MCSE.2011.37}

\bibitem[{{von Zeipel}(1924{\natexlab{a}})}]{von-zeipel_1924_aa}
{von Zeipel}, H. 1924{\natexlab{a}}, \mnras, 84, 684,
  \dodoi{10.1093/mnras/84.9.684}

\bibitem[{{von Zeipel}(1924{\natexlab{b}})}]{von-zeipel_1924_ab}
---. 1924{\natexlab{b}}, \mnras, 84, 665, \dodoi{10.1093/mnras/84.9.665}

\bibitem[{{Wang} {et~al.}(2019){Wang}, {Jones}, {Shporer}, {Fulton}, {Paredes},
  {Trifonov}, {Kossakowski}, {Eastman}, {Redfield}, {G{\"u}nther}, {Kreidberg},
  {Huang}, {Millholland}, {Seligman}, {Fischer}, {Brahm}, {Wang}, {Cruz},
  {Henry}, {James}, {Addison}, {Liang}, {Davis}, {Tronsgaard}, {Worku},
  {Brewer}, {K{\"u}rster}, {Zhang}, {Beichman}, {Bieryla}, {Brown},
  {Christiansen}, {Ciardi}, {Collins}, {Esquerdo}, {Howard}, {Isaacson},
  {Latham}, {Mazeh}, {Petigura}, {Quinn}, {Shahaf}, {Siverd}, {Rodler},
  {Reffert}, {Zakhozhay}, {Ricker}, {Vanderspek}, {Seager}, {Winn}, {Jenkins},
  {Boyd}, {F{\'{u}}r{\'e}sz}, {Henze}, {Levine}, {Morris}, {Paegert},
  {Stassun}, {Ting}, {Vezie}, \& {Laughlin}}]{wang_2019_aa}
{Wang}, S., {Jones}, M., {Shporer}, A., {et~al.} 2019, \aj, 157, 51,
  \dodoi{10.3847/1538-3881/aaf1b7}

\bibitem[{{Weiss} \& {Schlattl}(2008)}]{GARSTEC:2008}
{Weiss}, A., \& {Schlattl}, H. 2008, \apss, 316, 99,
  \dodoi{10.1007/s10509-007-9606-5}

\bibitem[{{Williams} \& {Kelley}(2015)}]{williams_2015_aa}
{Williams}, T., \& {Kelley}, C. 2015, "Gnuplot 5.0: an interactive plotting
  program"

\bibitem[{{Windhorst} {et~al.}(2018){Windhorst}, {Timmes}, {Wyithe},
  {Alpaslan}, {Andrews}, {Coe}, {Diego}, {Dijkstra}, {Driver}, {Kelly}, \&
  {Kim}}]{windhorst_2018_aa}
{Windhorst}, R.~A., {Timmes}, F.~X., {Wyithe}, J.~S.~B., {et~al.} 2018, \apjs,
  234, 41, \dodoi{10.3847/1538-4365/aaa760}

\bibitem[{Wu \& Li(2018)}]{wu_blap2018}
Wu, T., \& Li, Y. 2018, MNRAS, 487, 3871, \dodoi{10.1093/mnras/sty1347}

\bibitem[{{Wuchterl} \& {Feuchtinger}(1998)}]{rsp_wh98}
{Wuchterl}, G., \& {Feuchtinger}, M.~U. 1998, \aap, 340, 419

\bibitem[{{Yecko} {et~al.}(1998){Yecko}, {Kollath}, \& {Buchler}}]{rsp_FB1}
{Yecko}, P.~A., {Kollath}, Z., \& {Buchler}, J.~R. 1998, \aap, 336, 553

\bibitem[{{Zackrisson} {et~al.}(2011){Zackrisson}, {Rydberg}, {Schaerer},
  {{\"O}stlin}, \& {Tuli}}]{zackrisson_2011_aa}
{Zackrisson}, E., {Rydberg}, C.-E., {Schaerer}, D., {{\"O}stlin}, G., \&
  {Tuli}, M. 2011, \apj, 740, 13, \dodoi{10.1088/0004-637X/740/1/13}

\bibitem[{{Zahn}(1992)}]{Zahn1992}
{Zahn}, J.-P. 1992, \aap, 265, 115

\end{thebibliography}


\end{document}